\documentclass[twocolumn,trackchanges,tighten,resetfootnote]{aastex7}
\renewcommand{\figurename}{Fig.}
\usepackage{dsfont}
\usepackage{makecell}
\usepackage{adjustbox}
\usepackage{array}
\usepackage[none]{hyphenat}
\usepackage{comment}
\usepackage{rotating}
\usepackage{longtable}
\usepackage{bookmark}
\hypersetup{
  colorlinks=true,
  linkcolor=blue,
  citecolor=blue,
  urlcolor=blue,
  bookmarksnumbered=true
}
\usepackage{etoolbox}
\patchcmd{\thebibliography}
  {\hbox to\textwidth{\hss\normalsize REFERENCES\hss}}
  {}
  {}
  {}

\definecolor{myCol1}{RGB}{0, 11, 88} 
\definecolor{myCol2}{RGB}{0, 49, 97}
\definecolor{myCol3}{RGB}{0, 106, 103} 
\definecolor{myCol4}{RGB}{253, 235, 158}

\hypersetup{linkcolor=myCol2,citecolor=myCol3,filecolor=cyan,urlcolor=myCol1}
\graphicspath{{./}{images/}}

\hypersetup{pdfauthor={Name}}

\usepackage{enumerate}
\begin{document}
\title{{The Search for Technosignatures:\\
 A Review of Possibilities}}

\author[orcid=0000-0001-6689-5570]{Cl\'ement Vidal} 
\affiliation{Vrije Universiteit Brussel, Brussels, Belgium}
\affiliation{Blue Marble Space, Seattle, WA, USA}
\email[show]{contact@clemvidal.com}
\footnotetext{\footnotesize\raggedright Correspondence: contact@clemvidal.com, bnfields72@gmail.com, drsowinski@gmail.com, sbartlett@setiap.org, rich@terrile.com}


\author[orcid=0009-0008-7048-916X]{Benji L. Fields}
\affiliation{Blue Marble Space, Seattle, WA, USA}
\email[show]{bnfields72@gmail.com}

\author[orcid=0000-0003-3136-7449,gname=Damian,sname=Sowinski]{Damian R. Sowinski}
\affiliation{S.E.T.I. Institute, Mountain View, CA, USA}
\affiliation{University of Rochester, Rochester, NY, USA}
\email[show]{drsowinski@gmail.com}

\author[orcid=0000-0001-5680-476X]{Stuart Bartlett}
\affiliation{S.E.T.I. Institute, Mountain View, CA, USA}
\email[show]{sjbart@caltech.edu}

\author[orcid=0009-0004-1207-6018]{Richard J. Terrile}
\affiliation{Jet Propulsion Laboratory, retired, Pasadena, CA, USA}
\email[show]{rich@terrile.com}

\author[orcid=0000-0002-8120-7710]{Alex Ellery}
\affiliation{Carleton University, Ottawa, ON, Canada}
\email[show]{alexellery@cunet.carleton.ca}

\author[orcid=0009-0009-6500-0005]{Daliah Bibas}
\altaffiliation{SETI Post-Detection Hub, St. Andrews, UK}
\affiliation{Vrije Universiteit Brussel, Brussels, Belgium}
\email[show]{daliah.raquel.bibas@vub.be}

\author[orcid=0000-0002-9848-7915]{Armando M. Mastrogiovanni}
\altaffiliation{SETI Post-Detection Hub, St. Andrews, UK}
\affiliation{Baruch College, City University of New York, New York, NY, USA}
\email{a.m.mastrogiovanni@gmail.com}

\author[orcid=0000-0001-7935-727X]{Niklas Döbler}
\affiliation{University of Bamberg, Bamberg, Germany}
\email[show]{niklas.doebler@uni-bamberg.de}

\author[orcid=0000-0002-2858-9385]{Manika Singla}
\affiliation{Physical Research Laboratory, Ahmedabad, India}
\affiliation{Institute for Basic Science, Daejeon, South Korea}
\email[show]{manikasingla14@gmail.com}

\author[orcid=0000-0002-7304-8481]{Julia DeMarines}
\affiliation{University of California, Berkeley, CA, USA}
\email{julia.demarines@gmail.com}

\author[orcid=0000-0001-6910-5116]{Theresa Fisher}
\affiliation{University of Arizona, Tucson, AZ, USA}
\email[show]{theresafisher@arizona.edu}

\author[orcid=0000-0003-2792-4978]{Yuri Uno}
\affiliation{National Cheng Kung University, Tainan City, Taiwan}
\email{yuri.uno@gs.ncku.edu.tw}

\author[orcid=0000-0001-7836-1787]{Jake D. Turner}
\affiliation{Cornell University, Ithaca, NY, USA}
\email[show]{jaketurner@cornell.edu}

\author[orcid=0000-0001-5290-1001]{Evan L. Sneed}
\affiliation{University of California, Riverside, CA, USA}
\email[show]{evan.sneed@email.ucr.edu}

\author[orcid=0009-0001-5209-4518]{Advait Huggahalli}
\affiliation{University of California, Berkeley, CA, USA}
\email[show]{ahuggahalli@berkeley.edu}

\author[orcid=0000-0002-3012-4261]{Megan Grace Li}
\affiliation{University of California, Los Angeles, CA, USA}
\email{megangrace@ucla.edu}

\author[orcid=0000-0003-4285-4453]{Zhuofu (Chester) Li}
\affiliation{University of Washington, Seattle, WA, USA}
\email{zhuofu@uw.edu}

\author[orcid=0009-0006-0506-7345]{Ramiro Saide}
\affiliation{S.E.T.I. Institute, Mountain View, CA, USA}
\affiliation{University of Mauritius, Moka, Mauritius}
\email{mirosaide@gmail.com}

\begin{abstract}
This paper reviews the diverse range of technosignatures that have been proposed in the literature. 
We organize the review by spatial scales from the Earth and the solar system to the edge of the Universe, not tied to a specific observatory or technical capability. 
This provides a more comprehensive view of technosignature science beyond the search for radio signals that has dominated SETI since its inception. 
We introduce the Kardashev and Barrow scales before exploring exoplanetary technosignatures, from surface, atmospheric to orbital sources. 
Next, we consider stellar technosignatures that may involve massive energy utilization, stellar modification or stellar pollution, and end with a section about compact objects. 
We then review attempts to detect interstellar communication, and discuss many dimensions of the search space from first principles. 
We also consider interstellar travel technosignatures, and end with galactic, extragalactic and universal signatures. 
The paper argues for stronger synergies between biosignature and technosignature searches, and the importance of multimodal and multimessenger strategies.
Central to this is the importance of anomaly detection as a fundamental and agnostic search strategy.  
We highlight instruments for detecting technosignatures, as well as a science traceability matrix template. 

\end{abstract}

\maketitle
\makeatletter
\global\@firstsectionfalse
\makeatother
\clearpage
\phantomsection
\twocolumngrid
\tableofcontents
\clearpage
\twocolumngrid

\begin{flushright}
\textit{Leave no stone unturned.}\\
Euripides, \textit{Heraclid\ae} lyrics, 1002 (c 428 BC).
\end{flushright}

\section{Introduction}\label{sec: intro}

The question of the existence of life beyond Earth has fascinated minds from antiquity to the present \citep{dick1982PluralityWorlds,crowe1986ExtraterrestrialLife,crowe2008ExtraterrestrialLife,dick1996BiologicalUniverse}.
Astrobiology has taken off as a science \citep{cirkovic2012AstrobiologicalLandscape}.
While the existence of exoplanets was an unwarranted speculation just a few decades ago, today we have discovered over $6000$ of them.
The hopes and likelihood to find traces of life are high, helped with highly sensitive space telescopes such as the James Webb Space Telescope (JWST) that has recently found strong candidate biosignature gases \citep{madhusudhan2025NewConstraints}.
In strong analogy to the debate surrounding the putative detection of Phosphine on Venus \citep{vidal2025LifeClouds}, scientists have claimed that the similarly putative discovery of Dimethyl sulfide (DMS), serves as suggestive evidence of life on exoplanet K2-18b.
In this case, the presence of life is highly speculative because on Earth, while DMS overwhelmingly comes from marine biology via dimethylsulfoniopropionate (DMSP) produced by phytoplankton, potential abiotic synthetic pathways have also been suggested \citep{reed2024AbioticProduction}.
A dual-site follow-up for communicative technosignatures did not lead to good candidates \citep{tremblay2026NarrowbandTechnosignature}.

In astrobiology, most of the funding, resources, and activities are related to the search for \textit{biosignatures}, typically searching for microbial lifeforms in the solar system or for traces of biospheres in exoplanets.
But the universe might also host signs of exotic, intelligent or technological life.
There are pros and cons arguments between the search for bio- or techno- signatures, but technosignatures could be abundant, long-lived, highly detectable, and unambiguous compared to other biosignatures \citep{wright2022CaseTechnosignatures}.
In both cases, given the many layers of uncertainties, it remains very hard to guess which of bio- or techno- signatures will be confirmed and accepted first.

The term \textit{technosignature} was first coined by Jill \citet{tarter2007EvolutionLife} as a general umbrella term to include scattered efforts to search for intelligent life or technologically advanced civilizations.
These include search strategies and observational searches that have been conducted under the banner of {\it Search for Extraterrestrial Intelligence} (SETI), but also a wider search that has been called \textit{Dysonian SETI} \citep{cirkovic2006MacroengineeringGalactic, bradbury2011DysonianApproach}, which includes the \textit{Search for Extraterrestrial Artifacts} (SETA), the search for megastructures, macroengineering, astroarcheology, astroengineering, or stellar engineering.
	
Since the very beginning of the modern search for extraterrestrial intelligence in the 1960s, there have been many isolated technosignature search proposals and it is difficult to get a clear overview and picture of the landscape of possibilities.
Historically, it is worth mentioning that major strategies were articulated in the 1960s: looking for interstellar signals \citep{cocconi1959SearchingInterstellar}, considering interstellar probes \citep{bracewell1960CommunicationsSuperior} as well as stellar engineering \citep{dyson1960SearchArtificial}.
Although less known, there are many more technosignature possibilities, and this paper presents an organized review of them all.

\begin{figure*}[ht]
    \centering
    \includegraphics[width=0.8\linewidth]{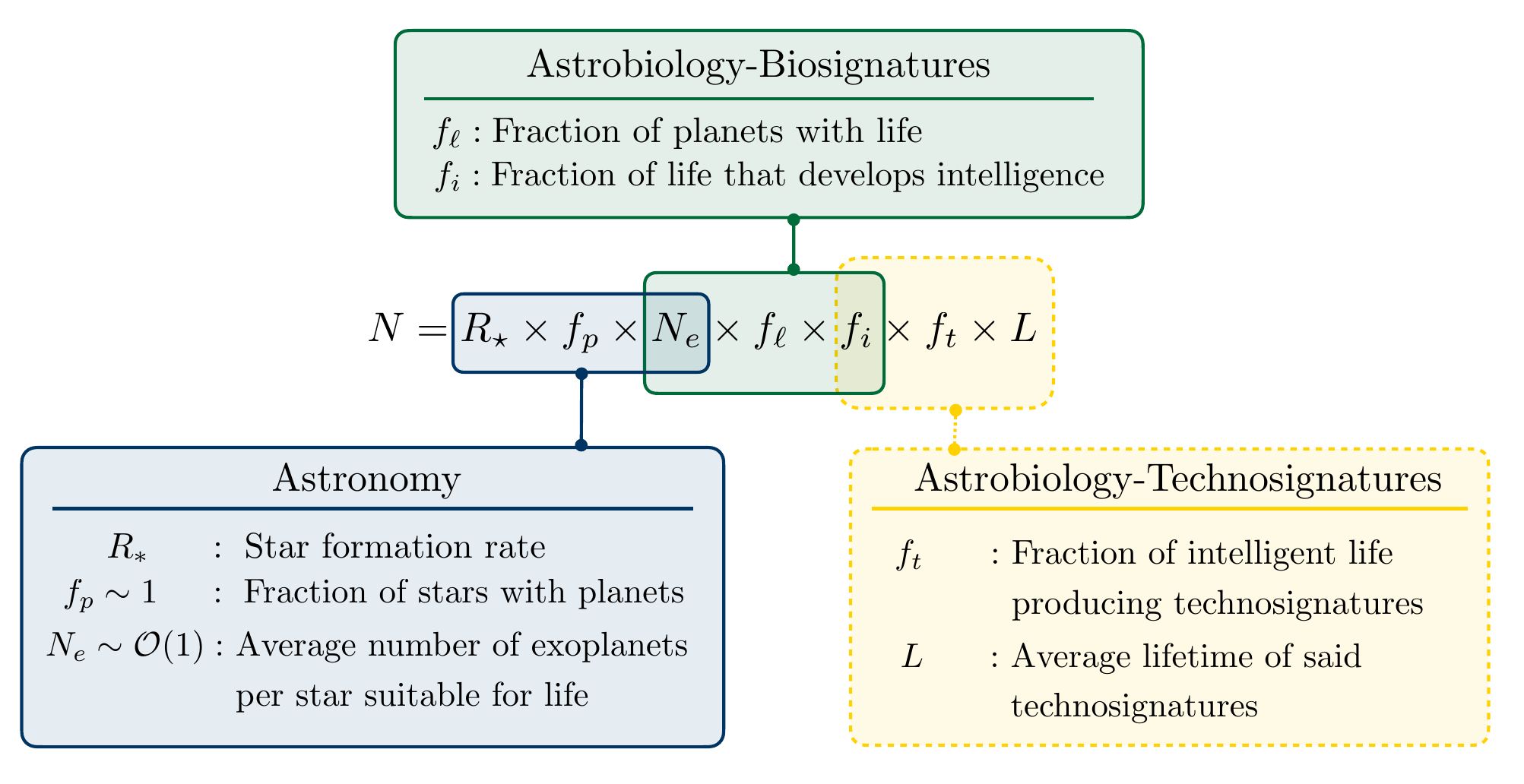}
    \caption{A Drake-like equation addresses the many factors used to estimate the number of civilizations producing technosignatures in our galaxy at any given time. Note that $R_*$ was initially set to 1 star per year by \citet{drake1965RadioSearch}, but recent estimates vary between 2-16 \citep{francois2015RateFormation}. A detailed and modern discussion of the various factors and their estimates can be found in \citep{vakoch2015DrakeEquation}.}
    \label{fig:drake equation}
\end{figure*}

Most of the search efforts have focused on the search for communicative signals (see e.g. \citealt{ekers2002SETI2020, tarter2010SETITurns}), guided by the famous \citet{drake1965RadioSearch} equation (see Fig. \ref{fig:drake equation}).

We should emphasize that the Drake equation was proposed as a way to structure the 1961 Green Bank meeting, and thus has always been seen as a guide and invitation to better understanding the various parameters leading to life and intelligent life.
Nevertheless, many authors have pointed out limitations and proposed variations to the Drake equation \citep{vakoch2015DrakeEquation}.
For our purposes of guiding the search for technosignatures, it is worth highlighting some blind spots, biases and implicit assumptions behind this equation.
For example, one might draw the conclusion that we have to know the first factors of the equation before being able to assess the later factors.
But it is clearly not the case: we do not need to know everything about star formation or to have complete statistics about all planets in the galaxy to start searching for biosignatures.
In much the same way, we don’t need to know everything about possible lifeforms and their evolutionary trajectories to start searching for technosignatures.
The equation also implicitly focuses on static planetary civilizations, i.e. Type I civilizations on Kardashev’s \citeyearpar{kardashev1964TransmissionInformation} scale.
For example a Type III galactic civilization could generate an exponential suppression on other factors by lasting very long and spreading across multiple stellar systems.
The Drake equation does not address the search for extraterrestrial artefacts either, which has led to the proposition of a Drake-like equation for artefacts \citep{benford2021DrakeEquation}.
Another major blind spot is that, at least in its original formulation, the equation attempts to estimate the number of \textit{communicative} civilizations.
But as we will expose in this review, there are many other ways to search for technosignatures beyond the search for interstellar communication.
We will thus include \textit{non-communicative} technosignatures, i.e. looking for traces of technological activity, even if there is no intention to communicate.
Technosignatures may also be passive in nature in the sense that an ET civilization could be long extinct, but its technosignatures could survive over geologically long timescales.

Another limitation of the original Drake equation is the meaning of the lifetime $L$ of a civilization.
What does it mean for an interstellar civilization seeding life or colonies, or for galactic colonization models?
Some colonies or technologies might go extinct, while others could transform so much that the link with parent civilization or others would be lost.
The picture could be much more dynamic and complicated, and this is why we defined $L$ to be the average lifetime of any technosignature.

We do not aim at a critical and detailed evaluation of each technosignature that we review: even the different authors of the present paper would certainly disagree about the different merits of the various technosignatures.
We would like to warn readers that we include both mature and speculative proposals, and invite readers to pay attention to distinguish them according to evidential status, detectability, and scientific ancillary value (for more on evaluation, see Section \ref{sec:discussion}).

As much as possible, we attempt to identify synergies between technosignature searches and other related fields.
In principle the strategy to look for synergies can go in both directions:
\begin{enumerate}
    \item a study in field $\mathcal F$ can also inform and be useful to search for a technosignature $\mathcal T$
    \item a search for a technosignature $\mathcal T$ can also be useful to study field $\mathcal F$.
\end{enumerate}

Strategy (1) is a cost-efficient strategy to search for technosignatures, for example by adding technosignature science goals when designing missions in the solar system \citep{haqq-misra2022OpportunitiesTechnosignature}, when observing exoplanets, or when running commensal SETI pipelines on some of Earth’s most capable radio telescopes \citep{worden2018PhilanthropicSpace,lebofsky2019BreakthroughListen}.
The bottom-line is that it would be a pity to miss a technosignature that would have been easy to detect within existing missions, adding just a little bit more effort and resources.
	
The strategy to synergize with other disciplines and efforts (2) goes the other way.
When designing a new SETI search strategy, it is recommended to think about what other non-SETI science goals it could achieve.
Indeed, we don’t know when -nor even if!- we’re going to find extraterrestrial life, so we have to design searches to also benefit other sciences at the same time.
As \citet{drake1965RadioSearch} put it, “any project aimed at the detection of intelligent extraterrestrial life should simultaneously conduct more conventional research”.
This is also known as the \textit{First Law of SETI Investigations}: “Every search for alien civilizations should be planned to give interesting results even when no aliens are discovered" \citep{nasatechnosignaturesworkshopparticipants2018NASASearch}.

For readers new to the field, an excellent recent textbook by \citet{wright2026SearchExtraterrestrial} covers much of the theory and practice of the field.
Its goal is to establish technosignature science and SETI as a teachable academic discipline.
Wright also covers much of "SETI Theory" topics such as the Fermi paradox, the Rare Earth hypothesis or the Doomsday argument, which we intentionally avoided in this review.
As a book-length work, it expands on whether we should send interstellar messages, on post-detection, and on societal and cultural dimensions.
Our focus by contrast is to produce a companion for researchers who want a comprehensive catalog of technosignatures that have been proposed, organized by scale, and to open observational paths to further pursue the search.

In this paper, we thus aim to answer the central question: What kinds of technosignatures could possibly be out there?
Our secondary focus is about detectability and instrumentation, i.e.: How can we detect them now or in the near future?

To organize the types of technosignatures, we simply follow \textit{spatial scales}.
We start very carefully with the controversial idea of looking for technosignatures on Earth (section \ref{sec: earth ts}), then we zoom out to Earth’s orbit, the Moon, the inner solar system, the asteroid belt, the outer solar system up to the Oort cloud (section \ref{sec: ss ts}); then we jump to planetary technosignatures, reviewing potential surface, atmospheric or orbital signatures (section \ref{sec: planet ts}); we discuss then stellar technosignatures which could reveal the existence of megastructures or the capability to directly use or modify stars (section \ref{sec: stellar ts}).
We continue with a review of interstellar signals, discussing key dimensions and strategies to search for interstellar communication (section \ref{sec: interstellar ts}).
Then we consider signatures of interstellar travel and migration (section \ref{sec: travel ts}), and we end with more exotic technosignatures that may be observable at a galactic, extragalactic or even universal scales (section \ref{sec: galactic ts}).
We end with a discussion on key issues, such as the lessons emerging from biosignatures and technosignatures searches, anomaly detection and machine learning, multimodal and multi-messenger searches, the issue of evaluation and prioritization as well as epistemological challenges (section \ref{sec:discussion}), and a summary of instruments relevant for technosignature as a \textit{science traceability matrix} in Appendix \ref{sec: STM}.

In choosing our references, whenever possible, we mention the first historical paper and review papers. We cite centrally peer-reviewed material and occasionally preprints when they fill gaps in the technosignature search space or make testable predictions.
We focus on organizing and reviewing \textit{all kinds of technosignatures} and as such we do not attempt to cite every technosignature paper: for this purpose, we release a bigger bibliography we used while researching this paper \footnote{\href{https://www.zotero.org/groups/5098525/technosignatures}{https://www.zotero.org/groups/5098525/technosignatures}}, which is itself based on \citep{reyes2019ComprehensiveBibliography,lafond2021FurtheringComprehensive}.

We hope that better organizing the wide array of possible technosignature strategies and targets will clear the way for strengthening, growing and legitimizing the field and the community.
We also hope to raise awareness of the breadth of possible technosignatures, that may be found by targeted or commensal programs, or just serendipitously.
We believe that pursuing both bio- and techno- signatures searches with an equal amount of resources should dramatically improve the overall chances of finding extraterrestrial life.
\clearpage
\section{Earth Technosignatures}\label{sec: earth ts}

Before going through the possibilities of past or present technosignatures on Earth, it is useful to consider the broader context of the potential expansion of civilizations in the galaxy.

Many arguments have been made showing that even at non-relativistic velocities, an advanced species could explore the galactic disk within $50$ million to one billion years \citep{newman1981GalacticCivilizations}.
It is worth emphasizing that the argument works with just one civilization embarking in such a program, thus it doesn’t assume that all civilizations should be doing space exploration and expansion.
The dynamics is very much non-linear, especially considering self-replicating probes, where they would go to the next star system, mine materials, and make more probes that would do the same, recursively.
This would lead to an amplifying dynamics that would cover the majority of the galaxy with such probes on a much faster scale than the age of the Milky Way, which is about $13$ billion years old \citep{nepal2024DiscoveryLocal}.

\citet{lineweaver2004GalacticHabitable} showed that 75\% of the stars in a defined galactic habitable zone are older than the Sun. A recent study showed that the Sun is indeed younger than $53\pm 2$\% of F8-K2 nearby stars in the galactic disk; more generally, our Sun was born after $86 \pm 5$\% of all stars born in the galaxy \citep{robles2008ComprehensiveComparison}.

This galactic and stellar context shows that there is ample time for advanced civilizations to have explored the galaxy systematically.
A long-lived civilization continuously traveling at a speed of $0.1$c could have crossed the galaxy 4500 times since the Earth has been around.
To see this, simply divide the age of the Earth ($4.5 \times 10^9$ years) by the galactic crossing time, which is $10^6$ years:
\begin{equation}
N_{vis} \approx \frac{T_\oplus}{L_{MW}/v}
= \frac{4.5\times 10^9\,\mathrm{yr}}{10^5\,\mathrm{ly}\,/\,0.1\,c} = 4500	    
\end{equation}
That civilization would have had the opportunity to visit Earth up to $4500$ maximally distinct times.
If there are $N$ civilizations exploring, this value is multiplied by $N$.

What can we conclude?
\citet{hart1975ExplanationAbsence} and \citet{tipler1980ExtraterrestrialIntelligent} used similar expansion models to conclude that we are alone in the universe.
The reasoning is that we haven’t seen evidence of probes or traces of past visitation on Earth, and given that galactic colonization should be fast, we are alone in the universe.
Of course, the argument is flawed because we have barely searched systematically for all kinds of probes or signatures of visitations.
Also, we may still be able to detect non-visiting civilizations.

This implies that we should search for more evidence.
Many of these arguments are classic justifications to do SETI in the universe or SETA in the solar system, but they also logically apply to Earth.
If ETI probes can do interstellar travel and get to our solar system, there is no reason to surmise that it’s somehow impossible for them to go through our own atmosphere.
If we put our shoes into a space-faring civilization, it would be quite absurd to travel for millennia, finally arrive to a solar system hosting life, and explore only its non-living parts.
This would be the implicit assumption if we would advocate to look only for technosignatures in our solar system, and not on Earth.

There is also a large uncertainty regarding how much matter falls into Earth, with estimates varying from $5$ to $300$ tons per day \citep{plane2012CosmicDust}.
This means that Earth is not closed to space, and that we are still far from a systematic, high resolution monitoring and awareness of everything that enters or exits the Earth.

At first sight, the strategy to look for technosignatures on Earth is the easiest one: It would be right here!
In practice however, the search for past or present traces of ETI on Earth may be the hardest search strategy because it has suffered and still suffers from numerous geological, sociological, psychological, and methodological obstacles.

\subsection{Searching for Past Visitation}

Astrobiologists have hypothesized the possibility of a \textit{shadow biosphere}, the existence of a separate genesis of life on Earth, with a different biochemistry that would still exist today \citep{cleland2005PossibilityAlternative, davies2009SignaturesShadow}.
We propose to extend this concept to technology, i.e. to also consider the possibility of a \textit{shadow technosphere}.
For example, Paul \citet{davies2012FootprintsAlien} did call for a general search for biological, geological and physical traces of extraterrestrial life or technology.
It means that sciences such as climatology, geology, archaeology or paleontology are relevant to search for pre-human technological artifacts or processed materials.

Even before searching for traces of past extraterrestrial visitation, we can ask whether we are the first advanced civilization in the geological history of Earth.
It turns out that traces of a civilization –even one going through an anthropocene phase– would be very challenging to detect because of erosion dynamics such as surface weathering or plate tectonics.
These constraints are discussed through the \textit{Silurian hypothesis} \citep{schmidt2019SilurianHypothesis}, and the authors conclude that artefacts or fossilized examples of a population older than 4 million years (Ma) would be unlikely to be found.

Still, what could be searched for in this 4 Ma window?
Here are a few examples.

We could look for evidence of large scale agriculture that would have led to a disruption of the soil nitrogen cycle.
Another line of research could look for evidence of mining, such as anomalous geological structures that would be indicative of large mining operations.
These could include anomalous depletion of minerals compared to the local standard, refined or processed ore, and evidence of large-scale digging sites.
Looking for plastics or microplastics deep in geological layers, or in ocean sediments would also be potential signatures.
Atmospheric pollution may be preserved in permafrost or sedimentary layers.
According to \citet{schmidt2019SilurianHypothesis}, all of the pollution of the anthropocene would fit within 1 cm of sediment layer, which makes sense given how short our industrial civilization has existed on geological timescales.
This explains why even if there was a pre-human civilization which went through its own anthropocene, we might not have noticed it in our sediment analysis yet, while also leaving open the possibility that we could discover such a layer in the future.
Other signatures could include persistent organic pollutants, which are organic molecules resistant on large time scales to degradation by chemical, photo-chemical or biological processes, such as Per- and poly- fluoroalkyl substances (PFAS).
There are ways to distinguish these from biological organics.
In particular, life-as-we-know-it generally obeys homochirality.
Finding racemic mixtures would indicate abiotic organic polymers as opposed to biological remains.
Such mixtures would still be ambiguous, as they could come from artificial or/and natural processes. 
Alternatively, finding a consistently opposed chirality to life-as-we-know-it might then be a trace of a shadow biosphere.

Another signature could be anomalous isotope ratios in CO$_2$, such as greater abundances of $^{12}$C than $^{13}$C, indicating mass burning of fossil fuels.
This may be investigated in correlation to markers of climate change such as a sudden increased temperature.

\citet{schmidt2019SilurianHypothesis} suggested paying attention to radioactive isotopes, especially Curium-247 and Plutonium-244 which have halflives of millions of years.
We could add Iodine-129, a byproduct of nuclear testing, very rare in nature, and that has a very long half-life too \citep{rao1999SourcesReservoirs}.

In this context, we can also mention an intriguing 1972 discovery of uranium ore from a mine in Oklo, Gabon \citep{bodu1972LexistenceDanomalies, neuilly1972LexistenceDans}.
Scientists discovered that the uranium had a lower concentration of the fissile isotope Uranium-235 ($U_{235}$), a characteristic similar to spent fuel from a man-made reactor.
While this initially led to some speculation about ancient technological civilizations, a thorough scientific investigation concluded that the reactor was a naturally occurring phenomenon \citep{cowan1976NaturalFission,meshik2004RecordCycling}.
The evidence overwhelmingly points to a natural origin rather than the product of a past civilization.
However, \citet{davies2010EerieSilence} points out that it's a kind of signature of a past civilization we could also look for in a technosignature context.
For example, if the reactor was dated younger, there wouldn’t have been time for a natural reactor.

We might also constrain the search within a broader geological, biological or astronomical context.
Technosignatures might be correlated with mass extinctions in the geological record, for example if a past anthropocene or nuclear war took place.

On the astronomical side, searching for artefacts of past interstellar visitors can be constrained by the knowledge of stellar flybys, happening about every $\sim100,000$ years.
This means that over the lifetime of the solar system, $45,000$ encounters happened within $1$ ly \citep{bailer-jones2022StarsThat}.
For example, Scholz’s star was inside the Oort cloud $70,000$ yrs ago \citep{mamajek2015ClosestKnown}!
The search for technosignatures in the geological record can thus be meaningfully focused around the time when such stellar flybys occurred.

Similar search strategies naturally apply to the search for pre-human indigenous technological species that would leave traces on dead and stable geological surfaces in the solar system \citep{wright2018PriorIndigenous}.
We could also search the Earth layers for nanomaterials such as nanotubes, nano fibers, or even nanomachines.
This makes sense if we look at the Barrow scale that proposes that civilizations progress by increasing their ability to manipulate, manufacture, and control smaller and smaller scales (see \citealt{barrow1998ImpossibilityLimits}, \citealt{vidal2014BeginningEnd}, and the Barrow scale section~\ref{sec: Barrow Scale}).

\begin{figure}
        \centering
    \includegraphics[width=0.95\linewidth]{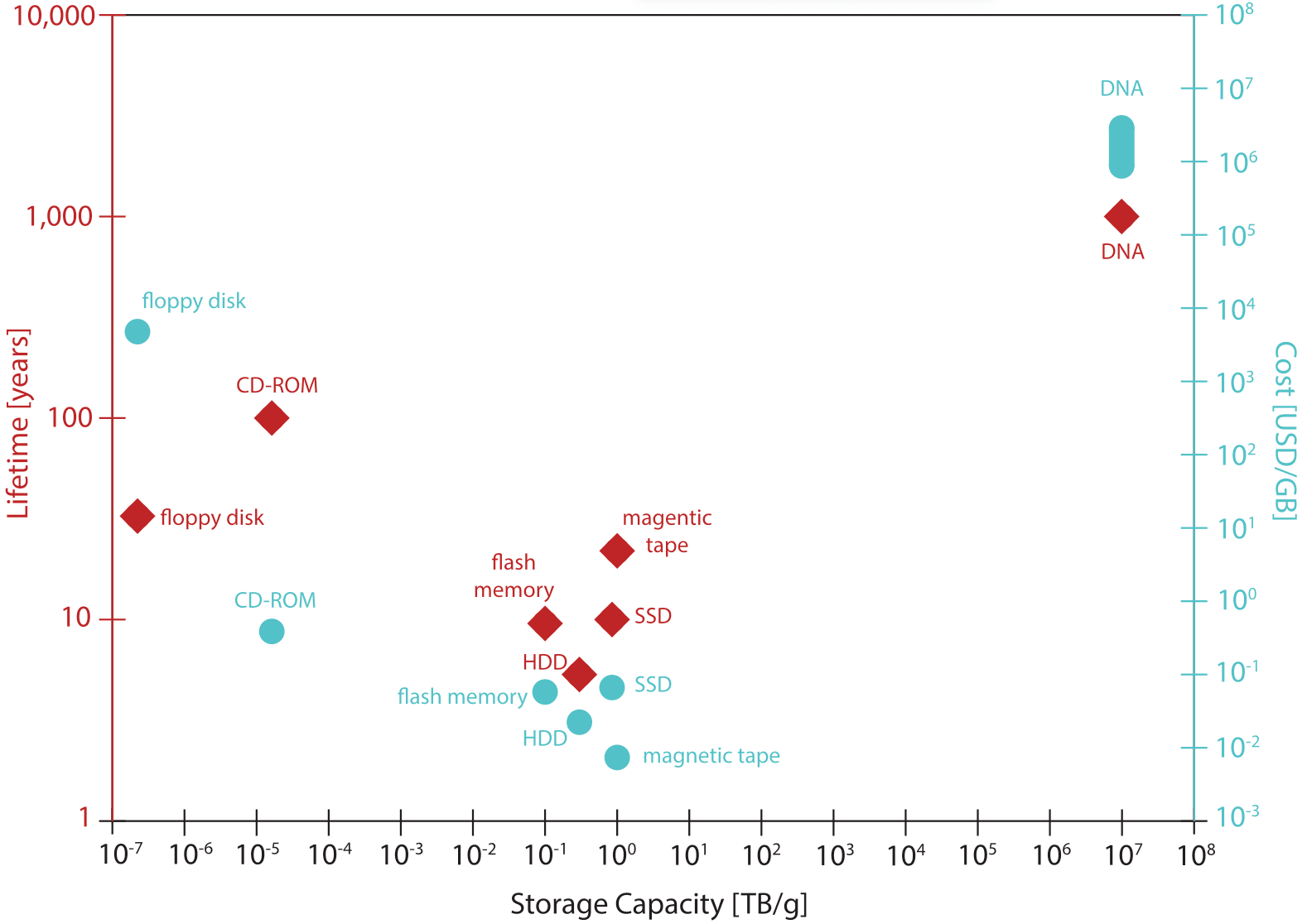}
    \caption{Lifetime, current storage capacity, and costs of various storage systems. Estimates for mainstream media were made with consumer pricing data. The authors of the study note that ``(1) these costs are an estimation at the point at which this work was written and (2) that the basis for calculation is not perfectly aligned, as to date DNA data storage cannot be purchased as a routine commercial electronic storage medium and read/write infrastructure has not been scaled". 
Fig. from \citet{meiser2022SyntheticDNA}}.
    \label{fig: information storage}
\end{figure}

Paul \citet{davies2010EerieSilence} did propose a ''Genomic SETI" program to leverage genome sequencing technologies and to find anomalies, to possibly interpret them as a message from ETI.
Although not very convincingly, some authors have speculated that there might be a message in the DNA of organisms \citep{yokoo1979BacteriophageFX174,nakamura1986SV40DNA}, or in the genetic code itself \citep{shcherbak2013WowSignal}.
DNA is an extremely dense information storage solution ($\sim10^{19}$ bits/g, \citealt{meiser2022SyntheticDNA}) in comparison to state-of-the-art solid state devices ($\sim 10^{12}$ bits/g, see Fig. \ref{fig: information storage}).
So it is still worth exploring whether there could be information content in the noncoding DNA of space resistant organisms.
\citet{vidal2019PulsarPositioning} suggested that a way to test directed panspermia would be to look specifically at the DNA of space-resistant organisms in the context of galactic navigation.
Millisecond pulsars are a prime candidate to provide stable and long-lived coordinate systems, so if ETI used the organism itself to store this navigation metadata, it might still be there.
These genomic SETI hypotheses assume that the Earth has been seeded, so they are part of directed panspermia hypotheses (\citealt{crick1973DirectedPanspermia}, see also Section~\ref{sec: travel ts} on travel technosignatures).

Another information system that might be a target of visit of an advanced ETI might be the internet itself, which led $80$ scientists to place an invitation to ETI on the web \citep{tough2001WideningRange}.

We can note that distinguishing between a past civilization from Earth and a past civilization coming from space would be challenging.
The distinction could be helped with an analysis of isotopic ratios of candidate artefacts, to estimate whether they come from our solar system or not.
Now that we have examined possibilities of past visitations, by extension it seems legitimate to also examine the possibility of present visitations.

\subsection{Unidentified Aerial Phenomena}

\subsubsection{Difficulties in the Past}
\begin{figure*}[t]
    \centering
    \includegraphics[width=1.0\textwidth]{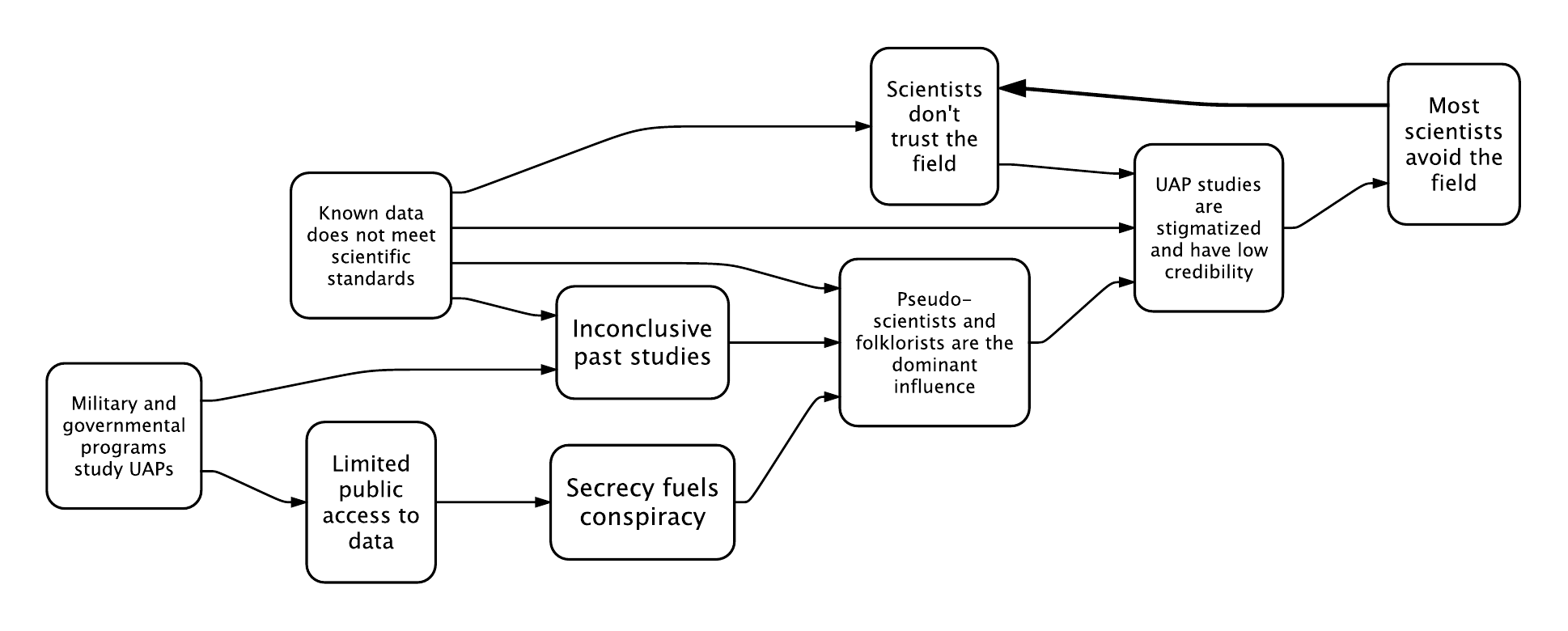}
    \caption{The UAP Taboo: vicious circles prevent serious studies of UAPs.}
    \label{fig:UAP_stigmatized}
\end{figure*}

Although there are reports of unidentified flying objects (UFO) dating back over millennia \citep{stothers2007UnidentifiedFlying}, the first modern sighting to popularize UFOs was reported by Kenneth Arnold in 1947: he saw nine crescent-shaped objects while flying his small plane.
His description of their movements as “saucers skipping on water” led journalists to coin the term “flying saucer”.
This sparked a litany of reported sightings as the media popularized the supposed phenomenon (see \citep{frank2023LittleBook} for more details of the history of modern UFO sightings).
Multiple U.S. Air Force programs began investigating this increasing number of reports, culminating in the extensive Project Blue Book \citep{pbba1951ProjectBlue}.
The head of the Project Blue Book investigation, Edward Ruppelt, replaced the term “flying saucer,” which by then was loaded with unscientific connotation, with the more neutral “Unidentified Flying Object”, or UFO.
UFO sightings grew into a strong social phenomenon, nurturing folklore and fascinating the general public; lack of substantive evidence drove the scientific community to ever higher skepticism.
Gradually, this created a UFO taboo that made serious study of any reports more and more difficult (Fig. \ref{fig:UAP_stigmatized}).

History repeated itself half a century later as UFO was changed to “Unidentified Aerial Phenomena” (UAP) in Project Condign, led by the UK Ministry of Defence \citep{defense-intelligence-agency2000ProjectCondign}.
The “A” of “Aerial” also shifted to “Anomalous”, for example by the “All-Domain Anomaly Resolution Office” \citep{aaro2024ReportHistorical} and therefore encompassing a greater variety of reports, such as those made underwater or in space.

\begin{figure}[h]
    \centering
    \includegraphics[width=0.95\linewidth]{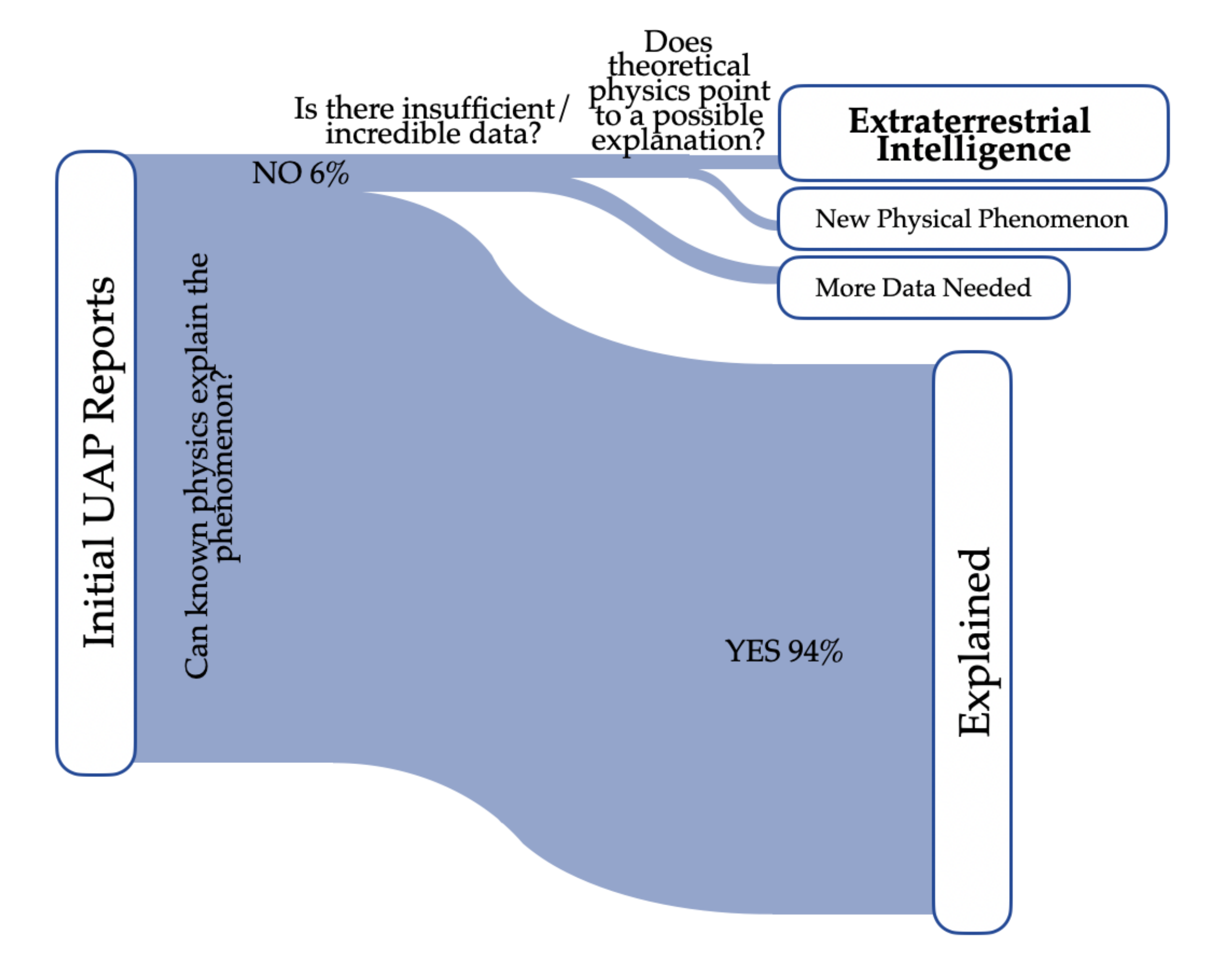}
    \caption{Decision tree illustrating the possible categorizations for UAPs. About 90-95\% of cases end up being explained, but about 5-10\% remain unexplained, and will likely eventually be explained in the future (see Table \ref{tab: EarthTS UAP false sources}). If not, they might remain unexplained due to their unrepeatable nature, or a lack of quality data. But this small unexplained percentage might also lead to the discovery of a new natural phenomenon, or to the discovery of extraterrestrial intelligence. 
}
    \label{fig: is it a UAP}
\end{figure}

How many reports are really unidentified or anomalous?
Out of 12,000 reports analyzed in Project Blue Book, 6\% of them remained unexplained.
How are UAP reports categorized? 90\%-95\% end up as (1) explained phenomenon.
The remaining 5-10\% of reports could end up (2) unexplainable due to lack of credible data.
Those that do have credible data imply either (3) an unknown physical mechanism, or (4) an unknown manifestation of extraterrestrial intelligence (see Fig. \ref{fig: is it a UAP}).

Explained UAP reports are attributed to multiple sources: celestial, meteorological, biological, instrumental, artificial, or psychological (see table \ref{tab: EarthTS UAP false sources}).
This means that ruling out systematically all the confounding sources requires expertise in different domains: astronomy, engineering, meteorology, but also biology or psychology.
It can be easy (e.g. Venus) or very hard if not impossible (e.g. a secret military device).
One should note that a typical astronomer interested in SETI would be expert in only a small subset of such sources, so explaining away UAPs is not within the typical expertise of scientists working in SETI.

An example of a phenomenon that remains unexplained are long-delayed radio echoes that were discovered in the 1920s and that are still considered anomalous.
It is a case where it could be an atmospheric phenomenon, we may also need more data, it might be a new physical phenomenon or even extraterrestrial intelligence as \citet{bracewell1960CommunicationsSuperior} suggested.

\begin{table*}[!ht]
    \centering
    \caption{Common sources causing UAP sightings. These need to be carefully considered and excluded before taking the ETI road.}
    \label{tab: EarthTS UAP false sources}
    \begin{tabular}{p{2.2cm}p{14.0cm}}
        \hline\hline
        \raggedleft Source & \raggedright\arraybackslash Example \\
        \hline\hline
        \raggedleft Celestial & Venus, Mars, Meteor, Meteorite\\
        \raggedleft Meteorological & Clouds (e.g. noctilucent, lenticular, punchhole), Reflected light off clouds, Mirages (e.g. fata morgana), Sprites, Ball Lightning, Ice crystals and sundogs\\
        \raggedleft Biological & Birds reflecting the Sun, Insects interfering with instruments\\
        \raggedleft Instrumental & Flares, Ghosting, Internal scattering of light inside the instrument, Other errors or distortions from observing instruments\\
        \raggedleft Artificial & Weather balloons, Airplanes, Contrails, Light on the ground, Sky lanterns, Military aircraft flying in formation, military and defense: secret military aircraft, experiment, or exercise, Satellites reflecting light, launching, cruising, or crashing, Space junk reflected or crashing, Rocket launch boom, Blimps, Search light or laser pointer, Artificial auroras, Hoax\\
        \raggedleft Psychological & Drugs, Hallucination, Fantasy prone personality, Altered states of consciousness, Mass hysteria\\
        \hline
    \end{tabular}
\end{table*}

To look for a UAP as a technosignature positively (i.e. not merely refuting alternative explanations) one could expect one or several anomalous properties (see table~\ref{tab:EarthTS_UAP_anomalies} and \citealt{siegel2024AskEthan}).
\begin{table}[h!]
    \centering
    \small  
    \begin{tabular}{>{\raggedright\arraybackslash}p{0.25\linewidth}>{\raggedright\arraybackslash}p{0.65\linewidth}}
        \hline
        \textbf{Property} & \textbf{Constraint} \\
        \hline
        \textbf{Size} & No known aircraft with a wingspan greater than about half a kilometer. \\[6pt]
        \textbf{Mass} & No aircraft heavier than a few thousand tons has ever been even proposed. \\[6pt]
        \textbf{Speed} & No aircraft or spacecraft in Earth's vicinity has exceeded more than around $17 \text{ km s}^{-1}$ ($\sim$11 miles/second) while near our planet. \\[6pt]
        \textbf{Acceleration} & No aircraft has sustained an acceleration or deceleration in any direction of more than $\sim 10g$ for more than a few seconds. \\[6pt]
        \textbf{Emission} & E.g.\ gamma-rays from antimatter propulsion (see also other exotic propulsion signatures in section \ref{sec: travel ts} on Interstellar travel). \\
        \hline
    \end{tabular}
    \caption{Properties of UAPs that would be considered anomalous \citep{siegel2024AskEthan}.}
    \label{tab:EarthTS_UAP_anomalies}
\end{table}
A study did analyze the accelerations of some past UAP cases, estimating minimum accelerations of 10s g’s, observed by the Japanese cargo flight JAL1628 in 1986, or even 100s or 1000s of g’s in the Bethune case of 1951 or the Nimitz case of 2004.
If confirmed by well calibrated instrumentation, these anomalous velocities and accelerations would be sufficient for interstellar travel, which may have important implications for an ETI origin hypothesis \citep{knuth2019EstimatingFlight}.

There are other historical examples of transient phenomena in the sky that turned into the discovery of new phenomena.
The core example is the discovery of meteorites that relied on eye-witness testimony after the fact, as opposed to real-time observation.
Historically, the discovery was pushed by the general public against an incredulous scientific community (see \citealt[chap.~2]{dick1996BiologicalUniverse}; \citealt{jacobs1975UFOControversy}).

Indeed, their existence contradicted the last remnants of the Aristotelian paradigm, assuming that there could not be rocks from space, because the cosmos was not made of the same matter as Earth (see also \citealt{watters2023ScientificInvestigation} for a discussion in the UAP context).
The rare phenomena of earthquake lights or ball lightning may also provide a good analog, and the latter has only recently been quantifiably observed \citep{cen2014ObservationOptical}.
The observing difficulties in all these cases - including UAPs - is that they are rare and transient phenomena.

How can we make scientific progress?
Unfortunately, the field is locked into a UAP taboo, where vicious circles prevent serious studies (see Fig. \ref{fig:UAP_stigmatized}).
There is a huge interest and fascination in popular culture about ETI coming to Earth, which is fueled by countless movies, TV shows, YouTube channels or blogs (see e.g. \citealt{denzler2003LureEdge}).
There are even a few UFO religions who may not want scientists to even approach the topic.

Many UFO enthusiasts want to further study the phenomena.
Unfortunately, they are not trained as scientists, so they end up doing pseudoscience as they have no or low scientific standards.
Indeed, the notion of UFOs epitomizes pseudoscience, and is associated with topics such as ancient aliens, abduction stories, conspiracy theories, out-of-body experiences, paranormal activities, or psychic channeling.
Trained scientists who dare to enter the field have their signal drowned in a loud crackpot noise and risk being ridiculed and rejected by their peers and the scientific community at large.
However, this is not an isolated sociological issue as there are pseudoscientists in all fields such as physics, medicine, etc. All scientific fields have to deal with the issue of drawing epistemic boundaries, deciding what qualifies as legitimate and proper scientific practice, and what to leave out.

It is no secret that governments have been and are studying UAPs (see \citealt{graff2023UFOStory}; \citealt{eghigian2024FlyingSaucers}; and \citealt{knuth2025NewScience} for a review).
However, another important obstacle comes from the way classified defense programs work: by definition, they don’t share much of their data and this can slow down scientific progress and discoveries.
For example, gamma-ray bursts were discovered on July 2, 1967 by the US Vela nuclear test detection satellites.
Although the data was not classified \citep{bonnell1996BriefHistory}, it took another six years before the scientific community interpreted the data as coming from the sky \citep{klebesadel1973ObservationsGammaRay}.
The instruments that the military builds have different purposes than astronomical instruments, and have in this example detected a novel phenomenon.
This is not an isolated example, and astronomer Harwit \citeyearpar[234-239]{harwit1981CosmicDiscovery} showed that out of 43 major astronomical discoveries, 26 of them were enabled thanks to instruments and technologies first built for military purposes.

When military data is released, it is often not scientifically actionable, for example because releasing the sensitivities of the instruments would reveal strategic information to adversary nations.
However, science cannot be done without transparency, data sharing, and reproducibility.
Such classified research is typically compartmentalized so no peer evaluation is possible, no consensus building dynamics can emerge, and thus very little progress can be made.
What is more, concealed data gives rise to conspiracy theories that are formulated using speculations instead of data.

Publicly available reports from past programs have failed to give conclusive scientific insights, which stifles current interest in the topic.
One historically significant example is the Condon Report \citeyearpar{condon1969FinalReport} which remains controversial.
The report states, “Our general conclusion is that nothing has come from the study of UFOs in the past 21 years that has added to scientific knowledge.”
However, the study was also criticized by several scientists at the time for lacking objectivity and taking a ‘conclusions first, data second’ approach \citep{mcdonald1972ScienceDefault, sturrock1999UFOEnigma, sagan1972UFOsScientific}.
Regardless of this particular debate, it underscores the difficulties of consensus building and peer review under such conditions.

All these aspects lead to an academic stigma against the UAP topic, so credible people often prefer to avoid its study (see again Fig. \ref{fig:UAP_stigmatized}).
In addition, the search for extraterrestrial life and intelligence has and is still suffering from a “giggle factor” as a seemingly not so serious field of research \citep{wright2018VisionsHuman}.
The technosignature community has faced this stigma repeatedly, and thus naturally wants to avoid an additional stigma from ufology.
So it makes sense that a wide portion of the community tends to reject the study of UAPs.

Although outside the scope of this paper, the current unfortunate situation could be transformed into opportunities for interdisciplinary education (e.g. refuting mundane explanations such as the ones in Table \ref{tab: EarthTS UAP false sources}); for teaching epistemology and the scientific method to debunk reports; and maybe even stimulating UAP enthusiasts towards solid citizen science projects related to UAPs, extraterrestrial life and intelligence.

\subsubsection{Present and future prospects}
Despite all these difficulties, we discuss UAPs in this review because they logically are a part of the technosignature search space, not because we believe that current UAP data provides strong support for an ETI interpretation.
Let us give an overview of the current research progress related to UAPs.

A recent survey showed that out of 1,460 American research university professors, $37\%$ of them agreed that the subject should be researched in academia -even if few were enthusiastic about the extraterrestrial explanation \citep{yingling2024AcademicFreedom}.
An impressive review of 20 historical government studies ranging from 1933 to 2025 has recently been published by \cite{knuth2025NewScience}.
The Society for UAP Studies aims to bring more academic credibility to the emerging field and publishes a journal, \textit{Limina} \citep{cifone2024EditorialFirst, rodeghier2024EditorialFirst}.
NASA commissioned an independent analysis to design a roadmap for getting better and more uniform data \citep{spergel2023NASAUnidentified}.
The Galileo project was founded in July 2021 as an effort to gather new high quality data with multiple modalities \citep{loeb2022OverviewGalileo,watters2023ScientificInvestigation}.
This is to be contrasted with most existing UAP reports that rely on eyewitness testimony, or very low resolution images.
As an illustration, if a UAP were to fly over the instruments of the Galileo project: All-sky infrared and optical cameras would capture it with their continuous panoramic views.
Machine learning algorithms would compare the detection with known confounding sources (such as the ones in Table \ref{tab: EarthTS UAP false sources} above).
The observational data would be checked against sources of false positives, such as the Automatic Dependent Surveillance-Broadcast (ADS-B) to check whether it was a known aircraft.
If the algorithm flags it as a genuine anomaly, then the Pan Tilt Zoom (PTZ) cameras can activate to get close-up imagery.
Passive radars would cross-check the reality of the object, to rule-out optical artefacts.
Microphones pick up a wide range of sounds, above and below what the human ear can perceive.
It is useful information to understand how the object is interacting (or not) with the surrounding air.
If the object is within range, the magnetometers would measure the electromagnetic activity around the object, and be able to compare with a local ambient baseline, as well as to the entire Earth’s magnetic field via the US geological survey (USGS).
Satellite images can complement all these ground-based observations by surveying more of the Earth’s surface than can be reached by a stationary observatory.

A team did a field expedition on an alleged UAP hotspot to collect reliable data, with a setup similar to the Galileo project, but adapted for a short term and mobile purpose, and found an ambiguous source of ionizing radiation picked up by an instrument capturing cosmic particle showers \citep{szydagis2025InitialResults}.

Getting physical access to UAP fragments would provide a lot of opportunities for continuous, exhaustive and repeated analyses by different independent teams.
For example, if we would find an unknown fragment, we could do advanced material analysis to determine whether its composition includes materials from interstellar origins.
Mass spectrometry can decipher elemental compositions and abundances, while a study of isotope ratios could determine whether the material did form in our solar system or not.
For an overview of these techniques and related ones, see \citet{nolan2022ImprovedInstrumental}.
Of course, finding evidence of interstellar origins would just be a first intriguing lead and not conclusive of ETI origin.
A next step would be to ascertain artificiality, for example, \citet{embaid2022PuzzleMeteoritic} suggested in a non peer-reviewed preprint paper that known Bustee, Kaidun and Tucson meteorites display minerals heideite and brezinaite that have no known natural occurrence on Earth, and suggests that they might be iron-based superconductors.
Another example is the meteor CNEOS 2014-01-08, and the debate about its potential interstellar origin.
\citet{siraj2022InterstellarMeteors} argued for an interstellar origin, calling it IM1 for “Interstellar Meteor 1”, and followed up with an underwater mission in Papua New Guinea to attempt to find and analyze the remains of the meteor \citep{loeb2023DiscoverySpherules, hyung2025InvestigationRelationship}.
However, the interstellar origin has been contested \citep{hajdukova2024NoEvidence}, and the debate is unlikely to be fully resolved from the CNEOS data, because it comes from a military instrument that does not disclose its uncertainties publicly.
More generally, at a planetary scale the study of ocean floor (bathymetry) remains incomplete and at low resolution, especially in the deeper water.
Following up on “transmedium” UAPs reports, one might want to map the ocean floor to a much higher accuracy and more systematically \citep{gallaudet2024SurfaceWe}.

Even if UAPs are not linked in any way to ETI, explaining away mysteries and anomalies is the core role of science, and as such they should be studied.
Other benefits include developing new instruments, new methods for continuous all sky monitoring, improving planetary defense, discovering new meteorological or natural phenomena.

\subsection{Earth Orbit}
\begin{figure*}
    \centering
    \includegraphics[width=0.9\linewidth]{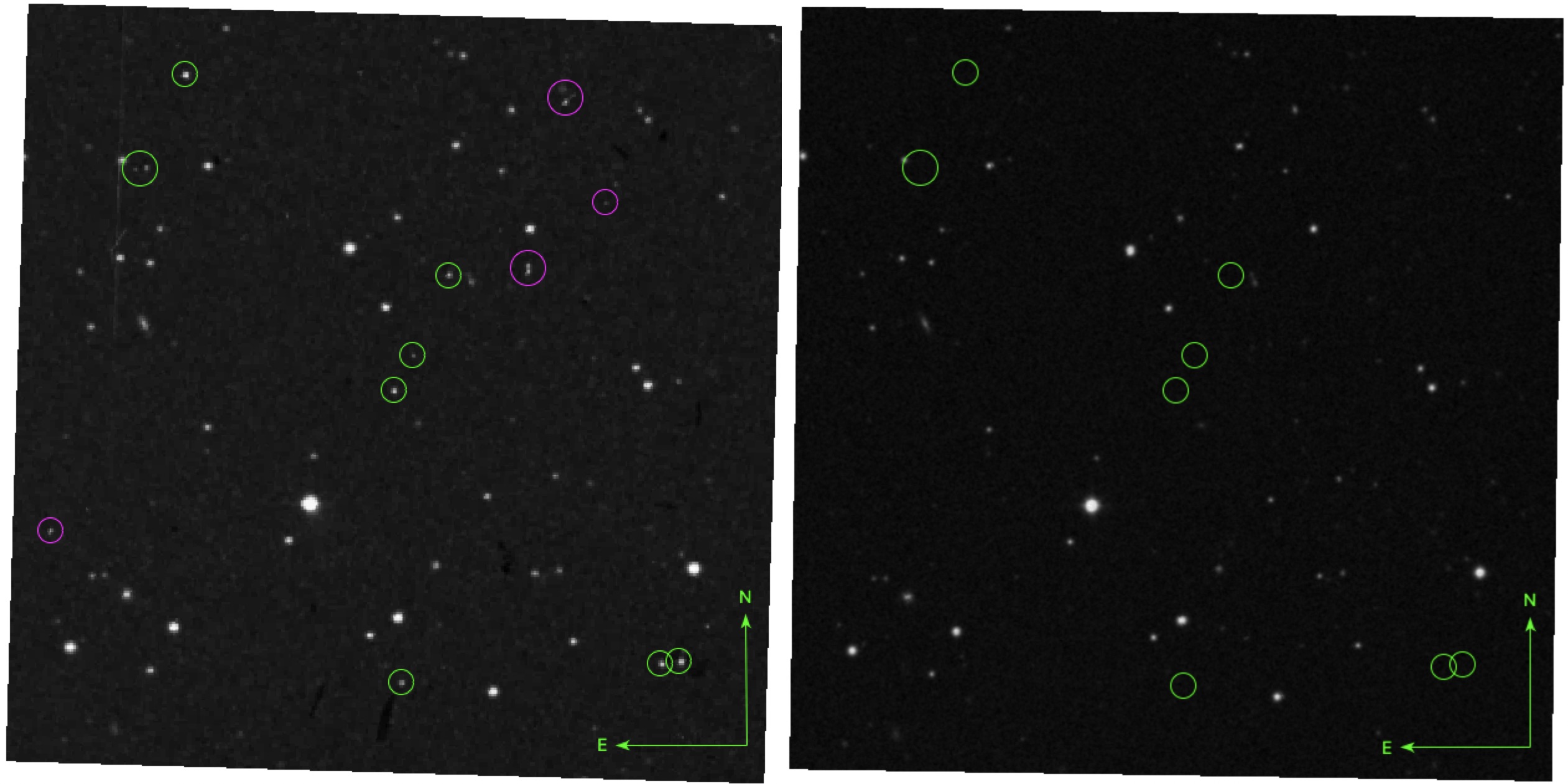}
    \caption{Nine simultaneously occurring transients on April 12th 1950, reproduced from \citet{villarroel2021ExploringNine}. Original caption: ``10 x 10 arcmin field shown in POSS-1 and POSS-2 red bands. In the POSS-1 image we see a number of objects that cannot be subsequently found, marked with green circles. Purple circles are artifacts during the scanning process. About 9 objects are present in the POSS-I E image (left) from the 12th of April, but not in the POSS-2 image (right) from 1996. One slightly larger circle host two transients. In addition, the 9 objects are neither visible in the blue POSS-1 taken half an hour earlier, nor in a second POSS-1 red image taken six days later on April 18th. The 9 transients are not caused by a difference in depth or spectral sensitivity. The images are based on the DSS digitizations of the Palomar plates''.}
    \label{fig: POSS-I}
\end{figure*}

\cite{villarroel2021ExploringNine,villarroel2022GlintEye} conducted an analysis of 1950s archival photographic plates from the Palomar Observatory Sky Surveys (POSS-I, 1948-1958 and POSS-II, 1980s-1990s).
The POSS-I survey was mostly before Sputnik, the first man-made satellite that was launched in 1957.
The team discovered intriguing multiple point sources that were not visible in subsequent observations taken only 30 mins later (see Fig. \ref{fig: POSS-I}).

This short window of time excludes far away sources, and nine objects would almost certainly not vanish within such a close interval of each other if they were located at interstellar distances.

They also considered radioactive dust from nuclear tests, and apparently found statistically significant correlations between the Palomar transients UAP sightings and nuclear tests \citep{bruehl2025TransientsPalomara}.
However, a recent study by \cite{watters2026CriticalEvaluation} argues that the correlation between the timing of observations and nuclear tests is completely determined by the observation schedule of the Palomar telescope.

A similar “triple transient” was found which appeared to vanish within a 50 minute window, though in that case it was hypothesized that it might have been a gravitational lensing event \citep{solano2024BrightTriple}.
One analysis argued that both transients could be spurious artefacts of the photographic emulsion (see \citealt{hambly2024NatureApparent}).
To further probe whether these transients are skybound or not, \citet{villarroel2025AlignedMultipletransient} did propose to explore their relation with the Earth's shadow, i.e. when the Earth is casting a shadow in the sky, like during a Lunar eclipse.
Intriguingly, they did find an apparently highly significant deficit of transients within Earth's shadow, arguing against the hypothesis that these were simply plate defects.
The transients might thus have been highly reflective objects in Earth's orbit that would seem to disappear when entering the Earth's shadow, simply because they would not be lit up by the Sun anymore.

However, the study by \cite{watters2026CriticalEvaluation} challenges the Earth shadow deficit argument and many others.
Yet, whatever the final word will be, this effort is a novel search strategy for extraterrestrial artifacts \citep[see also][]{villarroel2022ThereBackground}.
Future searches could also be done on older astronomical plates, going back even further before the advent of satellites.
We could also search for ETI satellites today, although the enormous number of satellites and space debris would make the endeavor challenging.
Nonetheless, searching currently could allow for multiple telescopes independently observing candidates (thereby ruling out artefacts), or more accurately localizing their orbits \cite[see also][]{villarroel2024SearchesNearEarth, villarroel2025CostEffectiveSearch}.

Other novel strategies could include searching the International Space Station (ISS) video feeds for anomalies, which would anyhow have ancillary benefits for Earth sciences or planetary defense.
We also suggest staying open to a SETI angle for anomalies that may be found during space domain awareness efforts in the cislunar region \citep[see e.g.][]{baker-mcevilly2024ComprehensiveReview}.

Let us now turn our attention further outward, looking for technosignatures in our solar system writ large, from the Moon to the Oort cloud.

\clearpage
\section{Solar System Technosignatures}\label{sec: ss ts}

There has been relatively little consideration of the Search for Extraterrestrial Artefacts (SETA) in the scientific literature, although various authors have called for a renewal in recent years (see e.g. \citealt{ellery2003SETIScientific,gertz2016ProbesLooking,benford2019LookingLurkersa,shostak2020SETIArgument} and \citealt{haqq-misra2012LikelihoodNonterrestrial} for a review).

As with Earth technosignatures (see section \ref{sec: earth ts}), the serious plausibility of solar system technosignatures is implied by the concept of \textit{Bracewell probes} \citep{bracewell1960CommunicationsSuperior,bracewell1974InterstellarProbes}, i.e. probes having communicating or monitoring purposes, or of \textit{self-replicating probes}, also called \citet{vonneumann1966TheorySelfreproducing} probes, focusing on galactic settlement \citep{freitas1980SelfreproducingInterstellar,ellery2022SelfreplicatingProbes}.
The two strategies are not mutually exclusive, and a Bracewell probe program could use a von Neumann strategy for fast propagation.

More precisely, there are two main strategies for a civilization to expand which can be characterized by r-K selection theory in biology \citep{boyce1984RestitutionKSelection}: r-selected species emphasize many offspring and fast reproduction (e.g. insects), while K-selected species have a slow reproduction and focus on having few offspring, but take greater care of them (e.g. mammals).
In the context of interstellar expansion, self-replicating probes are the optimal r-strategists \citep{jones1976ColonizationGalaxy} for practical \citep{valdes1980ComparisonReproducing}, economic \citep{ellery2017SpaceExploration}, reconnaissance and threat assessment reasons \citep{smith1995SituationAwareness}, or simply to make the most of astronomical resources \citep{bostrom2003AstronomicalWaste}.

There are two main dimensions we can look for in the solar system; the first is their position in space, whether it is in orbit, or on a surface; the second is whether the probe is active or passive in terms of energy use or communication \citep{lazio2023DataDrivenApproaches}.

Generally, active probes or artefacts may be less ambiguous, because one could readily assess whether the behavior seems anomalous, and one could look for transmission, but also look for waste heat or waste material signatures.
However, we could also have alternating periods of activity and dormancy, for example if a probe puts itself in a hibernation mode to save energy and limit wear.
Typically, in space engineering a rule of thumb is that an operating device undergoes ten times more wear when active compared to being dormant.
Regarding surface artefacts, the methods of archeology may be beneficial to explore SETA hypotheses, as explored in \citep{vakoch2014ArchaeologyAnthropology}.
One may object that if there were extraterrestrial technosignatures in the solar system we would have noticed it by now.
However the solar system is a huge place, and the vast majority of it has not been observed at a high resolution (\citealp{haqq-misra2012LikelihoodNonterrestrial}, see also Table \ref{SolarSystem_TS_resolution}).
What is more, technology tends to progress at ever-smaller scales, a trend known as the \textit{Barrow scale} \citep{barrow1998ImpossibilityLimits,vidal2014BeginningEnd}.
It implies that small and light probes are naturally preferred for fast interstellar missions as they require less fuel and can be propelled faster.
We illustrate this trend with human space technology and the concept of smart dust, which are millimeter cube computers \citep{warneke2001SmartDust}.
The point here is that we are far from having systematically sieved the solar system from micro- or nano- or smaller ETI technology that might be lurking.

\begin{table*}[ht]
    \centering
    \begin{tabular}{rcl}
        Solar System Body & Resolution (m) & Mapping Mission  \\ \hline\hline
hi        Moon & 0.5 & Lunar Reconnaissance Orbiter\\
        Mercury & $2.5\!\times\!10^2$ & MESSENGER\\
        Venus & $10^3$& Magellan radar\\
        Mars & 1m for $1\%$ of surface & Mars Reconnaissance Orbiter, \\ & 100 for global & {\it The Atlas of Mars}\\
        Main Belt Asteroids & $4.1\!\times\!10^2$ for Ceres& Dawn (best case example) \\
        Jovian Satellites & $\sim 6$ over narrow swath(s) & Europa\\
        Saturnian Satellites & $\sim 3\!\times\!10^3$& Iapetus, {\it Cassini} \\
        Uranian Satellites &$\sim 10^4$ & Voyager 2 flyby\\
        Neptunian Satellites&$\sim 3\!\times\!10^3$ over 1/3 of surface & Voyager 2 flyby
    \end{tabular}
    \caption{Missions have mapped solar system bodies at various resolutions, adapted from \citet{lazio2023DataDrivenApproaches} and \citet{fields2023InformationGain}.}
    \label{SolarSystem_TS_resolution}
\end{table*}

\subsection{The Moon}
\begin{figure*}
    \centering
    \includegraphics[width=0.99\linewidth]{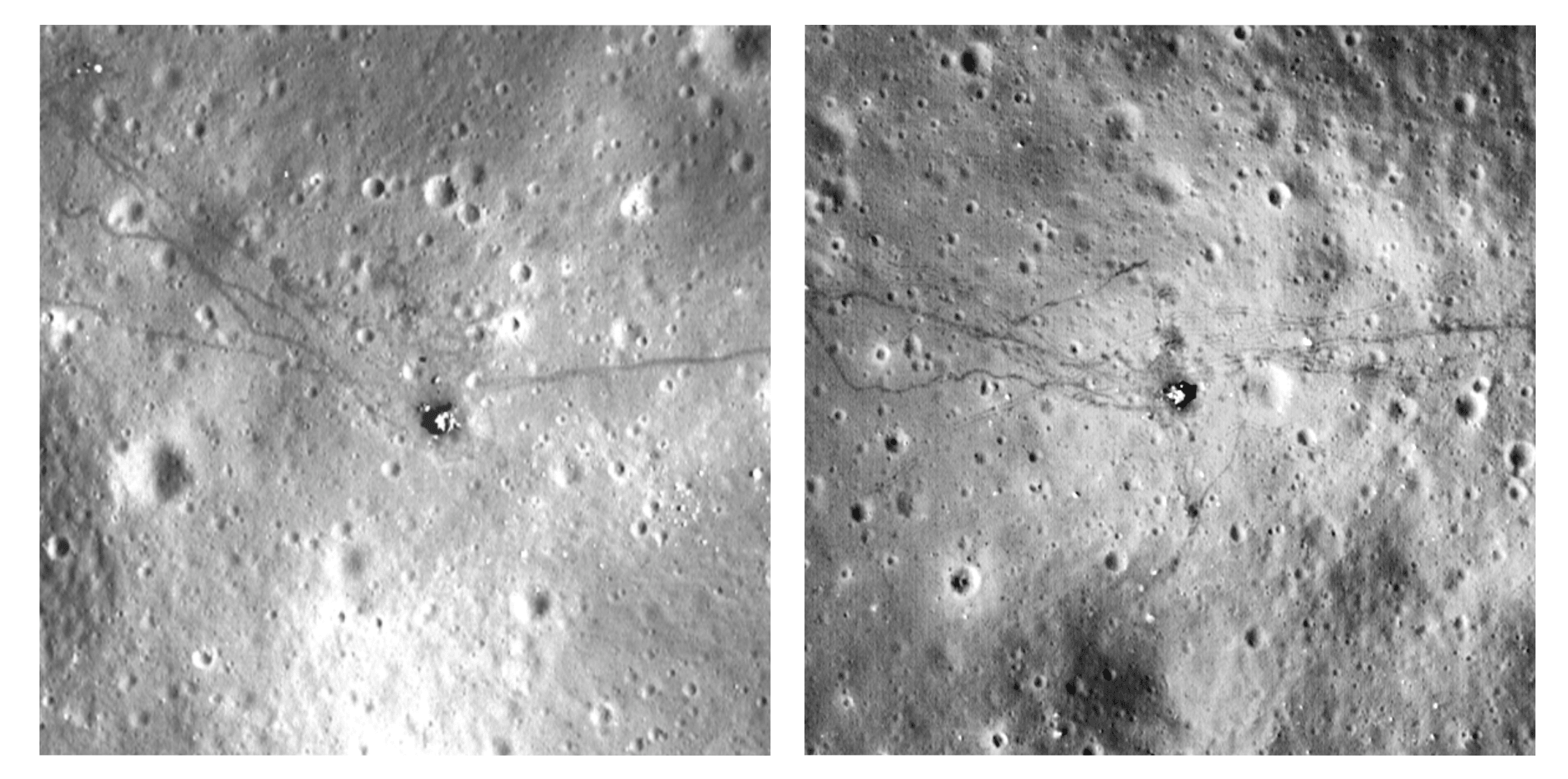}
    \caption{Left: An LRO/LROC image showing the Apollo 12 descent module and astronaut tracks located at selenographic coordinates 3.0128 S, 336.5781 E. Two extravehicular (EVA) activities were performed on the lunar surface totaling 7.75 hours, resulting in the collection of 35.34 kg of lunar samples. Right: An LROC image showing the Apollo 17 landing site. The sub-meter resolution image clearly shows the descent module and tracks. A few bright point-like features can be seen around the descent module, and are various experiment packages deployed on the lunar surface.}
    \label{fig:Apollo12}. Image credits: NASA/GSFC/ASU and \citep{robinson2010LunarReconnaissance}. 
\end{figure*}

The idea to search for extraterrestrial artefacts in the solar system has been popularized in science fiction most famously with the Monolith of \textit{2001: A Space Odyssey} \citep{clarke19682001Space}.
Compared to space, the Moon is a privileged place for placing long-term human or extraterrestrial artefacts and technology, because the Moon shields from many meteoroids, it shields from half of the ionizing radiation, no stabilization is required, lunar resources can be used for maintenance and repair \citep{arkhipov1995SearchAlien}.
Compared to the Earth, since the Moon is not geologically active, there is almost no erosion, it also doesn’t have an atmosphere so there are no atmospheric hazards, and since it’s dead there is no influence from biological activity.
\citet{arkhipov1998NewApproaches} also proposed a search for alien space debris on the Moon, and \citet{davies2013SearchingAlien} proposed to launch a citizen science project to systematically examine the Lunar Reconnaissance Orbiter (LRO) images that offer 0.5 m resolution.
This may be conducted using the methodology of similar citizen science projects (e.g. \citealp{bugiolacchi2016MoonZoo}).
The sub-meter spatial resolution of LRO’s imaging system has imaged various Apollo landing sites that clearly show evidence of past human activity on the Moon (Fig. \ref{fig:Apollo12}).

Preliminary studies of the Moon’s surface using machine learning are able to detect human activities on the Moon such as the Apollo 15 landing site \citep{lesnikowski2020UnsupervisedDistribution, loveland2024AnomalyDetection}, but a systematic search on the LRO data remains to be done.

\begin{figure}
    \centering
    \includegraphics[width=0.95\linewidth]{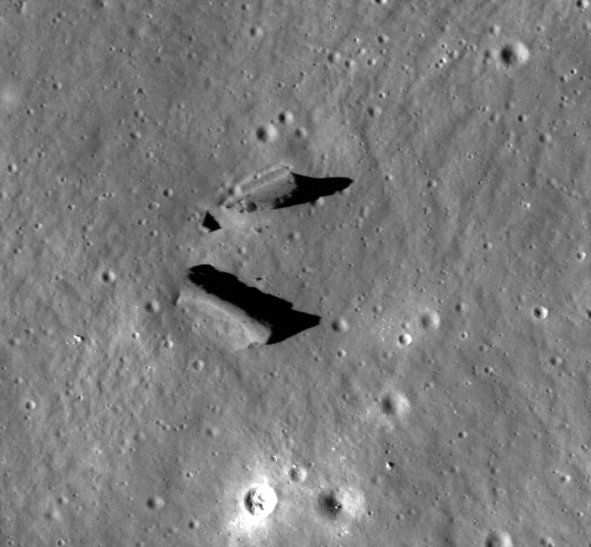}
    \caption{LRO LROC-NAC frame M118769870LC of the unusual structures located on the floor of the lunar farside crater Paracelsus C. The image shows geometrical structures and three-dimensional relief as evident by the cast shadows. The width of this whole image is approximately 382 meters. Image credits: NASA/GSFC/ASU and \citep{robinson2010LunarReconnaissance}.}
    \label{fig:Paracelsus C}
\end{figure}

One example are the structures observed on the floor of the farside crater Paracelsus C (Fig. \ref{fig:Paracelsus C}).

Since there is no weathering on the Moon due to the lack of an atmosphere, a natural explanation for these structures would be some type of non-trivial impact erosion or tectonic stresses that lead to contraction over time creating ridges and faults that can appear geometric.
These angular features could also result from impact-induced fracturing, ejecta deposition, or solidified melt pools.
Of course, visual inspection can offer clues at best, and humans are prone to “pareidolia”, a cognitive phenomenon where a superficial and isolated visual inspection can easily bias humans towards familiar shapes (e.g., faces, walls or other sharp boundaries) on random terrain, especially under low-resolution or shadowed conditions.
At least in this case, the high resolution images are sufficient to discern shadows cast thus revealing their three-dimensional structure.
While natural geological processes remain the primary explanation for these structures, some speculative theories suggest that Paracelsus C could be of interest for further exploration due to its atypical features \citep{carlotto2016ImageAnalysis}.
Future lunar missions may provide more insight into the true nature of these enigmatic formations.

One should also note that the Moon is a unique place for future astrobiological and SETI missions, radio telescopes on the Moon would enable to search for auroral radio emission from exoplanets (e.g. LuSEE-Night, FARSIDE; \citealt{burns2021LowRadio}) but also be capable to do radio SETI.
Actually, an exciting project to search for technosignatures from the Moon has been announced, the Lunar Farside Technosignature \& Transients Telescope (LFT3, \citealt{deboer2025LunarFarside,deboer2026LunarFarside}).
It proposes to seize the opportunity that the Moon is almost completely radio quiet now, before the launch of many missions that are planned for the next decade.

\subsection{Earth-Moon Lagrange Points}

\begin{figure*}[ht]
    \centering
    \includegraphics[width=0.75\linewidth]{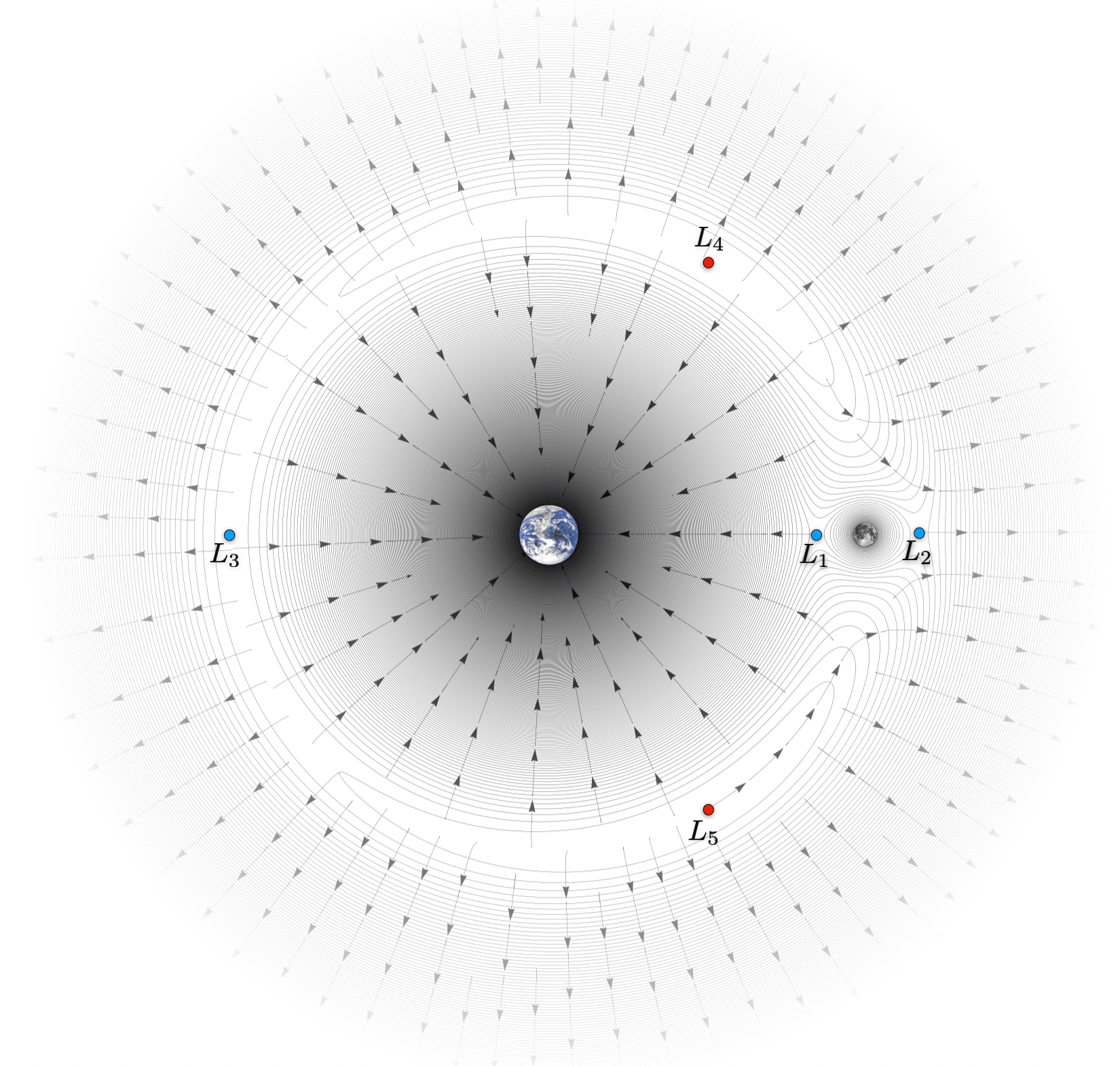}
    \caption{ The five Lagrange points of the Earth-Moon system in the co-rotating reference frame. Effective potential contours (logarithmic scaling) are superimposed on the field lines (streamlines). The first three Lagrange points occupy saddle points (blue dots) of the potential, while $L_4$ and $L_5$ local maxima (red points).  Counterintuitively, these are dynamically stable when Coriolis forces are taken into account. (Earth and Moon sizes not to scale).
}
    \label{fig: Earth-Moon Lagrange points}
\end{figure*}

The gravitational geometry of two orbiting bodies gives rise to stable regions in space, first discovered by the mathematician Joseph-Louis Lagrange in the 18th century.
Fig. \ref{fig: Earth-Moon Lagrange points} displays these aptly named {\it Lagrange points}.
Since objects in these regions require very little propellant to remain stationary, these regions are particularly useful for human (or extraterrestrial) artefacts in space.

This is why SETA researchers have focused on proposals to observe and examine the five Lagrangian points L1-L5, both for the Earth-Moon and the Earth-Sun system \citep{freitas1983IfThey, freitas1983SearchExtraterrestrial, freitas1985SearchExtraterrestrial, benford2019LookingLurkersa}.
In particular, it has been speculated that the Kordylewski clouds in L4 and L5 of the Earth-Moon system might host some kind of intelligent organism or “superbrain” \citep{temple2019KordylewskiDust}.
However, we may also speculate that amongst the cloud’s dust are small artificial probes that would be ideally placed for the purpose of long-term monitoring of planet Earth.

An ET could place a probe in a custom solar orbit, where it could be on a unique orbit that only occasionally brings it close to Earth, making it appear as just another near-Earth object that is difficult to track consistently.

Also, a horseshoe orbit would be an excellent candidate to place an ET probe for long-term monitoring of the Earth.
This type of orbit represents an optimal balance of concealment, long-term stability, and periodic opportunities for observation of the Earth.
The probe would slowly drift towards Earth from the L5 point (an example), but before it gets too close, the gravitational perturbation from Earth would nudge it, changing its orbit and causing it to reverse direction.
It could then spend decades or even hundreds of years travelling all the way around the Sun, passing through the L3 region.
Finally, it would eventually approach Earth from the other side (near the L4 point), where it gets another gravitational perturbation, sending it back the other way.
This 'cycle' or 'cyclical motion' repeats over very long timescales.
It is interesting to note that we have discovered asteroids (e.g. 2010SO16, 419624, ...) in this type of orbit around the Earth, and we might explore them from a technosignature perspective (e.g. metallic surface, light sail signature).

Co-orbital objects are also a potential target of interest, because they come very close to the Earth annually \citep{benford2019LookingLurkersa}.

\subsection{Inner Solar System}
In the solar system search for biosignatures, it is well-known that cross-contamination of living microorganisms --- planetary lithopanspermia --- is a plausible scenario over long timescales \citep{nicholson2009AncientMicronauts}.
Similarly one could imagine the possibility that an advanced civilization developing in our (or another) solar system might migrate away from a dying planet to a nearby one in the same stellar system.
In our solar system, we could frame this possibility as an extension of the silurian hypothesis \citep{wright2018PriorIndigenous} beyond the Earth’s boundary, and to search for extinct civilization coming from the nearby planets.

Historically, Percival Lowell thought he saw canals on Mars constructed by “Martians”, or a spider web on Venus, another typical example of pareidolia.
Another explanation of these signatures might simply be that the telescope was acting like an ophthalmoscope and that Lowell was describing the vascularization of his own eyeballs \citep{sheehan2003SpokesVenus}!

A general strategy is to look for outliers in the orbital parameter spaces of solar system objects \citep{lazio2023DataDrivenApproaches}, maybe similar to the ones of human’s technological solar-system missions.
These display technosignatures such as non-gravitational acceleration, fine-tuned gravitational assists, goal-directed behavior, or active communication.
Phobos, the innermost natural satellite of Mars was observed to have a non-gravitational acceleration in the 1940s \citep{sharpless1945SecularAccelerations}.
\citet[chap.~26]{shklovskii1966IntelligentLife} explored in detail the hypothesis that Phobos might be artificial. 
However a finer analysis of the data showed that the orbital anomaly could be attributed to systematics \citep{wilkins1969MotionPhobos}.
Anomaly detection using fractal dimension analysis has been applied to satellite images on Mars \citep{carlotto1990MethodSearching, carlotto2007DetectingPatterns}, finding intriguing but non-conclusive features.
Similar efforts could certainly be refined and more systematically studied with modern machine learning capabilities.

\subsection{Asteroid Belt}
The asteroid belt has raw material resources \citep{papagiannis1978AreWe,papagiannis1983ImportanceExploring}, and would be a good place to hide for lurkers because it has a lower gravitational well compared to the one of planets, which means it would be less energy costly to go in and out.
The worry about collisions ingrained through popular culture is actually unfounded because the density is low, and collisions are very rare now \citep{bottke2015CollisionalEvolution}.
Polarimetry and spectroscopy instruments can detect metal surfaces that would stand out from natural objects that rarely produce polarization \citep{socas-navarro2021ConceptsFuture}.
It has been proposed to look for forensic evidence of extrasolar asteroid mining \citep{forgan2011ExtrasolarAsteroid}, but obviously it would make even more sense to start such a search in our own solar system first.

\subsection{Interstellar Objects}
Interstellar objects (ISOs) going through our solar system have attracted attention through the interpretation that ‘Oumuamua might be an interstellar spacecraft propelled by a light sail \citep{bialy2018CouldSolar, loeb2022PossibilityArtificial}.
The tumbling motion of ‘Oumuamua led also to the idea that it was a passive probe, such as a debris of a megastructures like a Dyson sphere \citep{loeb2023InterstellarObjects}.
This ETI interpretation led to heated controversies as there are also natural explanations to explain its phenomenology \citep{bannister2019NaturalHistory, katz2021OumuamuaNot}.
One can also see positive outcomes of such debates \citep{elvis2022ResearchProgrammes}, although the lack of additional data is a key issue to solve this controversy.
In the near future, the Vera Rubin Observatory will be able to detect many more ISOs \citep{ezell2023DetectionRate}, and the JWST could help to further characterize them.
A recent review of ISOs in relation to technosignatures has been conducted in \citet{davenport2025TechnosignatureSearches}.
The authors suggest to look for anomalous trajectories, spectra, color, and shapes, and recommend detailed observational strategies during the whole observable timing of the ISO (see table \ref{tab:Solar_System_TS_ISO_obs}).

\begin{table*}
\centering
\caption{Observing recommendations for ISO technosignatures. Note that the timing here is a suggestion for the most likely detection, and almost all proposed technosignatures benefit from additional data throughout the ISO passage. Table from \citet{davenport2025TechnosignatureSearches}.}
\label{tab:Solar_System_TS_ISO_obs}
\begin{tabular}{lll}
\hline
Observing Mode & Technosignature Class & Timing \\
\hline
Astrometry \& tracking & Accelerations & Full arc of passage \\
Full sky photometry & Early detection \& origin & Earliest possible \\
Optical/IR spectroscopy & Spectral anomalies, lasers & Closest approach \\
Radio spectroscopy & Transmissions & Closest approach \\
Multi-band photometry & Color \& phase curve anomalies & Full arc of passage \\
High-cadence imaging & Rotational modulation & Multi-night campaign \\
Infrared photometry & Waste heat & Pre- and post-perihelion \\
Polarimetry & Unusual surface properties & Closest approach \\
Radar imaging & Shape anomalies & When within radar range \\
\hline
\end{tabular}
\end{table*}

\cite{davenport2025TechnosignatureSearches} suggest synergies between ISO SETI searches with studies of surface composition, comet activity and acceleration, shape, and rotation.

The Nancy Grace Roman Space Telescope, scheduled to be launched in 2027 is poised to be an exceptional tool for discovering interstellar objects due to its unparalleled combination of a wide field of view, deep-field imaging capabilities, and high-resolution optics.
By having a population to work with instead of a single anomalous case, one will have more material to do science, including statistics.
The ultimate data gathering may be to send a rendezvous mission to intercept and study such ISOs \citep{siraj2023PhysicalConsiderations, hein2022InterstellarNow}.
The recent discovery of dark comets \citep{seligman2024TwoDistinct} shows a population of solar system objects that have non-gravitational accelerations.
\cite{loeb2025DarkComet} argues that one of them matches the Venera 2 spacecraft, which implies that differentiating between a natural population of “dark comets” and artificial spacecraft may be nuanced, an example they use to argue that perhaps some of them are ETI.
\begin{figure*}
    \centering
    \includegraphics[width=0.95\linewidth]{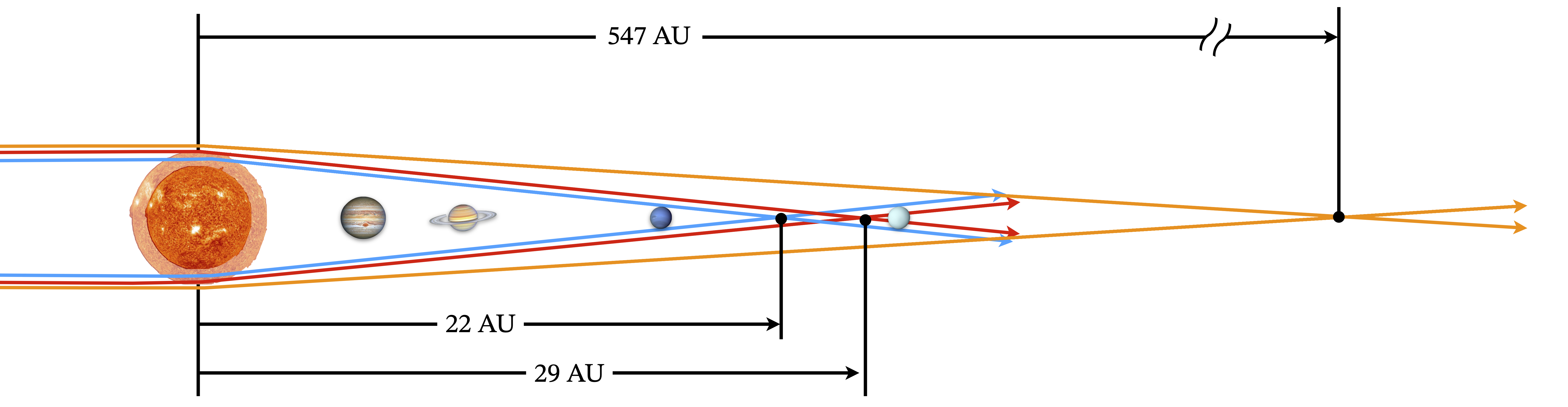}
    \caption{The Solar Gravitational Lens (SGL) is a region where gravitational and neutrino radiation starts to focus, respectively at 22.45 AUs (blue) and 29.6 AUs (red) while the focus of electromagnetic radiation starts from 547 AUs (orange). Fig. adapted from \citet[page xxxi]{maccone2009DeepSpace}. 
}
    \label{fig: SGL}
\end{figure*}
\subsection{Outer Solar System}
Recently, \cite{stanton2025UnexplainedStarlight} did find two fast optical pulses separated by 4.4s in the direction of HD 89389.
After ruling out other possibilities, Stanton suggested that the anomalous pulses might have been caused by an opaque solid ring-shaped object in our solar system, occulting the star sequentially.

\cite{matloff2004ProposedInfrared} proposed a search for artificial Kuiper belt objects, while \cite{loeb2012DetectionTechnique} showed that lights on Kuiper belt objects are in principle detectable with existing instruments.

Artificially illuminated objects might be detected in the outer solar system, but also on the surface of exoplanets \citep{loeb2012DetectionTechnique}.

\subsection{Solar Gravitational Lens}
\label{subsec:SGL}

As predicted by Einstein’s theory of relativity, massive objects such as our Sun bends and diffracts light waves from distant sources.
This produces a semi-infinite focal lines that extends beyond the minimal focal distance of 547 AU. 
Human or ETI probes, observational or transmitting, placed at these regions will benefit orders of magnitude in gains~\citep[see e.g.][]{maccone2013SunFocus,turyshev2024SearchGravitationally}.
Other modalities such as neutrinos and gravitational waves experience the same bending, each leading to a distinct focal distance (see Fig.\ref{fig: SGL}).
A radio search of SGL communication between our Sun and alpha centauri has been conducted \citep{tusay2022SearchRadio}, while \cite{kerby2021StellarGravitationalb} suggested that it would be a sound strategy to look for artifacts around the focal region.

\subsection{Oort Cloud}
The more we go towards the outskirts of our solar system, the more interstellar objects we can expect, and indeed interstellar objects outnumber solar system objects in the Oort cloud \citep{siraj2020CaseEarly}.
We can note that Oort clouds of different stellar systems are almost touching each other.
They might play the role of a boundary between stellar systems, where activities and exchanges occur, much as in a cell membrane, or as human activities mostly happen on the surface of planet Earth, at the interface of the planet and space.
From this viewpoint, they are important search targets for technosignatures, at the interface between our solar system and other stars.
\cite{romanovskaya2022MigratingExtraterrestrial} proposed that civilizations may be exiting their Oort cloud for migration purposes, and this might lead to signs of ejection, or associated communicative or heat technosignatures.

 To sum up, many present and future observational facilities such as the Vera Rubin Observatory or the Nancy Grace Roman Space Telescope as well as planned missions in the solar systems that together with machine learning techniques have the potential to enable a much more systematic and comprehensive search for anomalies and possibly artefacts in our own solar system (see also \citealt{haqq-misra2022OpportunitiesTechnosignature}).
\clearpage
\section{Exoplanetary Technosignatures}\label{sec: planet ts}
\subsection{Classification Schemes for Advanced Civilizations}
\subsubsection{The Extended Kardashev Scale}
As we shift our search for technosignatures beyond our home solar system, it is prudent to take a general perspective before looking at details.
From such a lofty perspective, energy utilization is perhaps the most important driving force behind civilization's technological development.
The greater that consumption, the greater the effect a civilization has on its environment, and the more possibilities there are for observing it.
\citet{kardashev1964TransmissionInformation} introduced his eponymous scale, outlining three main levels of civilizations, both in terms of spatial scale and energy (see table \ref{tab: kardashev scale table}).
An extension to the universal scale was proposed by \citet{gray2020ExtendedKardashev}, and a review made by \citet{cirkovic2015KardashevsClassification}.

\begin{table*}
    \centering
    \begin{tabular}{cccp{7cm}}
         Kardashev Index & { Spatial Extent} & { Power Usage} & { Technological Level} \\
        \hline\hline 
        0 & Planetary & $10^{-20}L_\odot\approx10^6$W & Continent spanning pre-Industrial to planet spanning industrial, using chemical energy sources \\
        I & Planetary & $10^{-10}L_\odot\approx10^{16}$W&  Planet spanning industrial with space-faring abilities for solar system exploration. Chemical, solar, and nuclear energy sources.\\
        II & Solar System& \hspace{20pt}$1 L_\odot\approx10^{26}$W  & Mega-structure engineering capacity and/or interstellar space-travel capacity. Extensive solar energy production.\\
        III & Interstellar & \hspace{6pt}$10^{10}L_\odot\approx10^{36}$W  & Mega-structure mass production capacity, exotic energy production capabilities\\
        IV & Intergalactic & \hspace{6pt}$10^{20}L_\odot\approx10^{46}$W  & Multi-galaxy spanning, with essentially omnipotent technology
    \end{tabular}
    \caption{The Kardashev scale describes potential civilizations at different spatial and energetic levels. Kardashev articulated the Types I, II and III, which we extend here one order of magnitude lower (Type 0), and higher (Type IV). }
    \label{tab: kardashev scale table}   
\end{table*}

Consider a civilization extracting energy from its environment at a rate of $dE/dt$.
Its Kardashev index $K$ is
\begin{equation}
K=2+\frac{1}{10}\log_{10}\frac{1}{L_\odot}\frac{dE}{dt}
\end{equation}
where $L_{\odot}$ is a solar luminosity.
A civilization that extracts energy at a rate equivalent to that produced by our Sun has a $K=2$, and is classified as a Type-II on the Kardashev scale.
Next, imagine a civilization that utilizes all the solar energy impinging on its planet.
If the planet has a radius r and an orbital radius R, then the civilization’s power extraction is $r^2/4 R^2$ times its stars luminosity, $L_\star$, giving

\begin{equation}
    K=2+\frac{1}{10}\log_{10}\left(\frac{r}{2R}\right)^2\frac{L_\star}{L_{\odot}}
\end{equation}

If human civilization reached this level of technological ability, it would have $K=1.066$, and be classified as a Type-I civilization.
We discuss technosignatures generated by such Type-I civilizations in this section, and of Type-II in section \ref{sec: stellar ts}.

The jump from a Type-I to a Type-II civilization requires the scaling up of energy harvesting technology from a planetary to a stellar scale, which would include a transition to the capability to harness resources at a stellar system scale.
Moving up another notch to Type-III requires further scaling to some 10 billion Sun-like stars, or a sizable portion of the output of an average spiral galaxy.
The impact of such an advanced civilization on its home galaxy could be observable, and are discussed in section \ref{sec: galactic ts}.
Moving up to cosmological scales, a civilization that has extended its reach across manifold galaxies is Type-IV (with some examples discussed in section \ref{subsec: Universal}).

\begin{figure}
    \centering
    \includegraphics[width=1\linewidth]{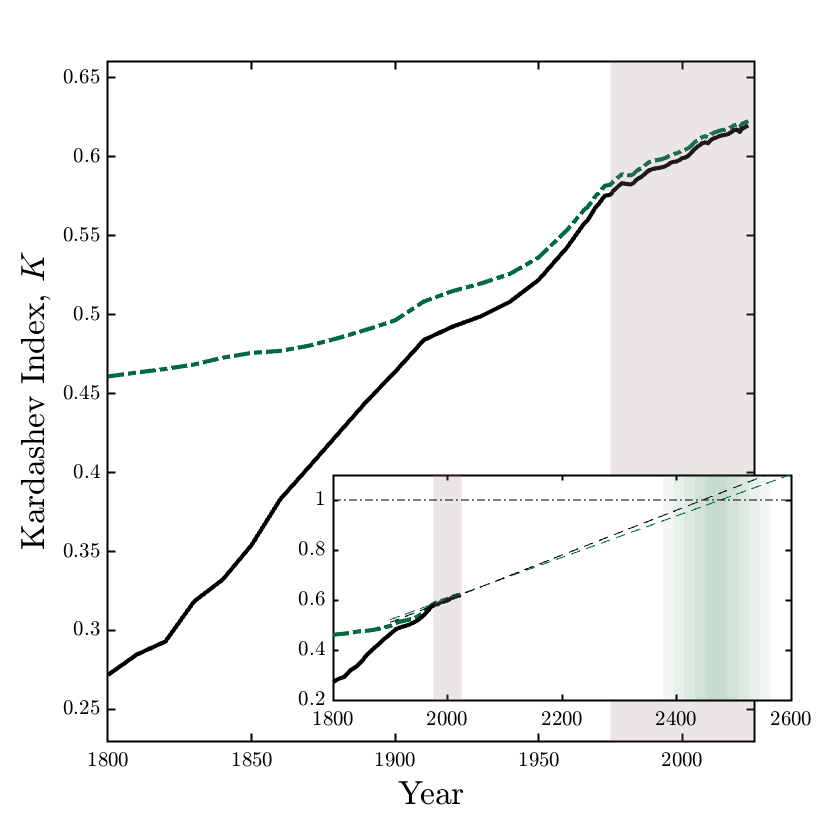}
    \caption{Total human energy use through time on the Kardashev scale. Depending on the definition of energy usage we have a lower-bound (black line) and an upper bound which adds the agricultural burning of biomass (green dot-dashed line).}
    \label{fig: exoplanetary_kardashev}
\end{figure}

To put things in perspective, in 2021 the total human energy usage from fossil fuels, hydroelectric, solar, and nuclear was estimated at 19.6 terawatts, or about $5.12\times 10^{-14}L_\odot$.
This puts human civilization at $K\approx 0.67$.
Slight variations from the literature on the definition of $K$ result in indices ranging from 0.55 to 0.76, classifying modern human civilization between a Type-0 and Type-I.
Fig. \ref{fig: exoplanetary_kardashev} shows the upper and lower bounds on the Kardashev index of human civilization since the 1800s \citep[data from ][]{ritchie2020EnergyProduction}.
The inset projects the rate attained since 1975 into the future, implying that at the current level of growth, humanity will achieve Type-I status around year 2500.
Projections for growth to Type-II and beyond are more speculative.

The remainder of the paper follows this spatial scale, and we will review possible technosignatures on planets, planetary systems, and stars up to the universal scale.
The Kardashev Type of civilizations engaging in producing interstellar signals (section \ref{sec: interstellar ts}) or sending probes would not be obvious to assess unless we know more about the originating civilization.

\subsubsection{The Barrow Scale}
\label{sec: Barrow Scale}
\begin{figure*}
    \centering
    \includegraphics[width=0.85\linewidth]{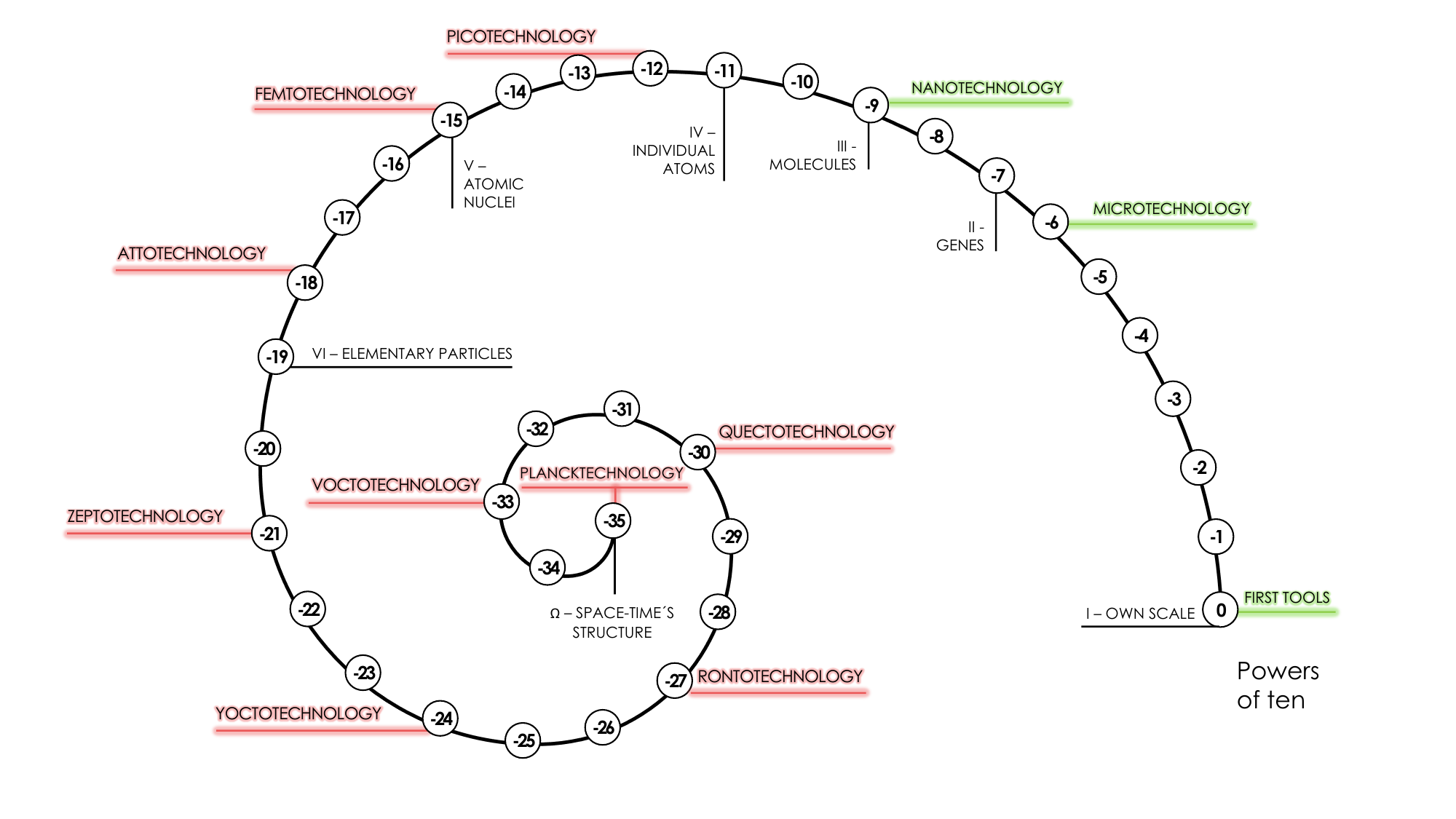}
    \caption{The Barrow scale highlights the ability to manipulate and control from our own scale (1m) to the minimal length-scale ($10^{-35}$ m) . Humanity has created many tools at our own human scale down to the nanotechnological realm  (green), but has almost no mastery below (red). \citet{barrow1998ImpossibilityLimits} gave examples of types: Type I-minus: The ability to manipulate objects on a human scale, such as building structures and mining, Type II-minus: The capacity to control and engineer genetics and living organisms; Type III-minus: The mastery of molecular and chemical bonds to create new materials; Type IV-minus: The ability to manipulate individual atoms, marking the advent of true nanotechnology; Type V-minus: The power to engineer atomic nuclei and their constituent nucleons; Type VI-minus: The capacity to manipulate elementary particles (quarks and leptons); Type Omega-minus ($\Omega$-minus): The theoretical apex, representing the ability to manipulate the fundamental structure of spacetime.}
    \label{fig: Barrow Scale}
\end{figure*}

The Barrow scale (Fig. \ref{fig: Barrow Scale}), proposed by cosmologist John D. \citet{barrow1998ImpossibilityLimits}, offers another compelling metric for classifying the technological level of a civilization.
In contrast to the Kardashev scale's focus on macro-level energy consumption, the Barrow scale measures a society's advancement in its ability to manipulate matter and environment on increasingly smaller physical scales, a concept we could call ``micro-dimensional mastery''.
This framework suggests that true technological sophistication lies not necessarily in harnessing stellar energies, but in the precise control over the building blocks of the universe.
The scale is delineated by ``minus'' types, signifying this inward focus (Fig. \ref{fig: Barrow Scale}).

The Barrow scale posits that the ultimate form of technological evolution may be the conquest of the small, achieving profound results with maximum efficiency and fundamental control.
Such a trajectory is also called the \textit{compression hypothesis}, often contrasted with the \textit{expansion hypothesis} \citep{last2017BigHistorical, smart2012TranscensionHypothesisa}.

However, manipulating small scales requires also a lot of energy ---think for example of the energy required for particle accelerators to break atoms--- so a civilization may not have to choose between developing along the Kardashev or the Barrow scale, but may instead co-develop along both axes \citep{vidal2016StellivoreExtraterrestrials}.
Contrary to intuition, more energy efficiency also leads to more energy usage, a paradox articulated by \citet{jevons1866CoalQuestion}, and known to economists as a kind of \textit{rebound effect} \citep{herring2008EnergyEfficiency}.

Let us now turn to the search for exoplanetary technosignatures that aims to identify signs of advanced technological civilizations on exoplanets.
This endeavor synergizes closely with exoplanet detection and characterization, as well as the study of debris disks, and can often be conducted commensally with other astronomical observations.
There are three main categories of planetary technosignatures: surface, atmospheric and orbital.

\subsection{Surface Technosignatures}
Surface technosignatures are features on a planet's surface indicative of technological activity.
Given the distances of exoplanets, detecting surface technosignatures is much more challenging than in our solar system, or than atmospheric signatures.
It was first proposed by archeologist John B. \citet{campbell2006ArchaeologyDirect}.
Agricultural activity has existed on our planet for over 10,000 years, so taking the Earth as a model, the observational window is arguably one of the most enduring human activities, and it has been proposed to look for proxies of agricultural societies, by looking at urban fires, landscape burning, deforestation, desertification due to overgrazing, sea canals, seasonal agriculture practices or irrigation \citep{lockley2021DetectionPreindustriala}.
The widespread use of solar panels would create a distinct spectral signature due to their high reflectivity in specific wavelengths, as explored by \citet{kopparapu2024DetectabilitySolar}.
Observations would remain challenging, as it would require hundreds of observation hours with an HWO-like observatory, assuming a 23\% land coverage with solar panels.

The surface composition of exoplanets affects the polarization phase curves due to the change in multiple scattering \citep{singla2023EffectMultiple}.
If there are artificial surfaces which may be composed of cement, plastic, or cloth, the change in surface albedo will be reflected in the polarization phase curves, so polarization studies will prove very useful to study surfaces.

Another compelling surface technosignature is artificial illumination, such as city lights on the night side of a planet \citep{beatty2022DetectabilityNightside}.
If a civilization routinely lights its habitats, cities, or transportation networks, such light could, in principle, be detected from interstellar distances.
The Earth's night side emits visible light from urban areas, observable from orbit.
A sufficiently luminous and widespread lighting network on an exoplanet could similarly produce detectable optical signatures.
The unique spectral energy distribution (spectral signature) of artificial lights differs markedly from natural sources.
These spectra differ from broadband thermal emission of planetary surfaces or atmospheres.

Furthermore, lighting is likely to be stable and predictable.
For example, on tidally locked planets orbiting M-dwarf stars, the constant night side of the planet offers optimal contrast and predictable location.

The JWST and two future space-based telescopes, HWO and LIFE, are capable of detecting night-time lights on nearby (30 ly) exoplanets.
For example, a study was done by \citet{tabor2021DetectabilityArtificial} on the detectability of artificial lights from Proxima b. The terrestrial planet is likely to be tidally locked to its M dwarf host star owing to its proximity to the star, which means that if it is inhabited, the night side would likely be illuminated.
The study simulated lightcurves from Proxima b to compare natural reflected starlight with artificial LED-like illumination.
It found that JWST could detect artificial lights if they are 500 times stronger than Earth's current LEDs, or if Earth-level illumination is emitted in a spectrum 1000 times narrower.

Nonetheless, the use of artificial lights as a technosignature is based on an anthropocentric assumption of the behavior of an alien civilization, and for example a different biological or technological evolution generalizing the use of NIR would render massive illumination unnecessary.

More generally, large-scale artificial structures, such as cities or industrial complexes, could also alter a planet's albedo or thermal profile, generating an urban heat island effect \citep{kuhn2015GlobalWarming}, i.e. a localized warming due to activity or the use of darker-colored materials in construction.
As an illustration, generating surface maps in different colors for the nearest exoplanet, Proxima b, may be achievable with reasonable ground-based telescopes/interferometers \citep{berdyugina2019SurfaceImaging}.
However, the observation remains challenging and requires high resolution, very large aperture telescopes, and a SGL telescope may be needed here too.

Exoplanet Surface Technosignatures could also benefit from a Solar Gravitational Lens (SGL) telescope (see section \ref{subsec:SGL}).
To characterize the planet, the spacecraft would move laterally to scan different portions of the lensed image, reconstructing a complete image pixel by pixel \citep{turyshev2020ImageFormation, turyshev2026DirectHighresolution}.
By performing spectroscopy on the light from individual pixels, scientists could analyze the composition of specific surface areas and overlying atmospheric regions, offering an unparalleled method for studying an exoplanet's global environment.

A 224 km resolution would allow the detection of massive, artificially created megastructures.
The key would be to look for anomalies within a pixel that are inconsistent with natural geology or vegetation.
For example, immense fields of solar panels covering thousands of square kilometers would have a distinct spectral signature.
They would be much darker and have lower reflectivity (albedo) than most natural surfaces like deserts, forests, or oceans, and could also create a thermal anomaly.
The SGL could even differentiate between bio- and techno- signatures: agriculture on a continental scale would create unnaturally uniform geometric patterns or vegetation types with a spectral signature different from native plant life, detectable across multiple adjacent pixels.

Furthermore, a sprawling network of cities or a single ecumenopolis (a planet-wide city, see ~\citep{doxiadis1962EcumenopolisUniversal}) would have a unique thermal and spectral signature.
At night, the collective heat from energy use and city lights could make a $224\times224$ km$^2$ area appear anomalously warm in the infrared spectrum.
During the day, the composite reflectivity of artificial materials like concrete and asphalt would differ from natural terrain.

\subsection{Atmospheric Technosignatures}
Advances in exoplanet detection methods, such as transit photometry, radial velocity, and direct imaging have paved the way for the identification of potentially habitable worlds.
Atmospheric spectroscopy offers a powerful tool for detecting both bio- and techno- signatures, which means that searches can be commensal for both sides, on the condition that the signature appears on the same wavelength range.

Ideally, technosignature gases should have a long residence time in the atmosphere and, if the purpose is to terraform, a very strong Global Warming Potential (GWP).
If we could detect an ozone layer that is seriously depleting over time, that would also be a strong lead of the presence of activity from a civilization.
Key candidates for technosignature gases include potent greenhouse gases such as sulfur hexafluoride SF$_6$, carbon tetrafluoride CF$_4$, and nitrogen trifluoride NF$_3$, which have high GWP and long atmospheric lifetimes, making them detectable across interstellar distances (Table \ref{tab: exoplanetary_species}).

\begin{table*}[htbp]
\centering
\caption{Atmospheric concentrations, Global warming potentials (GWPs), and lifetimes of selected species.}
\label{tab: exoplanetary_species}
\renewcommand{\arraystretch}{1.3}
\begin{tabular}{l c c c p{6cm}}
\hline
\textbf{Species} & \textbf{Concentration (ppt)} & \textbf{GWP (100 yr)} & \textbf{Lifetime (yr)} & \textbf{Sources} \\
\hline
CF\textsubscript{4} & $\sim$85 & 7380 & 50000 &
\citep{ipcc2014IPCCData, agage2025AGAGEAdvanced, droste2020TrendsEmissions}. Note $\sim$95~ppt in 2050. \\

SF\textsubscript{6} & 12.2 & 25200 & $\sim\!1000$ &
\citep{ipcc2014IPCCData, noaaglobalmonitoringlaboratory2025NOAAGlobal, thoning2022TrendsGloballyaveraged}. Note $\sim$22--24~ppt in 2050. \\

SF\textsubscript{5}CF\textsubscript{3} & 0.15--0.16 & 17700 & 800 &
\citep{sturges2012EmissionsHalted, ipcc2014IPCCData}. \\

NF\textsubscript{3} & 2.5 & 17400 & $>$500 &
\citep{weiss2008NitrogenTrifluoride, ipcc2014IPCCData}. Note $\sim$8--10~ppt in 2050. \\
\hline
\end{tabular}
\end{table*}

Other potential technosignatures include industrial pollutants \citep{dubey2025PolycyclicAromatic} and alterations to the nitrogen cycle caused by agricultural activity \citep{haqq-misra2022DisruptionPlanetary}.
Detecting these requires advanced spectroscopic techniques and high-resolution observations, achievable with next-generation telescopes like the JWST or future possibilities such as a SGL telescope.

However, distinguishing technosignatures from biological or abiotic processes poses significant challenges, as many gases, such as methane (CH$_4$) and oxygen (O$_2$), can be produced by both biological and technological activity.
Chlorofluorocarbons (CFCs), hydrochlorofluorocarbons (HCFCs), and perfluorocarbons (PFCs) are synthetic chemicals that are not naturally occurring, primarily used as refrigerants, having the potential to accumulate in planetary atmospheres \citep{haqq-misra2022DetectabilityChlorofluorocarbons}.
Sulfur hexafluoride (SF$_6$), Tetrafluoromethane (CF$_4$), and Trifluoromethyl Sulfur Pentafluoride (SF$_5$CF$_3$) are highly stable greenhouse gases used in various industrial applications \citep{schwieterman2024ArtificialGreenhouse}.
Nitrogen trifluoride (NF$_3$) also represents a promising technosignature gas, since it is not known to be produced naturally by any fluorine based minerals, photochemical, volcanic or other geological processes \citep{seager2023FullyFluorinated, petkowski2024ReasonsWhy}.

\begin{figure*}
    \centering
    \includegraphics[width=0.75\linewidth]{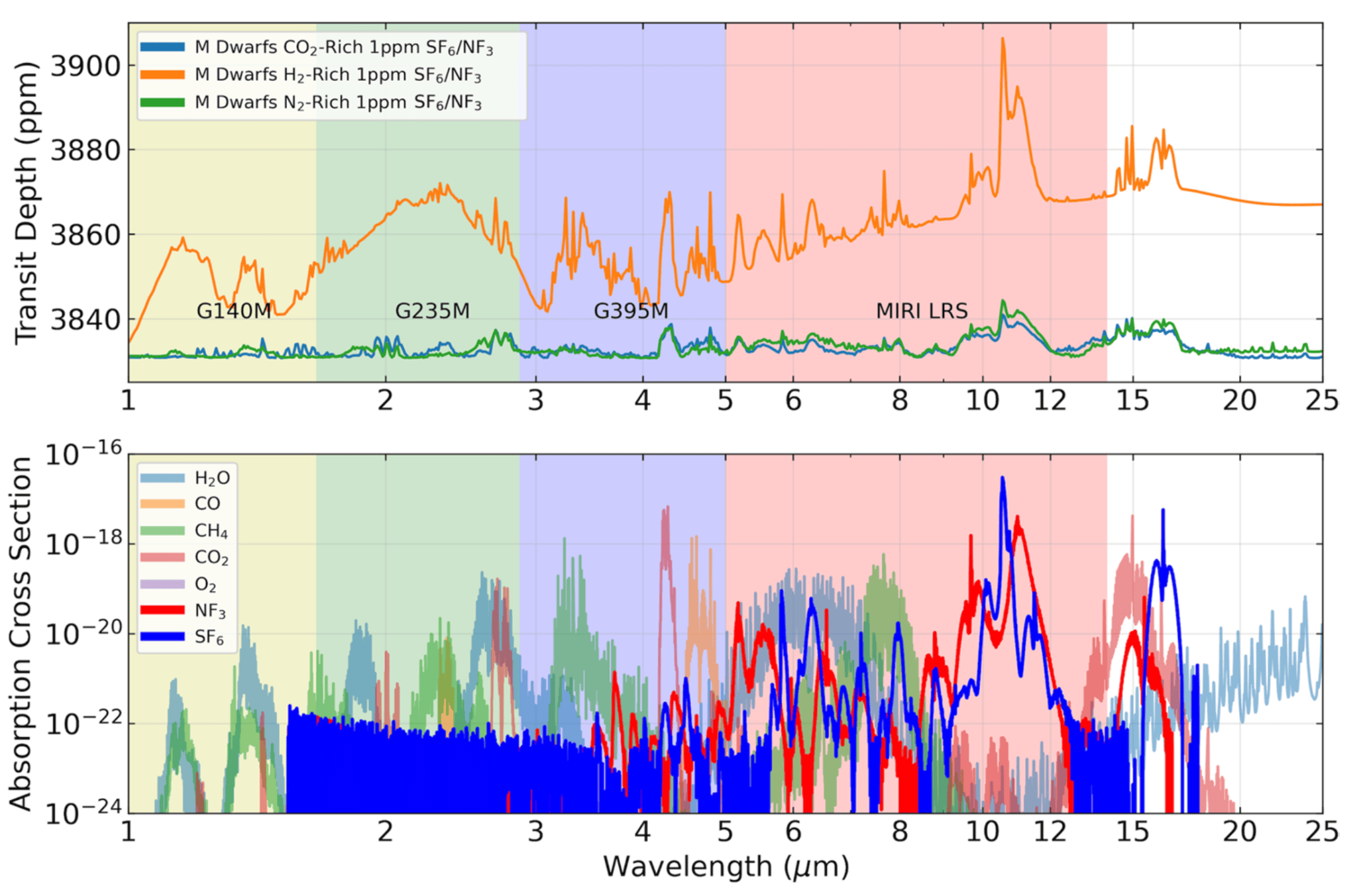}
    \caption{(Top) A simulated transit depth spectrum showing SF$_6$ and NF$_3$ in the atmosphere of terrestrial exoplanets of different atmospheric compositions orbiting an M Dwarf star. The concentrations of SF$_6$ and NF$_3$ are at 1 ppm. (Bottom) Absorption cross sections for various atmospheric molecules, including the industrial produced SF$_6$ and NF$_3$. It is noted that concentrations of 1 ppm for both SF$_6$ and NF$_3$ are orders of magnitude above that of present-day Earth’s abundances of these two industrial gases. The high concentrations of SF$_6$ and NF$_3$ could exist in so-called “service worlds". Fig. from \citet{seager2023FullyFluorinated}. }
    \label{fig: ExoPlanetaryTS_SF6_NF3}
\end{figure*}

\citet{seager2023FullyFluorinated} made a strong case to focus future atmospheric technosignature search efforts on SF$_6$, CF$_4$, and NF$_3$ (Fig. \ref{fig: ExoPlanetaryTS_SF6_NF3}).
The planned Large Interferometer For Exoplanets (LIFE) space observatory is a prime instrument to investigate these \citep{glauser2024LargeInterferometer}.
A search for these gases could be carried out by focusing on their fundamental vibrational modes within the wavelength range of 7.5 to 16 microns.

Nitrogen dioxide (NO$_2$) is also a good potential technosignature candidate because it is produced primarily by industrial activities, combustion processes (e.g., fossil fuel burning), and vehicle emissions \citep{kopparapu2021NitrogenDioxide, schwieterman2024ArtificialGreenhouse}.
On Earth, concentrations vary drastically from as low as 20 parts per trillion (ppt) in remote areas to over 45 parts per billion (ppb) in industrialized regions, highlighting its strong connection to technological pollution.
Natural processes such as volcanic activity or lightning also produce NO$_2$ but are more transient compared to a sustained industrial output.
To confidently claim NO$_2$ as a sign of technology on an exoplanet, scientists would need to detect it at concentrations that are significantly higher than what could be produced naturally.
The most effective method of detection is through visible reflectance spectroscopy; NO$_2$ absorbs specific wavelengths, particularly in the blue part of the spectrum between $400$ and $500$ nm. This absorption, and others in IR, creates a unique spectral fingerprint that could be detected on nearby exoplanets using the JWST and the future Habitable Worlds Observatory (HWO).

Trichlorofluoromethane (CCl$_3$F) and carbon tetrafluoride (CF$_4$) are also excellent technosignature candidates \citep{lin2014DetectingIndustrial}, as they are produced primarily from industrial processes, such as aluminum smelting and semiconductor manufacturing.
The low background concentration of CF$_4$ from volcanic activity or other geological processes would not be detectable at interstellar distance.
Thus, a detection of CF$_4$ based on its strong infrared absorption bands at $\sim$9.2~\textmu m~($\nu_1$) and $\sim$11.8~\textmu m~($\nu_4$) would most likely imply an industrial origin.
CCl$_3$F is entirely a synthetically produced gas on Earth, and is widely used historically in refrigeration, aerosols, and air conditioning.
There are no known natural processes (i.e., false positives) capable of producing CCl$_3$F, meaning that its detection via IR spectroscopy (e.g., $\sim11-13\mu$m) in the atmosphere of a terrestrial planet would strongly imply an artificial origin.

Another multi-gas technosignature are byproducts of incomplete combustion of fossil fuels called polycyclic aromatic hydrocarbons (PAH) \citep{dubey2025PolycyclicAromatic}.
Detecting large amounts of these would indicate a technological civilization at the same level as Earth's Industrial revolution.
The authors conclude that current observational abilities will not be sensitive enough to make such detections, though future observatories beyond the HWO may.

Other potential technosignature gases targets include "chemical fossils" left by past or present technology.
PFCs, such as carbon tetrafluoride (CF$_4$) and hexafluoroethane (C$_2$F$_6$), are purely artificial compounds with atmospheric lifetimes of thousands to tens of thousands of years.
Their extreme stability means they could serve as an extinct technosignature, a long-term record of industrial activity persisting long after a civilization might have vanished or transitioned to cleaner technology.

Reversing the observational logic as a proof of concept, \citet{lustig-yaeger2023EarthTransiting} considered Earth as another transiting exoplanet, and showed that JWST could detect an Earth-like planet both in terms of its biosignatures and its industrial pollutant technosignature.

However, significant detection challenges remain as the absorption signatures of some of these artificially produced gases can be difficult to discern in the mid-IR due to other strong features from water vapor and CO$_2$, and hydrocarbons.
Ideally, a large ($>15$ m) space-based telescope equipped with a high resolution mid-IR spectrometer would be needed that also integrate a stellar contamination control.
To mitigate these challenges, future space-based telescopes will need to observe in the least-confused spectral windows of specific technosignature gases, observe the spectrum in thermal emission mode, since this mode weights the vertical column differently than a limb transmission spectrum and can place the SF$_6$/NF$_3$/CH$_4$ bands on a planetary continuum instead of a stellar one.
Around $10-11\mu$m, CO$_2$ is less interfering than at $15\mu$m, and a sharp SF$_6$ Q-branch or NF$_3$ band can imprint on the N-band (centered around $10\mu$m) continuum.
Furthermore, post-processing techniques on observed thermal-IR spectrum (e.g. spectral deconvolution and/or derivative spectroscopy) can mitigate the challenges of detecting technosignature gas molecules in the atmospheres of terrestrial exoplanets.

An \textit{extant technosphere} refers to the physical and environmental traces of a technological civilization that persist even if the civilization itself is no longer active or has undergone significant transformation.
It encompasses industrial byproducts, infrastructure, atmospheric changes, and geological or isotopic anomalies.
As an example of isotopic anomalies, long-lived radioactive isotopes (e.g., plutonium-244) or anomalously low deuterium/hydrogen (D/H) ratios in planetary water could indicate the use of nuclear technologies like fission or fusion \citep{catling2025PotentialTechnosignature}.

Surface artifacts could also reveal evidence of megastructures, mining, or even disasters related to self-replicating nanotechnology (e.g. a "grey goo" scenario \citealt{stevens2016ObservationalSignatures}).
Stevens also argues that signs of photochemical byproducts from nuclear warfare such as gamma-rays may persist long after a civilization's demise.
Additionally, mass extinction events triggered by technological activity might leave behind decay products like methanethiol (CH$_3$SH) or other sulfur compounds \citep{stevens2016ObservationalSignatures}.

Of course, any of these atmospheric signatures would need to be complemented with a contextual analysis, in order to avoid false positives (see also section \ref{sec:discussion}).
As the discovery of several biosignature gases would strengthen the case for a biosphere, the presence of multiple industrial gases co-occurring would be a strong indication of a technosphere.

Temporal variations in potential technosignature gases would also be particularly indicative.
For example, following the example of humanity’s use of CFCs \citep{newman2009WhatWould}, one might expect to see an increase in the concentration of a given industrial product, followed by a relatively rapid decrease, indicating the initiation of a planetary regulation, or management of waste products \citep[see also][]{frank2022IntelligencePlanetary}.

As a side note, many radio telescopes are searching for auroral radio emission from exoplanets \citep{zarka2015MagnetosphericRadio, griessmeier2018SearchRadio, turner2021SearchRadio}: LOFAR, OVRO-LWA, NenuFAR, SKA.
There is thus a synergy possible between radio SETI in interstellar communication (section \ref{sec: interstellar ts}) and exoplanet studies: The collected data can be used both for studying auroral radio emission and for SETI search.
Indeed, all these telescopes have either intentional or piggyback SETI receivers.

\citet{fisher2023AtmosphericChemical} and \cite{fisher2023ComplexSystems} used complex systems modelling to distinguish between abiotic, biotic and anomalous sources of atmospheric gases.
The results of these works suggest that the impact of industrial pollutants on the topology of the atmospheric chemical reaction network could be a viable means of detection for exoplanets, regardless of the underlying atmospheric chemistry.
Their analysis focused on networks leading to the productions of CH$_4$ and CFC-12, but could also be applied to other technosignature gases identified above.
Other similar network-based studies looking for non-random, biology-like networks are also relevant for this endeavor \citep{wong2023NetworkBasedPlanetary}.

\subsection{Orbital Technosignatures}

The characterization of debris disks around stars gives insights into planetary system formation and dynamics, which may also inform the likelihood of technosignatures.
Indeed, unusual structures in debris disks, such as asymmetric features, could even hint at the presence of artificial megastructures or industrial activity \citep{jaiswal2023SpecularReflectionsa}.
This search for artificial structures or activities in orbit is also known as ‘Clarke exobelts’ \citep{socas-navarro2018PossiblePhotometric, sallmen2019ImprovedAnalysis}.
Since stellar evolution is highly predictable, and as the star heats up, it will change the habitable zones of planets.
One way to control the stellar flux is to place a \textit{starshade} at the L1 Lagrange point that could control how much light the planet receives \citep{gaidos2017TransitDetection, skoglund2025StarshadesTechnosignatures}.

\citet{lacki2019ShinyNew} has studied the signature, geometry and detectability of glints of light coming from artificial satellites.
Since we have not only satellites, but also mega-constellations of satellites, \citet{osmanov2021SpaceXStarlink} studied how an ETI equivalent of these could be detected with the Very Large Telescope Interferometer (VLTI).
In the early stages of stellar system formation, planetary accretion disks form.
If a system is not young (i.e. post zero-age main sequence) and still has a disk-like signature, this might be a lead to inquire whether the disk is artificial in nature \citep{socas-navarro2018PossiblePhotometric}.
Interestingly, the duration of such a technosignature could be long, as even non-active artificial debris disks could be long-lived and thus detectable during a long time window.

Solar power satellites in geocentric orbit would give power all day and night, in contrast to solar panels on the surface that give power only during daylight \citep{ellery2016SolarPower}.
The technosignature would be a strong beaming of energy from ‘exogeocentric’ orbit to the exoplanet.
As the energy demand grows, the solar-powered satellite band would grow too, making it more amenable to observations.
The limit being a kind of planetary Dyson sphere that would give a similar 10$\mu$m waste heat signature predicted for stellar Dyson spheres.

To sum up, atmospheric technosignatures—such as synthetic greenhouse gases and industrial pollutants—are the most accessible with current and near-future technologies.
Surface technosignatures, including nightside city lights and urban heat islands, offer additional, though technically demanding, potential evidence of technological activity.
Orbital technosignatures remain the hardest to detect due to their faint nature and the high resolution required to resolve orbital artifacts.

If a civilization has a strong impact in transforming its home planet, it seems logical and likely that over time its reach will broaden to having an impact on other planetary bodies in its planetary system.

\subsection{Exoplanetary System Technosignatures}

The search for exoplanet technosignatures should not necessarily be limited to planets within the star's habitable zone.
For example, water can also exist in micro-habitable zones such as those around gas giants where tidal friction induces geological activity in nearby moons.

Some advanced extraterrestrial civilizations may choose to use other terrestrial planets inside or outside of the habitable zone as \textit{Service world} briefly mentioned by \citet{lingam2023TechnosignaturesFrameworks} and defined by ~\citet[9-7]{wright2026SearchExtraterrestrial} as:
\begin{quote}
A planet with an extensive technosphere but no (or practically no) biosphere, for instance because it is the site of an interstellar settlement but has not been terraformed.
If interstellar settlement is common, most sites of technology might be on service worlds.
\end{quote}

These planets could be used solely for industrial processing, or they may be terraformed to make their environments suitable for habitability.
Such worlds may show abundances of industrial or greenhouse gases orders of magnitude greater than on present day earth.
In other words, a highly polluted and unlivable world by our standards might be a technosignature candidate.

The technosignature search strategies identified for our solar system (see section \ref{sec: ss ts}) also apply to other planetary systems, although the interstellar distances bring considerable challenges and re-assessments.
\citet{forgan2011ExtrasolarAsteroid} has proposed to look for active mining of many asteroids that would generate localized infrared excess.

Given a planetary system with a narrow habitable zone (M-K-G star), several terrestrial planets could be found in horseshoe constellations, forming a ring of planets \citep{raymond2023ConstellationsCoorbital}, detectable through a study of transit-timing variations.
Although by many orders of magnitude not energy efficient, a planetary system engineering civilization might also engineer the dynamics of the planets to form a beacon, producing mathematical sequences that would be highly unlikely to occur naturally \citep{clement2022MathematicalEncoding}.
\citet{ponnaganti2024MakingHabitable} proposed other examples of what they called strange exoplanetary architectures, and suggested that relocating planets using high-power lasers is feasible.

\subsection{Multiplanetary Systems and Terraforming}

A highly advanced terrestrial ET civilization may terraform the atmospheres of rocky planets in their planetary system to make them suitable to their form of life.
Terraforming is the mega engineering process of modifying a planet's atmosphere, temperature, surface topography, and ecology to be similar to Earth's environment, making it habitable for life forms that would otherwise not be able to survive there.
A typical example are explorations and models to terraform Mars \citep[see e.g.][]{beech2022TerraformingMars, ansari2024FeasibilityKeeping}.
To test this, one may try to observe infrared signatures of greenhouse gases with very high planetary warming potentials and long residence times in rocky terrestrial exoplanets.
The growth of an oxygen-rich atmosphere, or an adjustment of atmospheric pressure may all be indicative of such engineering.
The discovery of multiple exoplanets sharing very similar atmospheric compositions could be a potential indication of terraforming by an advanced extraterrestrial civilization \citep{romanovskaya2022MigratingExtraterrestrial, romanovskaya2023PlanetaryBiotechnospheres, tarter2024SearchingExtraterrestrial}.

This possibility suggests that the search for alien life should not be rigidly confined to the traditionally defined habitable zone of planetary systems.
An ETI might choose to terraform planets outside the habitable zone for strategic reasons, such as resource availability, lower risk from stellar events (for planets around volatile stars), or simply because they have the technology to make any suitable planetary body habitable.

The point of these wider considerations is that examining anomalous structural, dynamical and energetic patterns at the level of a planetary system may reveal such activities from ETIs.
Such feats bring us closer to consider even more ambitious stellar technosignatures.

\clearpage
\section{Stellar Technosignatures}\label{sec: stellar ts}
The idea that an ETI could become capable of significantly using energy from its home star, or even modify stars for altering stellar properties such as lifespan, luminosity, or mass composition has had various terminologies.
For example, “astroengineering” was used in the early days of the SETI (e.g. \citealt{dyson1973AstroengineeringActivity}) while \citealt{criswell1985SolarSystem} uses “stellar husbandry” or “star-lifting”, \citet{matloff2019StellarEngineering} wrote a book called “stellar engineering”, and \citet{beech2008RejuvenatingSun} speaks of stellar “rejuvenation”.
We recommend simply using “stellar technosignatures” as a general umbrella to denote these closely related notions.

\subsection{Stellar Megastructures}
The search for extraterrestrial transiting megastructures (e.g., Dyson swarms, or artificial structures designed to message) has the advantage that it can make use of well-established techniques that monitor the dimming of a star’s flux when an opaque object passes in front of it \citep{arnold2005TransitLightCurve, arnold2013TransmittingSignals, wright2016SearchExtraterrestrial}.
A transit light curve with an unusual or complex structure (e.g., asymmetries, irregular, jagged, non-uniform) may be indicative of artificial structures compared with the relatively smooth U-shaped transit curves produced by exoplanet transits (Fig. \ref{fig:ExoPlanetaryTS_transits}).
Indeed, a megastructure might cause transit depths that change or vary over time due to the modification (e.g. construction) of the structure, moving components of the structure, or an engineered design that varies in opacity.

\begin{figure}
    \centering
    \includegraphics[width=1\linewidth]{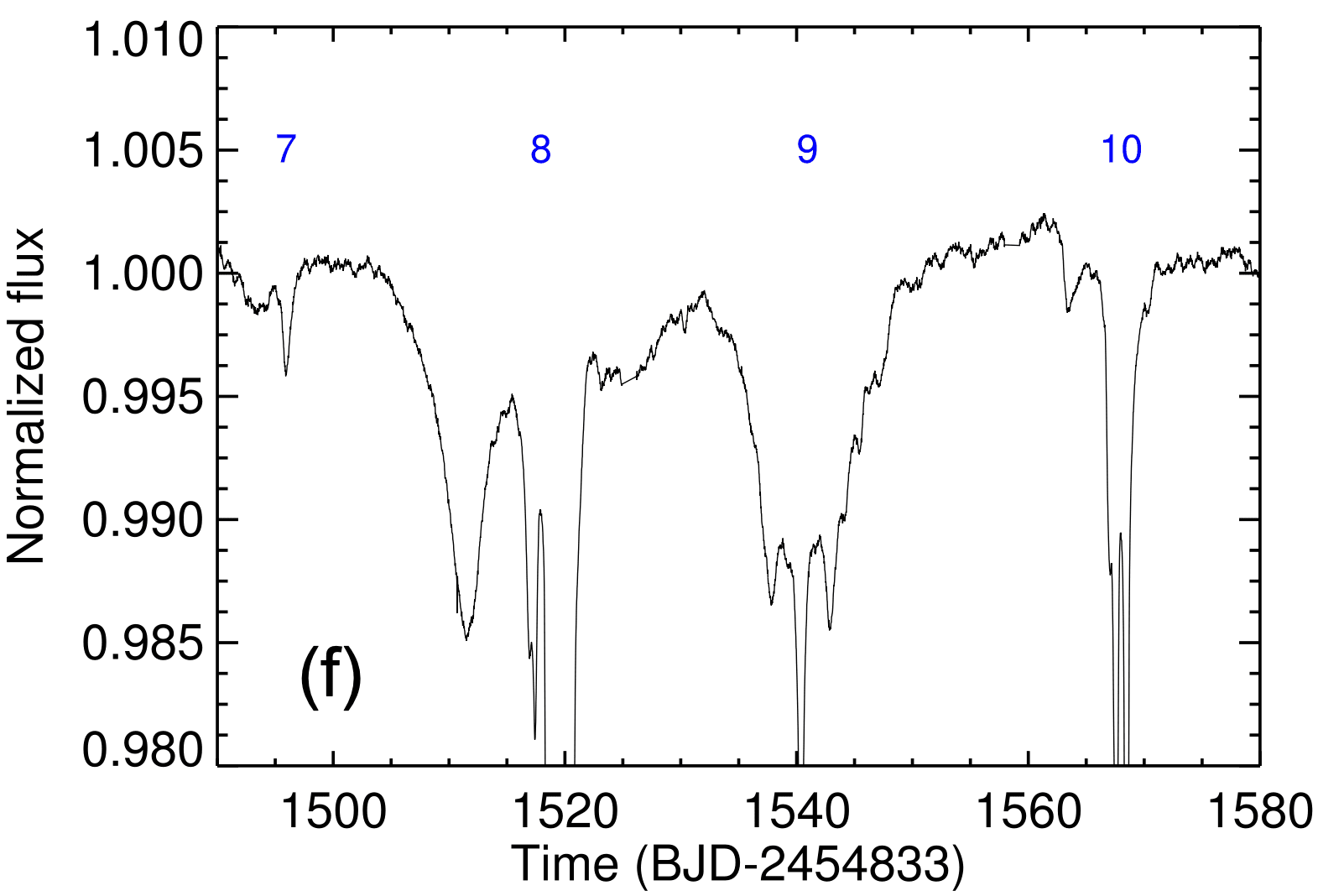}
    \caption{Kepler observations of Boyajian’s star (KIC 8462852) dips. Dip numbers (in blue) correspond to 4 of the 10 discrete dips analyzed in \citep{boyajian2016PlanetHunters}.}
    \label{fig: Boyajian}
\end{figure}

\begin{figure*}
    \centering
    \begin{minipage}{0.98\linewidth}
        \centering
        \includegraphics[width=\linewidth]{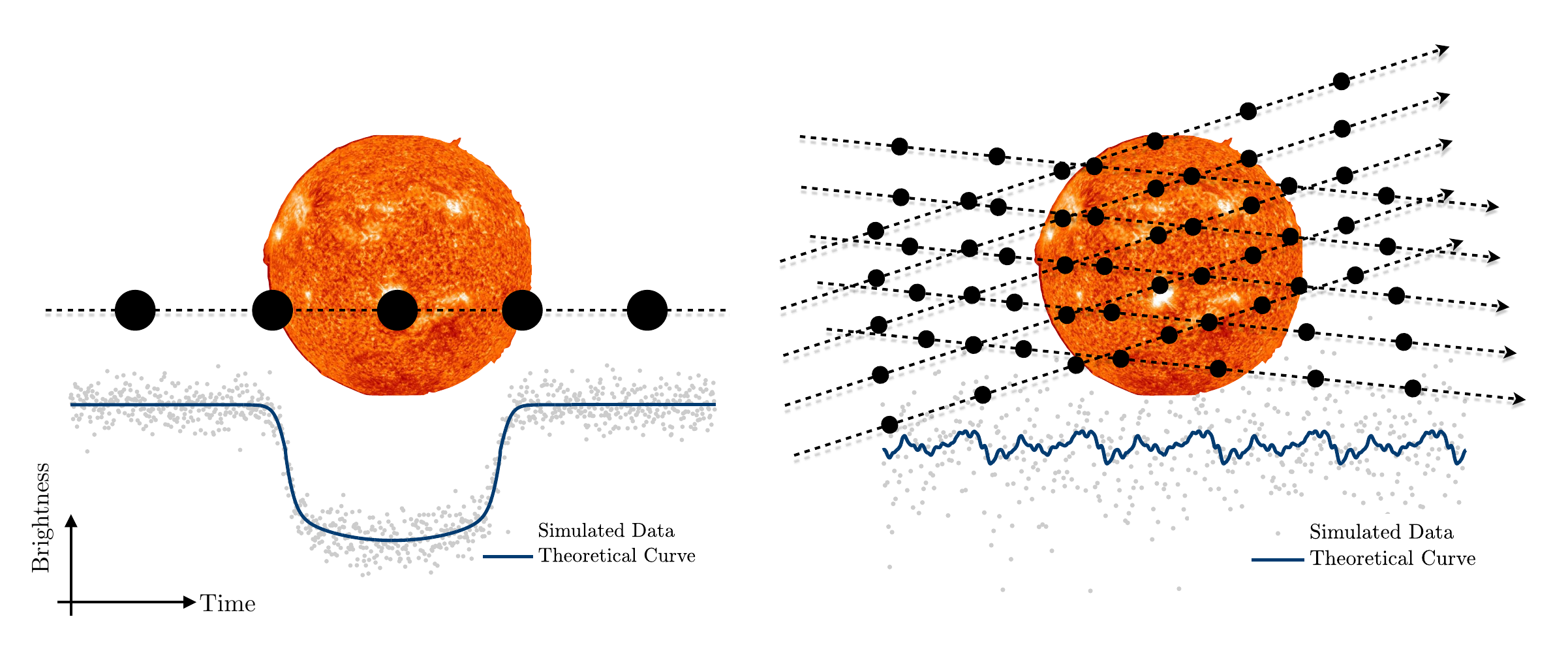}
    \end{minipage}
    \caption{The left panel displays a simulated transit light curve and a scaled visual inset of a natural, Earth-like exoplanet passing in front of a sun-like star, causing a brief dip in brightness. In contrast, the right panel illustrates a highly complex light curve produced by an artificial Dyson swarm as its various panels orbit and cross the star's disk. Comparing these two scenarios highlights how the intricate shape of a light curve can distinguish artificial cosmic structures from natural planets, though real-world observational noise will add further complexity to both signals.}
    \label{fig:ExoPlanetaryTS_transits}
\end{figure*}

Another strategy to look for candidate megastructures near a star is to use space-based transit observatories like the former Kepler mission or the currently operating TESS mission, looking for anomalous light curve signatures during transits.
For example, the highly unusual light curve of Boyajian’s star (Fig. \ref{fig: Boyajian}) has been intensely discussed as a candidate megastructure, although it is almost certainly dust \citep{boyajian2016PlanetHunters, wright2016FamiliesPlausible, ksanfomality2017HeritageKepler, lipman2019BreakthroughListen, schmidt2022SearchAnalogs, wright2018ReassessmentFamilies, wright2016SearchExtraterrestrial}.

Machine learning techniques are already used to search light curve data for signs of such unusual patterns \citep[for early results, see][]{tong2023RandomForest, giles2024AnomalyDetection}.

In the case of a Dyson swarm consisting of thousands of individually collecting “statites”, an advanced Type-II ETI would likely choose to live outside the radius of a Dyson swarm they constructed around their star (Fig. \ref{fig:ExoPlanetaryTS_transits}, right).
Indeed, the swarm would generate enormous amounts of waste heat, which could make inner planets uninhabitable.
Outside the swarm, the civilization would enjoy a cooler and more stable environment.
One can imagine that it could construct habitats on mined-out comets in the Oort Cloud or building free-floating, spin-gravity habitats that orbit the star beyond the swarm.
Such a civilization could wirelessly beam the immense energy collected by the swarm to power habitats, colonies, and projects throughout their star system.
The construction of the swarm would require a highly advanced interstellar mining and manufacturing infrastructure, likely involving the disassembly of inner planets, as originally proposed by \citet{dyson1966SearchExtraterrestrial}.
This stellar-system development would also serve as a staging ground for interstellar expansion, with the swarm's energy being able to power massive ships for traveling towards other star systems or building even more advanced megastructures.
If the civilization were to transition to a postbiological existence, its computational activities could be housed in a megascale computer, or "Matrioshka brain" \citep{bradbury2001MatrioshkaBrains}.
The point of highlighting these admittedly speculative possibilities is that the whole stellar system and the home star may be affected by intelligent life, and it makes sense to keep them in mind as a cluster of technosignatures.

\subsection{Energy Utilization}
The first proposal to harness the whole radiating energy of a star was famously articulated by \citet{dyson1960SearchArtificial, dyson1966SearchExtraterrestrial}.
Dyson initially spoke of a “spherical shell” and although it was later called a “Dyson sphere”, Dyson was aware that a solid sphere would be very unstable.
In an issue of the \textit{Science} magazine, Dyson replied to early critiques by Maddox, Anderson, and Sloane, and wrote that he was more thinking of a swarm of solar collectors in orbit around a star rather than a solid structure \citep{dyson1960ResponseArtificial}.
For historical and comprehensive reviews about Dyson spheres, see \cite{bradbury2001DysonShells} and \cite{wright2020DysonSpheres}.
The core point is that whatever the specific construction, such a project would leave an observable signature of waste heat.
Searches for such waste heat signatures have been conducted in the infrared, with the IRAS telescope \citep{jugaku2004SearchDyson, carrigan2009IRASBasedWholeSky}.
\cite{zackrisson2018SETIGaia} combined data from Gaia DR2, 2MASS, and WISE to perform an even more refined search.
Out of 5 million sources, \cite{suazo2024ProjectHephaistos, suazo2024SearchingDyson, suazo2022ProjectHephaistos, suazo2024ProjectHephaistosa} identified seven M-dwarfs exhibiting anomalous infrared excess, which could be interpreted as partially completed Dyson spheres.
These candidates deserve further analysis, although false positives could be circumstellar dust emission (which was also noted by \citealt{sagan1966InfraredDetectability}).
Another possibility is that while the Dyson sphere would capture back the energy, it would create a feedback process that would create a hot sphere \citep{osmanov2018PossibilityDyson,huston2022EvolutionaryObservational}.

Machine learning methods are allowing to detect more and more systematically such anomalous transits \citep{giles2019SystematicSerendipity,martinez-galarza2021MethodFinding}.
Simulations of transits \citep{sandford2019ShadowImaging, wheeler2019WeirdDetector} enable the exploration of potential shapes of megastructures and their light curves.
The use of Dyson spheres need not be limited to simply collecting energy, and speculation has been made for their use in communication and directed energy weapons.
Gerard Nordley proposed their use as a large baseline array, and speculated that they might be used as weapons that could destroy planets \cite[cited in][]{shostak2020SETIArgument}.
\subsection{Modification}
The first proposal to modify a star was made by Hubert \citet[120-122]{reeves1985AtomsSilence}, who argued that a way to delay the predictable red-giant phase of the Sun would be to pump the unburned hydrogen that sits between the core and the surface, thereby mixing fresh hydrogen into the core, extending ts lifetime by an order of magnitude.
He poetically proposed that humanity’s nuclear arsenal could be used for this purpose.
\citet{criswell1985SolarSystem} made a fascinating study on how the solar system could be industrialized, articulating in more detail how to prolong the lifetime of stars.
In a book-length essay, Martin \citet{beech2008RejuvenatingSun} proposed many observable signatures of stars being rejuvenated.
In particular, \citet{beech1990BlueStragglers,beech2008RejuvenatingSun} proposed an original hypothesis to puzzles related to the composition of blue stragglers, namely as being due to stellar engineering.
Recently, \cite{scoggins2023LazarusStars} made a contribution to the topic of stellar rejuvenation and developed analytical models that lead them to propose observational signatures.
\cite{huston2022EvolutionaryObservational} also proposed a model where a Dyson sphere with high reflectivity could extend the stellar lifetime of that star.

Besides energy utilization and stellar rejuvenation, another explorative engineering consists in modifying a star to control its motion in the galaxy by transforming it into a stellar engine (see section \ref{subsec:stellar_engines}).

\subsection{Stellar Pollution}
\begin{figure}
    \centering
    \includegraphics[width=1\linewidth]{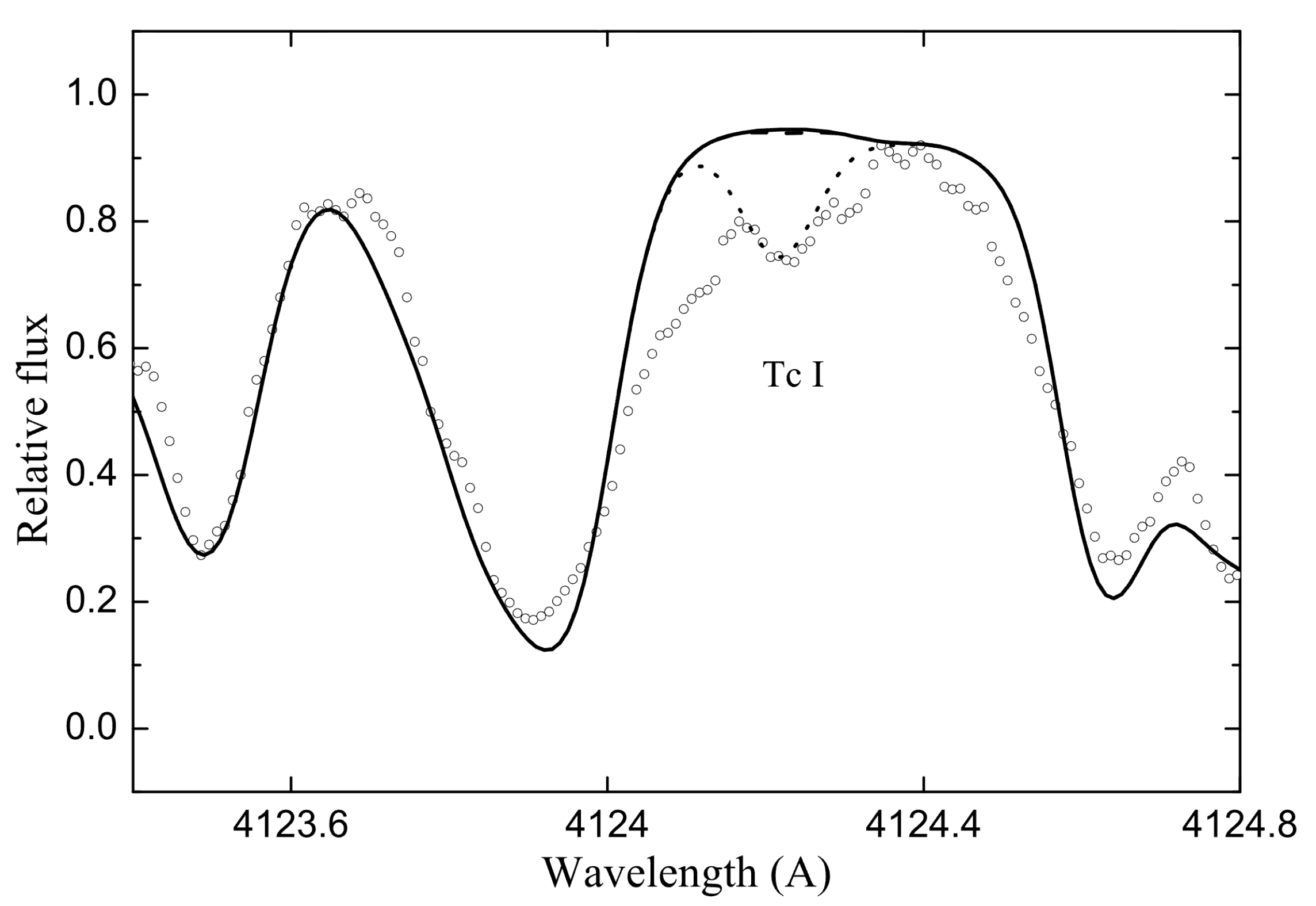}
    \caption{Original caption from Fig.2 in \citep{andrievsky2023AbundanceRadioactive}: ``Fragment of the observed spectrum of Przybylski’s star (open circles), covering the blend 4124.22 \AA. Synthetic spectra with technetium abundances $\log(\text{Tc}/\text{H})\approx4$ (dotted line), and with $\log(\text{Tc}/\text{H}) = 2.5$ (dashed line) are shown. Continuous line – no technetium in the atmosphere''.}
    \label{fig:StellarTS_Przybylsky}
\end{figure}

A Type II civilization may also end up polluting its home star, much as a Type I civilization pollutes its home planet.
Various authors proposed that a Type II civilization may dump radioactive waste into a star \citep{whitmire1980NuclearWaste, carrigan2010StarryMessages, tarter2024SearchingExtraterrestrial}, while \citet{stevens2016ObservationalSignatures} noted that high level nuclear fission waste products would remain in the photosphere of a star, and would be detectable in its stellar spectrum.
A candidate is Przybylski’s \citeyearpar{przybylski1961HD101065a} star (see Fig. \ref{fig:StellarTS_Przybylsky}), which has anomalies and peculiarities that might be consistent with such stellar pollution (\citealt{tarter2024SearchingExtraterrestrial}, see also \citealt{wright2017PrzybylskisStar}).

\subsection{Compact Objects}
Despite the fact that white dwarfs, neutron stars and black holes are often called the “stellar graveyard”, one can argue that these are attractive sites for advanced civilizations.
The extreme density affords immense computational capacity \citep{lloyd2000UltimatePhysical}; these compact objects are extremely long-lived compared to other stars, neutron stars can easily beam electromagnetic radiation throughout the galaxy; and black holes have many peculiarities that may make them attractors for intelligence \citep{vidal2011BlackHoles}.
The mastery of such exotic substrates is also consistent with a civilization climbing down the Barrow scale, being able to manipulate matter, energy and information at ever smaller and denser scales (see above).

\cite{matloff1986WorldShips} pointed out that the great energy released during the red-giant phase could provide a significant boost for any civilization attempting to migrate using solar sails.
He also noted that the remaining white dwarf would provide a much better gravitational assist than a main sequence star like our Sun.
\citet{gertz2019TheresNo} also made the case to look for technosignatures around white dwarfs, as it would be a site where a long-lived civilization would have survived the stellar evolution of its home star.
\citet{dyson1963GravitationalMachines} also analyzed the immense potential of compact objects for gravitational assists, discussing both the art of doing gravity assists around white dwarf and neutron star binaries.
He noted that if they were used as such, they should give off observable gravitational waves.
Building on Dyson’s work, \citet{kipping2018HaloDrive} proposed a propulsion model extracting kinetic energy from a black hole, while avoiding the strong gravitational tidal effects.

The notion of habitable zone has been extended to a narrow region very nearby white dwarfs \citep{agol2011TransitSurveys}.
In the context of technosignatures, the concept of Dyson spheres has also been expanded to apply to white dwarfs \citep{semiz2015DysonSpheres}, with observable infrared and optical signatures described in \citet{zuckerman2022InfraredOpticala}.
Dyson spheres might also be constructed around pulsars, leaving an infrared signature \citep{osmanov2016SearchArtificial}; or around black holes, which would be detectable in optical, UV and infrared \citep{hsiao2021DysonSphere}.
\citet{opatrny2017LifeBlack} also explored the possibility of a Dyson sphere that would absorb energy from the cosmic background radiation while dumping waste energy to a black hole.
Dyson spheres may even be detectable as a megastructure in X-ray binaries \citep{imara2018SearchingExoplanets}.
\citet{chennamangalam2015JumpingEnergetics} proposed a model where the pulsar radiation would be modulated by an artificial satellite, for the purpose of galactic-wide communication.
Sagan also proposed to analyze pulsars’ pulses for messages or traces of artificiality (in \citealt{dyson1973AstroengineeringActivity}).
The reasoning behind these proposals is that it is much easier to modulate an existing powerful natural beacon like a pulsar, than to build one from scratch.
A Dyson ring could modulate a pulsar’s equatorial emission that would give a distinct technosignature \citep{haliki2019BroadcastNetwork, kayali2025SearchDyson}.

The exceptional physics of black holes has led to various speculations on how advanced civilizations might use them.
For example, \citet{loeb2024IlluminationPlanet} proposed that a rogue planet might embark a “black hole Moon” as an energy source, and that, when used, the accretion from the planet to the Moon should give a particular gamma-ray signature.
The concept of “mini-Earths” was introduced in \citet{vukotic2022HabitableMiniEarths} and \citet{cirkovic2023NoteVukoticGordon}, defined as small rocky planets with small black holes inserted into their centers, allowing atmospheres to be retained.
Discovering such objects would be a strong suggestion of engineering.
Closer to stellar engineering, \citet{beech2008RejuvenatingSun} proposed a model where an artificial black hole would be inserted into a star to mix its layers and prolong its lifetime.
He noted that it would not be so profitable for a Sun-like star, but would be particularly advantageous for a star more than $2 M_\odot$.
Roger \citet[270--272]{penrose1969GravitationalCollapse} imagined a method to extract energy from a rotating (Kerr) black hole.
It consists of injecting matter into a black hole in a carefully chosen way, thereby extracting its rotational energy \citep[see also][908 for more details]{misner1973Gravitation}.
A signature here would be a black hole whose rotational rate is slowing down anomalously, i.e. a sudden change in angular momentum.
\citet{blandford1977ElectromagneticExtraction} suggested a similar process with electrically charged and rotating black holes.
Other proposals suggest collecting energy from gravitational waves of colliding black holes.
Misner imagined this in 1968 as a personal communication to Penrose (1969).
\citet{frautschi1982EntropyExpanding} also proposed that ETI could be merging black holes as a way to produce a power source.
\citet[pg.~197-198]{chaisson1988RelativelySpeaking} and \citet[pg.~370]{crane2010PossibleImplications} have suggested that black holes are the perfect waste-disposal device.
Since black holes are made of the densest state of matter, an advanced civilization could exploit them for quantum computing, and \citet{dvali2023BlackHoles} suggested that the signature of such artificial black holes would be a black hole of a mass of $10^9$g, for which there is no natural formation scenario.
Anticipating section \ref{sec: galactic ts}, supermassive black holes (SMBHs) may also be industrial or computational centers of choice for Type II-III civilizations \citep{dvali2023BlackHoles, inoue2011TypeIII}.

It may be counter-intuitive to even consider life or technology around the extreme physics surrounding these compact objects, but it remains plausible in the context of future extrapolations of life, if we argue that life could transition into a postbiological form that has completely different viability constraints \citep{dick2003CulturalEvolutiona, dick2008PostbiologicalUniverse, dick2009BringingCulture, cirkovic2006GalacticGradients}.

There are two main approaches when discussing such exotic lifeforms: first, assuming that all life follows a common developmental pattern, and thus that life would start as-we-know it before bifurcating towards other substrates, becoming life as-we-don't-know it.
Alternatively, there might be many developmental webs of life, and the one we know from Earth’s evolutionary history is just one particular instance.
In that case, one must consider that self-organization and complexification might happen through different pathways than life-as-we-know-it.
For example, \citet{anchordoqui2020CanSelfReplicating} did propose an early model of nuclear life in the interior of a star.
\citet{maldonado-lang2025EmergentComplexity} used the formalism of chemical organization theory to explore the potential of nuclear reactions to build complexity.

The topic of stellar consciousness was also explored by \citet{matloff2016StarlightStarbright, matloff2017StellarConsciousness}.
He proposes a test via a “volitional star hypothesis” to explain the Parenago discontinuity ---the observation that cooler, less massive, redder stars revolve around the center of the Milky Way galaxy a bit faster than hotter, more massive and bluer stars.
However, this interpretation competes with existing astrophysical models of age-velocity dispersion relations, or asymmetric drift.

\citet{vidal2014BeginningEnd, vidal2016StellivoreExtraterrestrials} and \citet{haqq-misra2025ProjectionsEarthsa} proposed to reinterpret some accreting binaries as Type-II civilizations feeding on stars, or \textit{stellivores}.
Indeed, their configuration satisfies the universal metabolic nature of living systems: the extraction of energy from the environment (here, the companion star), and the ejection of waste and entropy (in the form of jets or novas).
Tests still need to be developed to generate predictions and retrodictions that contrast with the standard astrophysical models that already explain many observations of these binary systems.

We have zoomed out from planets, to stellar systems, and stars, and we now reach the interstellar scale, which is the focal point of much past and present SETI searches.
\clearpage
\section{Interstellar Technosignatures}\label{sec: interstellar ts}

The search for interstellar communications marked the birth of modern SETI.
We structure the progression of our review with the following key questions: What communication theory do we rely on?
Should we attempt to Message to Extraterrestrial Intelligence (METI)?
What are the first principles and fundamental constraints of interstellar communication?
Then we consider questions related to the search space itself, such as where in physical space, when in time, where in the EM spectrum, and other fundamental strategies to look for signals.

\subsection{Communication theory}
\label{subsec:communication_theory}

\begin{figure*}
    \centering
    \includegraphics[width=0.6\linewidth]{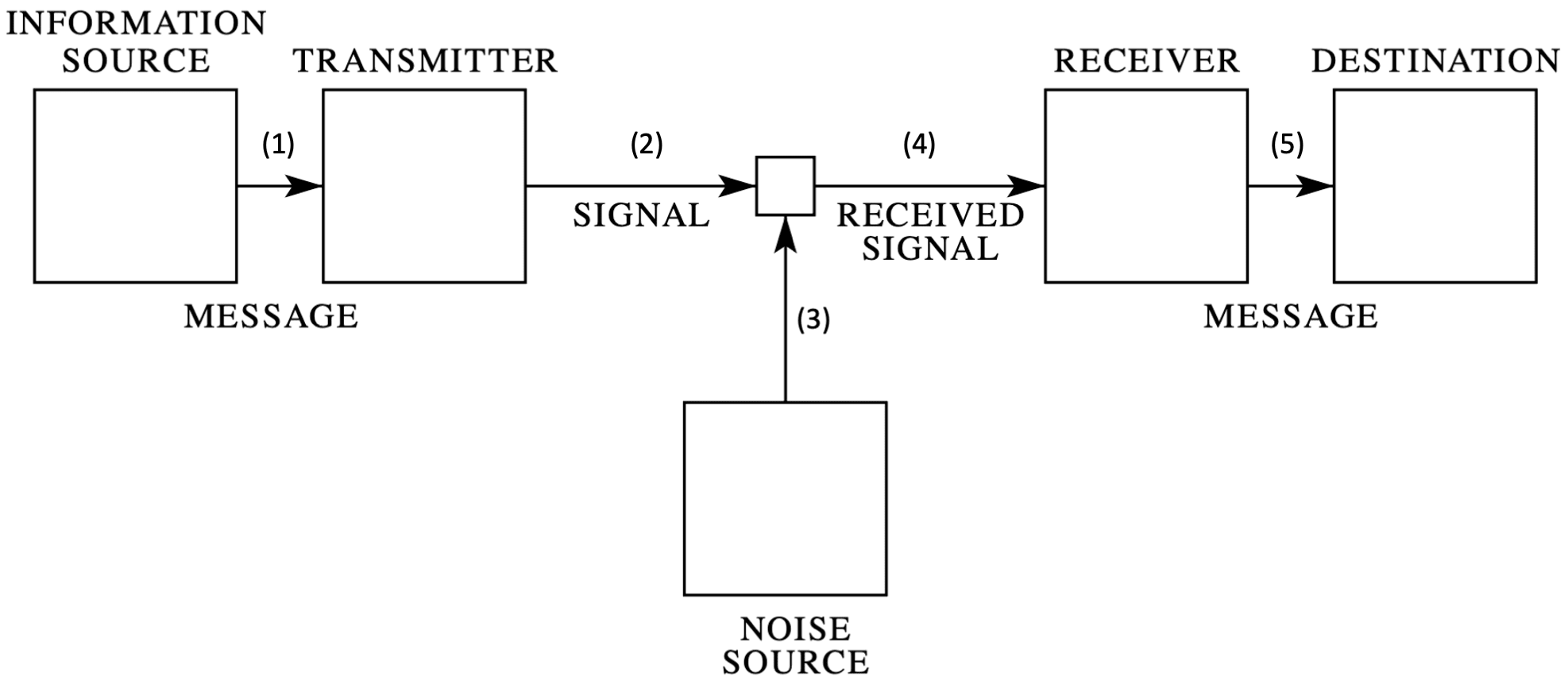}
    \caption{Schematic diagram of a general communication system after \citet{shannon1948MathematicalTheory}. }
    \label{fig:InterstellarTS_shannon}
\end{figure*}

Implicitly in the field of SETI, when we speak about communication, we mean cybernetic communication, i.e.\ the technical challenges related to information theory and the hardware implementation.
However, this is just one perspective on communication, as communication theory expert \citet{craig1999CommunicationTheory} distinguished seven families of communication theories: rhetorical, semiotic, phenomenological, cybernetic, sociopsychological, sociocultural, and critical.
We will stay with the scientific and technological tradition of communication from a cybernetic perspective, but we should remain aware of these multiple perspectives and disciplines---most of them from the humanities---having all to contribute to get a full picture of communication theory.
The most fundamental model is the one of \citet{shannon1948MathematicalTheory}, reproduced in Fig. ~\ref{fig:InterstellarTS_shannon}.

In the interstellar communication context, the left part of the Shannon model is the domain of Messaging to Extraterrestrial Intelligence (METI), while the right part is the central concern of SETI.
One first starts with an information source that one wants to transmit.
It needs to be encoded---arrow (1)---possibly with compression and/or encryption.
Then a transmitter can actually transmit a signal with a physical carrier (2).
Shannon's theory derives what the maximum rate of information transmission is, given the noise source (3), also known as the channel capacity.
More specifically, Shannon's noisy channel theorem says that whatever the level of noise is, there exists an encoding to transmit the message essentially error-free.
The receiver receives the signal (4), and decodes it (5).
SETI is primarily concerned with step (4), and not the decoding step (5).

Indeed step (5) is about decoding and interpreting a message, which is yet another chapter of SETI, sometimes called xenolinguistics \citep{vakoch2023XenolinguisticsScience}.
It is extremely challenging too, as when analysing signals that could have been generated by extraterrestrial civilisations, we do not know a priori what kind of encoding, encryption or compression scheme was employed, nor do we know whether the signal encodes a time series, an image, a 3D projection, or some other data object.
Algorithmic complexity theory has been shown to be a powerful tool in such a scenario.
It was found that local dips in algorithmic complexity correspond to the correct dimensionality, aspect ratio or partition for signals, encoded images and even 3D projections \citep{zenil2025OptimalUniversal}.
The authors simulated in particular the scenario of receiving the Arecibo message with no prior knowledge about the content or properties of the message encoded in the signal, and yet were able to infer the correct 2D nature of the message agnostically.

From this perspective, it is useful to conceptualize SETI and METI together.
SETI focuses on us as the receiver who attempts to de-encapsulate energy into meaning.
METI is the reverse exercise that encapsulates intention and meaning into physical energy.
This encapsulation and de-encapsulation process has been further standardized to clarify and help the development of Earth communication through the Open Systems Interconnection (OSI) standard, composed of seven layers: physical, data link, network, transport, session, presentation, and application \citep{ISO_IEC_7498_1_1994}.
However, we should note that METI is controversial.

\subsection{Should we attempt to Message ETI?}
\label{subsec:meti}

METI is the concept of intentionally signaling potential extraterrestrial onlookers via targeted EM transmissions, in hopes of a response.
Because of its proactive nature, it was also referred to as ``active SETI.''

\citet{vakoch2011AsymmetryActive} argues that we should transmit messages because potential ET onlookers may not wish to transmit unless we attempt an explicit contact first.
In other words, the strategy is about triggering potential lurkers in our solar system, to probe the zoo hypothesis \citep{ball1973ZooHypothesis}.
However, the same reasoning applies to attempts to trigger a response outside of the solar system.
The hope is to get a reply after at least $2x$ years, where $x$ is the distance of the target in light years.

METI transmissions, though few, have been ongoing since the first transmission to Venus in 1962 \citep{radulescu2025SovietInterstellar}.
METI is controversial for several reasons.
Most obviously, transmitting intentionally would expose our location to potential ET onlookers.
We have no idea of their intent, whether benevolent or malevolent.
A worst-case scenario would be that a malevolent ETI discovers us thanks to METI transmissions and then harms or destroys life on Earth, e.g.\ via ``deadly probes'' \citep{brin1983GreatSilence}.

Indeed, in the setting of sequential and incomplete information games, game theory \citep{yasser2020AliensFermi, fudenberg1983SequentialBargaining} tells us that an advanced civilization's best strategy is to remain undetected and therefore safe from destruction by a more capable civilization.
Alternative strategies are: (1) to signal their presence, exposing them to potential attack or benign communication; or (2) to remain silent while listening for other civilizations.
In game theory, under the assumption that the negative potential for complete annihilation outweighs any potential benefit from satisfying curiosity or collaboration, the Nash equilibrium implies that it is advantageous to immediately attack any source of a detected signal while remaining silent.
This gloomy conclusion is also known as the ``Dark Forest'' \citep{yu2015DarkForest} after the famous science fiction novel by Liu \citep{liu2015DarkForest}.

Another fundamental issue is when METI proponents transmit without having the consent of the majority of humans to craft and send a message that represents humanity, and by definition without being able to assume the responsibility especially because it could be relevant only in hundreds or thousands of years.

Currently there is no policy guiding or limiting METI transmissions, but efforts to address our more general cosmic footprint are arising \citep{normier2025CallAddressa}.
For more in-depth discussions about the METI debate, see \citet{haramia2026EthicsExtraterrestrial}, \citet{brin2019BarnDoor}, and \citet{haramia2019ImperativeDevelop}.

METI is not to be confused with Earth's unintentional EM leakage, which is not targeted to specific stellar systems.
We could monitor targets that have intercepted our leakage radiation.
For example, the deep space network radar signatures leak preferentially in the ecliptic, so we could narrow-down in space and time the targets that could respond to such leakage.
Inversely, \citet{fan2025DetectingExtraterrestrial} have analyzed how we might detect radar leakage radiation from ETI.

\subsection{First principles and fundamental constraints for interstellar communication}
\label{subsec:first_principles}

\subsubsection{The tyranny of the inverse square law}
\label{subsubsec:inverse_square}

The primary hurdle of interstellar communication is the natural weakening of the intensity of electromagnetic signals as they travel across vast interstellar distances.
This intensity falls very quickly, with the square of the distance ($1/r^2$).
This is simply due to the geometry of our universe that has 3 dimensions of space.

The inverse square law puts interstellar communication in a situation somehow analogous to the tyranny of the rocket equation where propellant mass explodes with a required $\Delta v$ (see section \ref{subsec:energy_sources}).
The common solution in the case of rockets is to launch multi-staged rockets, where the mass of the fuel tanks are disposed of.

In the case of the inverse square law, a rough analog to multi-stage rockets is to use amplifying signal relays so that the carrying of the signal is distributed.
For example, if one wants to send a signal across a distance $r$, the power required scales with $r^2$.
One relay at mid-length cuts the energy requirement in 2; 9 relays require only $1/10$th of the energy.
Of course, the physical relays need to be established first, which is no obvious feat and requires serious capacity for interstellar travel.

The tyranny of the inverse square law has strong implications for choosing a communication solution.
Some ways to mitigate it are, for example, on the sender side, to use concentrated power such as lasers, instead of omnidirectional transmissions.
Contrary to popular imagination, even lasers are cones growing with distance, not perfect cylinders.
Note that this choice saves energy but makes the problem of pointing to a desired target more sensitive.
On the receiver side, having bigger, more sensitive receiving antennas will always be beneficial.

Another way to magnify the very small signals coming from interstellar distances is to use the Sun itself as a lens, harnessing its mass as a gravitational telescope (see section \ref{subsec:SGL}). 
Observers would have to be located at 547~AUs, at the opposite side of the Sun to the target (see Fig. \ref{fig: SGL}).
This is a very large distance, requiring an immense fleet of instruments to cover the sky.
However, compact objects have a much stronger lensing effect.
The focal region for white dwarfs starts at 0.045~AU, for neutron stars at 1.9~km above their surface, and 60~km from the event horizon of a 10~$M_\odot$ black hole.
A relay-network made of compact objects would arguably be a very efficient architecture.

A last solution is to send data through inscribed matter, which is not subject to the inverse square law.

\subsubsection{Speed of light}
\label{subsubsec:speed_of_light}

Modern physics is founded on the finiteness of the speed of light, which is unfortunately still very slow when put in the perspective of the vastness of the universe.

In the context of interstellar communication, it means that immense patience or longevity is needed to exchange data.
One mitigation strategy would be to send not a short message like ``Hello'', but a massive amount of information at each iteration, such as a large language model (LLM), as suggested by Vidal \citep{alexandre2023GuerreIntelligences}.
Indeed, imagine if we would receive the chatbot of an extraterrestrial civilization: billions of humans would be able to ask billions of questions about that civilization during many years, which would make the waiting period for the next LLM more bearable.

Technically, these large distances require particular solutions, starting with rejecting network protocols that require sessions and many feedback cycles (e.g.\ acknowledging good receipt of a message, multiple handshake protocols, etc.).
Delay-Tolerant Networking (DTN) architecture has been developed precisely in situations of limited connectivity, and high error rates, with an application for an ``interplanetary internet'' \citep{cerf2007DelaytolerantNetworking, burleigh2003DelaytolerantNetworking, fall2003DelaytolerantNetwork}.
For example, the protocol can embed pre-emptive redundancy, to avoid slow and costly round-trip requests.

In recent years, delay-tolerant networking protocols have been developed to plant the seeds of an interplanetary internet \citep{burleigh2014SpaceInternet}.
This effort will confront us to real space network communication challenges, which would certainly also give insights in how to search for an interstellar internet.

\subsubsection{Heisenberg Uncertainty Principle}
\label{subsubsec:heisenberg}

The uncertainty principle places bounds on the three fundamental informational bottlenecks known in computer science: storage, processing, and transmission bandwidth.

In terms of storage, the fundamental limit of the amount of information that can be stored in a volume of space is defined by the \citet{bekenstein1981UniversalUpper} bound.
It is the number of Planck-length areas that can cover the surface of a black hole of mass $m$.
In other words, any material substrate of mass $m$ can not store more than this limit.
A one kilogram ``ultimate laptop'' would be able to store up to $10^{17}$~bits~kg$^{-1}$ \citep{lloyd2000UltimatePhysical}.

The \citet{bremermann1982MinimumEnergy} bound, on the other hand, places fundamental limits on the number of bit flip operations that can occur in a given amount of time, i.e.\ it is the maximal rate of information processing.
The related Margolus-Levitin theorem \citep{margolus1998MaximumSpeed} gives an upper bound for maximum processing rate, i.e.\ the number of operations per unit time per unit energy:
\begin{equation}
\frac{2}{\hbar \pi} = 6 \times 10^{33}\ \mathrm{s}^{-1}\cdot \mathrm{J}^{-1}
\end{equation}
It too is a consequence of the Heisenberg uncertainty principle.

Lastly, the transmission of information obeys the Pendry bound \citep{pendry1983QuantumLimits}, placing a minimum on the amount of energy needed to transmit a certain number of bits.
In other words, to send $10^9$~bits~s$^{-1}$ requires an energy flow of $E_+ \geq 10^{-16}$~Watts.

These bounds are not just curiosities; they give fundamental insights about how much data can ultimately be stored, processed, or transmitted, and at what energy cost.

\subsubsection{Galactic sources of noise}
\label{subsubsec:galactic_noise}

The sources of noise in the galaxy are varied, and will also widely vary depending on the distance of the two users communicating.
Typically, the greater the distance, the more power will be required.
Depending on the position of the source and destination of the users, the galactic medium will be very different.
For example, two civilizations wanting to communicate with each other from the opposite sides of the galaxy (100,000~ly), will have to deal with a lot of obstacles, delays, etc.\ and will devise a very different solution than two civilizations separated by 10~ly with no interstellar obstacle.
This means that we can expect widely varying communication channels and protocols depending on the respective locations of the transmitter and receiver.

Let us see the obstacles that a signal being sent from Earth would encounter.
First, it needs to go through the atmosphere of our planet.
In our case, the atmosphere is transparent to a narrow range of the EM spectrum.
However, it is hard to justify atmospheric planetary constraints as absolute ones, as we have already developed space-space communication solutions.
Furthermore, we do not want to assume that communication only happens within planets that have Earth-like atmospheres, as that would be unwarranted and anthropocentric.
Next, the signal needs to go through the interplanetary medium and solar system dust, which may be particularly relevant for observations in the ecliptic.

Then come the sources of interstellar attenuation, the scattering and absorption of light by the interstellar medium (ISM), which includes dust and plasma.
Additionally, nearly all astrophysical bodies have magnetic fields which can also interfere with communication efforts, from planets, stars, to the spiral arms of the galaxy.
Gravitational lensing is also well-known to affect light, here seen as a perturbation instead of a clever way to amplify communication as we discussed above.
The weakest constraint is to outshine the noise from the cosmic microwave background (CMB).

Dealing with these multiple interstellar constraints can require detailed modeling, and some communication solutions will clearly be better than others.
We dig deeper into the issue of choosing a part of the EM spectrum in Section~\ref{subsec:em_frequency}.

\subsection{Limitations with modern SETI}
\label{subsec:limitations_seti}

We outlined first-principles from which a search for extraterrestrial communication should start.
Historically, however, the birth of modern SETI started in a search for radio communications.
Since 2016, the Breakthrough Listen program dramatically increased the scale of the search \citep{worden2017BreakthroughListen}.
While a pragmatic starting point, this strategy is fraught with significant challenges.
Our chances of detection dramatically decrease if alien civilizations are not broadcasting powerful beacons in all directions.
We can assume that they would be sending highly targeted intentional messages towards us, but it is a strong assumption, and quite anthropocentric.
These different communication intents and their impacts on search strategies are compiled in Fig. \ref{fig: interstellar_TS_intent_topology}.

Finally, the search for radio signals rests on key assumptions about the technological development and behavior of other civilizations.
It is possible that advanced societies have moved beyond radio technology to other forms of communication, perhaps using other parts of the EM spectrum, or even other information carriers.

These factors combine to make radio SETI a highly complex and difficult task, and here we discuss strategies to address some of these concerns and to broaden the search.

The engineering constraints of sending and receiving interstellar messages are very hard to estimate, mostly because of the accelerating pace of technological progress.
For example, even a century ago, no human could have thought about quantum communication.
In addition, engineering is a practical endeavor that is strongly tied with economical constraints whether they are of human or extraterrestrial origin.

The economics of interstellar communication is rarely addressed \citep[see however][]{benford2010SearchingCostOptimizeda}, but is an important topic to put upper limits on communication solutions.
For example, if we consider two communication protocols, and one offers a higher number of bits per watt than the other, then it is a more promising candidate.
To avoid speculating about specific extraterrestrial economical structures and systems, one can stay with the general requirement that doing work requires energy, and energy is ultimately in finite supply.
So the domain of energy economics \citep{bhattacharyya2011EnergyEconomics} is also of relevance by its generality.

A detectable communication technosignature would only need to be evidence of an intelligent source, not necessarily a deciphered message from an ETI like in most science fiction movies (step (5) in Shannon's model in Fig. \ref{fig:InterstellarTS_shannon}).

A systematic galactic search would require an all-time, all-sky, all-wavelength monitoring for artificial signals, from all of the 1.8 billion stars in the currently biggest galactic catalog, Gaia DR3.
However, not only is this practically impossible to conduct, but such a strategy would be similar to the one of a butterfly collector, not a scientist.
Science progresses with theory and testable hypotheses, so the core issue here is to find strategies to clarify and constrain the search space, as well as to select targets.

To do so, we first structure our approach with two seemingly simple questions: Where should we look?
When should we look?
Then we consider the physical implementation of interstellar communication, which includes the transmission medium, i.e. the channel and the information carrier.
We focus on photons as information carriers, which begs the question of where to look in the electromagnetic (EM) spectrum.
Then we discuss how to deal with anthropogenic interferences such as Radio Frequency Interferences (RFIs), and consider alternative information carrier options.
Then we ask: What are the communication rules (i.e. the communication protocol)?
How is the signal shaped (i.e. which modulation)?
Regarding the topology of the communication network: Are we trying to find a communication that is one-to-one, one-to-many, many-to-one, or many-to-many?
Are there relays?

\subsection{Where should we look?}
\label{subsec:where_look}

\citet{lacki2021OneEverything} constructed a comprehensive review of a wide diversity of astrophysical targets to keep monitoring.
This ``exotica catalog'' includes not only Sun-like stars, but also a wide breadth of 816 distinct targets, including more ``exotic'' astrophysical phenomena spanning from Type I to Type III+ civilizations (see Table~\ref{tab:interstellarTS_lacki_exotica}).

\begin{deluxetable*}{p{0.05\linewidth}lp{0.5\linewidth}l}
\centerwidetable
\tabletypesize{\scriptsize}
\tablecaption{Phyla of astronomical phenomena: a high-level classification system used in the Exotic Target Catalogue \citep{lacki2021OneEverything}}
\tablehead{\colhead{Power on Kardashev Scale} & \colhead{Phylum} & \colhead{Characteristics} & \colhead{Example} }
\startdata
\onehalf    & Minor bodies	           & Solid, small, typically irregular shape, modification mostly from cratering after initial $^{26}$Al differentiation & 1I\'Oumuamua\\
I	        & Solid planetoids         & Hydrostatic equilibrium; solid{\bf, possibly with liquid oceans}; round; geology plays key role in interior evolution; {\bf thin atmosphere with insignificant mass compared to body} & Titan\\
I	        & Giant planets	           & Hydrostatic equilibrium; fluids dominate mass and evolution; {\bf larger than Earth}; typically high internal heat luminosity; formation by gas accretion {\bf onto} solid core & Jupiter\\
II	        & Stars	                   & Hydrostatic equilibrium; plasma; powered by nuclear fusion or {\bf sometimes gravitational contraction}; no solid cores; includes brown dwarfs and protostars by convention & Sun\\
II	        & Collapsed stars          & Supported by degeneracy pressure if at all; luminosity primarily from release of stored thermal, rotational, or magnetic energy & Sirius B\\
II          & Interacting binary stars & Evolution of component stars affected by mass transfer; substantial luminosity from accretion, surface nuclear burning, or shocked outflow; often compact object is mass recipient & SS 433\\
II\onehalf  & Stellar groups     	   & Gravitationally bound collections of stars, which do not otherwise interact for the most part; little to no dark matter & 47 Tuc\\
II\onehalf  & Nebulae and ISM          & Diffuse gases and plasmas; includes flows of matter to/from stars and diffuse galaxy; generally not self-bound & Orion GMC\\
III	        & Galaxies	               & Gravitationally bound, dominated by dark matter; generally contain vast numbers of stars, gas, and possibly a central black hole; gravitationally bound & {\bf M81}\\
III	        & Active galactic nuclei   & Powered by accretion onto a supermassive black hole; includes gas flows on and off SMBH, frequently with jets and particle-filled bubbles & Cygnus A\\
III	        & Galaxy associations      & Dark matter dominates internal gravitation; gravitationally bound collections of galaxies and intracluster medium & Virgo Cluster\\
III\onehalf & Large-scale structures   & Unbound or loosely bound structures on $\gg$Mpc scales; includes high-order arrangements of galaxies and diffuse gas (IGM); non-virialized & Shapley Supercluster\\
Any	        & Technology	             & Structures built intentionally, frequently for processing of matter, energy, and information; so far only known to exist on/near Earth & \emph{Voyager 1}\\
\hline
\nodata     & Reference	               & Sky locations only important relative to observer & Solar antipoint
\enddata
\tablecomments{The ``phyla'' are used to group objects in the Prototype and Superlative samples by shared physical traits.  From a SETI perspective, they indicate the need for very different techniques necessary for astroengineering, as reflected in the amount of used power measured by the \citet{kardashev1964TransmissionInformation} scale rating of on left.} Table reproduced from \citep{lacki2021OneEverything}.
\label{tab:interstellarTS_lacki_exotica}
\end{deluxetable*}

The motivation to closely examine known astrophysical objects is that we might keep missing ETI because we keep looking in the wrong ways or at the wrong places.
The idea to look cleverly at archival data makes sense since, after all, astronomers have observed the sky across many wavelengths over many years.
It could well be that there are technosignatures in existing data that await to be sifted.
To keep track of past searches since the 1960s, the SETI Institute has created a database called ``TechnoSearch''.\footnote{\url{https://technosearch.seti.org/}}

The opposite approach to archival search is to process real-time data only, and only save anomalous events that might correspond to an intelligent signal.
It has the advantage of being fast, and does not require infrastructure related to storage.
The disadvantage is that only the kinds of technosignatures that are considered worthwhile at a certain time are examined, and it is impossible to go back to deleted data.
Even if we cannot see the significance of it today, all data is potentially significant, and as technology progresses, future techniques and hardware may be able to extract patterns and data we cannot even think of today.
However, telescopes produce data very intensively, and storing such raw data remains prohibitively expensive.
A mixed approach between archiving a maximum of data and doing real-time searches would be to dedicate a percentage of the resources for a given SETI project to store at least some data for the long term.

\begin{figure}
    \centering
    \includegraphics[width=1\linewidth]{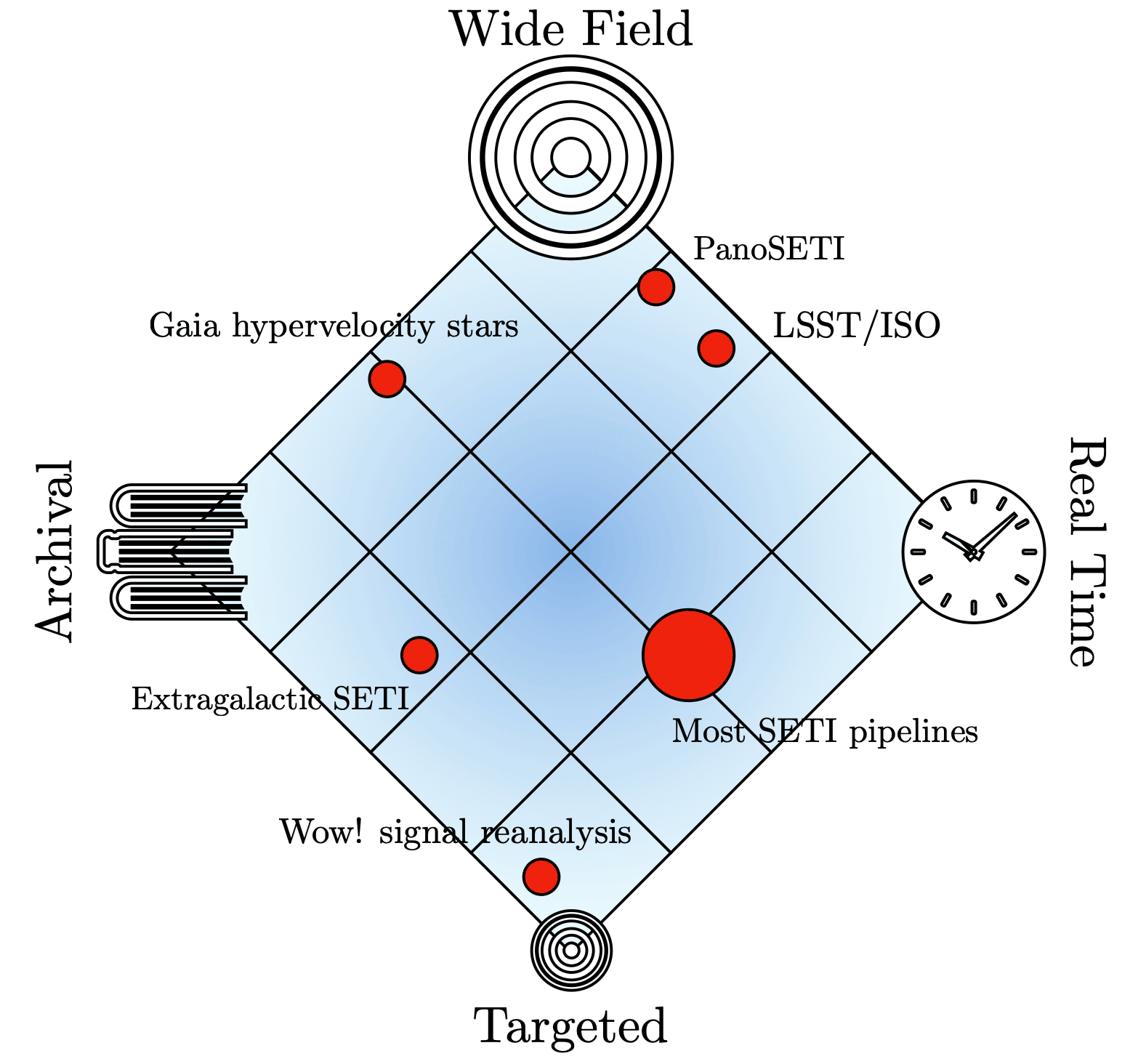}
    \caption{Strategies to look for signals. Two dimensions are represented here: archival vs. real-time in the x-axis, and targeted vs. wide-field in the y-axis.}
    \label{fig: InterstellarTS_2_dimensions}
\end{figure}

SETI researchers are very creative to select potential targets, an endeavor which is informed by theoretical expectations and constraints---sometimes called SETI Theory \citep{wright2022SETI2020}.
One way to organize the strategies to look for signals is through two dimensions: archival vs.\ real-time, and targeted vs.\ wide-field (see Fig. \ref{fig: InterstellarTS_2_dimensions}).

A strategy commonly adopted is to target stars with appropriate spectral types (M-K-G-F types); not A, B, or O because they have short lifespan and high UV radiation that would likely sterilize life-as-we-know-it.
\citet{tarter2001SETI2020} also proposed a roadmap with targets to focus on.
As \citet{turnbull2003TargetSelectiona, turnbull2003TargetSelection} argued, habitable stars like our Sun should also be a prime target, and the authors selected 17,129 targets around 140~pc from the Sun.
Stars with known exoplanets are also natural targets \citep{harp2016SETIObservations}, and the Breakthrough Listen program has also narrowed down targets for observation with the Green Bank Telescope, Parkes Telescope, and Automated Planet Finder \citep{isaacson2017BreakthroughListen}.

Taking the perspective of an extraterrestrial civilization looking for life signatures on Earth, \citet{castellano2004VisibilityEarth} noted that there is only a small set of stars that could observe the Earth transiting the Sun---which is the main method we use to find exoplanets.
This zone in the sky is called the Earth Transit Zone (ETZ).
The strategy is thus to assume that those stars in the ETZ might have detected the Earth easily, not only our technosphere but also our biosphere already billions of years ago.
Such stars are thus prime targets for co-detection and communication \citep[see also][]{heller2016SearchExtraterrestrial, sheikh2020BreakthroughListen, kaltenegger2020WhichStars, kaltenegger2021PresentFuture, sheikh2025EarthDetecting}.

Another way to converge is by looking for civilizations aligned with the solar gravitational lens (SGL, see section \ref{subsec:SGL}), \citep{voneshleman1979GravitationalLens, drake1988StarsGravitational, maccone2009DeepSpace, maccone2013GalacticInternet, tusay2022SearchRadio, turyshev2024SearchGravitationally}.
Telescopes placed at this region gain orders of magnitude in sensitivity, and can even image exoplanets \citep{turyshev2020ImageFormation}.
Similarly, power transmission sees huge increases in efficiency by taking advantage of precise alignment configurations \citep{turyshev2024GravitationalLensing}.

Another general strategy is to concentrate observations on high stellar density regions, simply because they offer more targets: not only more targets for us to search, but also more stars for ETI to live from.
Communication, interaction, and coordination are likely to be much faster and richer in a high-stellar density environment than between any two parts of the galaxy.
A logical place is the galactic plane where the vast majority of stars in our galaxy lie \citep{grimaldi2021DemographyGalactic}.
Even considering the trade-off between number of targets and distance to the stars, \citet{sullivan1984MilkyWay} argued that searching in the galactic plane is a superior strategy compared to nearby stars or all-sky surveys.
\citet{shostak1985SignalSearch} conducted a subsequent search in the galactic center with the Westerbork Synthesis Radio Telescope.
More recently, \citet{margot2021SearchTechnosignaturesa} conducted a search around 31 Sun-like stars in the galactic plane with the Green Bank Telescope, and \citet{gajjar2021BreakthroughListen} did a more thorough search towards the galactic center.

\citet{distefano2016GlobularClusters} also argued that since globular clusters (GCs) being extremely dense environments, it makes sense to prioritize searches in them.
They note that GCs contain many low mass stars that can burn during many billions or trillions of years, and that little gas and dust is present, which impedes star formation or core-collapse supernovae.
However, we can note that the metallicity of such stars is low, gas giants would be hard to form, and the formation of rocky planets remains a point of debate \citep[e.g.][]{sigurdsson1992PlanetsGlobular}.
If there are ETIs in such environments, we would be looking at a lifeform as we don't know it yet.

Finally, a way to observe many billions of stars at the same time is to conduct extragalactic SETI (see section \ref{sec: extragalactic}).

\subsection{When should we look?}
\label{subsec:when_look}

Without any prior communication, how could two civilizations know how to coordinate with each other?
Where and when could they hope to find each other?
This problem has been studied in game theory, and actors typically converge towards a focal point or Schelling point \citep{schelling1960StrategyConflict} in order to avoid coordination failure.
A key challenge that has been extensively discussed in the SETI literature is to guess what such spacetime Schelling points could be \citep{wright2020PlanckFrequencies}.
It is a mandatory exercise to prioritize searches given the huge multidimensional search space.

In particular, the question of when to look can be broken down into three: (1) At what particular time?
(2) How long should we look? and (3) How often?

\subsubsection{At what particular time?}
\label{subsubsec:when_particular}

The universe is not static, and astronomical bodies rotate, along their own axis, around their star, around the galaxy, and even galaxy pairs orbit a common center of mass.
In the context of communication with planet Earth, a good timing to transmit to Earth would be when it is very visible from an ETI point of view.
It would be so when it is midway through its transit.
This event is called the anti-solar point, and \citet{hort2024SETISearch} suggested that it would be an ideal moment to communicate.
The authors used the Hat Creek Observatory to observe the anti-solar point during $\sim$24 hours, but did not find signals of interest.

\citet{kipping2016CloakingDevice} suggested that advanced civilizations could alter their transit planetary signature with lasers, either to cloak themselves, or to broadcast their presence.
They proposed that if we would find a radial velocity signature of a planet, but not the planet itself, this might be a technosignature of a cloaking mechanism.
Such a search strategy is arguably contrived because by design, it may be nearly impossible to detect an advanced civilization wanting to hide.
However, this possibility has been made popular by science fiction author \citep{liu2015DarkForest}, with the ``dark forest'' hypothesis, in which most aliens hide from each other.
The opposite strategy from cloaking is where the ETI would deliberately signal its presence by altering the shape of the transit in a clearly artificial way.

Besides exoplanet transits, authors have proposed that ETI communication would likely be in relation to the transit of binary stars, especially during the apastron and the periastron \citep{pace1975TimeMarkers, pace1979UseBinary}.

Many have proposed that communication would be synchronized with major astronomical events, such as novas \citep{mclaughlin1977TimingInterstellar, mclaughlin1986SETIExperiment}, supernovas \citep{tang1976SupernovaeTime, seto2021SearchGalactic, nilipour2023SignalSynchronization}, Fast Radio Bursts (FRBs; \citealt{zhang2020QuantitativeAssessment}), Gamma-ray bursts (GRBs; \citealt{corbet1999UseGammaRay}), or gravitational wave events \citep{nishino2018SearchExtraGalactic}.

Cordes and Sullivan \citep{cordes1995AstrophysicalCoding, sullivan1995AstrophysicalCoding} also introduced the notion of \textit{astrophysical coding}, arguing that intentional signals are likely to be encoded and detectable with the same techniques used to conduct regular astrophysics.
In particular, they noted that timing analysis can be used to determine (1) the location of the source, (2) its distance via its dispersion measure, (3) orbital motion of the host planet or host satellite, and (4) orbital perturbations, precession, and relativistic effects, and mentioned pulsars as a prime example of astrophysical coding.

Since millisecond pulsars (MSPs) provide stable, continuous, high cadence timing reference points, assuming that civilizations use them as a time standard \citep{vidal2017MillisecondPulsars}, it implies that there is no privileged timing, because timing metadata is available since the existence of MSPs, and pulsars are theorized to last for billions of years.

\subsubsection{How long should we look?}
\label{subsubsec:how_long}

Assuming we are looking at the right target, how long should we look to detect a technosignature?
This is a general issue concerning any bio- or techno- signature.
Some technosignatures may appear very quickly such as laser pulses at a nanosecond timescale, while others very rarely, every year, decade, century, millennium, etc.\ (see Table~\ref{tab:timescales}).

\begin{table}[ht]
\centering
\caption{Observational timescales illustrated with potential technosignatures, from the Planck timescale to the billion-year timescale. Any technosignature requiring more than years or decades of data-long effort is likely to exhaust the patience of humans.}
\label{tab:timescales}
\begin{tabular}{rp{6cm}}
\hline
Timescale & Potential technosignature \\
\hline
Planck time   & Theoretical limit \\
Femtosecond   & Laser pulses (future technology?) \\
Nanosecond    & Laser pulses (Optical SETI) \\
Microsecond   & Fast radio bursts (FRBs) \\
Millisecond   & Signals in sync with millisecond pulsars \\
Second        & Active ongoing communication \\
Minute        & Transient communication (e.g.\ the Wow! signal) \\
Hour          & Transit timescale (e.g.\ 13 hours for Earth) \\
Day           & Light illumination (exoplanet day) \\
Year          & Anomalous motion of a spacecraft; grey goo scenario \\
Decade        & Progressive construction of megastructures; industrial pollutant concentration variations; disappearing stars \\
Century       & Close stellar encounters \\
Millennium    & Terraforming \\
Million-year  & Major biosphere/technosphere changes \\
Billion-year  & Long-term tracking of stellar evolution and migration of civilizations \\
\hline
\end{tabular}
\end{table}

Looking at biology, lifespans vary from mayflies that live only 5 minutes, to hundreds of years in the case of clams, and up to 5000 years for trees.
For the purpose of doing science, the timescale issue can largely be mitigated by having many samples.
For example, a botanist may not be able to study the dynamics of a 5000-year-old tree in isolation, but could get a quite comprehensive picture of the species by studying many samples, from seeds to trees of all ages up to 5000-year-old ones.
Similarly, the universe is full of different samples of diverse astrophysical objects, which is why we can still reconstruct a general picture.

We can note that technological change happens much more rapidly than biological change, and major changes observed in a few decades could be quite unambiguous (e.g.\ a gradual reddening of a star over decades could be a signature of a megastructure being built).

Studying timescales going beyond human lifespans is still conceivable, either by a consistent human and economical investment \citep{longnowfoundation2025LongNow}; or by building self-maintaining machines that can operate reliably during decades, centuries or millennia.
Other strategies such as hibernation, or storing only surprising, anomalous observations could enhance the energetic and memory autonomy of such multigenerational projects.
The point is that technosignatures have no reason to appear in an anthropocentric timescale window (from seconds to decades).
Finding a strong proof of a technosignature might require a decentering in timescale, such as a long-term effort with a multigenerational commitment to looking at large temporal scales; or by developing technologies that can probe shorter and shorter timescales.

The ideal of doing ``all time'' observation obviously means that the longer the observation the better, but pragmatically, telescope time is very expensive and competitive.
The Breakthrough Listen initiative is a game-changer in this regard, including many SETI observational hours at the Green Bank Observatory (GBT), many backends on major telescopes such as MeerKAT, Parkes, Lick Observatory, in addition to the Hat Creek Radio Observatory (HCRO) that is managed by the SETI Institute and can thus be prioritized towards SETI targets.

An important development of SETI searches is thus the use of commensal strategies, i.e.\ installing back-ends on the raw data of observations done by major telescopes, and running SETI pipelines on them.
It has the advantage that it does not require competition for telescope time, but the disadvantage is that targets cannot be chosen directly.

\subsubsection{How often?}
\label{subsubsec:how_often}

The cadence at which messages would be sent depends strongly on the kind of communication aimed at.
Warning messages might be very rare, for example, only when a civilization-threatening supernova or GRBs would be about to occur.
Leakage radiation may happen often or not, depending on whether the leakage is accidental, and depending on how carefully civilizations cloak their EM footprint, if they do it at all.
The highest communication cadence would be between two civilizations communicating continuously between each other, or the expectation to find a continually broadcasting beacon.

A very slow communication method would be to encode artificially mathematical sequences within planetary period ratios \citep{clement2022MathematicalEncoding}.
However, as we noted earlier, this would be extremely expensive and inefficient in terms of the number of bits per energy unit.

Even assuming very powerful civilizations, economical analyses can constrain the search.
Continuous broadcasting is extremely expensive on the sender end, and continuous observation is also expensive on the receiving end (at least on present Earth).
\citet{benford2010SearchingCostOptimizeda} made an economical assessment for building and maintaining interstellar beacons, and concluded that it leads to different strategies than most SETI searches, namely steady searches in the galactic plane on the scale of years.

\citet{zenil2025FractalSpatiotemporal} took seriously the problem of timescale and its major underlying uncertainties.
They proposed a fractal messaging scheme, using the Weierstrass function that is scale invariant, as a means to design a spacetime invariant communication protocol.

Yet, the galaxy is not in the timing dark, and millisecond pulsars (MSPs) provide extremely stable, accurate, and durable timing information.
For this reason, MSPs are very likely to be used as a timing, navigation, and communication metadata standard \citep{vidal2017MillisecondPulsars}.
This means that we can infer a lower-limit on the time resolution of any kind of communication that requires reliable galactic timestamping.
This lower-limit is then in the range of fast MSPs, i.e.\ milliseconds, and it thus constrains the time resolution of communication signals.
Thus if ETIs indeed use MSPs as a metadata time standard, then we can predict that nanosecond optical messages are unlikely, and thus that Optical SETI at that timescale is unlikely to succeed \citep{vidal2025PulsarPositioning}.

\subsection{At what electromagnetic frequency?}
\label{subsec:em_frequency}

Choosing a spacetime region to look at is just one decision to make in the search space.
The next question is to choose where to look in the space of electromagnetic (EM) frequencies.

We noted that communicating at different distances requires different strategies, and \citet{hippke2017InterstellarCommunicationb} made various proposals to deal with interstellar extinction at various distances, ranging from short distances ($<100$~pc), medium distances (100--500~pc) to large distances (500--8000~pc).
For example at $>480$~pc distances, a step appears where the optimal low-energy wavelength moves from UV (160~nm) to near-infrared (690~nm).
These detailed considerations lead to the strategy to search at different wavelengths depending on the distance of the target.

In their foundational paper that set the modern SETI era, \citet{cocconi1959SearchingInterstellar} argued from the perspective of an extraterrestrial civilization wanting to make deliberate contact with Earth-like planets.
This required that the frequency would be transparent to the planetary atmospheres.
Writing in the technological time of the 1960s, they considered optical up to gamma-rays but chose not to pursue that direction because such wavelengths would require more power at the source or complicated techniques.
They famously proposed to target the emission line of neutral hydrogen, the most abundant element in the universe, at 1.42~GHz.

This frequency was later extended to a range from 1.42~GHz (H, hydrogen) to 1.67~GHz (OH, the hydroxyl radical).
This region of the EM spectrum has an extremely low interstellar noise background.
What is more, H and OH form the components of water (H$_2$O), the molecule essential for life-as-we-know-it.
This range has been called the ``cosmic water hole'', playing both on the meaning of water, and a water hole as a place where animals end up meeting in the wild \citep{oliver1971ProjectCyclops}.

\citet{schwartz1961InterstellarInterplanetary} made the case for looking for laser transmissions, both in the optical part and infrared part of the spectrum (see Table~\ref{tab:radio_vs_laser}).
They noticed that focusing light on a tight optical beam considerably increases the bandwidth of communication compared to radio.
This suggestion has since grown into ``Optical SETI,'' or OSETI.
Notably, just as Tesla was a pioneer of radio communication and quickly considered using them to communicate with planets \citep{tesla1901TalkingPlanets, raulin-cerceau2010PioneersInterplanetary}, Townes was a pioneer of laser and maser technologies, and even co-received the Nobel Prize in physics for building the first maser.

\begin{table}[ht]
\centering
\caption{Comparison of radio and laser communication modalities for SETI.}
\label{tab:radio_vs_laser}
\begin{tabular}{>{\raggedright\arraybackslash}p{0.4\linewidth}>{\raggedright\arraybackslash}p{0.5\linewidth}}
\hline
\textbf{Radio} & \textbf{Laser} \\
\hline
\textit{Pros:} & \textit{Pros:} \\
Suitable for omnidirectional beacons, to communicate widely; wide window for observation. & Suitable for point-to-point communication; a collimated beam implies less leakage (privacy), more energy efficiency, and more bandwidth. \\[6pt]
\textit{Cons:} & \textit{Cons:} \\
Less bandwidth; more energy needed due to the omnidirectional nature. & Can only be detected if the signal is directed toward us, or if our solar system is in the path of two communicating civilizations; more affected by interstellar dust \citep[see][]{hippke2018InterstellarCommunicationa}. \\
\hline
\end{tabular}
\end{table}

As with any observing modality, laser searches can synergize with other strategies or programs.
For example, \citet{marcy2021SearchOptical} searched for signals coming from the SGL, assuming a probe would transmit towards us or Alpha Centauri.
Also, it was through an optical SETI search directed towards HD~89389 that \citet{stanton2025UnexplainedStarlight} suggested that the pulsed source came from a potential ring-shaped artefact in our solar system (see section \ref{sec: ss ts}).

To solve the interstellar dust problem that would affect laser communication, a more robust network of carrier nodes could be deployed, possibly also leveraging the gains at the solar gravitational lens \citep{maccone2013GalacticInternet}.
\citet{hippke2018InterstellarCommunicationa} also suggested that to minimize extinction by interstellar dust, a wavelength at 1064~nm and its second harmonic 532.1~nm would be optimal from a laser engineering point of view.

In addition to interstellar communication, we could also detect laser communication going on inside an extrasolar system, for the purpose of communicating with spacecraft, or between planets.
Nowadays, much of our own space communication is migrating towards laser solutions \citep{wu2017OverviewDeep}.

There are also non-communicative uses of lasers, such as using them as a directed energy system to accelerate small lightsail probes \citep{forward1984RoundtripInterstellar, guillochon2015SETILeakage, kulkarni2018RelativisticSpacecraft}.
The concept of a Solar Wind Power Satellite that would harness the stellar wind to power a civilization may also use high power lasers to beam energy where needed \citep{harrop2010SolarWind}.
Lasers might also be sporadically used for planetary defense purposes such as deflecting asteroids \citep{lubin2016DirectedEnergy} or even as weapons.

Due to the phenomenon of thermal broadening, most astrophysical phenomena will be spread around the electromagnetic spectrum, while lasers are very narrow and monochromatic.
This means that a laser can outshine even a star at a given wavelength and that they can be readily distinguished from most natural emission lines.
Two strategies can be used to detect laser emissions: spectroscopy and photometry.
In spectroscopy, lasers are defined by linewidths that are narrower than natural emission lines, and can be pinpointed by wavelength.
In photometry, lasers are detected by looking at the time-variability of brightness, particularly at flashes of short durations.
Each strategy has its own pros and cons.
Photometry is sensitive to time variability, and can detect very short duration pulses, but cannot reveal much information about the wavelength.
Spectroscopy, on the other hand, can reveal information about the wavelength and thus the nature of the laser, but is insensitive to time variability, and can only show the result of the whole integration time.
It means that spectroscopy is better suited for detecting continuous lasers and focusing on precise targets; while photometry is better suited for detecting pulsed lasers and conducting all-sky observations.

Following the spectroscopy strategy, the Keck observatory observed 5600 stars, but did not find traces of laser emissions \citep{reines2002OpticalSearch, tellis2015SearchOptical, tellis2017SearchLaser}.
As a follow-up to the anomalies of Tabby's star, a program was conducted to search for laser transmissions, without positive results \citep{lipman2019BreakthroughListen}.
Recently, \citet{marcy2023SearchTransient} looked for laser emissions in the galactic plane, using the objective prism Schmidt telescope, which uses low-resolution short exposure time spectra to look for time variability down to about 1~sec, and thus leverages both advantages of spectroscopy and photometry.
No candidates were found.
\citet{fields20252821Star} used archival data from the HARPS spectroscope to analyze 2821 stars, and found 3 ambiguous candidates from 2 red dwarfs and 1 red giant that are currently under further investigation.

Turning to photometric instruments, these allow going down to the nanosecond timescale resolution, at which laser pulses would be unambiguous because such rapid variability is rare or nonexistent in nature \citep{howard2004SearchNanosecond}.
SETI researchers used the Lick and Leuschner Observatories to conduct a near-infrared search \citep{maire2014NearinfraredSETI}.
\citet{stone2005LickObservatory} targeted 4605 stars of spectral types F--M within 200 light-years of Earth; they found 14 events of interest, all of them being explained as false positives, except one, HD~212107.
Further observations did not reveal any repeat detections.
More recently, PanoSETI is the main consistent effort for an all-sky observation program \citep{maire2022PanoramicSETI}.

\citet{howard2004SearchNanosecond} observed 13,000 Sun-like stars during 5 years, and found one particularly interesting candidate, HIP~107395, which is part of the habitable catalog \citep{turnbull2003TargetSelectiona}.
Furthermore, it showed up at two independent observatories, the Harvard/Smithsonian Oak Ridge and Princeton's Fitz-Randolph.
The paper reports that the probability of any pair of triggers randomly occurring within 1~ms is about $3 \times 10^{-5}$.
Part of the observational strategy was to sync the two observatories using GPS clocks, because this allowed the team to know that if both telescopes picked up a candidate, it was from the same source.
Unfortunately, the GPS clock at Princeton was not working at the time, making it difficult to confirm that the two instances happened at exactly the same time and leaving only a 2\% chance that it was actually the same signal.
They did follow-up this star without subsequent pulse detections, but it remains a candidate worthy of further follow-up.

\citet{hanna2009OSETISTACEE} did a search for nanosecond optical transients via the Solar Tower Atmospheric Cherenkov Effect Experiment (STACEE), but did not find candidates.
\citet{korzoun2024PeVGammaray} have shown that a multimessenger synergy is possible between PanoSETI and the Cherenkov telescope VERITAS to search for laser signatures.
VERITAS was also used to conduct a search for optical flashes associated with Tabby's star, with null results \citep{abeysekara2016SearchBrief}.
Laser SETI is an ongoing program led by the SETI Institute \citep{klesman2017NowYour} using a wide angle grating-based spectrograph, designed to find narrow, monochromatic emission lines characteristic of a laser.

However, only small parts of the electromagnetic spectrum have been the focus of SETI programs.
Already in 1983, \citet{townes1983WhatWavelengths} cautioned that considerable uncertainty remains when betting on a particular wavelength.
For example, low frequency SETI has only recently been debuted, notably using simultaneously multi-site LOFAR stations in Europe \citep{johnson2023SimultaneousDualsite}.
This region of the EM spectrum has not been probed before, and the multi-site configuration allows for RFI rejection.

\citet{hippke2017InterstellarCommunicationb} did a study of optimal frequency to maximize data rate in interstellar communication, and concluded that high-energy photons present serious advantages.
Indeed, X-ray lasers can carry a high data transmission rate compared to radio, infrared and optical due to their high frequency.

\begin{figure*}
    \centering
    \includegraphics[width=0.6\linewidth]{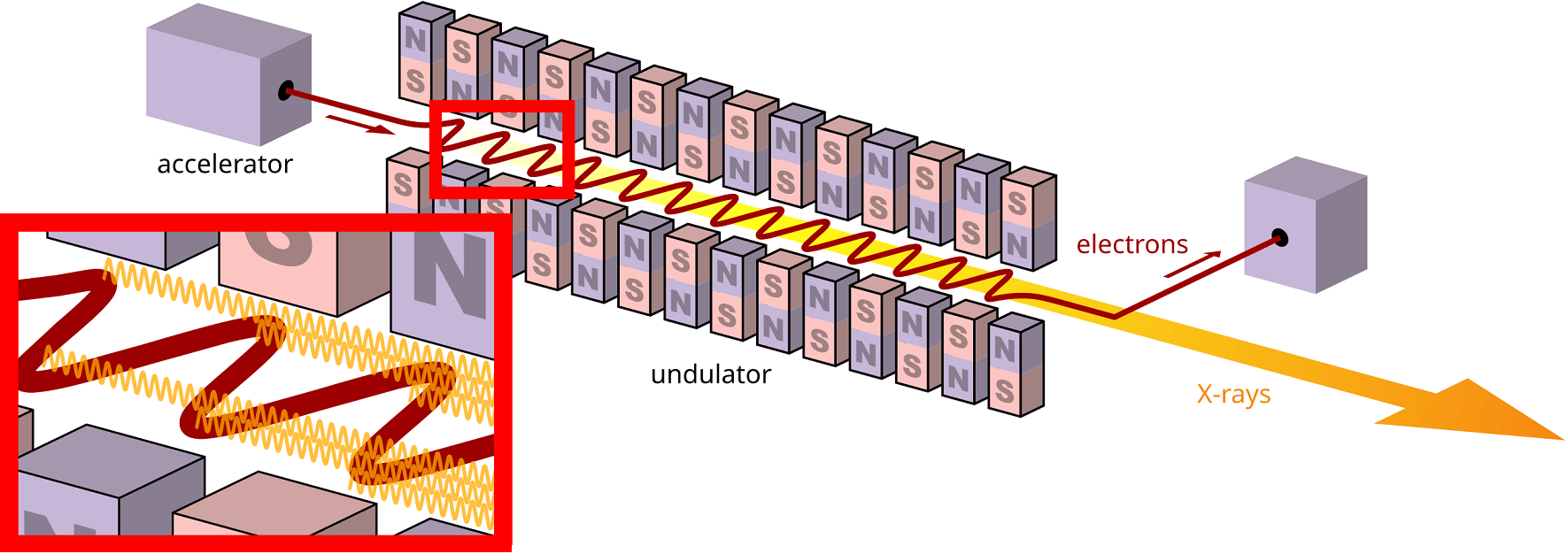}
    \caption{A schematic illustration of a X-ray Free Electron Laser (XFEL).  An electron gun fires a beam of electrons that are directed through an undulator after being accelerated through a particle accelerator. The beam of electrons then passes through an undulator, which is a periodic arrangement of magnets whose function is to produce the highly coherent X-ray pulses/beam. Diagram courtesy of Wikipedia, based on \citep{patterson2010NovelOpportunities}.}
    \label{fig: interstellarTS_XEFL}
\end{figure*}

In particular, X-ray lasers are capable of producing highly focused and intense X-ray beams with a very narrow divergence angle which allows for highly energy-efficient interstellar communication.
While natural astrophysical sources of X-ray emissions are generally characterized by specific spectral lines, we could search for free electron lasers, which accelerate free electrons to nearly the speed of light, directing them through an alternating magnetic field in a way that produces highly coherent X-ray pulses (see Fig. \ref{fig: interstellarTS_XEFL}).
Although above our present technological capabilities, fusion-powered X-ray lasers are another possibility to generate X-ray pulses.

In space, background radiation is lower in the X-ray spectrum compared to radio waves \citep{carstairs2002SpreadingNet}.
This reduced noise makes it easier to detect X-ray signals against the natural background, which is advantageous for transmitting and receiving clear and distinct signals over vast distances.
There are still natural X-ray sources coming from synchrotron radiation, but the sources are generally known.
If we would find patterns of X-rays that are outside the range of known X-ray sources, that would be an interesting anomaly to clarify.

Long wavelengths such as radio waves are susceptible to magnetic fields in the galaxy (inducing the Faraday effect or scintillation), whereas X-rays are much less susceptible to these effects \citep[see][]{berera2022ViabilityQuantum}.

\citet{hippke2017InterstellarCommunicationa} did an archival X-ray spectra search, looking for narrowband X-ray signals.
They found 19 monochromatic candidate signals, but concluded that they are most likely due to natural astrophysical origin.
G-type stars would represent good candidates to search for X-ray spectral lines since these stars do not produce strong X-ray emission lines.
X-ray SETI remains challenging with present X-ray observatories, notably because signal to noise ratio is very low: for example, it takes about 5 days to build a spectrum with the Chandra X-ray observatory.
A drawback of X-ray observatories is their lack of high spatial resolution, but they are still very good in the time domain.

In another study, \citet{berera2022ViabilityQuantum} investigated the feasibility of quantum communication across interstellar distances.
The authors concluded that X-ray photons might allow quantum coherence to be maintained over interstellar distances.
These studies contribute to the theoretical foundation for considering X-ray lasers as a possible mode of interstellar communication.

Gamma-ray communication is also a logical option rarely explored \citep[see however][]{harris1991SETIGammaray, corbet1999UseGammaRay}, and has been mostly discussed as signatures of advanced propulsion (see section \ref{subsec:energy_sources}).
\begin{table*}[ht]
\centering
\caption{Extinction effects of the interstellar medium (ISM) at various frequencies.}
\label{tab:ism_attenuation}
\begin{tabular}{lcll}
\hline
EM Region & Wavelength/Energy & Status in ISM \\
\hline
Radio      & $> 10$~cm            & Transparent      & \\
Microwave  & 1~mm -- 10~cm        & Transparent      &  \\
Infrared   & 700~nm -- 1~mm       & Transparent      & \\
Visible    & 400~nm -- 700~nm     & Hazy            &  \\
Near UV    & 300~nm -- 400~nm     & Opaque           &  \\
Extreme UV & 10~nm -- 121~nm      & Very opaque   &  \\
Soft X-Ray & 0.1~keV -- 2~keV    & Opaque           &  \\
Hard X-Ray & 2~keV -- 100~keV    & Transparent      &  \\
Gamma Ray  & $> 100$~keV          & Transparent      &  \\
\hline
\end{tabular}
\end{table*}

Even when a particular part of the spectrum is chosen, there remain many spectral channels to look into.
Since the inception of SETI, researchers have proposed particular Schelling points related to mathematical constants, physical and cosmological constants, and astronomical references.

\subsubsection{Mathematical constant Schelling points}
\label{subsubsec:math_schelling}

The idea of modulating light according to universal mathematical regularities to provide a clear signature of an artificial source dates back to Charles \citet{cros1869EtudeMoyens}, who proposed that optimal signals towards Mars or Venus should follow simple periodic laws, and not be random.
Mathematical scientists have since considered that many Schelling points could be chosen; for example, \citet{makovetskii1976StructureCall} proposed that calls could be composed starting with the waterhole frequency at 1420~MHz and multiplying it with a mathematical constant such as $\pi$, $e$, or $\sqrt{2}$.
Staying with the 1420~MHz window, \citet{shklovskii1966IntelligentLife} suggested its first or second overtone.

Another common suggestion is to bet on prime numbers, because of their unique mathematical properties, although one can argue that they have simple algorithmic (Kolmogorov) complexity and might only be a good beacon, but are not a good trace of a complex signal.

\begin{figure}
    \centering
    \includegraphics[width=0.85\linewidth]{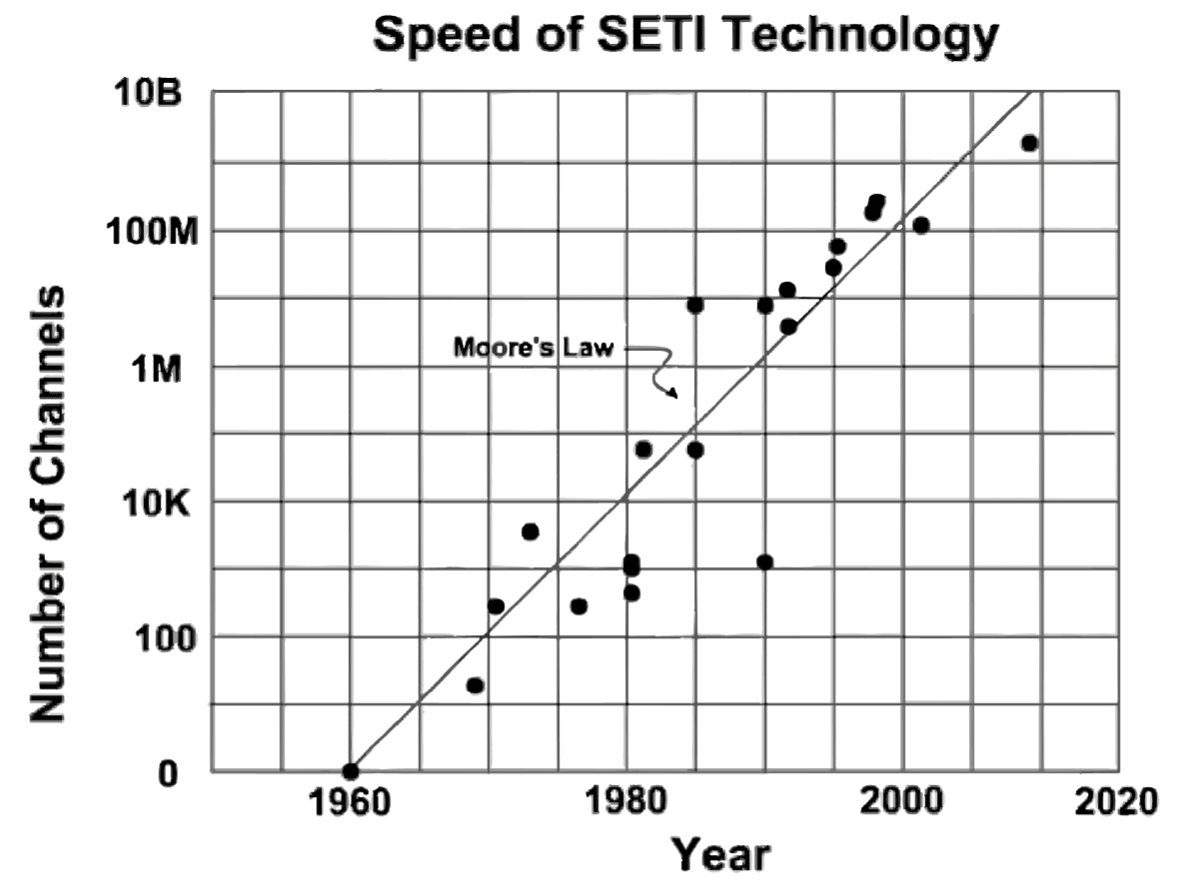}
    \caption{Moore's law in spectral channels, from \citep{tarter2009AdvancingSearch}.}
    \label{fig:moores_law_channels}
\end{figure}

One should note that many of these proposals were made in a time where not many frequencies could be analyzed, and thus stringent choices had to be made.
Nowadays, massive computing and spectrometer technological progress follow a kind of Moore's law (see Fig.~\ref{fig:moores_law_channels}), and now measure billions of frequencies simultaneously.

\subsubsection{Physical \& cosmological constants Schelling points}
\label{subsubsec:physical_schelling}

Using the cosmic water hole as a foundation, and adding other fundamental constants \citet{blair1993ListPossible} derived a list of 55 possible frequency choices.

Fundamental physical and cosmological constants also provide natural Schelling points.
Historically, this goes back to \citet{planck1900UeberIrreversible} when he defined his eponymous units, and wrote in passing that they "remain meaningful for all times and also for extraterrestrial and non-human cultures".
\citet{wright2020PlanckFrequencies} developed this suggestion and derived a "Planck Frequency Comb".
He also emphasized that there are still inevitable anthropocentric choices when making such suggestions.

\citet{kardashev1979OptimalWavelength} did also consider cosmological parameters, namely the range of maximum intensity of the cosmic microwave background.

\subsubsection{Astronomical Schelling points}
\label{subsubsec:astro_schelling}

Astronomical phenomena can also provide natural timing and frequency references.
For example \citet{heidmann1988PulsaraidedStrategies, heidmann1992PulsaraidedSETI} and \citet{heidmann1992PulsaraidedSETIa} suggested to use multiples of pulsar frequencies, and conducted searches accordingly.
\citet{vidal2017MillisecondPulsars} argued that millisecond pulsars are prime and readily usable galactic standards for timing, navigation, and metadata communication purposes.
This convenience also led \cite{vidal2019PulsarPositioning} to suggest the hypothesis that stable and regular MSPs functioning like a "pulsar positioning system" might have been engineered to a certain degree (see also section \ref{sec: galactic}).

\subsection{How to deal with anthropogenic interferences?}
\label{subsec:rfi}

Adding to the difficulty is the abundance of both natural and human-made interference.
The interstellar medium (ISM) itself creates background noise that can obscure faint signals.
Furthermore, radio frequency interference (RFI) from our own technology, such as satellites and mobile phones, present a constant challenge for researchers trying to distinguish a genuine extraterrestrial signal from terrestrial noise.

Back in 1999, a major goal of the SETI@Home project was to use citizens' idle computers to search for artificial signals, but it also included algorithms to reject false positives such as RFIs \citep{korpela2025SETIhomeData, anderson2025SETIhomeData}.

RFIs give rise to many false positives, sometimes not easy to reject, as in the case of Breakthrough Candidate 1 (BLC1; \citealt{smith2021RadioTechnosignature, sheikh2021AnalysisBreakthrough}).

Through the years, SETI scientists have devised sophisticated RFI mitigation and rejection techniques, for example by conducting multi-site observations \citep[e.g.\ LOFAR;][]{johnson2023SimultaneousDualsite}; by pointing the telescope ON or OFF its target to see if the signal persists \citep[e.g.][]{enriquez2017BreakthroughListen}; by looking at the drift rate of EM waves since local RFI generally does not drift; by using polarization properties \citep{li2023PolarizationCriteria}; or by using machine learning algorithms that can learn the patterns of RFI, and thus reject them \citep{jacobson-bell2025AnomalyDetection}.

Another mitigation strategy is to prioritize the search on higher frequencies that are not yet contaminated with our technology.
Unfortunately, the Earth may become too contaminated for technosignature searches---and for astronomy at large---which motivates the deployment of telescopes on the farside of the Moon which is almost completely radio-quiet (e.g.\ \citealt{mumma1990AstrophysicsMoon}), including the recently announced Lunar Farside Technosignature \& Transients Telescope \citep[LFT3;][]{deboer2025LunarFarside}.

\subsection{What are the non-EM information carrier options?}
\label{subsec:non_em}

\citet{hippke2018BenchmarkingInformation} reviewed information carrier options for interstellar communication (Table~\ref{tab:carrier_performance}), confirming that photons are still a premium carrier, but also that other carriers have advantages in certain scenarios.

\begin{table*}
\caption{Carrier performance summary. All values for $E={\rm keV}$, transmitter aperture 1\,m, distance 1\,pc. Masses and lifetimes from \citep{mohr2008CODATARecommended}. Text in \textcolor{red}{red color} indicates problematic properties. Adapted from \citep{hippke2018BenchmarkingInformation}.}
\label{tab:carrier_performance}
\footnotesize
\centering
\begin{tabular}{lcccccccccc}
\hline
Carrier 
  & Mass & Lifetime & Velocity & Charged & Beam & Extinct. & Lensing & Diff. & Init. Cost & Op. Cost \\
  & (MeV/$c^2$) & (s) & ($c$) & & (arcsec) & & & & & \\
\hline
Photon   & 0 & stable  & 1  & no & $10^{-4}$ & 0.001  & good & low & medium & medium \\
Neutrino & ${\approx}0.001$ & oscillations & $\lessapprox 1$ & no & \textcolor{red}{$10^6$} & ${\approx}0$ & very good & medium & \textcolor{red}{high} & medium \\
Electron & 0.51     & $>10^{36}$  & \textcolor{red}{0.0630} & \textcolor{red}{yes} & \textcolor{red}{$10^6$}  & medium & medium & \textcolor{red}{${\approx}1$} & medium & medium \\
Proton   & 938.27   & $>10^{41}$& \textcolor{red}{0.0015} & \textcolor{red}{yes} & \textcolor{red}{$10^6$} & \textcolor{red}{${\approx}1$} & very good & low & medium & medium \\
Neutron  & 939.56   & \textcolor{red}{$881.5$}    & \textcolor{red}{0.0015} & no & \textcolor{red}{$10^6$} & \textcolor{red}{${\approx}1$} & very good & low & medium & medium \\
Muon     & 105.66   & \textcolor{red}{$10^{-6}$}  & \textcolor{red}{0.0044} & \textcolor{red}{yes} & \textcolor{red}{$10^6$} & \textcolor{red}{${\approx}1$} & very good & low & \textcolor{red}{high} & medium \\
Tau      & 1776.82  & \textcolor{red}{$10^{-13}$} & \textcolor{red}{0.0011} & \textcolor{red}{yes} & \textcolor{red}{$10^6$} & \textcolor{red}{${\approx}1$} & very good & low & \textcolor{red}{high} & medium \\
Higgs    & $1.25\times10^5$ & \textcolor{red}{$10^{-22}$} & \textcolor{red}{$10^{-6}$} & no & \textcolor{red}{$10^6$} & ${\approx}0$ & very good & \textcolor{red}{high} & \textcolor{red}{high} & \textcolor{red}{high} \\
\hline
Inscribed matter & $>0$ & $>10^{13}$ & $0\dots1$ & no & ${\approx}0$ & ${\approx}0$ & no & low & \textcolor{red}{high} & zero \\
Occulter & $>0$ & $>10^{13}$ & 1 & possible & \textcolor{red}{$10^3$} & 0.001 & unlikely & low & \textcolor{red}{high} & very low \\
Gravitational wave & 0 & stable & 1 & no & \textcolor{red}{$4\pi$} & 0 & low & \textcolor{red}{high} & \textcolor{red}{high} & \textcolor{red}{high} \\
\hline
Axion   & ${\approx}10^{-9}$ & stable & $\lessapprox 1$ & no & $10^{-4}$ & ${\approx}0$ & good & \textcolor{red}{high} & unclear & unclear \\
Tachyon & unclear & stable & $>1$ & no & unclear & unclear & unclear & \textcolor{red}{high} & unclear & unclear \\
\hline
\end{tabular}%
\end{table*}

\subsubsection{Occulters}
\label{subsubsec:occulters}

In many ways, the simplest solution to signal over interstellar distances is to use an occulter in front of an already existing, natural bright source.
Indeed, this solution dispenses with the engineering and economic burden of creating the energy source, and focuses only on modulating the signal.
An analog on Earth is the smoke signaling that has been used for centuries across ancient China, North America, and Australia.

\citet{morrison1962InterstellarCommunication} proposed that an occulter could be used by advanced ETI to communicate from Andromeda to our galaxy.
He proposed an opaque screen of $10^{20}$~grams that could beam algebraic equations.

In SETI, \citet{arnold2005TransitLightCurve} suggested that transiting megastructures themselves could be used as information carriers to encode prime numbers.
\citet{chennamangalam2015JumpingEnergetics} suggested that pulsar radiation could be modified with an artificial satellite, a proposal that \citet{vidal2019PulsarPositioning} kept as a possibility regarding the hypothesis that millisecond pulsars would have been modified for the purpose of engineering a ``pulsar positioning system''.

\subsubsection{Neutrinos}
\label{subsubsec:neutrinos}

As already noted by \citet{morrison1962InterstellarCommunication}, neutrino communication is a plausible way to engage in interstellar communication.
Neutrinos are nearly massless, electrically neutral particles that have extremely low interaction cross-sections with matter.
As such, they go through many layers of the Sun's composition, rarely absorbed by even the densest core layers.
In relation to searching for probes in the solar system, this implies that neutrino radiation converges through a solar gravitational lens effect at 29.6~AU \citep{maccone2009DeepSpace}, so searching for probes in this area is much nearer and feasible than searching for a photon-based probe at 547~AU (see again section \ref{subsec:SGL} and Fig. \ref{fig: SGL}).

Neutrinos are produced in nuclear fusion reactions, high-energy astrophysical phenomena such as supernovae and active galactic nuclei (AGN), and nuclear reactors.
These high-energy particles have been proposed for global communications with submarines, planetary-wide communications, and interstellar communications \citep{stancil2012DemonstrationCommunication, huber2010SubmarineNeutrino}.
Their extremely low interaction cross-section makes them good candidates for interstellar communication, since they rarely interact with matter: they can travel through interstellar dust, gas clouds, planetary and stellar objects, or even strong magnetic fields surrounding pulsars and neutron stars with negligible attenuation.
In other words, neutrino emissions are immune to common interstellar communication issues like dispersion, scattering, absorption, or polarization rotation (problems prevalent with electromagnetic signals).
This enables neutrino signals to propagate across interstellar distances while maintaining coherence and fidelity.

Natural neutrinos typically exhibit broad spectra and random arrival directions.
In contrast, artificial neutrino signals from extraterrestrial civilizations could be designed to appear distinctly structured (e.g., coherent, pulsed signals with specific temporal patterns or unusual energy distributions, or another modulation scheme that we use with EM frequencies, see Table \ref{tab:modulation}).
Neutrinos communication could be modulated according to a specific neutrino flavor (electron, muon, or tau); or by switching between neutrinos and antineutrinos.

They persist undistorted over timescales spanning millennia or even millions of years, which means that they would provide a reliable beacon for ET civilizations separated across vast temporal scales.
The generation and detection of neutrinos on a directed, coherent scale requires enormous energy resources and technological advancements far beyond current human capability.
At high energies, neutrinos are rare from any given direction, and in contrast to photons, they are essentially noise-free in the galactic plane where galactic civilizations may be located \citep{learned2009GalacticNeutrino}.

\citet{learned1994TimingData} also conducted a theoretical exploration of neutrinos as a high cadence timing standard.
However, the location of neutrino sources is hard to pinpoint, and timing and navigation require many stable sources.

\subsubsection{Gravitational Waves (GW)}
\label{subsubsec:grav_waves}

We saw that the energy of electromagnetic radiation follows an inverse-square law ($1/r^2$) in intensity as it propagates through space.
However, even if the energy flux of gravitational waves decays as $1/r^2$, the detection of gravitational waves is done through measuring gravitational wave amplitude ($h$), representing strain (relative changes in distances), and diminishes only as $1/r$.
\citet{wright2026SearchExtraterrestrial} also notes that radio telescopes cannot measure the $1/r^2$ power amplitude because one does not know the waveform of radio signals in advance, as we know the waveform of gravitational waves.

Much like neutrinos, gravitational waves also offer the advantage that they propagate virtually unaffected through ordinary matter \citep{wang2025GravitationalCommunication}.
Unlike electromagnetic signals, gravitational waves interact extremely weakly with dust, gas, magnetic fields, or charged plasmas in the interstellar medium, making them ideal for penetrating dense environments without signal degradation.

\begin{table*}[ht]
\centering
\caption{A comparison of the advantages and disadvantages of electromagnetic waves versus gravitational waves for interstellar communication.}
\label{tab:em_vs_gw}

\begin{tabular}{>{\raggedright\arraybackslash}p{0.1\linewidth}>{\raggedright\arraybackslash}p{0.32\linewidth}>{\raggedright\arraybackslash}p{0.32\linewidth}}
\hline
Feature & Electromagnetic Waves (Radio/Laser) & Gravitational Waves \\
\hline
Obstacles & Blocked/absorbed by stars, planets, dust & Pass through everything unimpeded \\
Noise     & High (stars, CMB, dust, electronics)     & Extremely low (at high frequencies) \\
Jamming   & Easy to jam or shield                    & Nearly impossible to jam or shield \\
\hline
\end{tabular}
\end{table*}

Gravitational wave communication offers a lower natural background noise level.
For example, electromagnetic channels (radio, optical, infrared) are crowded with natural and artificial sources of interference (stars, interstellar medium, cosmic microwave background, human-made signals).
In contrast, GW channels have comparatively lower natural background noise in certain frequency ranges, especially in the high-frequency regime above stellar binaries and below cosmological GW backgrounds (see Table \ref{tab:em_vs_gw}).

The primary reason we do not use this technology yet is the generation problem: gravity is the weakest of the four fundamental forces, $10^{40}$ times weaker than electromagnetism.
To generate a detectable gravitational wave currently requires accelerating masses the size of black holes.
To use this for communication, a civilization would need to master High-Frequency Gravitational Wave technology, which requires energy scales far beyond our current capabilities.
However, \citet{jackson2018GravitationalWave} proposed a concept for a gravitational wave transmitter by a civilization using small black holes.
\citet{abramowicz2020GalacticCentre} argued that a Jupiter-mass GW emitter at the center of our galaxy would make sense to be built and would be detectable by LISA-type GW detectors.

As with other carriers, artificial GW signals, if engineered at characteristic frequencies or modulations, could stand out distinctly from natural sources.

\subsubsection{Inscribed matter}
\label{subsubsec:inscribed_matter}

If the desired data transfer is very high, inscribed matter may be a better choice than sending information in the EM spectrum \citep{rose2004InscribedMatter, hippke2018BenchmarkingInscribed, lingam2021LifeCosmos}.
Think of slow postal services versus fast internet transfer.
However, what is faster to transfer 2 petabytes of data?
Maybe counterintuitively, sending a box with one hundred 20~TB hard drives to the other side of the world goes faster than attempting to transmit it via the internet, even with the best optical fiber consumer-grade connection.
An adage by Andrew Tanenbaum in computer science illustrates this: ``Never underestimate the bandwidth of a station wagon full of tapes hurtling down the highway.''

This train of thought involves a tradeoff between the desired latency versus bandwidth.
As argued by \citet{rose2004InscribedMatter}, this advantage provides additional strong arguments to look for solar system artefacts that might be extremely information-rich (see section \ref{sec: ss ts}).

Also, inscribed matter does not need the recipient to be listening at the exact moment when the data is being transmitted.
A civilization could discover a message with inscribed matter millions of years after it actually arrived, much like in the famous science fiction novel \textit{2001: A Space Odyssey}.

We saw that the inverse square law makes the data rate of EM radiation fall off proportionally to $1/r^2$.
How does it fall off for inscribed matter?
Consider that the total bits delivered with one package of inscribed matter is a constant ($B$), the time ($t$) it takes to arrive is simply distance ($r$) divided by velocity ($v$), i.e.\ $t=r/v$.
Therefore, the data rate (bits per second) is:
\begin{equation}
\text{Data Rate} = \frac{B}{t} = \frac{B}{r/v} = \frac{Bv}{r}
\end{equation}
Because the distance ($r$) is in the denominator, the data rate for inscribed matter falls off proportionally to $1/r$.
Naturally, the equation shows that the velocity $v$ at which the package is sent matters too.
This means that EM radiation degrades much faster (inverse-square law, $1/r^2$) due to beam diffraction, or requires exponential energy increases to maintain the rate.
To sum up, inscribed matter is favored over longer distances, unless the inscribed matter is travelling at a low velocity, or unless the inscribed matter device is not very information-dense.

From the perspective of a civilization broadcasting, inscribed matter also has many advantages.
Indeed, for a civilization sending EM photons, it only gains information if it sends messages, receives a reply, and is able to decode it.
By contrast, a civilization that sends a probe could have a hybrid EM and inscribed matter strategy.
Indeed, a probe could host massive amounts of inscribed information onboard, but could simultaneously carry antennas that could engage in real-time conversations with local lifeforms, and could also transmit information back to the host.
Even if no ETI is found by the probe, it could still serve as a potent galactic scouting strategy.

\subsection{What signal modulation scheme?}
\label{subsec:modulation}

We already dealt with the questions of choosing targets in space, time and frequency, but what remains is the question of the modulation scheme.
This aspect is important not only for the goal of deciphering a message, but also because different modulation schemes imply different communication choices and thus search strategies.
A communication protocol is a system of rules that allows two or more entities of a communications system to transmit information.
The protocol defines the rules, syntax, semantics, and synchronization of communication and error recovery methods.
Even if we guess the right frequency and information carrier, to conclusively determine if a signal is artificial, we cannot assume that it would always be a simple carrier wave, and we need to dig into possible and suitable modulation schemes.

At first sight, we might want to systematically explore the whole space of possible modulation schemes, at least the ones known to humanity \citep[for a textbook see e.g.][]{haykin2009CommunicationSystems}.

Delay-Tolerant Networking protocols, already mentioned in the context of speed-of-light delays are particularly designed for space applications \citep{cerf2007DelaytolerantNetworking, burleigh2003DelaytolerantNetworking}, and have also been successfully tested on the Moon \citep{wang2025DelayDisruption}.

The most common way engineers implement modulation is by varying the amplitude (e.g.\ AM Radio), the frequency (e.g.\ FM radio), or the phase (phase modulation) of a carrier wave.
Modern modulation solutions vary both amplitude and phase, a method known as Quadrature Amplitude Modulation (QAM).
These modulations can be done both in analog or digital settings (see Table~\ref{tab:modulation}).

\begin{table*}[ht]
\centering
\caption{Examples of common analog and digital modulation techniques. 
These techniques aim to encode information by varying parameters of the basic equation of a sinusoid carrier wave: $x(t)=A\cdot\sin(2\pi ft+\phi)$, i.e.\ varying either amplitude, frequency, phase, or both amplitude and phase in the case of QAM.}
\label{tab:modulation}
\footnotesize
\begin{tabular}{>{\raggedright\arraybackslash}p{0.22\linewidth}>{\raggedright\arraybackslash}p{0.23\linewidth}>{\raggedright\arraybackslash}p{0.2\linewidth}>{\raggedright\arraybackslash}p{0.23\linewidth}}
\hline
Analog Modulation & Digital Modulation & Parameter Changed & Behavior \\
\hline
AM (Amplitude Modulation) & ASK (Amplitude Shift Keying) & Amplitude ($A$) & Changes signal strength \\
FM (Frequency Modulation) & FSK (Frequency Shift Keying) & Frequency ($f$) & Changes oscillations per second \\
PM (Phase Modulation)     & PSK (Phase Shift Keying)     & Phase ($\phi$)  & Changes the starting angle of the wave \\
Analog QAM & Digital QAM & Amplitude ($A$) \& Phase ($\phi$) & Changes both strength and starting angle simultaneously \\
\hline
\end{tabular}
\end{table*}

Another engineering consideration is the requirement for robustness.
A common solution in modern digital wireless engineering is to use several parts of the EM spectrum, a technique known as \textit{spread spectrum}.
It is actually so common in communication engineering that \citet{messerschmitt2012InterstellarCommunication} argued that we should search for such modulation patterns in SETI.

\subsubsection{Polarization}
\label{subsubsec:polarization}

In astronomy, light polarization refers to the geometry of the electric field of the EM wave.
There are multiple ways of characterizing an EM wave's polarization, including elliptical, linear, and circular.
That polarization can be stable (fully polarized), unstable and random (unpolarized), or in between (partially polarized).
Polarization constitutes a continuous degree of freedom, which can be characterized through the Stokes parameters \citep{stokes1851CompositionResolution}.
In optical communication engineering, polarization states can be modulated continuously through this state space (also commonly represented as a Poincaré sphere).
Polarization could thus be used as an additional degree of freedom in interstellar communication, although it has rarely been considered in SETI.

One can even go further and combine quantum communication with the Stokes parameters, which gives a quantum version of the parameter space, and in that case we speak of Stokes operators.
The polarization on the Poincar\'{e} sphere is then represented by cells instead of points, due to the Heisenberg uncertainty principle.
This opens even more channel capacity, up to 33\% compared to coherent beam combining, with a squeezed polarization technique \citep{schnabel2003StokesoperatorsqueezedContinuousvariable}.

Polarization is associated with \textit{Spin Angular Momentum} (SAM).
However, there is another way to encode much more information with light, using \textit{Orbital Angular Momentum} (OAM), by modulating the shape of the wavefront and creating vortex beams \citep{allen1992OrbitalAngular}.
It has been shown that a wireless network using such encoding could achieve 2.5~Tb~s$^{-1}$ \citep{wang2012TerabitFreespace}.
In the context of SETI, \citet{hippke2021SearchingInterstellar} noted that OAM degrades more than $1/r^2$ over interstellar distances, as beam divergence increases at least with the square root of the number of OAM modes.
OAM also occurs in natural sources such as our Sun or rotating black holes, but an artificially generated OAM wavefront could have distinguishing features, a topic not yet explored in a SETI context.

Classical communication uses bits, but quantum communication uses qubits, which could also be used for interstellar communication \citep{hippke2021SearchingInterstellar, berera2022ViabilityQuantum}.
Given that we might be just decades away from generalized quantum communication and quantum computers, this axis of research should be considered seriously.
Indeed, as \citet{hippke2021SearchingInterstellar} argues, quantum communication (1) might be a way to gate-keep less advanced civilizations who do not know how to use it, (2) quantum computers can solve some problems faster than classical computers (quantum supremacy), (3) quantum key distribution and cryptography are very secure and enable the detection of eavesdropping, and (4) it offers higher information efficiency.

However, qubits are just one way to use quantum properties to transmit information.
In quantum communication, the concept of \textit{qudit} is used, a quantum $d$-ary digit that is a unit of quantum information with $d$ levels.
This means that a qubit is a particular case where $d=2$, but other systems can be built with much larger values of $d$; for example, a 25-dimensional qudit was demonstrated recently \citep{dong2023HighlyEfficient}.

\subsubsection{Exotic protocols}
\label{subsubsec:exotic_protocols}

All of these considerations might also be outdated or inferior to other more advanced, exotic communication solutions.
An exotic communication protocol might use properties of both quantum and relativity theories.
On the quantum side, a major property of quantum systems is that they can display entangled states, also known as Einstein Podolsky Rosen (EPR) correlations \citep{einstein1935CanQuantumMechanical}.
On the relativity side, wormholes were first proposed as an Einstein-Rosen (ER) bridge \citep{einstein1935ParticleProblem}.

\citet{maldacena2013CoolHorizons} formulated the ER=EPR conjecture, proposing that an ER bridge is the geometric manifestation of EPR entanglement.
More specifically, two black holes would be entangled via a wormhole.
Crucially, faster-than-light communication would still be forbidden, but this geometry, with the addition of a classical channel, might be a way to implement the quantum teleportation protocol.

Another exotic possibility would be the ability to harness dark matter or dark energy as an energy source or for communication, which has been evoked by \citet{garrett2025BlinkYoull}, although the possible anomalies or technosignatures have not been elaborated yet.

Many thinkers have argued or assumed that ETI would converge towards one particular solution for interstellar communication.
If taken too dogmatically, these ways of restricting the search space might create lock-in situations, where alternatives would be forgotten.

To summarize, if we combine our insights, the ultimate communication solution for two advanced civilizations might be an X-ray laser modulated hybridly with quantum and classical protocols, including polarization and orbital angular momentum encoding, and pulsed at the millisecond timescale (to be congruent with millisecond pulsars for galactic spacetime timestamping).
As of today, no one has searched for such signals.

\subsection{What is the communication transmission mode and purpose?}
\label{subsec:transmission_mode}

We distinguish four main communication network topologies: a beacon broadcast (one-to-all); point-to-point unicast (one-to-one); multicast (one-to-many); and many-to-one (e.g.\ a warning signal).
Note that an intentional communication to Earth would be a particular case of a one-to-one communication, and that there are many multicast options: any given communicating civilization might communicate with $n$ other civilizations.

As we argued earlier, the issue of inverse square attenuation with distance ($1/r^2$) would arguably be counteracted with a relay network that repeats or re-amplifies signals.
Searching for such a relay network is thus also a valid strategy, e.g. by trying to find similar signals at different intensities.

\begin{sidewaysfigure*}
    \centering
    \includegraphics[width=1\linewidth]{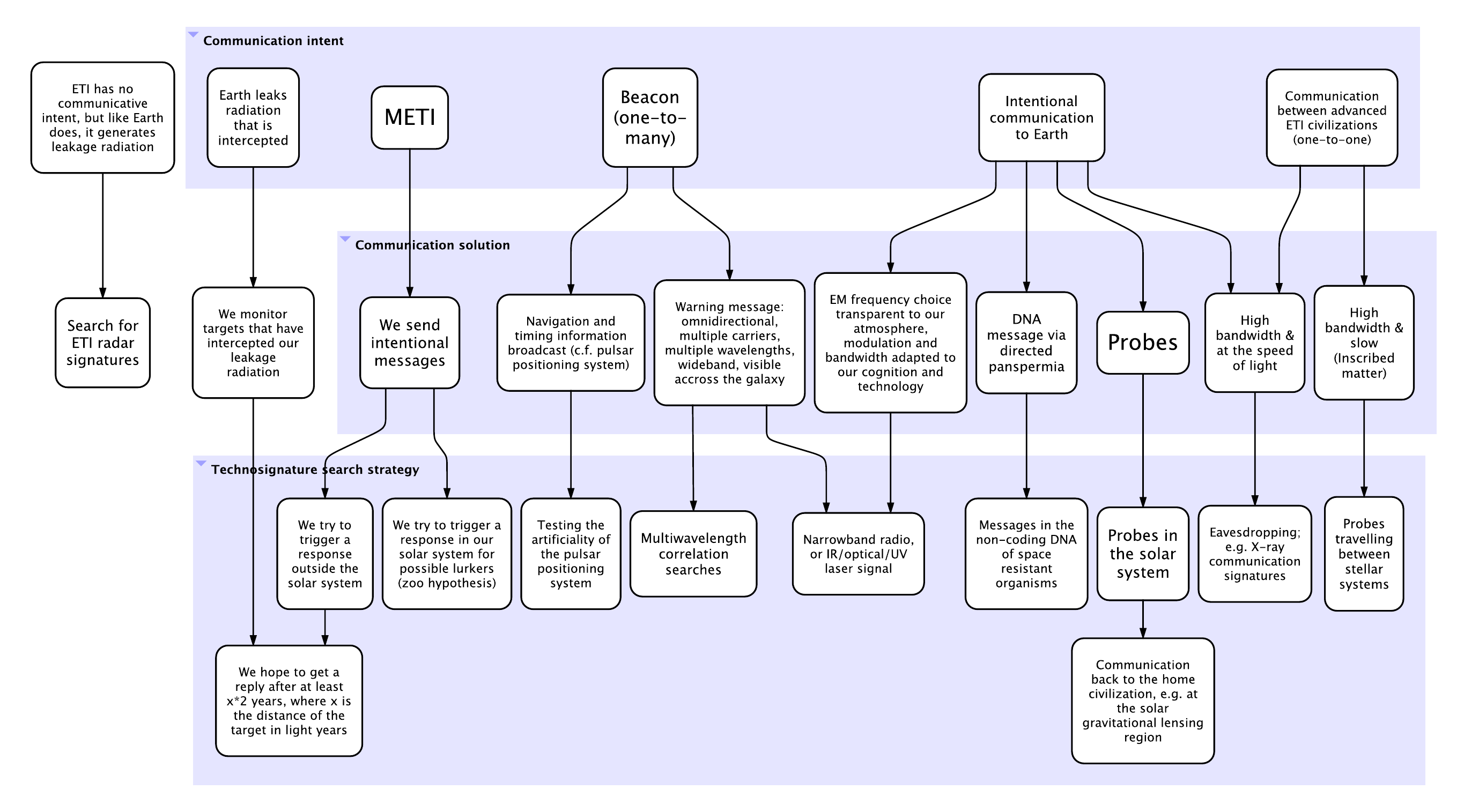}
    \caption{From communication intents to communication solutions, and their associated technosignature search strategies.}
    \label{fig: interstellar_TS_intent_topology}
\end{sidewaysfigure*}

These communication network topologies imply different preferred modulation schemes, and thus different kinds of search strategies that we can employ to detect them (see Fig. \ref{fig: interstellar_TS_intent_topology} for an illustration).

\subsubsection{One-to-one}
\label{subsubsec:one_to_one}

A direct one-to-one communication could take several forms.
A common scenario coming from popular culture is a direct communication to Earth, with a high signal/noise ratio, adapted to our cognition/technology, and anti-cryptographic, so humans are able to decode it.
But one-to-one communication could also happen between two communicating ETIs, and for us to detect this case, we would be eavesdropping on them.
The decoding step (5) in Shannon's diagram (Fig. \ref{fig:InterstellarTS_shannon}) would most likely be nearly impossible to achieve, but we may still be able to detect that communication is going on.
The more secure and secret the communication channel is, the more the signal would look like noise, and thus the harder it would be for us to receive and decode it.
Highly compressed signals also look like noise, which is deeply problematic for SETI \citet[p.~836]{wolfram2002NewKind}.

\subsubsection{One-to-many}
\label{subsubsec:one_to_many}

A beacon broadcast targeting dense stellar regions such as the galactic center or globular clusters would constitute a one-to-many communication.
As a one-to-many example, the Arecibo message was targeted to a globular cluster \citep{staffatthenationalastronomy1975AreciboMessage}.
Expected features of a one-to-many communication could include: a strong signal to cover large distances, long-lasting, isotropic, multiwavelength, and anti-cryptographic.
Interestingly, over long operational timescales, broadband pulsed beacons are more energetically efficient compared to narrowband beacons, which is another reason to also search for broadband communication \citep{gajjar2022SearchingBroadband}.

A beacon could also be used as a warning message, and if it were to warn the whole galaxy, it would be an instance of a one-to-all communication effort.
In that case, much as warning messages on Earth use multimodal solutions (sound sirens, radio, TV, internet, etc.), we might detect interstellar warning signals via a correlation across wavelengths or communication channel options.

\subsubsection{Many-to-one}
\label{subsubsec:many_to_one}

The topology of many-to-one requires specific and speculative assumptions, where several civilizations would want to warn us with a similar message.
For example, a warning of a bad event such as a supernova risking to wipe life from planet Earth.
Note that it would be a very peculiar impact scenario: it would be climactic for SETI as a field, and anticlimactic for humanity.

\subsubsection{Many-to-many}
\label{subsubsec:many_to_many}

A many-to-many communication system could include a distributed system analogous to Global Navigation Satellite Systems on Earth, which actually has a direct analog as the pulsar positioning system \citep{vidal2019PulsarPositioning}.
An amplification system made of distributed relays is also a candidate.

\subsubsection{No communication intent}
\label{subsubsec:no_communication_intent}

Note that signals may be detectable outside of the possibilities of these communication network distinctions.
Indeed, humanity does leak radiation in space, through military radars or the Deep Space Network.
\citet{fan2025DetectingExtraterrestrial} has proposed ways to search for such Earth analogs of leakage radiation as a SETI search strategy.
Assuming a solar-system spanning civilization, \citet{sneed2023SearchSpillover} proposed to search for overhearing spillover of signals between planets, much like the Deep Space Network leaks powerful signals as we communicate with space missions in our solar system.

Some communication solutions are close to travel signatures (e.g.\ sending probes or inscribed matter), which will be the topic of the next section \ref{sec: travel ts}.

\subsection{A cosmic haystack}
\label{subsec:cosmic_haystack}

We can summarize the main dimensions of the search for signals by mapping the ``cosmic haystack problem'' \citep{wright2019OriginTerm}.
In Table~\ref{tab:cosmic_haystack} we compile core dimensions we reviewed that are relevant to the search.
One should note that each time one adds a dimension to the search space, this leads to an exponential growth in the search space volume, a phenomenon known as a ``curse of dimensionality.''

\begin{table*}[ht]
\centering
\caption{A cosmic haystack: interstellar messages might be embedded in various dimensions. Partially inspired from \citet{hippke2018BenchmarkingInformation, tarter2010SETITurns, wright2018HowMuch}. Note that it is ``a'' haystack and not ``the'' haystack as we might be missing dimensions.}
\label{tab:cosmic_haystack}
\begin{tabular}{>{\raggedright\arraybackslash}p{0.2\linewidth}>{\raggedright\arraybackslash}p{0.35\linewidth}>{\raggedright\arraybackslash}p{0.35\linewidth}}
\hline
Dimension & Options & Notes \\
\hline
Target in space & Targeted searches; wide-field surveys; archival data mining & The three options are complementary and are regularly analyzed by various studies. \\[4pt]
Target in time & Various timescales from Planck timescale to billions of years (see Table~\ref{tab:timescales}) & Time dimensions address: (1) At what particular time? (2) How long? (3) How often? \\[4pt]
Information carrier & Photon; Occulter; Neutrino; Gravitational waves; Inscribed matter; Electron; Proton; Neutron; Muon; Tau; Higgs & Photons are still a premium choice. See Table~\ref{tab:carrier_performance} and \citet{hippke2018BenchmarkingInformation}. \\[4pt]
Frequency & Radio, Optical, X-ray, etc. & Radio has been the most searched. \\[4pt]
Frequency range & Narrowband, wideband, broadband & Narrowband signals have been most searched for. \\[4pt]
Sampling rate & Frequency intervals chosen for the detection algorithms (e.g.\ every 1~Hz) & Also often referred to as ``channels'' in radio astronomy. \\[4pt]
Modulation \& encoding scheme & Analog: AM, FM, phase modulation; Digital: spread spectrum, pulse-amplitude modulation, etc.; Quantum communication modulation techniques; Polarization; Orbital Angular Momentum (vortex beams), see Table \ref{tab:modulation}. & Polarization provides degrees of freedom that could be used to encode more information. Quantum communication is an underexplored frontier in SETI.\citep{hippke2021SearchingInterstellar}. \\[4pt]
Transmission modes and purposes & One-to-one; One-to-many (beacon); Many-to-one; Many-to-many (e.g.\ warning signal) & \\
\hline
\end{tabular}
\end{table*}

\addtocontents{toc}{\protect\pagebreak}
\clearpage
\section{Travel Technosignatures}\label{sec: travel ts}
\subsection{Motivations}
\label{subsec:travel_motivations}

Why would a long-lived civilization even engage in interstellar travel?
After all, interstellar distances are immense, travel times are very long, and require significant economical and political investment to secure the required technology and energy.
In addition, travel is often associated with leisure or the luxury of exploration for its own sake.
However, two universal evolutionary motivations will make interstellar travel a necessity for any long-lived civilization: \textit{survival} and \textit{reproduction} \citep[see][]{vidal2024SpiderStellar}.

Survival motivations include avoiding a death threatening supernova or migrating towards a nearby star as the home star fades away \citep{zuckerman1985StellarEvolution, hansen2021MinimalConditions}.
A pioneering study by \citet{hansen2022UnboundClose} looked for close stellar encounters in the solar neighborhood.
The strategy is then to look for active interstellar migration, where ``generation ships'' are sent during a close encounter window, hitting this window because it would cost orders of magnitude less time and energy than crossing the otherwise vast interstellar spaces.
Hansen proposes this method as a way to constrain search targets because a lot of heat or communication signatures might be associated with such migration.

By a similar logic, two civilizations inhabiting two planets orbiting each star of the same binary system (S-Type) would have a strong incentive to visit or communicate with each other.
Such systems are thus good targets for technosignature searches, and a first candidate system, TOI-2267 has recently been discovered \citep{zuniga-fernandez2025TwoWarm}.
Note that this is distinct from a circumbinary configuration where planets orbit the center of mass of two stars (P-Type).

To avoid frequent migrations, an advanced civilization might also choose to target high stellar density environments such as globular clusters or the galactic center.
If this is true, it implies that we can expect more travel signatures from and towards GCs and the galactic center.

The reproduction motivations may include the settlement of a second planet in a given extrasolar system.
This would arguably imply intense traveling routes between the two inhabited planets to exchange goods and materials.

Another possibility is directed panspermia, which was proposed by \citet{crick1973DirectedPanspermia}.
This is plausible as life is space resistant \citep{olsson-francis2010ExperimentalMethods}.
It is actually so much space resistant that it is extremely hard and expensive to confidently sterilize any space mission.
This field of study is called \textit{planetary protection} \citep{coustenis2023PlanetaryProtection}, with the objective to avoid contaminating solar system targets from Earth-life (forward contamination), but also to prevent potential extraterrestrial microorganisms from entering the Earth (backward contamination).

The question of whether we ourselves should engage in a directed panspermia program is another debate that raises ethical questions, much like the METI debate but this time sending matter and life, and not only electromagnetic radiation.
We should note that in this context the meaning of ``directed panspermia'' makes a 180 degree turn, i.e.\ the idea that humans would send life in space \citep{mautner1979DirectedPanspermia}.

Directed panspermia initiated by humans is not just an abstract idea, as the Lunar lander ``Beresheet'' \citep{aharonson2020ScienceMission} sent information to the Moon: the Wikipedia encyclopedia, the PanLex database, and an artefact designed by the Long Now foundation called the Wearable Rosetta disc.
More controversial was the decision to add tardigrades as part of the launch of the lander \citep{oberhaus2019CrashedIsraeli}.

Search strategies for interstellar travel technosignatures are obviously very similar to the search for ISO technosignatures (see Table \ref{tab:Solar_System_TS_ISO_obs}).
The difference being that some cases, such as radar searches, are much more limited because of the huge distances involved.
The observational time window of travel technosignatures might also be narrow, visible only during maneuvers like accelerating, decelerating, or direction changes.
Still, some general strategies and anomalies to look for have to do with the origin and destination of spacecraft.
Indeed, if we were to find a high velocity object whose extrapolated trajectory begins and ends at precisely two stars, it would be extremely unlikely to be a natural occurrence.
Such studies are possible in principle by sifting through the Gaia mission catalog \citep{prusti2016GaiaMission}.

This strategy works to detect the longest cruising phase of interstellar travel but fails while a spacecraft is in a maneuvering phase near either the point of departure or destination, or during some intermediary gravitational assist.
Other anomalous trajectories include three body interactions that generate high-velocity ejections, as explored in detail by \citet{breeden2013GravitationalAssist, breeden2014SETIImplications}.
Another logical signature to look for is an interstellar contrails \citep{kitajima2023OriginNarrow}.
Indeed, it can be seen as an instance of a waste signature, that has the advantage of being particularly extended in space, and thus detectable.
We could look more specifically whether candidate contrails are going towards or away from habitable stars or planets.

\citet{zubrin1995DetectionExtraterrestrial} studied the radio spectral signatures of advanced interstellar spacecraft, and concluded that the most detectable form of starship radiation would be low frequency radio emissions of cyclotron radiation caused by interaction of the interstellar medium with a magnetic sail.

Independent of the propulsion technology, at extreme velocities, a relativistic spacecraft would interact with CMB photons, which would produce specific signatures \citep{yurtsever2018LimitsSignatures}.
These may be interpreted as technosignatures if not associated with other natural inverse-compton radiation such as Sunyaev-Zeldovich effect in cosmology.


\subsection{Energy sources for propulsion}
\label{subsec:energy_sources}

Considering any spacecraft relying on onboard fuel, using a high energy density source of fuel is almost a prerequisite for interstellar travel.

In comparison to chemical rockets, a nuclear fission source has $\sim10^{6.8}$ more specific energy, nuclear fusion $\sim10^{7.4}$ and the absolute theoretical maximum, matter-antimatter annihilation is $\sim10^{9.8}$ times as much.
This means that crossing the threshold from chemical rockets to nuclear fission propulsion leads to a gain of almost 7 orders of magnitude, while going from fission to fusion is a gain of less than one order or magnitude, and from fusion to matter-antimatter annihilation means 'only' gaining 2 orders of magnitude.

\cite{harris1986DetectabilityAntimatter} argued that matter-antimatter propulsion would produce a contrail of neutrinos that may be detectable.
Also pushing the theoretical limits, \citet{crane2009AreBlack} made a theoretical exploration of a spacecraft powered by an artificial black hole.
They derived a technosignature consisting of gravitational radiation emitted from such a spacecraft  of wavelengths at nuclear scales. 
This would require major advances in gravitational wave detector sensitivity.

Other technosignatures might be derived from work on advanced propulsion methods, for example, the edited book by \cite{millis2009FrontiersPropulsion} explores frontiers propulsion science and proposals, from nuclear propulsion to exploiting vacuum energy, and many others.

As we mentioned, the core hard physics constraint to interstellar travel is known as the \textit{tyranny of the rocket equation}: A spacecraft only carries a certain amount of weight.
The further one wants to travel, the more propellant one has to carry.
However, doing so increases the overall weight, thus also increases the propellant needed to accelerate it: a tyrannous feedback loop.

\subsection{Light sails}
\label{subsec:light_sails}

One way to bypass the tyranny of the rocket equation is to propel a spacecraft not with onboard fuel, but by beaming directed energy to it \citep{lubin2022PathTransformational}.

\begin{figure}[h!]
    \centering
    \includegraphics[width=0.95\linewidth]{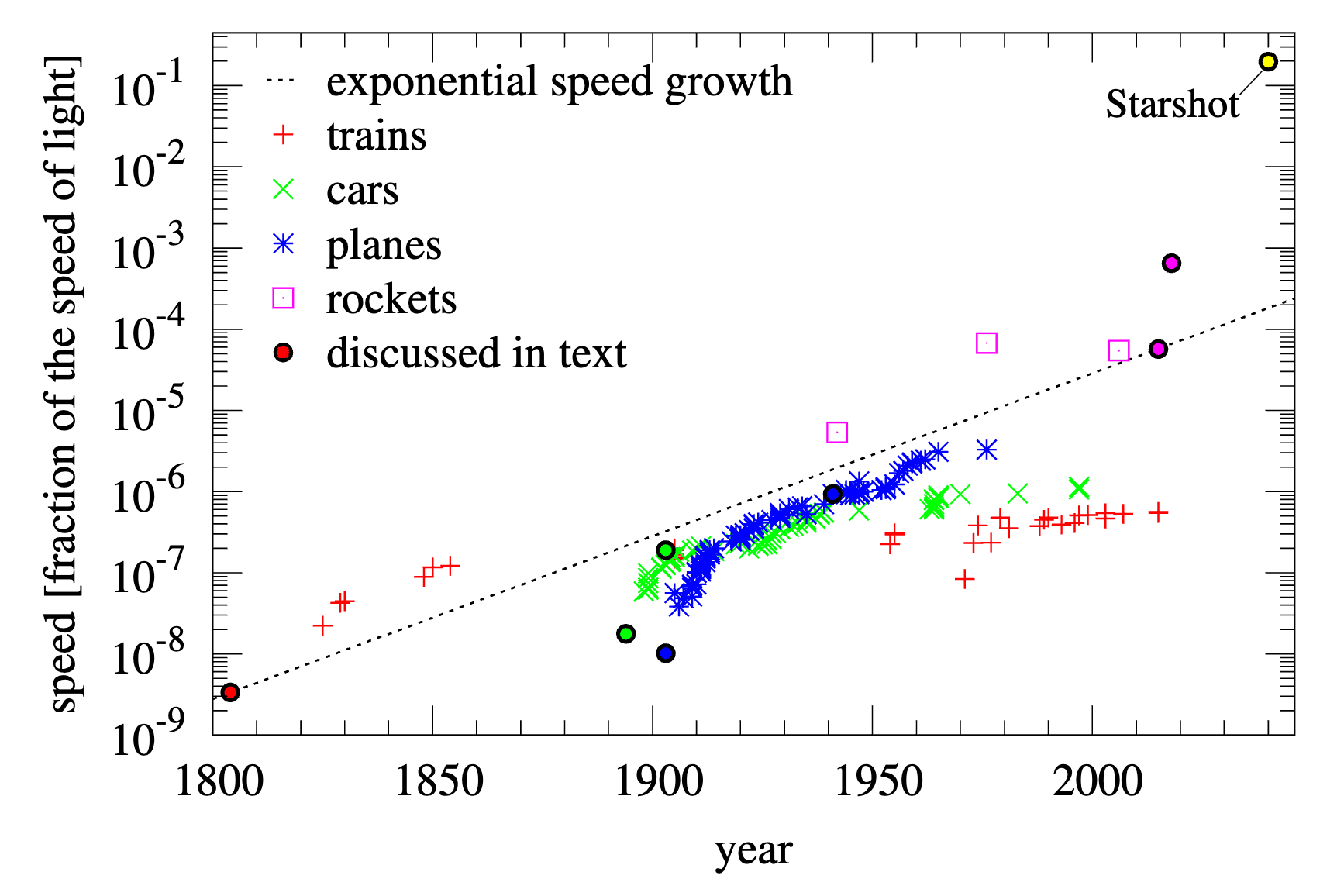}
    \caption{Historical speed records of vehicles or probes, with the Starshot project in the upper-right. 
    Fig. from \citet{heller2017RelativisticGeneralization}.}
    \label{fig: historical rocket speed}
\end{figure}

A recent example is the Breakthrough Starshot program \citep{parkin2018BreakthroughStarshot}, that explored the engineering, scientific and economical constraints and challenges of launching probes towards Alpha Centauri.
Unfortunately, the project has recently been discontinued \citep{scoles2025VoyageNowhere}.
However, its success would have meant breaking a trend of an already exponential growth of maximum speed of vehicles (see Fig. \ref{fig: historical rocket speed}).

As \citet{heller2017RelativisticGeneralization} noted, reaching $0.1c$ would not happen before 150 years from now, assuming this exponential growth continues unabated.
In that sense, the project might have been a few centuries ahead of its time!

Of course, even if we do not use directed energy propulsion solutions yet, ETI might already be doing it.
This reasoning can be called the \textit{cosmic reversibility principle} that \citet{crick1973DirectedPanspermia} proposed in the context of directed panspermia, and that was generalized in \citep{vidal2019PulsarPositioning}:

\begin{quote}
``If we are capable of doing X in the future, then, given that the time was available, another extraterrestrial civilization might well have done X already.''
\end{quote}

This principle is often at play implicitly in many SETI speculations: whatever technological project we can do now or in the near future, ETIs might already have done it better, more efficiently or at a larger scale.
As we saw in the context of Earth Technosignatures (section \ref{sec: earth ts}), there is ample time for this to happen given the age of the galaxy.
So, by attempting more and more ambitious space or interstellar travels ourselves \citep[see e.g.][]{mallove1989StarflightHandbook}, we will have better and better ideas about how ETIs might be doing it.

In the case of directed energy, \citet{lubin2016SearchDirected} proposed a general search for directed energy beaming activities, and \citet{guillochon2015SETILeakage} proposed to look for leakage from a light sail spacecraft traveling between planets of a given stellar system.
This search can be done in synergy with optical laser SETI searches.
However, note that if the beam matches perfectly with the size of the sail, then there is no leakage to detect, so we would be looking for leakage from a system designed to minimize it, which may be hard.

\citet{lingam2017FastRadio} also explored a directed energy signature, namely that Fast Radio Bursts (FRBs) might actually be powering light sails, arguing that the properties of FRBs are particularly well-suited for this purpose.
One might also try looking for the light sail itself; for example, when Avi Loeb argued that `Oumuamua might be a light sail, or if the next generation coronagraphs such as the forthcoming HWO would be used to detect reflected light from huge light sails, possibly with similar strategies to look for starshades \citet{skoglund2025StarshadesTechnosignatures}.

Clifford E.\ \citet{singer1980InterstellarPropulsion, singer1981QuestionsConcerning} proposed an analogous solution to directed energy that we could call \textit{directed mass}.
He proposed that a home world could send pellets at high speed that would bounce off a receiving area of the spacecraft and thereby transfer the momentum.
He argued that it would offer one to two orders of magnitude improvement over a fusion rocket in terms of propulsion system mass, reaction mass requirements, and power consumption.
We are not aware of associated observable technosignatures, but a stream of high-energy particles going out from a planet, hitting a spacecraft and heating it might be such a signature.

A few ultra-high energy cosmic rays (UHECRs) have been detected, such as the Amaterasu particle, or the "Oh-My-God" particle.
Cosmic rays are generally very hard to trace, but UHECRs are more promising to trace because their high kinetic energies makes them less susceptible to be re-routed by magnetic fields.
The acceleration to such energies produces synchrotron radiation which is a secondary signature from another modality.
In this perspective, this makes UHECRs at the boundary of extreme natural objects and technosignature candidates (see section \ref{subsec:anomaly}).
As with directed energy leakage searches, this search also suffers from an attempt to detect "failure modes", i.e. leakage from particles of a directed mass propulsion system.

\subsection{Ramjet}
\label{subsec:ramjet}

Another approach to avoid carrying all the fuel is to scoop material from the ISM and use it as fuel.
This is the proposal of the \textit{Bussard Ramjet} \citep{bussard1960GalacticMatter, fishback1969RelativisticInterstellar}.

Carl \citet{sagan1963DirectContact} explored the consequences of such a spacecraft and concluded that with accelerations and decelerations of only 1 g, the whole Galaxy would be accessible within a human lifetime of the traveler, due to relativistic time dilation.
Kevin \citet{knuth2019ConstraintsSocieties} has further explored the consequences and mindset of nomadic civilizations traveling at relativistic velocities.

Peter Schattschneider and Albert A. Jackson \citeyearpar{schattschneider2022FishbackRamjet} re-analyzed the ramjet design and found not only that travel within a human lifetime would in fact not be possible, but also that the ramjet would require a 1 AU wide magnetic scooping area.
If such a megastructure is actually built, it makes the ramjet actually more observable.
We are not aware of specific observable signature proposals, but we could imagine that the combination of a 1-AU collecting ISM structure, a high-velocity or accelerating object, leaving behind a fusion exhaust and a void in the ISM would be a set of intriguing properties.

\subsection{Planet engines}
\label{subsec:planet_engines}

The more payload a spacecraft can carry, the longer its crew can survive interstellar space.
The limit to this logic is to travel with one's planet to form a \textit{planet engine}, or one's home star by designing a \textit{stellar engine}.

We already saw that \citet{ponnaganti2024MakingHabitable} proposed a mechanism to move planets within a stellar system thanks to high-powered laser.
The notion of a controlled planet engine for interstellar travel has been touched upon by \citet{romanovskaya2022CosmicHitchhikers} who considered ejected rogue planets as potential technosignatures.
Indeed, an active iron core could provide energy for a long time, even for a rogue planet without a stellar energy source.
A technosignature search strategy could try to examine the trajectory of such rogue planets, and see whether they are directed away and towards nearby stars.

Another example of an observable technosignature from a planet-mass spacecraft would be when it accelerates or decelerates abruptly \citep{sellers2022SearchingIntelligent}.
Changes in velocity generate time-varying quadrupole moments, a source of gravitational waves (GWs).
Detectors in our Solar System would notice periodic GW pulses that would look artificial, being distinct from astrophysical merger chirps.
\citet{sellers2022SearchingIntelligent} concluded that a Jupiter-mass RAMAcraft (Rapidly Massive Accelerating craft) undergoing a change in velocity of 30 percent the speed of light would be detectable from 10-100 kpc. This range is decreased to 1-10 pc when the mass of the craft is lowered to the mass of the Moon.

\begin{figure}
    \centering
    \includegraphics[width=0.99\linewidth]{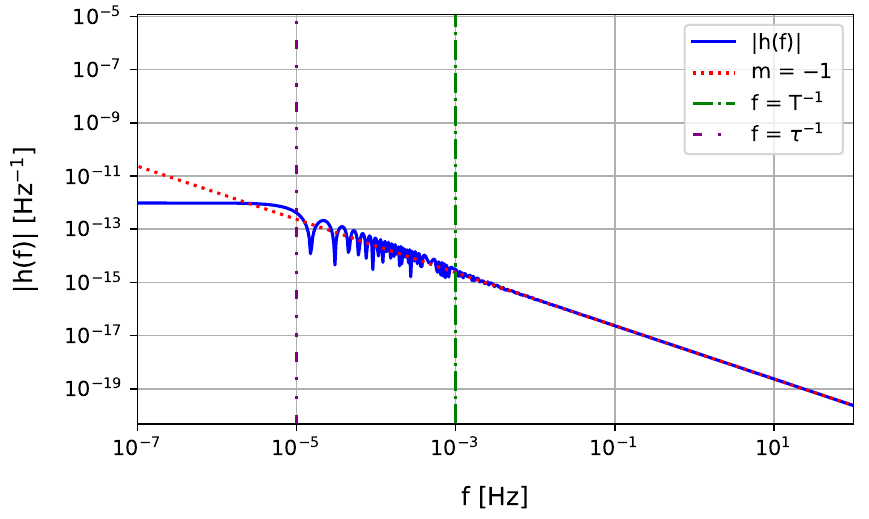}
    \caption{The gravitational wave power spectrum for an acceleration period of $T=10^3$s sampled for $10^5$s. Figure from \citep{sellers2022SearchingIntelligent}.}
    \label{fig: travelts_gw}
\end{figure}

As another example, a massively decelerating spacecraft approaching our solar system would also produce an unusual gravitational waveform (see simulation in Fig. \ref{fig: travelts_gw}).
The Fig. plots the gravitational wave spectrum observed for $\tau=10^5$s of a spacecraft accelerating for $T=10^3$s. This spectrum was derived by considering a lunar mass RAMAcraft at 1-10pc or a Jovian mass one at up to 100kpc, accelerating to $30\%$ of the speed of light.

The main technical challenge for detecting any gravitational wave signals from a massive decelerating spacecraft is the extremely small characteristic strains.
Future gravitational wave detectors are expected to achieve a characteristic strain of ($10^{-25}$) or better, as proposed for next-generation ground-based observatories like the Einstein Telescope and Cosmic Explorer.
This will be accomplished through new technologies such as longer arm lengths, laser stabilization, and quantum optics to increase sensitivity and overcome noise limitations.

\subsection{Stellar engines}
\label{subsec:stellar_engines}

After a planet engine comes the possibility of stellar engines that Type~II civilizations could drive.
The idea of stellar engines was pioneered by the great astronomer Fritz Zwicky (1957/\citeyear{zwicky2012MorphologicalAstronomy}, 260) who also considered using the Sun and planets as nuclear propellants.
The first detailed model of a stellar engine was proposed by \cite{shkadov1987PossibilityControlling}, where a giant parabolic mirror is held at fixed distance from a star in order to create thrust.
\cite{fogg1989SolarExchange} proposed another design building on Criswell’s \citeyearpar{criswell1985SolarSystem} mechanism to extract stellar mass for the purpose of extending the Sun’s lifetime.
More sophisticated designs and analytical studies were developed by \cite{caplan2019StellarEngines} and \cite{svoronos2020StarTug}, focusing on achieving high-acceleration capabilities.
\cite{vidal2024SpiderStellar} proposed the first binary stellar engine design, highlighting potential candidates in spider pulsars that may have anomalous velocities, trajectories, accelerations, or active steering signatures.

As a pioneering observational search of stellar engines, \citet{lingam2020ConstraintsAbundance} searched the Gaia catalog for hypervelocity stars, but without finding good candidates.

An advanced Type III civilization that has developed during billions of years might also want to preserve its galactic structure and integrity, and try to gain some control over the merging of its galaxy with another.
We know that our galaxy will merge with Andromeda, which will slowly but surely affect many stars, gas regions, before the merger stabilizes.
Galaxy pairs are actually common in the universe, and so are galaxy mergers.
The merging of two supermassive black holes will lead to about 5-10\% of their mass being converted to gravitational energy in the form of gravitational waves.
A Type III civilization could try to make the two SMBHs end up in a wide orbit to delay the ultimate merger, or to align their spin, this in order to avoid turning the resulting merger into an active quasar that might destroy many ETIs across a galaxy.

\subsection{Newtonian gravitation for propulsion}
\label{subsec:newtonian_grav}

\begin{figure}
    \centering
    \includegraphics[width=0.85\linewidth]{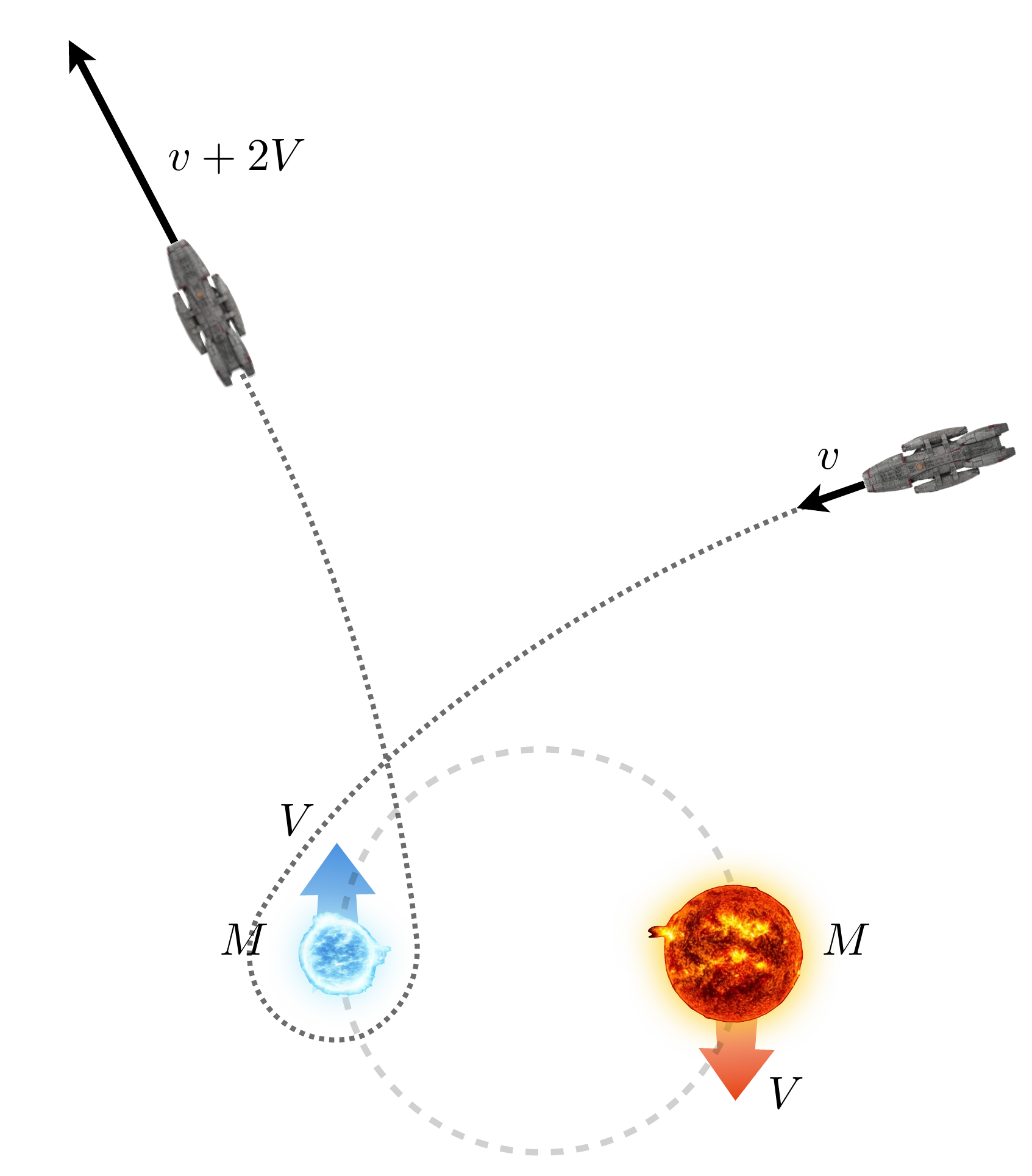}
    \caption{Illustration of a gravitational machine \citep{dyson1963GravitationalMachines} for accelerating spacecraft using binary star orbital energy. Diagram based on \citet[p.~141]{mallove1989StarflightHandbook}.}
    \label{fig: Travel_TS_Dyson_GRAV_MACH}
\end{figure}

A universal strategy to gain energy and momentum is to harness gravitational energy by doing gravity assists.
In this line of thinking, \citet{dyson1963GravitationalMachines} proposed the concept of a \textit{gravitational machine} which offers a mechanism for extreme acceleration.
As illustrated in Fig.~\ref{fig: Travel_TS_Dyson_GRAV_MACH}, a spacecraft approaches a tight binary of equal masses $M$ that orbit each other with speed $V$ about the barycenter.
The craft dives past one star on a near-parabolic swing-by so that, in the star's rest frame, its velocity reverses direction ($\approx 180^\circ$ turn).
Transforming back to the barycentric frame, that reversal adds roughly $2V$ along the star's instantaneous motion, so the outbound speed becomes $v + 2V$ in the ideal case.
The extra kinetic energy comes from the binary's orbital energy, exactly like a planetary gravity assist but with far faster ``primaries.''
With careful phasing and safe closest approach, the craft can gain of order thousands of km~s$^{-1}$; repeated passes could add more, limited by geometry, tides, and the finite energy of the binary.

Let us try to translate this possibility into an observable technosignature.
If many or very massive spacecraft would use binaries as a gravitational machine, the inspiral rate would be faster than the effect of gravitational wave radiation alone.
If our timing is lucky enough, we might even observe propulsion signatures.
Following this lead, binary neutron stars, and by extension binary compact objects, should be part of technosignature target lists.

As with interstellar communication, our imagination may be failing, and other more exotic propulsion solutions might be used by ETI.
Let us review some of them.
\citet{kipping2018HaloDrive} proposed the halo drive concept, using a moving black hole as a gravitational mirror.
The idea is to shoot light towards the photosphere of the black hole, and collecting the photons coming back to the spacecraft with more energy, blueshifted.
Such a burst of light from an isolated moving black hole would have its light rapidly shifting in wavelengths, and would be a candidate signature.
Kipping proposes other observable technosignatures for binary black holes, a slightly eccentric orbit at the extraction point at periapsis, or changes in the inspiral rates in binary black holes leading to higher merger rates, which might be observable as a gravitational wave signal.

\subsection{Propulsion through spacetime manipulation}
\label{subsec:einst_gravitation}

\subsubsection{Warp drives and spacetime bubbles}
\label{subsubsec:warp_drive}

\begin{figure}
    \centering
    \includegraphics[width=1\linewidth]{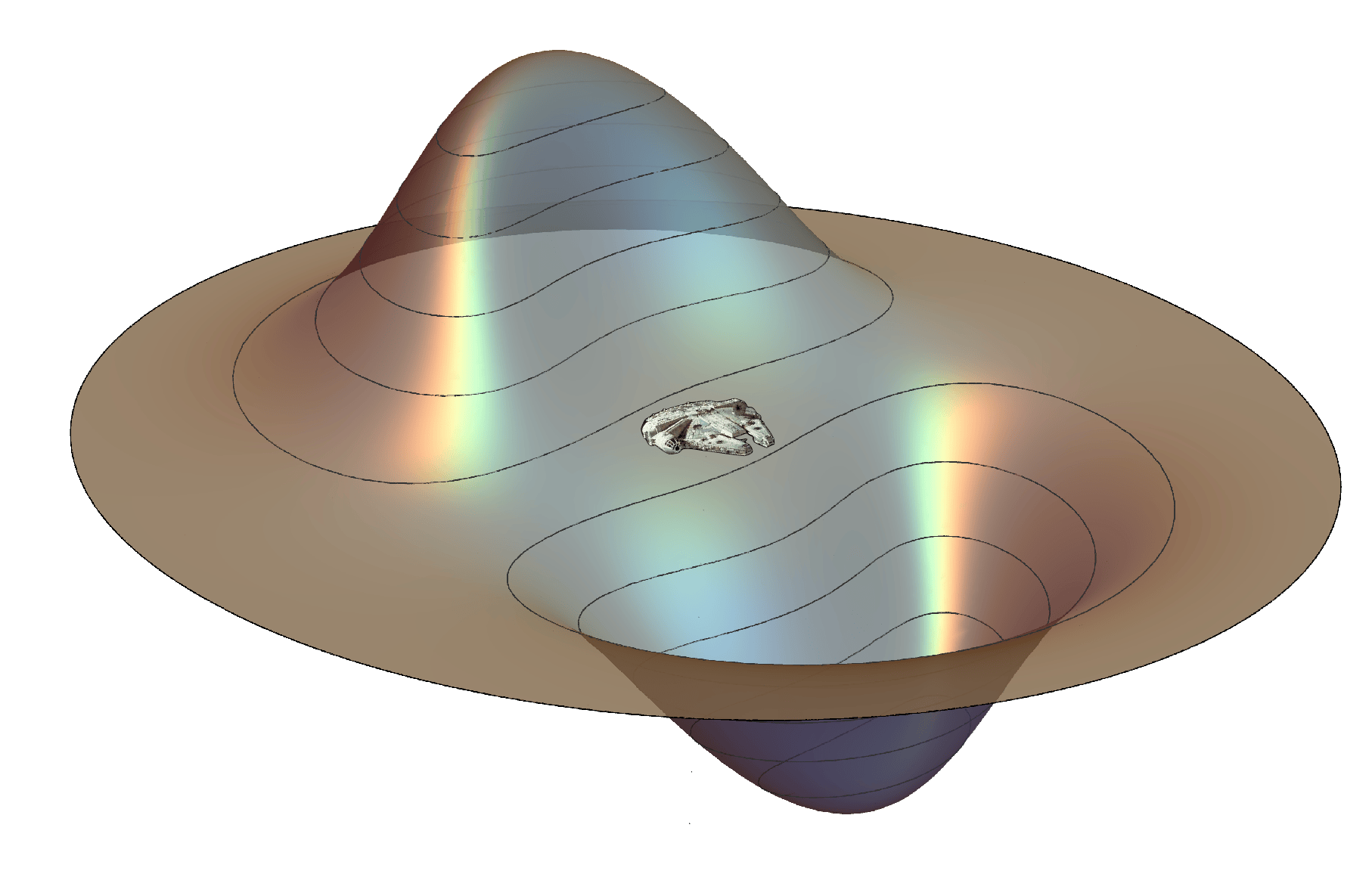}
    \caption{The York-time representation of an Alcubierre spacetime bubble, showing a localized region of warped space with contracted space ahead and expanded space behind.}
    \label{fig: Travel_TS_aclubierre}
\end{figure}

Warp drives are fascinating, albeit still theoretical, possibilities that could, arguably, be harnessed by sufficiently advanced technology.
Inspired by science fiction, \citet{alcubierre1994WarpDrive} discovered the first of these strange solutions of Einstein's general relativity.
His bubble solution consists of a highly warped shell of spacetime surrounding a run-of-the-mill flat region (see Fig. \ref{fig: Travel_TS_aclubierre}).
The warping contracts spacetime ahead of the vessel, and expands it behind, leading to external observers measuring the bubble motion as super-luminal.
\citet{natario2002WarpDrive} showed that this compression/expansion is not a necessary feature of warp drives, constructing a super-luminal bubble solution with no ``warp''.

Warped spacetime or not, these super-luminal solutions violate the Weak/Strong Energy Condition (WEC/SEC) is a major challenge to the creation of warp bubbles.
The Alcubierre metric requires a stress-energy tensor with negative energy density in the walls of the bubble, violating the WEC.
While quantum effects like the Casimir effect \citep{casimir1948AttractionTwo, bordag2001NewDevelopments} produce localized negative energy density, the vast quantities required for macroscopic warp bubbles are well beyond the technological capability of human civilization.
This may not be the case for highly advanced civilizations.
More recently, metric solutions for spacetime warp bubbles that require only positive energy sources have been discovered \citep{lentz2021BreakingWarp, bobrick2021IntroducingPhysical, fuchs2024ConstantVelocity}.
Remarkable if true, there is disagreement in the field as to the validity of these positive energy solutions\citep{santiago2022GenericWarp}.

Another hypothetical, and contested, challenge with warp drives is that particles with an initial positive velocity in the spacetime bubble's path are swept up and accelerated.
During the collapse of the bubble, these particles produce a concentrated beam of extremely high energy, effectively irradiating any destination and its inhabitants \citep{mcmonigal2012AlcubierreWarp}.
Though unfortunate, such high energy events could provide a discernible technosignature.

A bi-modal signal \citep{lentz2024MotivatingEmissions} is a unique technosignature predicted to occur only when the spacetime bubble's speed is greater than the speed of the emissions it is broadcasting.
This mechanism is a purely craft motion effect, since the craft is moving super-luminally, essentially outrunning the signals it produced earlier in its path.
Thus, a distant observatory would record emissions that occurred at two different times simultaneously.
One signal would move in the apparent direction of the craft's motion, showing the emissions occurring in the correct, forward order in time.
The second highly unusual signal would move in the opposite apparent direction, presenting the craft's emissions in a reversed temporal order.
This technosignature is considered a key observable \citep{lentz2024MotivatingEmissions} because there is no known natural phenomenon that could produce such a signal.

The collapse of a warp bubble is predicted to also release a very distinct gravitational wave signature---short, high-frequency bursts unlike the slowly in-spiraling ``chirps'' of merging black holes or neutron stars \citep{clough2024WhatNo}.
A non-electromagnetic technosignature from a warp bubble collapse would be an intense gravitational wave burst signal with a characteristic frequency proportional to the bubble's size (e.g., $\sim$300~kHz for a 1~km size bubble).
Unfortunately, such gravitational wave signals are outside the range of current detectors, but potentially measurable by future gravitational wave detectors.

\subsubsection{Traversable Wormholes}
\label{subsubsec:wormholes}

While at CalTech, Michael S. Morris and Kip S. Thorne, popularized the idea of traversable wormholes to the physics community \citep{morris1988WormholesTime}.
Traversable wormholes, in theory, allow material to move between different regions of spacetime within the universe, or from one universe to another.
Wormholes, traversable or not, are solutions of classical general relativity, having been discovered soon after the theory was proposed \citep{einstein1935ParticleProblem}.
In contrast to black holes or their non-traversable brethren, traversable wormholes do not have event horizons.
Einstein and Rosen's original solution turned out to be non-traversable - the spacetime throat pinched off before anything could pass through\citep{fuller1962CausalityMultiply}.
The first traversable solutions were found soon after, simultaneously by Homer G. Ellis at the University of Colorado and Kirill Alexandrovich Bronnikov at Moscow's RUDN University ~\citep{bronnikov1973ScalartensorTheory, ellis1973EtherFlow}.
In these, light passes through, permitting observations of cosmic sources on the other side.
The resulting lensing pattern is a clear signature of the type of wormhole, so that, if the hole is produced by an ETI, gives a distinctive techno-signature.

\begin{figure}
    \centering
    \includegraphics[width=0.95\linewidth]{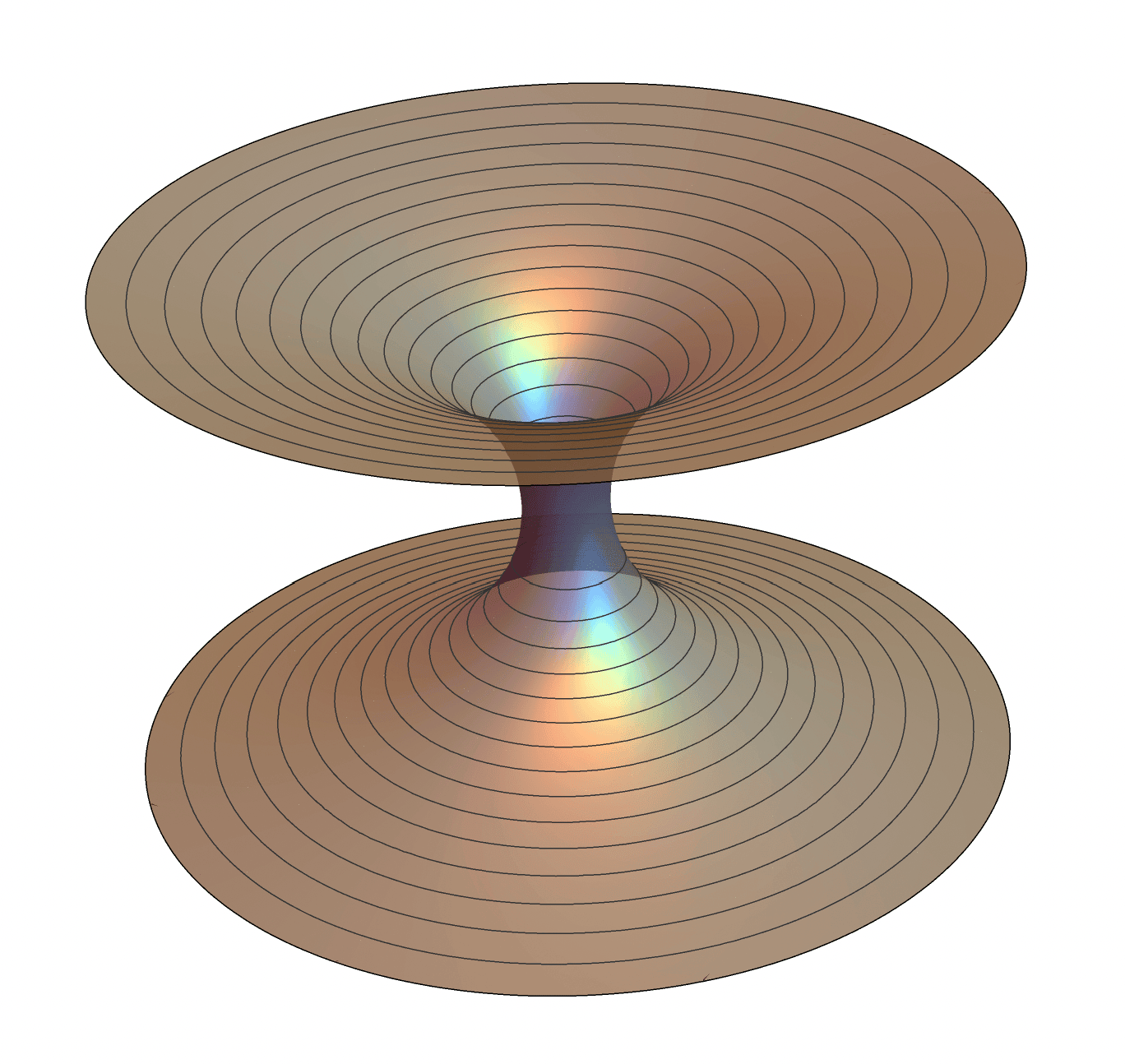}
    \caption{An embedding diagram of the $t=0$, $\theta=\pi/2$ spatial slice of the Ellis wormhole. }
    \label{fig: Ellis Wormhole}
\end{figure}

The two traversable solutions, and those to follow faced the problem of requiring the presence of negative energy density or ``exotic matter,'' violating the WEP.
To show how this problem arises explicitly, consider the simplest wormhole metric having line element
\begin{equation}
ds^2 = -c^2dt^2 + dr^2 + \left(r^2 + \ell^2\right)(d\theta^2+\sin^2\theta d \phi^2),
\end{equation}
where $\ell$ is the radius of the throat.
Fig. \ref{fig: Ellis Wormhole} The total energy of the matter needed to curve spacetime in such a way is
\begin{equation}
    E=-\frac{\pi}{4}\frac{c^4\ell}{G}=-9.51\times 10^{43} J\left(\frac{\ell}{1\text{m}}\right).
\end{equation}
The negative sign means that the matter is exotic --- nothing like the atoms we are used to.
Even a Planckian Ellis wormhole ($\ell\sim 10^{-35}$m) requires a Gigajoule of negative energy.

Visser attempted to minimize the volume of exotic matter needed to make a wormhole ~\citep{visser1989TraversableWormholes, visser1996LorentzianWormholes}.
His ``cut-and-paste'' construction method describes wormholes with polytope (many flat faces glued together), instead of spherical mouths.
This results in a multiply-connected spacetime that localizes the exotic matter to the edges between the faces.
Edward Teo generalized the Morris-Thorne construction to include the effects of rotation ~\citep{teo1998RotatingTraversable}.
This opened up a new can of worms, pun intended, exposing the potential to create closed timelike curves (CTCs) and all the associated time-travel paradoxes ~\citep{morris1988WormholesTime}.

Eventually solutions were found that do not violate the WEC, but still require us to push our imaginations to the extremes of theoretical physics.
\citet{woodward2011MakingStargates}, for example requires the bare mass of the electron to be negative, while \citet{klinkhamer2023DefectWormhole} requires defects in the fabric of spacetime.
As strange sounding as some of these ideas are, they do not violate the known laws of physics, so could, if realized in our universe, be harnessed by a suitably advanced civilization.

\subsection{Fundamental forces for interstellar travel}

To sum up this section, it makes sense to harness the four fundamental forces of physics for the difficult task of interstellar travel.
We saw that nuclear forces are a prime choice for high-density fuel, electromagnetism can be used as directed energy, while Newtonian gravitation allows the well-known gravitational assist maneuvers, and Einsteinian gravitation allows in principle to alter the fabric of spacetime.

A realistic propulsion solution would employ multiple strategies using all available forces.
For example, a spacecraft could be launched with rockets, perform a gravitational assist around its home star, deploy sails and get directed energy (or matter) from the home system.
While cruising far away from the home system, it could use a ramjet to scoop matter from the ISM, perform gravitational assists on massive objects along the way, and while approaching the final destination use onboard fuel for critical maneuvers.
\clearpage
\section{Galactic and beyond}\label{sec: galactic ts}
\subsection{Galactic}\label{sec: galactic}
\begin{figure*}
    \centering
    \begin{minipage}{0.48\linewidth}
        \centering
        \includegraphics[width=\linewidth]{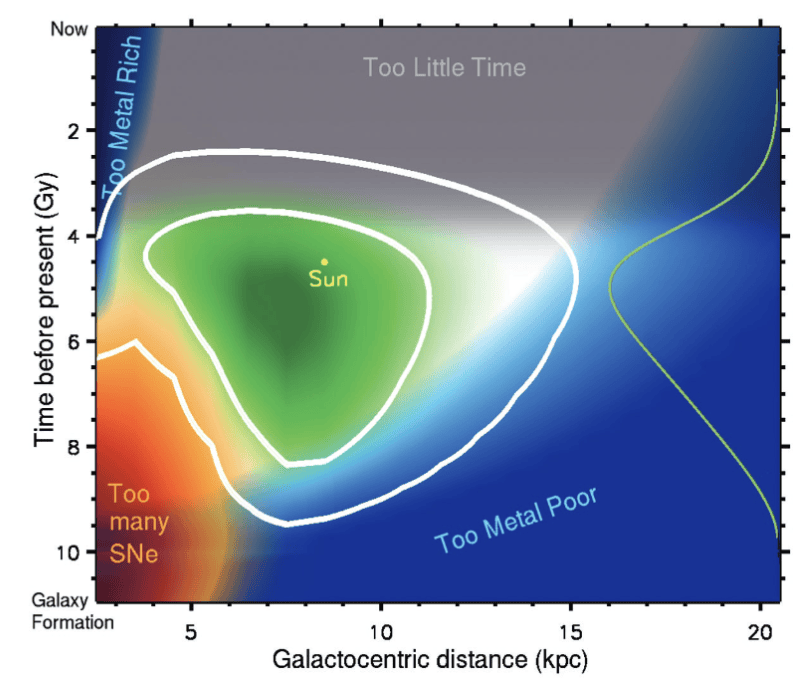}
    \end{minipage}
    \hfill
    \begin{minipage}{0.48\linewidth}
        \centering
        \includegraphics[width=\linewidth]{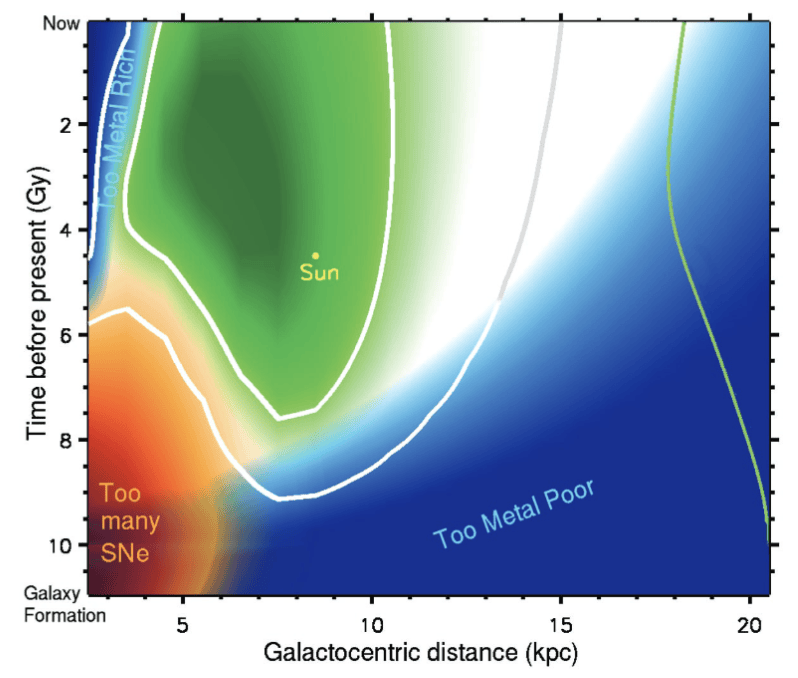}
    \end{minipage}
    \caption{The GHZ (Green) as a function of radial distance from the galactic center (GC) and time before the present for the Milky Way Galaxy. The left model is for complex life, for which it assumes life requires $4\pm1$ Gyr to form, while the right model considers all life. The green curve on the right of the left Fig. is the age distribution of complex life, and in the right Fig. the age distribution of all life. The GHZ begins as a thin annulus some ~8 kpc from the GC around ~8 Gyr, and widens. Its inner radius eventually reaches a minimal distance to the GC, and, due to the higher metallicity rates in the core, halts and shrinks. Meanwhile, the outer radius continues to expand as the Galaxy ages. Fig. from \citet{lineweaver2004GalacticHabitable}.
}
    \label{fig:Galactic_TS_GHZ}
\end{figure*}
In our galaxy, the Galactic Habitable Zone (GHZ) depicted in Fig. \ref{fig:Galactic_TS_GHZ} is the region within the galaxy where conditions are thought to be most favorable for the emergence and long-term sustainability of complex life \citep{lineweaver2004GalacticHabitable}.

It is based on the metallicity gradient of the galaxy, its temporal evolution, the location of potentially hazardous astrophysical phenomena including supernovae explosions and gamma ray bursts, as well as relative motions and locations of stars in the spiral density waves (also known as the corotation circle).

Even though it is biased towards parts of the galaxy conducive to life-as-we-know-it, the GHZ provides an excellent starting point in a search for technosignatures from Kardashev Type II civilizations.
Such civilizations might have built either a massive structure around a single star or multiple smaller structures around many stars.

In the case of the latter, communication within the galaxy-spanning civilization is subject to delays and galactic time synchronization issues due to the finite speed of light.
Broadcast hubs might continually transmit important information, making it available to even the most distant colonies.
One galactic center radio transient (GCRT J1745−3009) has been suspected as a potential candidate \citep{benford2010SearchingCostOptimizeda}.

With an estimated $10^6-10^8$ pulsars in the Milky Way \citep{rozwadowska2021RateCore}, we have already mentioned that these stellar remnants could be used for both navigation and timing \citep{vidal2017MillisecondPulsars}.
\citet{edmondson2003UtilizationPulsars} and \citet{edmondson2010TargetsSETI} envisioned a SETI-strategy based on pulsars seeking alignments between Earth, a pulsar and a habitable star.
If pulsars would be artificially modulated, they might also form a galaxy spanning information network \citep{chennamangalam2015JumpingEnergetics, vidal2019PulsarPositioning, haliki2019BroadcastNetwork, davis2024FindingSignal}.

Given the vast distances between planets and the resource demands of an interstellar civilization, planets suitable for colonization, resource extraction, and even terraforming may be distributed very differently than those in the GHZ. In such scenarios, the distribution of planets with expected technosignatures would differ greatly from those harboring biosignatures.
Recent models of galactic settlement \citep{carroll-nellenback2019FermiParadox, wright2021DynamicsTransition} distinguish between habitable and settleable worlds, and assess the transition of a Type I civilization to a Type II.
Settlement driven by minimal distances and maximal energy extraction results in settlement wavefronts strongly biased towards the galactic center.
Since this is where less habitable planetary systems are to be found, technosignatures for Type II civilizations decouple from those of Type I or early Type II.
 
Getting to a galactic core increases the energy available to a Type II civilization not just due to the higher density of stars there; supermassive black holes (SMBHs) residing in the cores of galaxies are prime candidates for energy exploitation.
A SMBH of the mass of about four million suns ($M=4\times 10^6M_{\odot}$), and low Eddington ratio ($10^{-3}$), should form an accretion disk with a luminosity of 100 million suns ($L=10^8 L_{\odot}$), and a total corona and jet luminosity an order of magnitude higher \citep{hsiao2021DysonSphere}.
\citet{inoue2011TypeIII} showed that a Dyson swarm around such a SMBH provides the energy equivalent of smaller swarms around $\sim10\%$ of a galaxy’s stars.
It means that rather than spreading out across a galaxy, to satisfy their energy needs, some Type II civilizations may instead concentrate at the cores of galaxies.
The energy harvest there is enough to transition a low Type II into a high Type II, and possibly even a Type III civilization \citep{hsiao2021DysonSphere}.
A civilization near a SMBH could also make the most of extreme gravitational lensing for galactic and intergalactic observation or communication \citep{maccone2012SETIGalaxies}.

Another speculative source of energy utilization by a Type III civilization is the concept of a ``Black Hole Bomb" \citep{cardoso2004BlackholeBomb}.
It is a theoretical device that exploits a phenomenon called superradiance to extract the rotational energy of a spinning black hole.
Thus extracting the rotational energy stored in a supermassive black hole at the center of a galaxy would constitute a power plant more powerful than output of billions of stars combined.

\subsection{Extragalactic}\label{sec: extragalactic}

Galaxies outside of the Milky Way might also harbor extraterrestrial civilizations.
The Breakthrough Listen program has put constraints on extragalactic transmitters, using existing observations \citep{garrett2022ConstraintsExtragalactic,tremblay2024ExtragalacticWidefield, uno2023UpperLimits}.
Given the tremendous distances involved, the magnitude of energy usage that could feasibly be observed by astronomers here on Earth would have to be immense, implying that such technosignatures would have to be produced by Type III civilizations or beyond.
\citet{kardashev1964TransmissionInformation} speculated that quasars might be artificial sources.
This was before the current understanding (which emerged in the 1970s) that quasars are active SMBH.
Simultaneously, the farther out we look, the farther back in time we observe galaxies, implying that any technosignatures must come from civilizations that arose and advanced very early in the history of the universe.

\begin{figure*}
    \centering
    \includegraphics[width=1\linewidth]{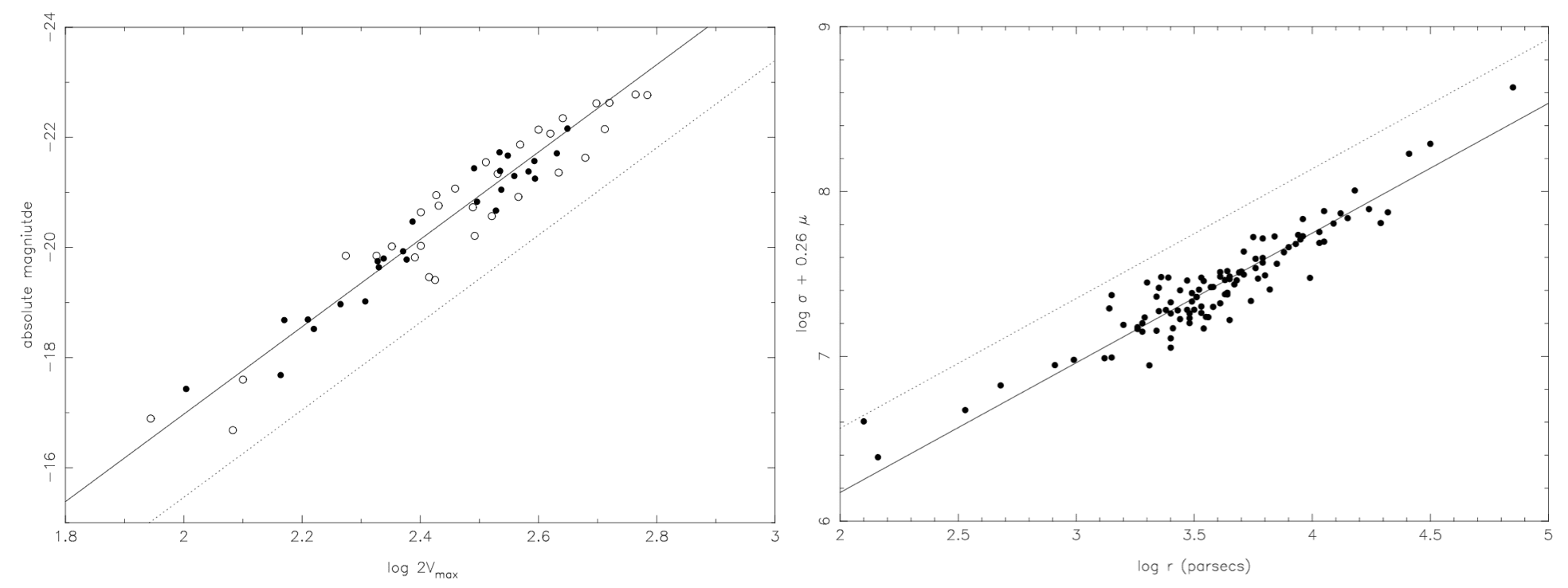}
    \caption{Galaxies display astrophysical correlations: the Tully-Fisher for spiral galaxies (left) and the Faber-Jackson for elliptical galaxies (right). Fig. from \citet{annis1999PlacingLimit}.}
    \label{fig: extragalactic_ts_tully}
\end{figure*}

What are possible observables of galaxy spanning civilizations?
Astronomers studying galaxies have discovered several empirical scaling relations between galaxy sizes (R), luminosity (L), and dispersion of stellar velocities ($\sigma$), that are well-motivated by simple gravitational physics.
For spiral galaxies the Tully-Fisher relation and for ellipticals the Faber-Jackson and Fundamental Plane show that when you plot the logarithms of these variables against one another for a great many galaxies, you find that they are highly correlated (Fig. \ref{fig: extragalactic_ts_tully}).
A galaxy hosting a Type III+ civilization might be observed as a large deviation from one of these scaling relations, if the luminosity of the galaxy is altered by megastructures.
\citet{annis1999PlacingLimit} performed an observational search of dimming effects due to Dyson swarms on dim galaxies.

The section \ref{sec: stellar ts} “Stellar technosignatures” discusses Dyson spheres but looking for them doesn't have to be confined around stars. Some searches looked for the cumulative effect of many Dyson spheres at galactic scales \citep{wright2014InfraredSearch, wright2014InfraredSearcha, griffith2015InfraredSearch}.
\citet{zackrisson2015ExtragalacticSETI} selected optically underluminous galaxies for followup infrared observations that would violate the Tully-Fisher correlation.

If megastructures are built around black holes instead of stars, ~\citet{hsiao2021DysonSphere} argued that they could still be detectable through their waste heat.
\citet{lacki2016TypeIII} went a step further arguing that if a whole galaxy could be enclosed in a black box, it could still be observable in the Planck catalog.
None of these approaches resulted in any promising candidates.

Galaxy wide dimming negative results thus far point towards Type III civilizations typically concentrating instead of spreading out.
Technosignatures could still be observed if these civilizations created very powerful point sources.
Manipulating Planck length volumes of spacetime requires a particle accelerator whose size can range from several AU to hundreds of AU. Such accelerators could harness ~100 solar masses worth of energy; the collisions in them produce neutrinos with energies measured in YeV ($10^{24}$ eV!).
Thus, the scientific endeavors of Type-III civilizations could provide point sources with distinct signatures \citep{lacki2015SETIPlanck}.
The energy infrastructure could play a similar role.
Energy harvested from a SMBH would have to be distributed if the Type III civilization wants to expand.
Power stations at the SMBH producing highly collimated beams pointing at receiving stations in distant star systems were proposed by \citet{inoue2011TypeIII} as a possible technosignature coming from this kind of activity.

\citet{voros2013GalacticscaleMacroengineering} argued that the very peculiar morphology of Hoag’s object might be the result of galactic engineering.
The signature remains ambiguous as galactic collision simulations can also reproduce such a ring \citep[e.g.][]{brosch1985NatureHoags}, which means that more stringent tests need to be developed.

\subsection{Universal}\label{subsec: Universal}
	
Moving up in scale to swathes of spacetime encompassing many galaxies over cosmological timescales, could civilizations exist that are no longer bound to their home galaxies?
A Type-III civilization which has expanded to its local group would only be a fraction of the way to a Type-IV.
Typical distances between galaxies are an order of magnitude larger than their sizes, so the communication difficulties faced are significantly greater than those of a merely galactic Type-III civilization.
Ambitious technological life, life that seeks to maximize resource availability, has been proposed as motivation for the existence of multi-galactic civilizations \citep{olson2017EstimatesNumber}.
All it takes is for a single civilization across hundreds of galaxies to be ambitious for the onset of aggressive expansion \citep{olson2015HomogeneousCosmology, olson2016VisibleSize, olson2018ExpandingCosmological}.
As we will see, this behavior can keep pushing a civilization up the Kardashev scale.

\begin{figure}
    \centering
    \includegraphics[width=1\linewidth]{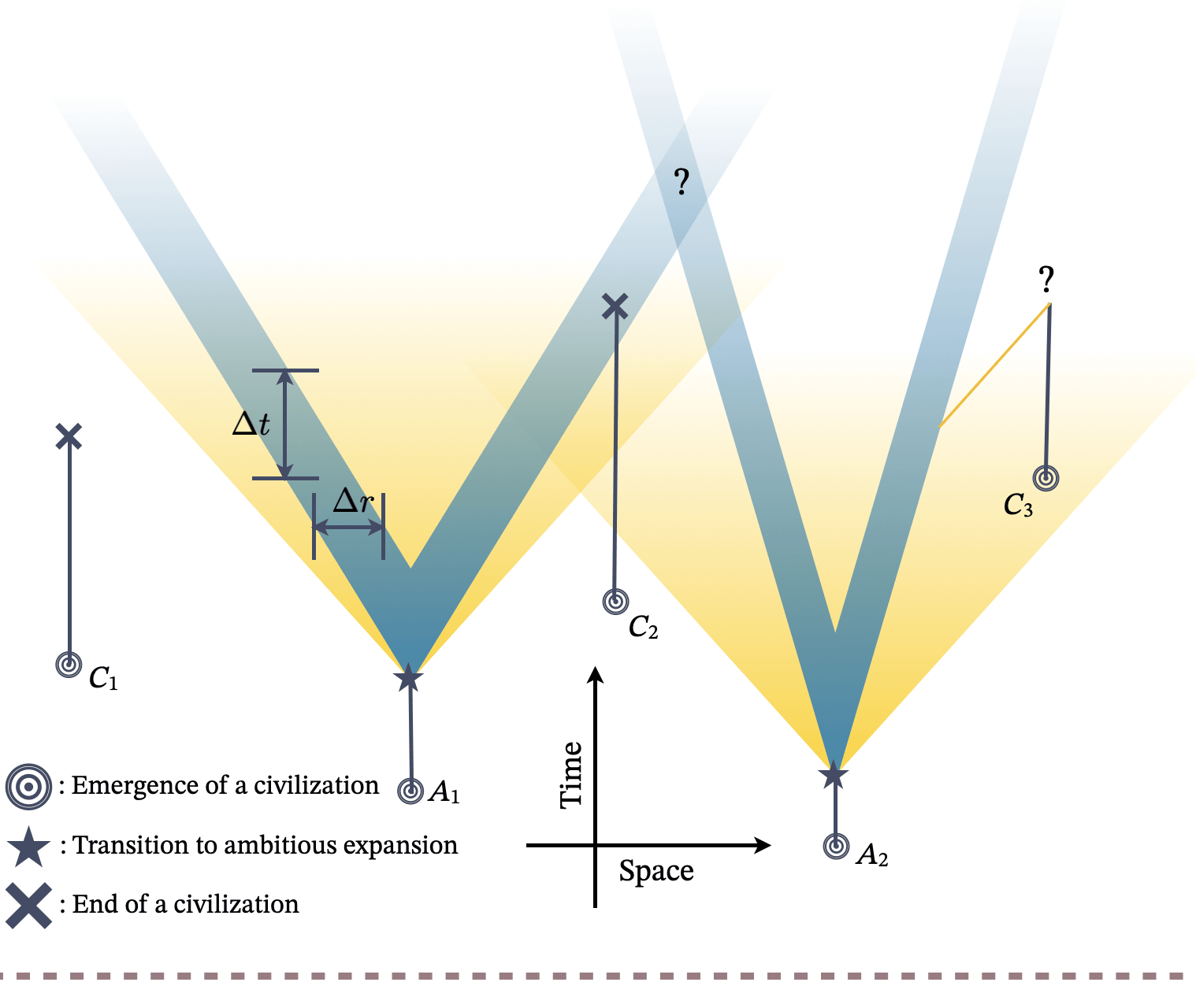}
    \caption{A spacetime diagram of the emergence of ambitious ($A_1$, $A_2$) and non-ambitious civilizations ($C_1$, $C_2$, $C_3$). 
    The horizontal direction represents space, and the vertical time; light cones are represented in yellow. }
    \label{fig: GalacticTS_Olson}
\end{figure}

An ambitious civilization utilizes free energy in their domain at a pace comparable to the rate of expansion.
As they spread out they use all available resources in the galaxies they encounter and then continue on their way.
A mathematical treatment models the emergence of civilizations as rare point processes in spacetime, represented by the bullseyes in Fig. \ref{fig: GalacticTS_Olson}.

Those which achieve ambitious expansion (stars) begin to spread out from their home galaxy at a speed $v=\Delta r/\Delta t$.
This expansion produces a wavefront propagating outward.
This wavefront speed depends on both the technological limits of the civilization’s spacecraft, as well as the timescales associated with energy harvesting, $\Delta t$, though its geometry can be approximated as spherical.
Energy harvesting also determines the thickness of the wavefronts, $\Delta r$, describing the spatial extent of colonized galaxies.
Taking into account factors such as the appearance of human civilization relative to others on cosmological timescales, gamma ray burst suppression of galactic habitability, and the improbability of such life, these models place limits on what we, here on Earth, might expect to see as the extent of such a civilization in our night sky.
Rather modest assumptions imply that one such civilization would cover a bubble taking up between $\sim0.1\%-1\%$ of the sky.
This may seem small until one realizes that the Moon takes up $\sim0.0008\%$ of the sky; even on the smaller end of the estimate, an ambitious civilization would cover a circle with a diameter $\sim10$ times that of the Moon in the sky!
The distances to such bubbles are on the order of a billion light years, making these intergalactic civilizations both incredibly large, close to Type-IV, and incredibly far away from us \citep{olson2015HomogeneousCosmology}.

If ambitious technological life exists then the universe is peppered with civilizations nearing Type-IV.
Detecting them, however, requires leaping farther into the unknown, and hypothesizing about technology at the limits of our knowledge of physics.
For example, if such a civilization has an efficient way of converting matter into energy, then one would expect the average matter density within its domain to be lower.
Accordingly the radiation density would be higher.
Since the expansion of the universe is sensitive to the ratio of matter to radiation, such activity would produce a backreaction on the galactic filaments of the cosmic web \citep{olson2015HomogeneousCosmology}.
Significant local deviations in the statistical distribution of voids could herald the presence of an ambitious civilization.

\subsection{The Simulation Hypothesis}\label{sec: simulation}
The simulation hypothesis \citep{bostrom2003AreWe} posits that our universe could be an artificial simulation operated by an advanced civilization\footnote{More precisely, Bostrom's argument defines an advanced civilization as one that is capable of running an {\it ancestor simulation}. Then either (a) advanced civilizations do not exist, (b) advanced civilizations have the capability but do not run ancestor simulations, or (c) we are almost surely in an ancestor simulation.}, possibly even our future selves \citep{wolpert2024ImplicationsComputer}.
Bostrom concluded from reasonable assumptions that the number of simulated conscious beings would vastly outnumber the original biological population.
Without the ability to distinguish between the simulated and biological, it is far more likely that you, a conscious being, are in fact simulated.

Such a possibility is congruent with digital and computational philosophy and physics, that attempt to model and understand the world according to the principles of information and computation theories \citep{floridi2003BlackwellGuide, lloyd2000UltimatePhysical, lloyd2005ProgrammingUniverse}.
In our context of the search for technosignatures, it means searching for empirical indicators or anomalies within the universe that could reveal its simulated nature by an intelligent creature \citep{campbell2017TestingSimulation}.

In our context, the technosignature search is geared towards empirical phenomena that serve as fingerprints of an underlying computational substrate.
These signatures would not be intentional communications but rather inherent artifacts of the universe's code, such as: Structural limitations imposed by a finite computational grid; ``glitches" or rounding errors in the execution of physical laws; or resource-saving optimizations that manifest as observable quantum phenomena.

\subsubsection{Discrete Spacetime and the GZK Cutoff}
A computer simulation would likely operate on a discrete grid rather than a continuous spacetime, implying a minimum possible length or ``resolution" to the universe.
\citet{beane2014ConstraintsUniverse} argue that a spacetime lattice would violate perfect rotational symmetry.
This anisotropy would be detectable in the behavior of ultra-high-energy cosmic rays (UHECRs).
In standard physics, there is a theoretical energy limit for cosmic ray protons known as the Greisen–Zatsepin–Kuzmin (GZK) cutoff (around $5 \times 10^{19}$ eV).
Protons above this energy should interact with cosmic microwave background (CMB) photons and lose energy.
\citet{beane2014ConstraintsUniverse} predict that on a discrete lattice, this energy cutoff would become direction-dependent, varying based on the cosmic ray's path relative to the lattice axes.
The detectable technosignature would be a statistically significant anisotropy in the GZK cutoff energy across the sky, providing a lower-limit on the size of the lattice that we would be simulated on.

\subsubsection{Physical Law ``Glitches," Anomalies, and Varying Constants}
A simulation running on finite computational resources might exhibit artifacts analogous to rounding errors or processing shortcuts.
These could manifest as slight deviations from what we consider fundamental laws.
\citet{campbell2017TestingSimulation} suggest searching for such inconsistencies.
This includes unexplained drift in fundamental constants like the fine-structure constant ($\alpha$) or momentary, localized violations of conservation laws; such phenomena could constitute candidate technosignatures worth examining.
\citet{whitworth2008PhysicalWorld} further speculates that the inherent "weirdness" of quantum mechanics (like non-locality and wavefunction collapse) might itself be an artifact of a computational substrate.

\subsubsection{Quantum Indeterminacy as Computational Efficiency}
Quantum mechanics posits that a particle's state is indeterminate until measured.
Hamieh \citeyearpar{hamieh2021SimulationHypothesis} has drawn an analogy to video games in computing, where a system only renders details when they are needed or observed.
The act of observation, which collapses the wavefunction, might be a sign that the universe is conserving computational resources by not calculating definite states for unobserved systems.
A potential technosignature avenue of research would be the observer effect in quantum mechanics interpreted as a resource-saving mechanism.

\subsubsection{Computational Complexity Limits}
Certain physical systems are so complex that simulating them with classical computers is believed to be intractable.
\citet{ringel2017QuantizedGravitational} showed that simulating a specific quantum Hall effect phenomenon would be exponentially difficult.
The fact that our universe appears to handle such complexity could be seen as evidence against a classical simulation.
Conversely, if we discover that nature systematically avoids states of intractable computational complexity, it might imply that the universe operates under computational constraints.
Therefore the potential technosignature would be evidence that the universe avoids physical states that are computationally prohibitive to simulate.
The existence of hypercomputational possibilities such as Malament-Hogarth spacetimes \citep{pitowsky1990PhysicalChurch, hogarth1992DoesGeneral}, e.g. a Kerr black hole could imply that the computational power of our universe is much higher than what we can imagine.
This could mean either that advanced civilizations have access to even higher computational capabilities, or that the simulation hypothesis is disfavored.

\subsubsection{Cosmological Fine-Tuning and Embedded Messages}
The simulation hypothesis offers a potential explanation for the ``fine-tuning problem", i.e. why the fundamental constants of our universe appear precisely calibrated to allow for complex structures and life \citep{barrow2007LivingSimulated}.

A related idea is that the simulators might have embedded an intentional message.
This ``Easter egg" could be hidden in the numerical values of physical or mathematical constants.
Some have speculated that a message might be in patterns of the CMB \citep{hsu2006MessageSky}, which was later rebutted by \citet{hippke2020SearchingMessage}.
The technosignature here would be the discovery of a recognizably intelligent and non-natural pattern or message encoded in fundamental physics.

Leaving the idea of simulation aside, some authors have proposed that intelligent life in another universe might have fine-tuned our universe to some degree.
This is not a wild speculation, but an effort to tackle the fine-tuning problem in an evolutionary framework, starting with Lee \citeauthor{smolin1992DidUniverse}’s \citeyearpar{smolin1992DidUniverse} cosmological natural selection theory.
Indeed, questions of apparent design in biology have been successfully explained with evolutionary theory, and Smolin hypothesized that an evolutionary process may have given rise to the peculiar, complexity-creating values of the constants of our universe.
Cosmologists Louis \citeauthor{crane2010PossibleImplications} (1994\slash\citeyear{crane2010PossibleImplications}) and Edward \citet{harrison1995NaturalSelection} have proposed that the selection of universes might not be just “natural” but also “artificial” involving intelligent life in the process.
\citet{gardner2004PhysicalConstants} argued that the fine-tuned physical constants themselves are the life signature!
In other words, that the bio/techno signature lies within our reach, in cosmological and fundamental physics models.
\citet{vidal2014BeginningEnd} developed a book-length argument of this scenario from a big picture perspective, and evolutionary thinkers have also argued for the plausibility and cogency of this view \citep{stewart2010MeaningLife, smart2009EvoDevo, gardner2003BiocosmNew, floresmartinez2014SETILight, price2019CosmologicalNatural,azarian2022RomanceReality}.
\clearpage

\section{Discussion}
\label{sec:discussion}

\subsection{Biosignatures and technosignatures}
\label{subsec:bio_techno}

The frontier between biosignatures and technosignatures may be fuzzy.
If we detect a technosignature on an exoplanet, it might also have biosignatures, much like present Earth displays both.
\citet{romanovskaya2023PlanetaryBiotechnospheres} did introduce the notions of \textit{biotechnosignatures} and \textit{biotechnospheres}, and explored possible observables.
Let us give a few examples along this line of thinking.

\citet{raup1992NonconsciousIntelligence} argued that since some animals can not only detect radio signals, but also generate electrical pulses, we could envision an exoplanet whose evolution develops animals using radio communication that would leak into space and thus be detectable.
The technosignature is in fact a biosignature.

As another example, imagine an ETI analyzing an Earth satellite image of an agricultural field.
It would notice a biosignature of the vegetation, marked by the red edge feature in its reflectance spectrum, but also a technosignature through its geometrical design, as a pattern with the purpose of delineating a boundary.

Now imagine an exoplanet with genetically modified organisms used in the upper atmosphere to regulate its climate.
The organisms would have a spectral biosignature, and since they are being used as a technological deployment to control the planet's climate, they could also be classified as a technosignature.
In a similar vein, advanced ET civilizations could harness bioluminescent organisms to create light, motivated by the desire for low-energy solutions for nighttime illumination.
The observation would be a combination of bioluminescent biosignatures and structured lighting patterns associated with technosignatures.

\begin{figure}
    \centering
    \includegraphics[width=0.8\linewidth]{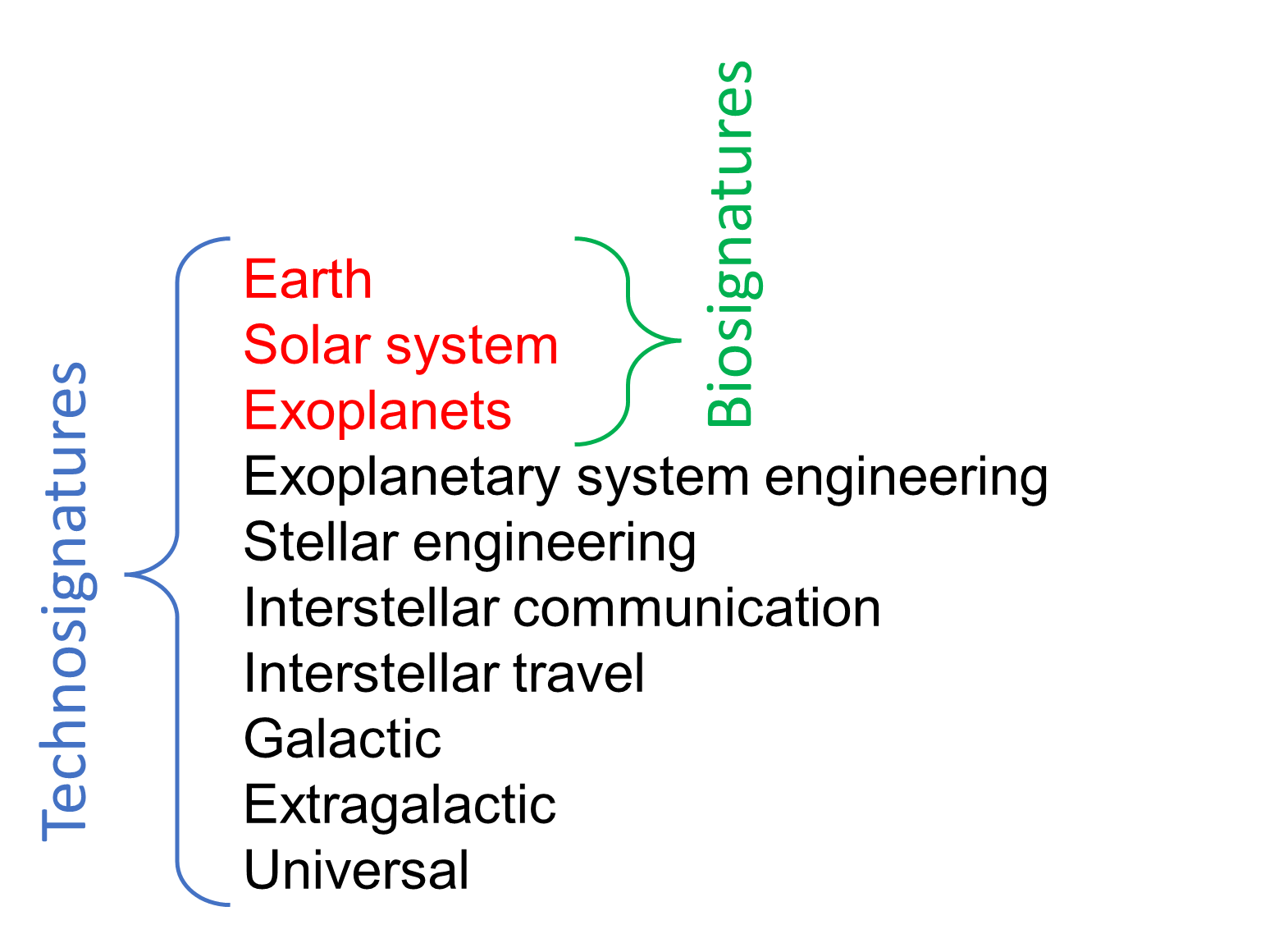}
    \caption{Synergies between bio- and techno- signatures overlap for targets at the Earth, solar system and exoplanet scales (in red). Technosignatures searches extend beyond these scales where they have different synergies with astronomy, astrophysics or engineering (in black).}
    \label{fig: Discussion_bio_techno}
\end{figure}

A key difference between biosignatures and technosignatures is the sheer number of possible targets.
It is very low for biosignatures, with a handful in the solar system and a few thousand exoplanets known; while it is very high for technosignatures, with billions of potential targets and a huge search space.
The number of targets is so high that the main issue for technosignatures is to find strategies to constrain observations and select targets.

An argument regarding the duration of signatures is that it is long for biosignatures (after all, Earth has had a biosignature for 4 billion years) and unknown for technosignatures.
Even if the existence of technosignatures is not recent, or short, given the exponential dynamics of technological progress, the observational effects of civilizations expanding in the galaxy or affecting larger scales could still be ubiquitous.

We also do not know how typical the development of life and technology on Earth is.
To the extreme, one could speculate that Earth was exceptionally slow to develop life and technology, in which case the 4 billion year observational window of biospheres would not be typical nor guaranteed.

We can expect biological life to exist in many places in the universe more so than technological life, due to the many more steps required to reach technology.
However, the chances of us finding technological life could be much higher than biological life, due to technosignatures being more distinct from background noise than biosignatures \citep{desmarais2008NASAAstrobiology, wright2022CaseTechnosignatures, smith2023LifeDetection} and the fact that technological life may have intentions to find other technological life, while biological life would not.

Taking the Drake equation too seriously, one may also naively conclude that the more we go towards the right of the equation, the more constrained the outcome will be, so the less likely and the less targets there will be.
This is not the case.
Planets outnumber stars, as many of them can form from a single circumstellar disk. 
In the same way, many technologies can form from a single world, so the number of technosignatures may outnumber the number of biosignatures.

Technological fossils---traces of a previous civilization on Earth (see Section~\ref{sec: earth ts})---or technological trash, such as inactive, broken probes in our solar system, broken Dyson spheres \citep{loeb2023InterstellarObjects}, and as \citet{holmes1991ArchaeologySpace} noted more generally, rubbish, debris, defunct equipment, and defunct spacecraft are also potential technosignatures.
For attempts to quantify this longevity factor of technosignatures, see \citet{lingam2019RelativeLikelihood} and \citet{cirkovic2019PersistenceTechnosignatures}.

Putting all these considerations of number of targets, duration, and detectability together, it remains uncertain what the best chances are to find either bio- or techno- signatures \citep[see also][]{wright2022CaseTechnosignatures}.
One reason that testing technosignatures may be perceived as more difficult is that it requires studying and extrapolating not just biological agency, but also cognitive, intelligent, and technological agency.
This is a blind spot in traditional natural sciences that seeks to study causal effects in a detached and objective way, and thus neglects or avoids the complexities of modelling agents \citep[see][]{frank2024BlindSpot}.

\subsection{Anomaly detection and agnostic searches}
\label{subsec:anomaly}

Historically, many anomalous astrophysical phenomena were hypothesized to be due to activity from ETI.
For example, we mentioned that \citet{shklovskii1966IntelligentLife} examined in detail whether Mars' moon Phobos could have been an artificial satellite based on observed anomalies in its orbit.

\citet{kardashev1964TransmissionInformation} suspected that two recently discovered strong radio sources might be artificial, but later were identified as quasars.
Pulsars were first suspected to be ETIs; the discovery team, not wanting to jump to such conclusions too hastily, did not release the news until after developing a physical mechanism \citep{penny2013SETIEpisode}.
Likely influenced by the Breakthrough Starshot project, \citet{lingam2017FastRadio} speculated that Fast Radio Bursts might be beams powering large sail spacecraft.

We want to highlight that the diversity of possible technosignatures we reviewed share a common thread: searching for technosignatures agnostically is ultimately searching for anomalies.
The bottom line is that anomalies cry out for explanation, so even when they end up not being ETI, their resolution contributes direct benefits for science.
The topic of anomalies in epistemology is central to the philosopher of science Thomas \citet{kuhn1996StructureScientific}, who wrote in his foundational \textit{The Structure of Scientific Revolutions}:

\begin{quote}
``If all members of a community responded to each anomaly as a source of crisis or embraced each new theory advanced by a colleague, science would cease.
If, on the other hand, no one reacted to anomalies or to brand-new theories in high-risk ways, there would be few or no revolutions.''
\end{quote}

This quote may translate into two attitudes regarding the search for ET life.
Amateur or pseudo-scientist enthusiasts about UAPs may overreact to anomalies, while the natural philosophers who refused to accept heliocentrism underreacted to anomalies.

Another way to express this discrepancy is that two attitudes are possible in the face of anomalies.
First, a too optimistic ``alien-of-the-gaps'' attitude, where, much as with the god-of-the-gaps, if there is a new phenomenon not understood, it is assumed that it must be extraterrestrial life.
This leads to many false positives.
The opposite attitude is one of extreme skepticism, or a ``nature-of-the-gaps'' attitude, where whatever anomaly is found, it is never ETI.
This leads to false negatives.
Carl Sagan \citeyearpar{sagan1993SearchLife} famously wrote that ``life is the hypothesis of last resort.''
But on careful examination, there is an infinite number of physical mechanisms that could attempt to explain a given phenomenon, which means that under this attitude, we are never going to exhaust all the possibilities, and thus never consider the life or ETI hypothesis.
In sum, a healthy solution is to be aware of and to carefully navigate between these two symmetrically biased attitudes, and to choose the hypotheses that best explain the past through retrodictions, and that best anticipate the future through predictions \citep{vidal2014BeginningEnd}.

Modern astronomical surveys such as Gaia \citep{vallenari2023GaiaData} or LSST \citep{ivezic2019LSSTScience} give rise to petabytes of data, containing high-dimensional datasets with significant potential for extraterrestrial anomaly detection.
Analyzing the data of large SETI searches can be done by crowdsourcing \citep{li2024AreWe}, or if the data to process is too big for humans, machine learning (ML) becomes a key tool.

The algorithms can be tested by injecting anomalies in test datasets, and checking whether the SETI pipelines can detect them.
Recently, \citet{ma2023DeeplearningSearch} set a complete ML pipeline to reanalyze 120 TB of data from the Green Bank Telescope; their method using a state-of-the-art variational autoencoder saw fewer false positives than previous analyses, as well as 8 new signals of interest.

Information theory has provided a panoply of tools for agnostically analyzing data, and offers prospects to do so in a less biased way than was possible before.
The diversity of animal communication is of particular relevance to study with information theory \citep{doyle2011InformationTheory}.
The problem of communicating information is also always relative to a background, and \citet{brooks2025FireflyinspiredModel} took inspiration from fireflies to model how ETI communication might proceed.

\begin{figure}
    \centering
    \includegraphics[width=1\linewidth]{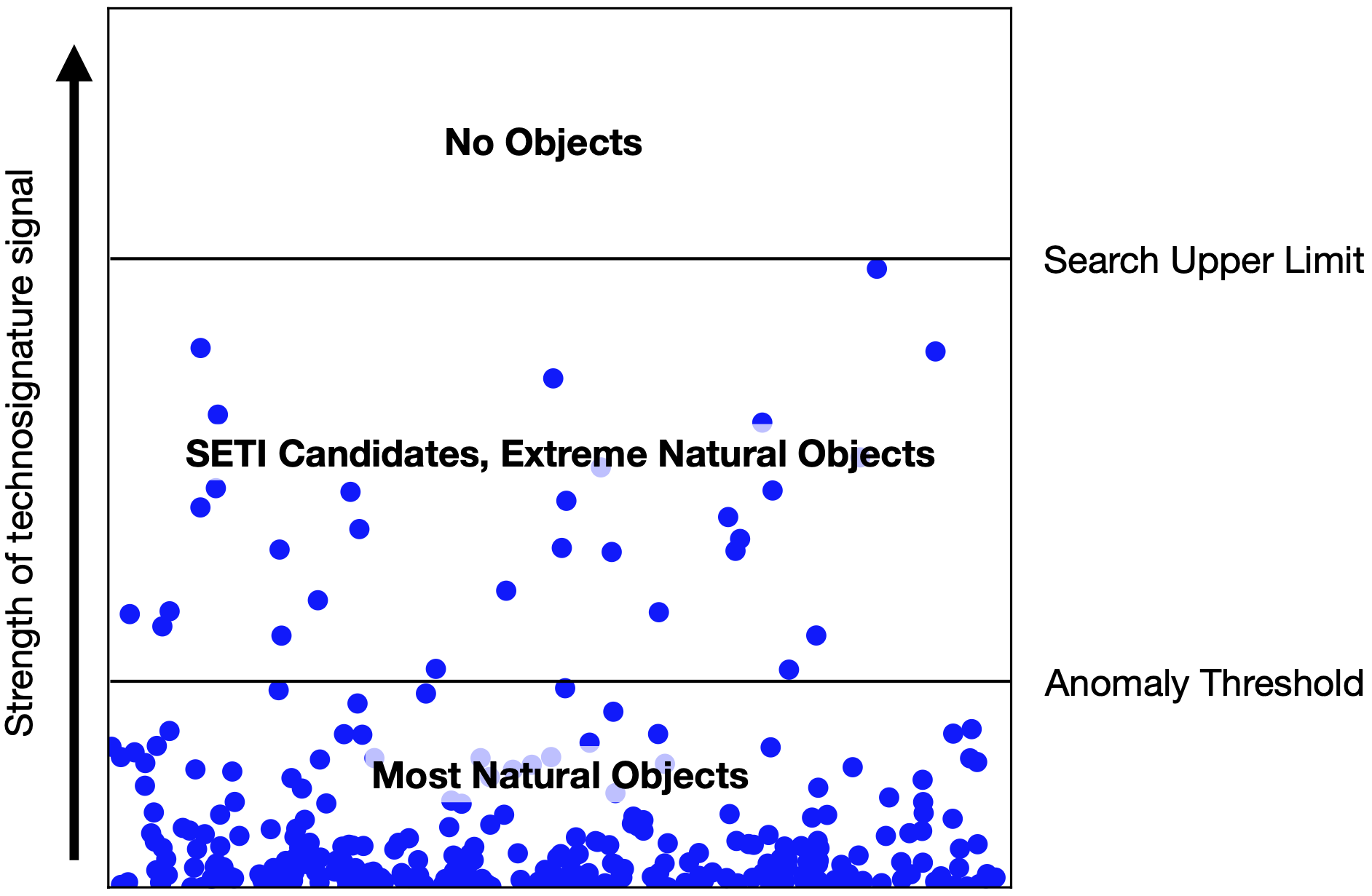}
    \caption{Schema of an anomaly-driven search for technosignatures. This method helps to identify SETI candidates, but also to detect and study extreme natural objects, between a defined upper-limit and an anomaly threshold. As such, it naturally produces ancillary science, even if no ETI is found. Fig. from \citet{lazio2023DataDrivenApproaches}.}
    \label{fig:Discussion_anomaly_threshold}
\end{figure}

Still, we should keep in mind that we cannot simply feed big data into an ML algorithm and hope to get as output all possible anomalies.
A great deal of data preparation and processing is needed, along with extensive training, and more or less implicit or explicit underlying theories and assumptions.
There is also the question of relying on big data from terrestrial systems.
This approach may overfit to life-as-we-know-it, creating a high-dimensional black box attuned to life on Earth that might not be able to detect extraterrestrial life.
What we call an ``anomaly'' is always relative to a chosen baseline, a threshold for anomalies, but also an existing theory, a scientific paradigm, and additional assumptions (see Fig. \ref{fig:Discussion_anomaly_threshold}).

Another approach is to use network theory to characterize whether a network has properties that are either universal (shared with biotic and abiotic systems), or whether some network structures and dynamics are uniquely present in bio- and techno- logical systems.
Such a dynamic, process-centered approach to searching for life is a sound foundation for a general agnostic search for life, even life-as-we-don't-know-it, because it assumes only general living functions such as dissipation, autocatalysis, homeostasis, or learning \citep{wong2024ProcessWe, bartlett2020DefiningLyfe}.

Techniques such as network reconstruction from noisy data \citep{newman2018NetworkStructure} may be particularly useful for remote characterizations.
For example, characterizing reaction network topologies in exoplanet atmospheres, in combination with other measurements, can help to assess the degree of thermal disequilibrium of an exoplanet \citep{fisher2022InferringExoplanet}.

Below we discuss some promising machine learning, information and complexity metrics that are especially relevant for SETI.
We saw that information theoretic tools are a key foundation for technosignature research (see section \ref{sec: interstellar ts}).
The most important metric at the heart of modern information theory is the entropy, as defined by \citet{shannon1948MathematicalTheory}:
\begin{equation}
H[X] = -\sum_{i}^{N} p_i \log_2 p_i
\label{eq:H_ent}    
\end{equation}
Entropy represents the uncertainty when dealing with a set of $N$ possible outcomes; these outcomes are quantified by the probability distribution $p$, where $p_i=p(X=x_i)$ is the probability of the $i^{\text{th}}$ outcome, $x_i$, occuring.
The more uncertain an observer is in the outcome, the higher the entropy.
Shannon showed that $H[X]$ is equivalent to the number of yes/no questions an observer would need to have answered in order to become certain of an outcome.
Maximum entropy is associated with maximum randomness (at least, from the perspective of the observer with the given prior distribution $p$).

As can be seen from \autoref{eq:H_ent}, it is very straightforward to calculate the entropy of a discrete distribution, such as a discretized signal received from the cosmos.
But we cannot simply associate high or low signal entropy with a high likelihood that that signal encodes meaningful information from a technological civilization.
Shannon entropy alone is insufficient to assess whether a signal is artificial, because non-living and non-technological sources can generate low-entropy signals.
A pulsar, for example, creates a periodic signal with low entropy, whereas white noise creates a signal with maximal entropy.

Single symbol entropy \autoref{eq:H_ent} does not factor in any kind of temporal correlations (if you scrambled the sequence of symbols in your string, the single symbol entropy would be unaffected).
Therefore, more subtle and sophisticated methods must be wielded \citep{gleiser2018ConfigurationalInformation}.
A common approach, used heavily in linguistic analysis, is to compute block entropies, where one uses the equation above but applied to finite-sized blocks of symbols, as opposed to just single symbols.
One can then assess how the entropy varies as a function of block size, and compare the scaling of these metrics with known information-rich signals.
Crucially, if the block entropy rate decreases below that of the single symbol entropy as a function of block size, then this implies higher level structure in the strings of symbols, e.g., a grammar \citep{doyle2002InformationTheory}.

One step further into the realm of intra-signal dependencies takes us to conditional entropies.
First, the joint distribution of seeing two symbols adjacent to one another is $p_{ij}=p(X=x_i, Y=x_j)$, with the single letter probability related to it by $p_i=\sum_j p_{ij}$.
Conditional entropy uses an analogous expression to $H[X]$, but invokes the conditional probability $p_{i|j}=p_{ij}/p_j$, the probability of encountering symbol instance $i$, given that $j$ was just encountered:
\begin{equation}
H[X|Y] = -\sum_{i,j}^{n} p_{ij} \log_2 p_{i|j}
\end{equation}
Note that one can also explore conditional probability distributions of blocks (words) of individual symbols, e.g., the probability that a given block of 3 symbols such as “ere”, follows a given block of 2 symbols such as “th” \citep{shannon1951RedundancyEnglish}.
Conditional probabilities and entropies introduce the idea of context-dependence, also a fundamental component of formal linguistic analysis, and the foundation for constructing so-called Hidden Markov Models (HMMs).
HMMs are abstract, statistical generative models for any process that produces strings of symbols from finite alphabets (e.g., the bits 0 and 1, or the Roman characters used in the English language).
They are generated through an inference process that begins with observations of whatever physical, biological or technological process produced the data of interest in the first place.
One example class of HMMs is the $\epsilon$-machine from computational mechanics, discussed below \citep{crutchfield2012OrderChaos}.

Additional information theoretic metrics use the concept of ‘how many yes/no questions must be asked to ascertain a given signal’ applied to multiple probability distributions.
For example, the KLD is a similarity metric for probability distributions, based on the information entropy concept \cite{kullback1951InformationSufficiency}.
It measures the excess uncertainty when one assumes the distribution is $q$ when it is in fact $p$:
\begin{equation}
D_{\mathrm{KL}}(p \| q) = \sum_{i} p_i \log \frac{p_i}{q_i}.
\end{equation}
Meaning, how many more YES-NO questions must one ask, on average, when working under a wrong assumption.
This metric was developed in order to be able to generalize the Shannon entropy to continuous events.

Since the KLD between $p$ and $q$ is not the same as the KLD between $q$ and $p$, one can also employ the symmetrized version known as the \textit{Jensen-Shannon (JS) divergence},
\begin{equation}
    JSD(p,q)=\frac{1}{2}(D_{KL}(p\|\frac{p+q}{2})+D_{KL}(q\|\frac{p+q}{2})
\end{equation}
The JS divergence has recently been used as the basis of a putative life detection method, where planetary atmospheric transmission spectra are treated as probability distributions \citep{vannah2024InformationTheory,vannah2025InformationalEntropic}.
Those exoplanet spectra carry crucial information about the chemical composition of a given planet’s atmosphere.
Despite the intrinsically physical meaning of transmission spectra, one can equally treat them as probability distributions, rendering them amenable to all the above techniques of information theory.
In this vein, \citet{vannah2024InformationTheory} compared different planetary spectra using the JS divergence.
This permitted an entropic comparison of a given spectrum with that of Earth, hence yielding a measure of ‘Earthlikeness’ grounded in information theory.

Another promising entropy metric for biosignatures and technosignatures is a measure of time series complexity called \textit{permutation entropy} \citep{bandt2002PermutationEntropy}.
It captures how unpredictable a sequence of values is by analyzing the orderings of neighboring points rather than their magnitudes: low values indicate regular or deterministic behavior, high values indicate randomness, and intermediate values reflect structured but non-random dynamics typical of complex systems.
It was recently integrated into a software package for time series analysis, including astrophysical phenomena \citep{hyman2025PECCARYNovel}.
Given its strengths as a complexity measure, the permutation entropy is particularly promising for assessing putative biosignatures and technosignatures: the expectation is that life will produce signals in the middle ground between perfectly ordered and purely random, which the permutation entropy can capture.

Recent developments in information theory also probe the concept of meaning or semantics, a feature that \citet{shannon1948MathematicalTheory} admitted was absent from his framework.
This burgeoning field of Semantic Information Theory \citep{kolchinsky2018SemanticInformation, sowinski2023semantic} proposes new ways to search for life in samples, such as might be returned from, or analyzed at solar system bodies \citep{bartlett2025PhysicsLife}.
The basic concept involves interventional causal experiments, where information that might be relevant to the putative living system is systematically scrambled, and the resulting effects then tracked.
A terrestrial example would be scrambling genes of an organism and then evaluating the impact that such gene loss has on its viability (i.e. its ability to survive).

However, extraterrestrial life might use different molecules or systems to store information, therefore scrambling other information stores or correlations between the agent and its environment would likely be necessary.
The so-called \textit{causal leverage density} attempts to measure semantic information of any kind (not just that related to the viability of an agent) \citep{sowinski2025CausalLeverage}.
Hence information that is meaningful to the agent in any sense, e.g., influencing its behavior, can in principle be measured by this technique.

Beyond information theory, there are several powerful sub-disciplines at the intersection of information theory, computer science and complexity science, which hold great promise for discerning signs of life and technology.
While the majority of complexity metrics invoke some form of entropy measure, there are other techniques that go even further and try to automate the process of generative model-building.
Most such techniques were formulated before machine learning became a mainstream tool of science and could hence be regarded as ‘ahead of their time’.

One such method, \textit{epsilon machine reconstruction}, takes as input a time series of measurements (e.g., a radio signal), and uses it to ‘reconstruct’ an HMM model (called an epsilon machine) capable of reproducing time series that are statistically equivalent to the input \citep{crutchfield2012OrderChaos}.
Crutchfield highlights several broad families of processes: 1) simple and deterministic, e.g., bit flipping $010101010101\dots$, 2) simple and random, e.g., white noise, 3) complex and deterministic, e.g., the Lorenz attractor, 4) complex and stochastic, e.g., a planetary atmosphere.
(see Fig. 1 in \citep{crutchfield2012OrderChaos}).

Epsilon machines have been used as the basis for a new approach to life detection, wherein planetary reflectance light curve time series are assessed in terms of statistical complexity and Shannon entropy \citep{bartlett2022AssessingPlanetary}.
It was found that synthetic versions of Earth with reduced atmospheric and surface features clustered at incrementally lower complexity and entropy values (see also \citep{segal2024PlanetaryComplexity}, and Earth was shown to have a higher complexity than Jupiter \citep{bartlett2022AssessingPlanetary}, as well as Mars.
Given these foundational results, the method has the potential to apply to a range of signals from both biotic and technological sources, and even establish a benchmark that can quantify distinctions between physical, biological, cultural or technological processes.

Arguably the ‘purest’ approach to signal analysis involves the use of Turing machines \citep{turing1937ComputableNumbers} that represent the most general and universal of all computational devices.
Algorithmic Complexity Theory \citep{zenil2023AlgorithmicInformation} builds on this foundation and uses Kolmogorov Complexity (KC), essentially measuring the size of the smallest Turing machine program that can reproduce a given string of symbols of a data stream.
A purely random string is by definition incompressible, and hence its KC is simply equal to its length.
In contrast, highly compressible signals, such as a sine wave or the digits of $\pi$, have very small minimal Turing machine programs and hence low KC.
Although there is no universal method to infer the smallest Turing machine program for a given string of symbols (due to the halting problem), there are various approximation methods available \citep{zenil2020ReviewMethods}.
This method has been applied in the context of decoding an extraterrestrial intelligent signal (see section \ref{subsec:communication_theory} and \citep{zenil2025OptimalUniversal}).

\subsection{Instruments for detecting technosignatures}
\label{subsec:instruments}

Our focus in this paper was on possibilities, but many targets require more instrumentation and search strategy development to be actually testable.
There are ongoing efforts in this direction, e.g.\ NASA's technosignature gap list \citep{timmaraju2022TechnosignaturesGapa}.
Without aiming to be comprehensive, we propose in Appendix~\ref{sec: STM} a Science Traceability Matrix template that illustrates the main categories of technosignatures we reviewed.

\subsection{Evaluation \& Prioritization}
\label{subsec:eval_prioritize}

Reviewing all these possible technosignatures can bring a sense of anxiety: which one should we prioritize, and how?
The pragmatic and economic answer is simply that the research that gets funded gets prioritized.

More fundamentally, as a result of a 2018 NASA Technosignature Workshop at the Lunar and Planetary Institute (LPI), nine axes of merit emerged for technosignature searches \citep[see][]{sheikh2020NineAxes}.
Of course, assessments of search strategies along these axes is itself a matter of debate, in the sense that different researchers would rate a given strategy differently on the different axes.
They may also apply a different subjective weight to the axes themselves.
One way to aggregate expert opinions and identify the most salient points of disagreements or agreements is to use the axes in combination with the Delphi method \citep{profitiliotis2022DelphiApproach}.
Another way to aggregate expert opinions is to use fuzzy adaptations of Multi-Criteria Decision-Making \citep{sanchez-lozano2024DecidingTechnosignature}.

\begin{figure}
    \centering
    \includegraphics[width=1\linewidth]{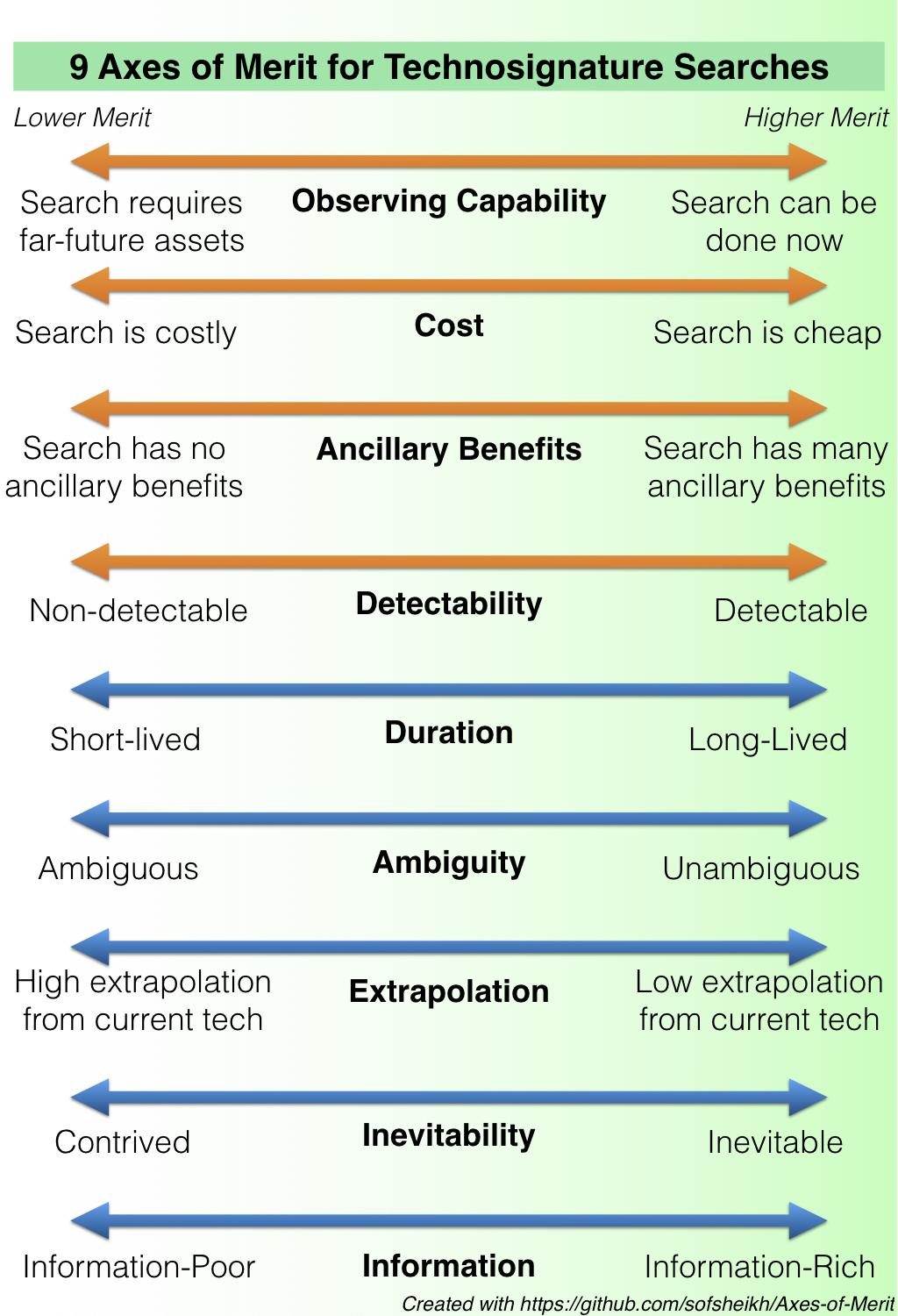}
    \caption{Axes of merit for technosignature searches, from \citep{sheikh2020NineAxes}. Note that axes 1--4 (in orange) cover the more practical aspects of any technosignature searches \citep{sheikh2020NineAxes}.}
    \label{fig:Discussion_AoM}
\end{figure}

\begin{figure}
    \centering
    \includegraphics[width=1\linewidth]{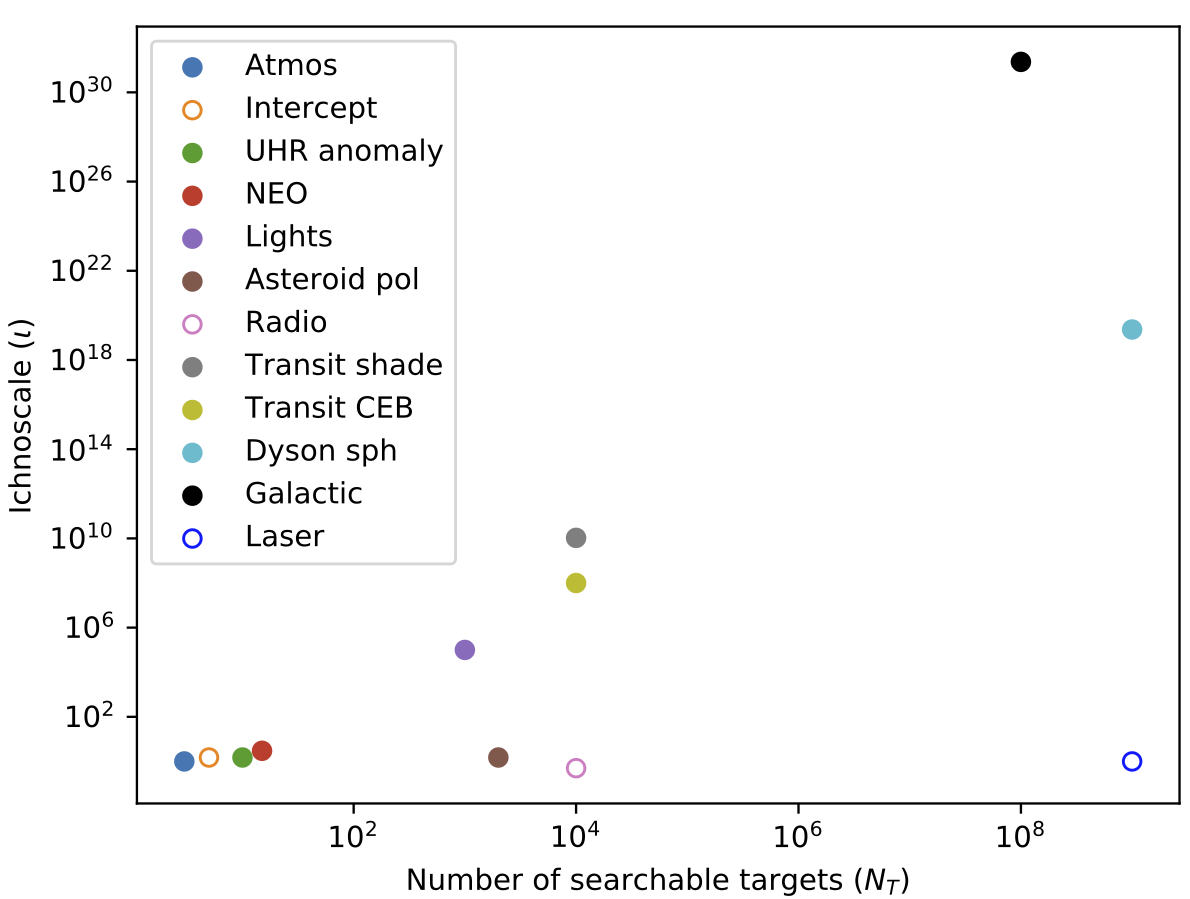}
    \caption{Ichnoscale (relative footprint of a given technosignature in units of current Earth technology) vs.\ number of targets for several possible technosignatures. Filled (empty) circles represent continuous (discontinuous) observables. Fig. from \citet{socas-navarro2021ConceptsFuture}.}
    \label{fig: Discussion_ichnoscale}
\end{figure}

The axes of merit do not explicitly cover the number of targets, but another scale, the \textit{ichnoscale}, does (see Fig. \ref{fig: Discussion_ichnoscale}).
This scale attempts to quantify a technosignature footprint relative to current Earth technology \citep[see][]{socas-navarro2021ConceptsFuture}.

We note that other assessment scales have been proposed to rank potential biosignatures \citep{neveu2018LadderLife, green2021SearchingTechnosignatures, meadows2022CommunityReport}.
However, further work is needed to create an encompassing framework for assessing and comparing both biosignatures and technosignatures.

\subsection{Epistemological issues}
\label{subsec:epistemological}

\subsubsection{Defining technology}
\label{subsubsec:defining_technology}

Like the effort to define ``life,'' reflection on the nature of technology goes back to antiquity.
It is a prominent element not only in ancient Greek mythology, notably in the Prometheus myth \citep{hesiod2017TheogonyWorks, aeschylus1961PrometheusBound}, but there are also ancient roots of technological thought in Asia \citep{hui2022QuestionConcerning}, as well as in the foundations of western philosophy in Plato's \textit{Timaeus}, \textit{Protagoras}, and \textit{Phaedrus}.

Aristotle introduced the foundational opposition between \textit{physis} (nature) and \textit{techne}, which is the category of everything artificial, everything whose production is outside it: all the objects that only exist thanks to human intervention within nature.
We find in his distinction between \textit{physis} and \textit{techne} a criterion that persists in the contemporary search for technosignatures.
The presence of technology is betrayed by motions, arrangements, or distributions of matter and energy that suggest the intervention of a factor external to the pertinent physical regime.
In SETI, we call that factor ``intelligence.''

Our implicit or explicit assumptions about human consciousness and its relation to technology may place unnecessary limits on our reasoning about technology, and therefore technosignatures.
Setting those assumptions aside, as in the agnostic search for anomalies, may allow us to imagine more technosignatures, and perhaps increase our chances of detection.

If technology becomes part of more general informational capabilities of life, the possibilities for technodiversity at the cosmic scale begin to open up.
Indeed, life on Earth might itself be on the verge of a technodiversity explosion \citep{vidal2015BiodiversityTechnodiversity}.
This review attempts to capture a glimpse of what such a technodiversity might be in the universe.

In fine, we can easily transpose the lessons from efforts to define ``life'' \citep{cleland2019QuestUniversal} to efforts to define ``technology,'' and conclude that what matters is not to focus on finding a short, ideal definition of technology and technosignatures, but rather to develop theories of technology that make predictions about observable technosignatures.
This has been our approach throughout this review.

\subsubsection{N=1 and the Wow signal}
\label{subsubsec:wow}

The $N=1$ problem in astrobiology refers to the fact that we have only one data point, life and technology from Earth, from which we attempt to extrapolate other lifeforms and intelligence in the universe \citep[see][]{cleland2019QuestUniversal}.
Detecting a technosignature only once thus puts us in a similar epistemological situation of having only one data point.
An event that is non-repeating makes scientific progress and verification extremely challenging if not impossible.

\begin{figure}
     \centering
     \includegraphics[width=1\linewidth]{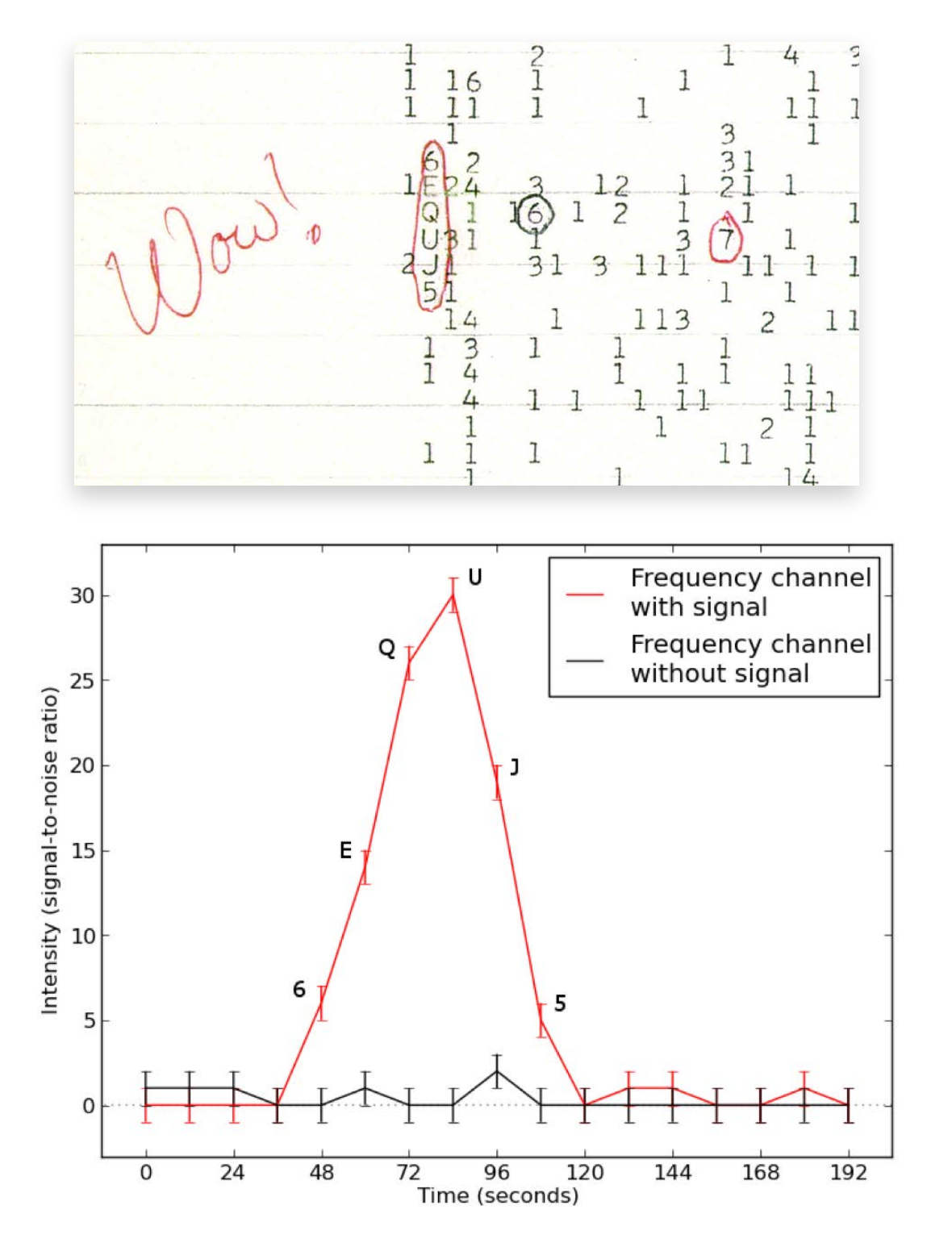}
     \caption{The “Wow” signal recorded on August 15, 1977 at the Big Ear Ohio State Radio Observatory, immortalized by Jerry Ehman's annotation upon reviewing the data \citep{Charbonneau2018Wow}.}
     \label{fig: Discussion_wow}
 \end{figure}

The ``Wow!'' signal detected on August 15, 1977 is such an example of a non-repeating event (see Fig. \ref{fig: Discussion_wow}).
Its strength, narrow band, and location near the 21-centimeter hydrogen line all point toward a possible artificial origin.
The signal's 72-second duration perfectly matched the observation window of the Big Ear radio telescope, suggesting it came from a fixed point in deep space.
However, the signal has never been detected again despite follow-up searches.
Of course, scientists have also considered natural, though rare, phenomena such as an astronomical maser flare as a potential source (for a recent update, see \citealt{caballero2022ApproximationDetermine}).

The fact that the signal was detected in only one of the telescope's two beams might support the hypothesis of a sweeping beacon that a telescope could have just caught once, on a brief pass.
On the side of the Earth, humanity itself sent the powerful but brief Arecibo message in 1974 only once, as a symbolic gesture.
Similarly, one can speculate that the ``Wow!'' signal could have been a similar ``message in a bottle,'' a momentary announcement from another civilization in the cosmos ... before METI became also controversial!

\subsubsection{Science fiction}
\label{subsubsec:scifi}

The role of imagination is key to the scientific process.
The core difference between science and science fiction is that science fiction aims to create emotional and engaging stories for human entertainment, while science tries to gain new insights, knowledge, and understanding, highly constrained by its methods and criteria.
A systematic study of major science fiction novels to derive technosignature strategies would be worthwhile, although outside the scope of this paper.
There is a rich interplay and synergy between science and science fiction \citep[see][]{kowal2023LifeUs}: many new ideas start in science fiction and inspire scientists, while new scientific theories and discoveries inspire hard science fiction authors.

However, science fiction is a double-edged sword for academic SETI.
On the negative side, it contributes to the ``giggle factor,'' creating implicit associations between entertainment and serious science.
On the positive side, science fiction addresses the question of extraterrestrial life and intelligence, which is so popular and fascinating that it is a huge opportunity for science education and outreach to draw people of all ages towards science.

\subsubsection{Epistemology}
\label{subsubsec:epistemology_sub}

Most technosignatures reviewed here involve observing remotely with the help of astronomical instruments.
Instruments like telescopes do not retrieve information unfiltered \citep{duhem1954AimStructure}.
Instead, they have their very own way of accessing the world, i.e., are sensitive to a specific wavelength, and modify photons in specific ways.
This means that any scientific instrument must ``translate'' its ``perception'' to modalities comprehensible to humans.
Such a translation ``mediates'' our experience, in a way that constitutes a particular understanding of the external world \citep{ihde2011StretchingInbetween}.

Let us look again at the Wow! signal.
What you see is the output of the ``Big Ear'' telescope which transformed the incoming information to a human-readable form, in which numbers, letters, and curves indicate signal strength.
To further understand this output, humans need a particular set of skills, attunement to the way the instrument presents its data, and knowledge of the scientific paradigm.
For any layperson, the symbolic output may be relatively meaningless, but for Jerry R.\ Ehman, this output generated at that time was truly astonishing.

This example shows that technosignature detection hinges upon the mastery of scientific instruments by individuals whose ability to make sense of the output is collectively acknowledged \citep{dobler2023AffordancesOutera}.
In other words, since we cannot directly visually perceive most processes in the universe, we rely instead on inferring their presence from the mediated information whose visual appearance may simply be a graph or diagram instead of direct visual evidence of the extraterrestrial.

Yet, given the uncertainty revolving around scientific observation, inference, and communication, converging on exactly one interpretation of the data requires time to confront and integrate hard work, independent validation, criticism, and debate amongst scientists.

A classical objection against astrobiology is \citeauthor{simpson1964NonprevalenceHumanoids}'s (\citeyear{simpson1964NonprevalenceHumanoids}) statement that ``this `science' has yet to demonstrate that its subject matter exists!''
While it is true that it lacks a concrete object of study, one can reply that astrobiology is a scientifically informed search, an exploratory scientific effort \citep{tarter2001SearchExtraterrestrial}.
We can draw a parallel with the discovery of gravitational waves.
They were theorized to exist by \citet{einstein1915FeldgleichungenGravitation}, but it took another 60 years before indirect evidence of gravitational wave radiation would be found \citep{weisberg1981GravitationalWaves}, and 40 more years before direct evidence would be brought thanks to the construction of gravitational wave detectors \citep{ligoscientificcollaborationandvirgocollaboration2016ObservationGravitational}.
From theory to strong observational proof, a full century has elapsed!
\citet{chyba2005AstrobiologyStudy} give other historical examples of searching for non-existent objects such as black holes, the Higgs boson, or proton decay.
As \citet{wright2026SearchExtraterrestrial} notes, the study of SETI is even better motivated than the search for black holes or gravitational waves, because technological life on Earth means $N=1$ and not $N=0$.

One can reasonably hope that the search for bio- and techno- signatures will follow a similar path.
The theoretical part is the sum of interdisciplinary knowledge that astrobiologists have gathered through cosmology, astronomy, the origin of life, and evolutionary theory, leading to the hypothesis that life is common and detectable.
We have no agreed-upon evidence today, but might find only indirect or insufficient evidence at first, that would only be confirmed by the construction of next-generation observatories later.

Even if the search space is large, each SETI experiment provides constraints and upper limits on the prevalence of a particular technosignature.
Such a hypothesis-driven search is ruling out and closing the net on possibilities.
As such, astrobiology is like any other science, following the regular hypothetico-deductive method \citep{popper1959LogicScientific}, and the null results that are found simply falsify hypotheses.

What is more, philosophers of science have a broader understanding of what science does beyond studying concrete observable objects, and understand that it has various stages, from pre-paradigm, to normal science and revolutionary science \citep{kuhn1996StructureScientific}.
\citet{lakatos1970FalsificationMethodology} argued that at heart scientists conduct research programs that can either be progressive if they continue to make new predictions and these predictions are corroborated, or degenerative if they rely too heavily on ad-hoc hypotheses.
The undeniable creativity and diversity of predictions we reviewed clearly open avenues for testing and corroboration, and in that sense technosignature research programs are progressive.

In addition, if a given SETI experiment is designed to bring results and synergies with other efforts, then again, there is benefit for ancillary science, and thus continuous progress.
This can also happen as a technological driver, for example when the NSA and SETI researchers shared technological know-how to detect artificial signals during the Cold War \citep{charbonneau2024MixedSignals}.
\section{Conclusion}
\label{sec:conclusion}

A few decades ago, the very existence of exoplanets was a highly speculative topic.
Today, it is normal science, at the heart of astrobiology.
Advances in technology and observational techniques are hard to predict, and it could well be that many of the technosignatures reviewed in this paper will become easily and routinely observable.

We are aware that we have cast a very wide net on possible technosignatures.
Since the search is ultimately poorly constrained, it is the most prudent approach.
Some technosignatures like UAPs, Genomic SETI or wormholes are in a state of either high controversy, low maturity, or with low observational support, but we included them because they are still potentially testable.

 What emerges from this review is that technosignature science need not be defined by communication alone.
 We believe more progress can be made by attempting to test a variety of well-articulated proposals, rather than trying to lock the search into a narrow set of assumptions and search strategies.

We saw that biosignature and technosignature searches have natural synergies in the cases of solar system exploration and exoplanet observations.
Technosignature searches are thus likely to be prioritized around these scales and targets.
We have also identified synergies between technosignature searches and other fields such as astronomy.
This means that even if technosignature searches fail, we should still learn something new.
As with many major discoveries of the past, we should keep in mind that technosignatures could be found serendipitously, which is why the synergies should also include an interdisciplinary dimension.

An emerging conclusion is that multiscale, multimodal, multimessenger, and multisite searches are the most future-proof search strategies.
Indeed, if an anomalous event is detected with a given modality or technique (e.g.\ transit), it can be cross-searched with other observational modalities (e.g.\ radio, heat signatures).
Imagine for example that we would find a strong atmospheric biosignature; it would be natural to follow up with searches for waste heat, communication signals, and transit anomalies.
Such multimodal follow-ups are already being practiced, for example the Hat Creek Radio Observatory telescope has been following up candidate megastructures such as KIC~8462852 \citep{harp2016RadioSETI} and `Oumuamua \citep{harp2019RadioSETI}, albeit both with null results.
The multisite aspect also helps to confirm the skybound nature of sources, and RFI mitigation in general.

More fundamentally, we argued that machine learning, as well as information and complexity theories, make us see the search for biosignatures and technosignatures as a unified search for anomalies.
This makes sense because both biological and technological systems are complex organizations that cannot be fully described as simple physical systems.
Whatever bio- or techno- signature will be found first, we must also remain aware that it is likely just one of many, and the search for others must continue.

\section*{Acknowledgments}
This paper was started as a collective workshop at the Penn State SETI Symposium (PSETI 2023).
We thank Macy Huston for their valuable input during that workshop.
We thank Mark Elowitz for his many contributions to an earlier version of this paper.
We thank Paolo Oliveira for insightful and relevant feedback on an earlier version of this paper.
Manika Singla was supported by the Institute for Basic Science (IBS-R035-C1).
\newpage
\section{References}
\bibliography{references}
\bibliographystyle{apaurldoi3}

\clearpage
\section{Appendices}\label{sec: Appendix}
\subsection{Technosignature Science Traceability Matrix Template}\label{sec: STM}
\begin{longtable}[l]{|>{\raggedright\arraybackslash}p{2.3cm}|>{\raggedright\arraybackslash}p{4.4cm}|>{\raggedright\arraybackslash}p{4.4cm}|>{\raggedright\arraybackslash}p{4.4cm}|}
\caption{Technosignature Science Traceability Matrix (STM). Note that this is a template, and an actual STM requires additional measurement and instrument requirements.}\label{tab:tstm}\\
\hline
\textbf{Science Goal} & \textbf{Technosignature / Artifact Type} & \textbf{Observational Feasibility (today and/or near-future)} & \textbf{Required Instruments and methods} \\
\hline
\endfirsthead
\multicolumn{4}{c}%
{\tablename\ \thetable\ -- \textit{Continued from previous page}} \\
\hline
\textbf{Science Goal} & \textbf{Technosignature / Artifact Type} & \textbf{Observational Feasibility (today and/or near-future)} & \textbf{Required Instruments and methods} \\
\hline
\endhead
\hline \multicolumn{4}{r}{\textit{Continued on next page}} \\
\endfoot
\hline
\endlastfoot
\multicolumn{4}{|l|}{\textbf{Earth Technosignatures (Section \ref{sec: earth ts})}} \\
\hline
Searching for past visitation & Long-lived artifacts or time capsules in the geological, fossil, or archaeological record; a ``shadow technosphere'' of past extraterrestrial activity on Earth (extension of the ``shadow biosphere'' concept). & A dedicated survey of the fossil and geological record would demand massive resources; however, it may be serendipitously discovered by archeologists and geologists in the field. Climatology, geology, archaeology and paleontology become relevant disciplines. & Methods and instruments used by archeologists, paleontologists, and geologists; isotopic and chemical analysis of strata; anomaly detection on geological databases. \\
\hline
UAP studies  & Active spacecraft or probes in Earth's atmosphere, ocean, or near-space. & Closest possible technosignature, so has a high potential visibility. The phenomenon is transient and unpredictable, which demands ideally 24/7 multimodal all-sky coverage. $\sim$5--10\% of reports remain unexplained. See also the UAP science traceability matrix in \citet{watters2023ScientificInvestigation}. & Galileo-Project-style multimodal observatories: all-sky fish-eye cameras, passive radar, acoustic sensors, high-resolution/high-zoom camera with ML outlier detection, magnetometers, infrared cameras, cosmic-ray / high-energy particle detectors, and high-resolution remote-sensing satellites (Landsat, Planet Labs, Maxar, etc.). \\
\hline
Earth orbit & Probes and artefacts in Earth orbit; high-albedo objects in geosynchronous orbits; non-terrestrial artefacts on archival photographic plates \citep{villarroel2022GlintEye}. & Very low distance and high observability, but distinguishing an alien artefact from human-made space junk or satellites is not obvious. & Historical astronomical plates (pre-satellite age / Sputnik); dedicated telescopes and radars with algorithms to distinguish from satellites; VASCO and ExoProbe-style searches; photographic-plate archives (e.g. POSS1-E). \\
\hline
\multicolumn{4}{|l|}{\textbf{Solar System Technosignatures (Section \ref{sec: ss ts})}} \\
\hline
General solar-system SETA & Active or passive probes, artificial structures, Bracewell / Von Neumann probes; surface artefacts and orbital artefacts. & Most of the solar system has not been observed at high resolution (see Table 3 of the paper). The Barrow scale implies that technology tends towards ever-smaller scales, making detection harder. Active probes are easier to detect if they communicate or produce waste heat and waste materials. & High-resolution imaging from existing planetary missions; archeological methods applied to surface artefacts; commensal SETI on archival data; future deep-space telescopes and probes. \\
\hline
The Moon & Lunar artefacts, debris, monoliths, unusual structures (e.g. the unusual geometric features in Paracelsus C crater); evidence of past activity at sub-meter scales. & The Moon is a privileged place for long-term artefacts (shields from meteoroids and half of ionizing radiation, no erosion, no atmosphere, no biology). LRO has imaged the surface at 0.5 m resolution. Machine-learning analyses can detect human activities, but a systematic search on LRO data remains to be done. & Lunar Reconnaissance Orbiter (LRO) high-resolution imagery with machine learning anomaly detection; citizen-science image searches; future lunar missions. \\
\hline
Lagrange points & Probes at Lagrange points L1--L5 (Earth--Moon and Earth--Sun); artificial probes embedded in the Kordylewski dust clouds at L4/L5; co-orbital / horseshoe-orbit objects that would be used as long-term Earth monitors. & Targeted multiwavelength observations and in-situ missions. The L2 Earth--Moon Lagrangian point requires a space mission as it is permanently occulted by the Earth. Horseshoe orbits provide concealment, long-term stability and periodic observation opportunities. & Targeted multiwavelength observations covering the Lagrange regions; dedicated in-situ spacecraft; surveys of co-orbital asteroids and horseshoe-orbit objects with anomaly detection. \\
\hline
Inner solar system (Mercury, Venus, Mars and their moons) & Probes from migrating civilizations (an extension of the silurian hypothesis); anomalous surface or orbital features; outliers in orbital-parameter space (non-gravitational acceleration, fine-tuned gravity assists, goal-directed behaviour). & Beware of pareidolia in low-resolution imagery (e.g. Lowell's ``canals''). Anomaly detection on existing planetary imagery is a promising route. & Anomaly detection on Mars Reconnaissance Orbiter, MESSENGER, Magellan, Mars Global Surveyor, and other planetary-mission archives; orbital-parameter outlier surveys. \\
\hline
Asteroid belt & Lurkers hiding in low-gravity wells; forensic evidence of extrasolar asteroid mining; metallic surfaces standing out from natural objects. & Polarimetry and spectroscopy to detect metallic surfaces uncommon among natural objects. & Polarimeters and spectroscopes on existing/future surveys; asteroid mining-signature searches; high-resolution imaging missions (Dawn-class). \\
\hline
Interstellar objects (ISOs) & Anomalous ISOs interpreted as passive probes, light sails or megastructure debris (e.g. 'Oumuamua-like candidates); ``dark comets'' with non-gravitational accelerations. & Detection rates will rise sharply with the Vera C. Rubin Observatory and the Nancy Grace Roman Space Telescope. \citet{davenport2025TechnosignatureSearches} recommend a multi-mode observing protocol during ISO passages (see Table 4): astrometry/tracking, full-sky photometry, optical/IR spectroscopy, radio spectroscopy, multi-band photometry, high-cadence imaging, IR photometry, polarimetry, and radar imaging. & Vera C. Rubin Observatory (LSST); Nancy Grace Roman Space Telescope; JWST; planetary radar (Goldstone and successors); rendezvous missions for ISO interception \citep{siraj2023PhysicalConsiderations,hein2022InterstellarNow}. \\
\hline
Outer solar system & Anomalous occulting objects & Detectable via high-cadence optical photometry; rare events that demand cross-instrument validation. & High-cadence optical photometric surveys; archival pulse searches; multi-wavelength follow-up. \\
\hline
Kuiper belt & Artificial Kuiper belt objects; artificially illuminated surfaces. & Lights on Kuiper belt objects are in principle detectable with existing instruments; same capabilities can also be applied to exoplanet surfaces. & Sensitive optical/IR telescopes capable of detecting faint illumination (HST, JWST, Subaru, LSST/Rubin). \\
\hline
Solar gravitational lens (SGL) region & Probes or communication relays at the SGL focal region (EM focus from $\sim$547 AU; neutrino focus from $\sim$29.6 AU; gravitational-wave focus from $\sim$22.45 AU). & A radio search of SGL communication between the Sun and $\alpha$ Centauri has been conducted \citep{tusay2022SearchRadio}; \citet{kerby2021StellarGravitationalb} suggested searching for artefacts around the focal region. & Targeted radio and optical searches towards the SGL regions; future probes to the SGL focal region. \\
\hline
Oort cloud & Civilizations migrating across or operating at the boundary between stellar systems; ejection signatures; communicative or waste heat signatures at the periphery \citep{romanovskaya2022MigratingExtraterrestrial}. & Difficult given enormous distances; neighbouring Oort clouds are almost touching, which makes the periphery a natural interface for ETI activities. & Wide-field optical/IR surveys (Vera Rubin / Roman); IR and communicative anomaly surveys at the solar-system periphery. \\
\hline
\multicolumn{4}{|l|}{\textbf{Exoplanetary Technosignatures (Section \ref{sec: planet ts})}} \\
\hline
Surface technosignatures & Artificial night-side illumination; surface megastructures (solar-panel fields, ecumenopolis, large industrial complexes); polarization signatures of artificial surfaces (cement, plastic, cloth); urban heat-island effect. & JWST could detect artificial lights on Proxima b if 500$\times$ stronger than Earth LEDs or in a spectrum 1000$\times$ narrower. Solar-panel detection demands hundreds of HWO-class observation hours assuming $\sim$23\% land coverage. SGL telescope could yield $\sim$224 km/pixel at 10 pc, enough to detect for megastructures. & JWST; future Habitable HWO; LIFE space mission; SGL-positioned telescope; coronagraph and starshade equipped large telescopes; polarimeters; ELTs. \\
\hline
Atmospheric technosignatures & Industrial pollutants and engineered gases: CFCs, HCFCs, PFCs (CF$_{4}$, C$_{2}$F$_{6}$, CCl$_{3}$F); high-GWP greenhouse gases (SF$_{6}$, NF$_{3}$, SF$_{5}$CF$_{3}$); NO$_{2}$ from combustion; chemical fossils; isotopic anomalies (plutonium-244, anomalous D/H ratios); methanethiol or other markers of nuclear/biotechnological catastrophes; ozone-layer depletion. & Long residence times and high GWPs make many of these detectable across interstellar distances. Best targets: SF$_{6}$, CF$_{4}$, NF$_{3}$ (vibrational modes 7.5--16 $\mu$m); NO$_{2}$ in visible reflectance (400--500 nm); CCl$_{3}$F in mid-IR $\sim$11--13 $\mu$m. Detection ideally requires a $>$15 m space-based mid-IR telescope with high-resolution spectroscopy and stellar-contamination control.  & JWST; planned LIFE mission; future HWO; high-resolution mid-IR spectrometers; ELTs. \\
\hline
Orbital technosignatures & Clarke exobelts (artificial satellite belts); mega-constellations and glints from artificial satellites; asymmetric or anomalous debris-disk features \citep{jaiswal2023SpecularReflectionsa}; persistent disk-like signatures around old (post-zero-age) stars; geocentric solar-power satellite beaming; planetary Dyson sphere with 10 $\mu$m waste-heat signature. & Detectable via transit/photometric signatures and infrared characterization of debris disks. Mega-constellations could be detected with VLTI \citep{osmanov2021SpaceXStarlink}. The duration of debris-disk technosignatures could be long, because still observable if ETI is inactive or extinct. & Kepler-class transit photometry (TESS, PLATO); Very Large Telescope Interferometer (VLTI); debris-disk imaging with ALMA / JWST / HST; future high-resolution coronagraphs. \\
\hline
Exoplanetary- system technosignatures & Service worlds (highly polluted, terraformed planets outside the habitable zone); active asteroid mining generating localized infrared excess; engineered planetary systems. & Detectable through transit-timing variations, dynamical anomalies, and infrared excess from in-system mining. Search strategies developed for our solar system (Section \ref{sec: ss ts}) generalize to other systems, with greater distance challenges. & Transit/timing photometry (Kepler/TESS/PLATO); IR photometry (Spitzer/JWST); ALMA for debris-disk imaging; dynamical/orbital surveys. \\
\hline
Multiplanetary systems and terraforming & Multiple planets in one system with engineered, suspiciously similar atmospheric compositions; high GWP greenhouse gas signatures across planets; engineered planetary pressure, O$_{2}$ growth or temperature. & Searches should not be limited to the conventional habitable zone, as ETIs could terraform inside or outside it. A cluster of planets with anomalously similar atmospheres would be the key indicator. & Transit/atmospheric spectroscopy with JWST, HWO, LIFE, ELTs; comparative multi-planet atmospheric studies. \\
\hline
\multicolumn{4}{|l|}{\textbf{Stellar Technosignatures (Section \ref{sec: stellar ts})}} \\
\hline
Stellar megastructures (transit detection) & Anomalous transit light curves, e.g. asymmetric, irregular, jagged, non-uniform, or time-varying, indicating Dyson swarms, ring structures, or messaging megastructures. Boyajian's star (KIC 8462852) was a candidate, although now attributed to dust). & Well-established transit-photometric techniques. Machine learning to flag unusual light curves systematically. Simulations of megastructure transits help guide pattern recognition. & Archival Kepler; TESS; PLATO; LSST/Rubin; ML-based anomalous-transit pipelines. \\
\hline
Energy utilization (Dyson spheres/swarms) & Infrared waste-heat signature of partial or complete Dyson spheres/swarms ($\sim$300 K, $\sim$10 $\mu$m at 1 AU); anomalous IR excess inconsistent with circumstellar dust. & Has been studied with infrared observatories (IRAS, WISE). \citet{zackrisson2018SETIGaia} and Suazo \citet{suazo2024SearchingDyson} cross-matched Gaia DR3, 2MASS and WISE on $\sim$5 million sources, identifying seven M-dwarfs with anomalous IR excess as candidates. & IRAS, WISE, 2MASS, Spitzer, Gaia DR3 cross-matches; JWST; future IR observatories. \\
\hline
Stellar modification / rejuvenation & Stars with anomalously extended main-sequence lifetimes (e.g. anomalously composed blue stragglers as proposed by \citeauthor{beech1990BlueStragglers} \citeyear{beech1990BlueStragglers}); high-reflectivity Dyson spheres that prolong stellar lifetime; stellar mixing signatures. & Models propose observable signatures \citep{scoggins2023LazarusStars, huston2022EvolutionaryObservational, beech2008RejuvenatingSun}; requires precise stellar parameters and population studies. & High-resolution stellar spectroscopy; stellar-evolution modelling; photometric monitoring; precise asteroseismology. \\
\hline
Stellar pollution & Photospheric spectral signatures of dumped nuclear-fission waste products; anomalous heavy-element abundances. Candidate: Przybylski's star (HD 101065). & Detectable via high-resolution stellar spectroscopy & High-resolution stellar spectrographs (e.g. ESPRESSO, HARPS, Keck HIRES, ELTs). \\
\hline
Compact objects & Dyson spheres around white dwarfs, neutron stars/pulsars or black holes (IR/UV/optical/X-ray signatures); engineered pulsar modulation patterns; ``mini-Earths'' (rocky planets with central small black holes); ``stellivores'' (reinterpretation of some accreting binaries) anomalously slowing black-hole rotation (Penrose process). & Detectable through anomalous IR/X-ray excess, pulsar timing anomalies, and X-ray binary peculiarities.  & X-ray observatories (Chandra, XMM-Newton, future Athena); pulsar timing arrays (NANOGrav, EPTA, IPTA); IR observatories; gravitational-wave observatories. \\
\hline
\multicolumn{4}{|l|}{\textbf{Interstellar Technosignatures --- Communication (Section \ref{sec: interstellar ts})}} \\
\hline
Artificial radio signals & Narrowband or broadband radio signals, which may be beacons, modulated pulses, intentional or leakage radiation (e.g. deep-space-network analogs). The ``cosmic water hole'' region (1.42--1.67 GHz) and low-frequency bands are key search regions. & Many radio telescopes already scan for signals, but requires long observation times and sophisticated RFI mitigation. Commensal SETI on major-telescope raw data is increasingly productive. Earth contamination motivates lunar-farside deployment. & Breakthrough Listen program and observatories such as Green Bank Telescope; Parkes; Automated Planet Finder; MeerKAT; FAST; Hat Creek Radio Observatory (Allen Telescope Array); SKA; LOFAR; OVRO-LWA; NenuFAR; GMRT; future Lunar Farside Technosignature \& Transients Telescope (LFT3). \\
\hline
Pulsed (short - duration) optical laser transmissions & Nanosecond-scale laser pulses producing flashes far brighter than the host star at the pulse wavelength. & Requires high time-resolution sensitivity and long observation windows. & PanoSETI; Laser SETI program at the SETI Institute; VERITAS; STACEE; Lick/Leuschner Observatories; STACEE. \\
\hline
Continuous optical laser transmissions & Narrow monochromatic emission lines from continuous-wave lasers, distinguishable from natural emission by their narrow linewidth. & Requires high spectral resolution to distinguish narrow monochromatic emission lines from background starlight. Many high-resolution spectrographs are being used for exoplanet detection, so there's abundant archival data to survey. & Keck HIRES; HARPS; ESPRESSO; archival exoplanet-spectroscopy data mining. \\
\hline
X-ray laser signals & Artificial pulsed or continuous X-ray laser emissions producing narrow X-ray emission lines distinct from natural astrophysical sources. & Requires future high spectral resolution and sensitivity to detect narrow X-ray emission lines from a laser. Currently challenging: Chandra requires $\sim$5 days to build a spectrum; SNR is the main limitation. & Current X-ray observatories (Chandra, XMM-Newton); future high-resolution/sensitivity X-ray spectrometers (e.g. ESA Athena); Lynx X-ray Observatory \\
\hline
Gamma-ray communication & Narrowband or pulsed gamma-ray signals; gamma-ray signatures from antimatter propulsion or warp-bubble snowplow ejections. & Rarely explored. Could synergize with advanced-propulsion technosignatures. & Fermi-LAT; INTEGRAL; future MeV--GeV missions. \\
\hline
Neutrino communication & Coherent, pulsed, or unusually structured artificial neutrino signals. & Neutrinos can penetrate the ISM and dense objects essentially unimpeded; near-noise-free in the galactic plane. Solar gravitational neutrino focus starts at $\sim$29.6 AU, much closer than the EM focus. & IceCube; KM3NeT; future MEMPHYS- or 1-Mton-class water-Cherenkov detectors. \\
\hline
Gravitational wave communication & Engineered gravitational-wave signals at characteristic frequencies / modulations distinguishable from natural mergers. & \citet{abramowicz2020GalacticCentre} argue a Jupiter-mass GW emitter at the galactic centre would be LISA-detectable. & LIGO, Virgo, KAGRA; future LISA, Einstein Telescope, Cosmic Explorer; high-frequency GW concepts. \\
\hline
Inscribed matter & Physical messages encoded in matter: e.g. messages in non-coding DNA of space-proof microorganisms; ET probes hosting onboard archives. & Genomic sequencing; in-situ examination of solar-system artefacts. 
Inscribed-matter data rate falls only as 1/r (vs 1/r$^{2}$ for EM), giving an advantage at large distances; particularly attractive for searches inside our solar system. & DNA sequencing of extremophiles followed by anomaly detection (e.g. millisecond-pulsar spacetime timestamps); high-resolution imaging of solar-system bodies; rendezvous/sample-return missions. \\
\hline
Occulters & Artificial opaque objects modulating an already-existing bright background source; ETI-modulated pulsar emission \citep{chennamangalam2015JumpingEnergetics, vidal2019PulsarPositioning}; modulated Cepheids \citep{learned2008CepheidGalactic}. & Detectable as anomalies in transits or in pulsar timing patterns. & Transit photometry (Kepler/TESS/PLATO); pulsar timing arrays; high-cadence photometric and spectroscopic monitoring. \\
\hline
Polarization, OAM, and quantum communication & Signals encoded in polarization states, orbital angular momentum (vortex beams), or quantum protocols (qubits/qudits). & Underexplored frontier. Polarimetry already feasible; OAM degrades faster than 1/r$^{2}$ so favours moderate distances. Quantum coherence may be maintainable at X-ray wavelengths over interstellar distances \citep{berera2022ViabilityQuantum}. & Polarimeters; future quantum-receiving systems; high-resolution spectroscopy supplemented with polarimetric capability. \\
\hline
\multicolumn{4}{|l|}{\textbf{Travel Technosignatures (Section \ref{sec: travel ts})}} \\
\hline
Interstellar migration via close stellar encounters & Generation ships travelling during stellar-encounter windows; associated heat or communication signatures along close-encounter trajectories. & Cross-matching Gaia data for close stellar encounters with heat/communication anomalies \citep{hansen2022UnboundClose}. Binary (S-type) systems offer strong incentives for short interstellar travels. & Gaia astrometry; IR and radio follow-up of close-encounter pairs; targeted S-type binaries. \\
\hline
Light sails (directed-energy propulsion) & Direct detection of reflected light from large light sails; leakage radiation from high-power laser beams between planets \citep{guillochon2015SETILeakage}; Fast Radio Bursts hypothesized as sail-powering beams \citep{lingam2017FastRadio}; starshade-like signatures. & Detectable via radial-velocity wobbles, photometric reflections, leakage radiation, and FRB statistics. Leakage detection limited if beams are well-matched to sail size. & Optical/IR coronagraphs (HWO, JWST); time-domain optical surveys; FRB detectors (CHIME, ASKAP, DSA-2000). \\
\hline
Bussard ramjet & Combined signatures: $\sim$1 AU magnetic scoop megastructure; high-velocity or accelerating object; fusion-exhaust contrail; void carved through the ISM behind the spacecraft. & \citet{schattschneider2022FishbackRamjet} argue that even a megastructure-class scoop cannot achieve relativistic flight; but the very existence of the megastructure makes it potentially observable. & Multi-wavelength surveys including IR, radio, and plasma diagnostics targeting ISM voids and exhaust signatures. \\
\hline
Spacecraft --- interstellar contrails & Low-frequency cyclotron radio emission from interaction between a magnetic sail and the ISM (Zubrin 1995); inverse-Compton interaction signatures of relativistic spacecraft with the CMB \citep{yurtsever2018LimitsSignatures}; spatially extended exhaust trails. & Contrails are extended in space and thus relatively detectable, but inverse-Compton signatures are degenerate with the Sunyaev--Zel'dovich effect and other natural emission. Could check whether candidate contrails point towards or away from habitable systems. & Low-frequency radio arrays (LOFAR, SKA); multi-wavelength contrail-tracing pipelines; Gaia for tracing origin/destination. \\
\hline
Spacecraft --- gravitational-wave signatures & Gravitational-wave signatures from massive accelerating spacecraft (RAMAcraft, \citeauthor{sellers2022SearchingIntelligent} \citeyear{sellers2022SearchingIntelligent}) or from collapsing / accelerating Alcubierre warp bubbles. & Demands very large masses and accelerations for the spacecraft to be detectable; physics of Alcubierre warp drives may or may not be possible. LIGO would be sensitive to some of these scenarios, and higher-frequency gravitational-wave observatories are being planned. & Gravitational-wave observatories (LIGO, Virgo, KAGRA); future LISA, Einstein Telescope, Cosmic Explorer; high-frequency-GW concepts. \\
\hline
Matter--antimatter propulsion & Exhaust leakage from matter--antimatter annihilation producing copious neutrinos and gamma-rays; possible neutrino contrails \citep{harris1986DetectabilityAntimatter}. & Requires high energies to be detectable; if achieved, abundant neutrinos and gamma-rays would be produced. & Neutrino detectors (IceCube, KM3NeT); gamma-ray telescopes (Fermi-LAT, INTEGRAL). \\
\hline
Directed-mass propulsion (Singer pellet stream) & High-energy particle streams (akin to ultra-high-energy cosmic rays, UHECRs) directed from a home world to a spacecraft; secondary synchrotron radiation from re-routed particles. & A few UHECRs (Amaterasu particle, ``Oh-My-God'' particle) sit at the boundary of natural and technological origin. Tracing UHECRs is hard but possible at very high energies due to reduced magnetic deflection. & Pierre Auger Observatory; Telescope Array; multi-messenger correlation with synchrotron sources. \\
\hline
Planet engine & Controlled motion or goal-directed trajectory of rogue planets, with internal-iron-core energy sources providing long-duration thrust \citep{romanovskaya2022CosmicHitchhikers}; abrupt accelerations producing time-varying quadrupole gravitational-wave signatures \citep{sellers2022SearchingIntelligent}. & Rogue planets are extremely difficult to detect, and tracking them is even harder. A Jupiter-mass RAMAcraft with $\Delta$v = 0.3 c is detectable from 10--100 kpc with proposed future GW detectors. & Wide-field astrometric surveys (Gaia, Vera Rubin, Nancy Grace Roman); microlensing surveys; future high-strain GW detectors. \\
\hline
Stellar engine & Stars moving in goal-directed or anomalous ways (Shkadov thrusters, Caplan-style high-acceleration engines, binary stellar engines as spider pulsars from \citet{vidal2024SpiderStellar}. & Requires precise astrometry over long baselines. & High-precision astrometry (Gaia and successors) complemented with VLBI or radial-velocity observations; targeted searches around spider pulsars and other compact binaries. \\
\hline
Newtonian gravitational assists & Anomalously fast inspiral rates of compact binary systems; propulsion signatures at periapsis (slight residual eccentricity, anomalous mass loss); changes in inspiral rate increasing merger rates of binary black holes \citep{dyson1963GravitationalMachines, kipping2018HaloDrive}. & Detectable if many or very massive spacecraft use binaries as gravitational machines; binary compact objects should be on technosignature target lists. & Gravitational-wave observatories (LIGO/Virgo/KAGRA, future LISA); pulsar timing arrays; optical/X-ray monitoring of compact binaries. \\
\hline
Einsteinian propulsion --- warp bubbles (Alcubierre) & Gravitational-lensing signatures of warp bubbles (front: defocusing/dimming/redshift; rear: focusing/brightening/ Einstein-ring arcs/blueshift); ``bi-modal'' signal \citep{lentz2024MotivatingEmissions} when bubble outruns its own emission; short-duration gamma-ray bursts from particle ``snowplow'' upon deceleration; characteristic high-frequency GW bursts ($\sim$300 kHz for a 1 km bubble) at bubble collapse. & Outside the range of current GW detectors but potentially measurable by future detectors; lensing/light-curve anomalies could be searched for in time-domain surveys. & Future high-frequency GW detectors; time-domain optical surveys; gamma-ray observatories. \\
\hline
Einsteinian propulsion --- traversable wormholes & Concentric ring/arc gravitational-lensing patterns with no event-horizon shadow; distinct redshifts between inner and outer rings (light from a different region of the universe); asymmetric/elliptical distortions depending on viewing angle; duplicate or time-delayed background images. & Distinguishable from black-hole lensing; requires very high resolution and sensitivity. & Large space-based telescopes; very-long-baseline interferometers; Event-Horizon-Telescope-class imaging. \\
\hline
\multicolumn{4}{|l|}{\textbf{Galactic and Beyond (Section \ref{sec: galactic ts})}} \\
\hline
Galactic & Engineered pulsar positioning system, i.e. millisecond pulsars used as galactic timing / navigation / communication standards \citep{vidal2019PulsarPositioning}; Dyson swarm around the central SMBH \citep{inoue2011TypeIII}; ``Black-Hole Bomb'' exploiting SMBH rotational energy \citep{cardoso2004BlackholeBomb}; modulated Cepheid variables \citep{learned2008CepheidGalactic}; galactic-centre radio transients (e.g. GCRT J1745$-$3009). & Kinematics and distribution of millisecond pulsars tuned for engineering purposes; anomalous IR excess and beamed signals from the galactic centre; transient surveys may reveal candidate megabeam pulses. & Pulsar timing arrays (NANOGrav, EPTA, IPTA); IR observatories (WISE, Spitzer, JWST); radio transient surveys (CHIME, MeerKAT, ASKAP, SKA). \\
\hline
Extragalactic & Galaxy-wide infrared excess from Dyson-spheres-at-scale \citep{griffith2015InfraredSearch, wright2014InfraredSearcha}; deviations from Tully--Fisher, Faber--Jackson and Fundamental Plane scaling relations (under-luminous galaxies, \citeauthor{zackrisson2015ExtragalacticSETI} \citeyear{zackrisson2015ExtragalacticSETI}); morphological anomalies (e.g. Hoag's object; \citeauthor{voros2013GalacticscaleMacroengineering} \citeyear{voros2013GalacticscaleMacroengineering}); YeV neutrino point sources from Type-III particle accelerators \citep{lacki2015SETIPlanck}; SMBH-collimated power beams to receiving stations; Fast Radio Bursts with unnatural periodicity or modulation. & Most galaxy-wide dimming searches so far have yielded no candidates, suggesting Type III civilizations may concentrate rather than spread out. Extragalactic SETI lets us probe billions of stars simultaneously. & WISE / Spitzer / IRAS for IR excess; Planck for far-IR fully-enclosed galaxies; FRB detectors (CHIME, ASKAP, DSA-2000); large radio telescopes (Green Bank, FAST, MeerKAT); Breakthrough Listen; GMRT. \\
\hline
Universal & Bubbles of ambitious (near-Type-IV) civilizations covering $\sim$0.1--1\% of the sky; locally depressed matter density / elevated radiation density; significant statistical deviations in cosmic-web void distributions \citep{olson2015HomogeneousCosmology}. & Requires statistical anomalies in cosmic-scale matter/radiation distributions; large surveys needed to confirm. & High-resolution spectroscopy (HWO, Keck HIRES, HARPS); large-scale-structure surveys (Euclid, DESI, LSST/Rubin); CMB experiments. \\
\hline
Simulation hypothesis & Anisotropy of the Greisen--Zatsepin--Kuzmin (GZK) cutoff energy in UHECRs (signature of a discrete spacetime lattice; \citeauthor{beane2014ConstraintsUniverse} \citeyear{beane2014ConstraintsUniverse}); drift in fundamental constants (e.g. fine-structure constant); momentary localized violations of conservation laws. & Each candidate signature is independently testable but at the edge of current observational capability; statistically significant deviations are required. & Pierre Auger Observatory and Telescope Array (GZK anisotropy); ultra-stable atomic-clock networks (constant drift). \\
\hline
\end{longtable}
\pagebreak

\subsection{Abbreviations}
\begin{longtable}[l]{rl}
  \hline
  \endfirsthead
  \hline
  \endhead
  \hline
  \endfoot
    
AARO     & All-Domain Anomaly Resolution Office \\
AAS      & American Astronomical Society \\
ADS-B    & Automatic Dependent Surveillance--Broadcast \\
AMD      & Arnowitt–Deser–Misner \\
AGN      & Active Galactic Nucleus \\
AM       & Amplitude Modulation \\
ASK      & Amplitude Shift Keying \\
AU       & Astronomical Unit \\
BLC1     & Breakthrough Listen Candidate 1 \\
CETI     & Communication with Extraterrestrial Intelligence \\
CFCs     & Chlorofluorocarbons \\
CMB      & Cosmic Microwave Background \\
CNEOS    & Center for Near-Earth Object Studies \\
DMS      & Dimethyl Sulfide \\
DMSP     & Dimethylsulfoniopropionate \\
DNA      & Deoxyribonucleic Acid \\
Gaia DR3 & Gaia Data Release 3 \\
DTN      & Delay-Tolerant Networking \\
ELT      & Extremely Large Telescope \\
EM       & Electromagnetic \\
EPR      & Einstein--Podolsky--Rosen \\
ER       & Einstein--Rosen (bridge) \\
ESA      & European Space Agency \\
ESPRESSO & Echelle Spectrograph for Rocky Exoplanet and Stable Spectroscopic Observations \\
ET       & Extraterrestrial \\
ETI      & Extraterrestrial Intelligence \\
ETZ      & Earth Transit Zone \\
EVA      & Extravehicular Activity \\
FM       & Frequency Modulation \\
FRBs     & Fast Radio Bursts \\
FSK      & Frequency Shift Keying \\
GBT      & Green Bank Telescope \\
GC       & Galactic Center \\
GHZ      & Galactic Habitable Zone \\
GRB      & Gamma-Ray Burst \\
GW       & Gravitational Waves \\
GWP      & Global Warming Potential \\
GZK      & Greisen--Zatsepin--Kuzmin (cutoff) \\
HabEx    & Habitable Exoplanet Observatory \\
HCFCs    & Hydrochlorofluorocarbons \\
HCRO     & Hat Creek Radio Observatory \\
HMMs     & Hidden Markov Models \\
HWO      & Habitable Worlds Observatory \\
IM1      & Interstellar Meteor 1 \\
IPCC     & Intergovernmental Panel on Climate Change \\
IR       & Infrared \\
IRAS     & Infrared Astronomical Satellite \\
ISM      & Interstellar Medium \\
ISO      & Interstellar Object \\
ISS      & International Space Station \\
JS       & Jensen--Shannon (divergence) \\
JWST     & James Webb Space Telescope \\
KC       & Kolmogorov Complexity \\
KIC      & Kepler Input Catalog \\
KL       & Kullback--Leibler (divergence) \\
LED      & Light Emitting Diode \\
LFT3     & Lunar Farside Technosignature \& Transients Telescope \\
LIFE     & Large Interferometer For Exoplanets \\
LIGO     & Laser Interferometer Gravitational-Wave Observatory \\
LISA     & Laser Interferometer Space Antenna \\
LOFAR    & Low-Frequency Array \\
LPI      & Lunar and Planetary Institute \\
LRO      & Lunar Reconnaissance Orbiter \\
LROC     & Lunar Reconnaissance Orbiter Camera \\
ly       & light-year \\
LSST     & Legacy Survey of Space and Time \\
METI     & Messaging to Extraterrestrial Intelligence \\
MIT      & Massachusetts Institute of Technology \\
ML       & Machine Learning \\
MSPs     & Millisecond Pulsars \\
NASA     & National Aeronautics and Space Administration \\
NEC      & Null Energy Condition \\
NOAA     & National Oceanic and Atmospheric Administration \\
OAM      & Orbital Angular Momentum \\
OSETI    & Optical Search for Extraterrestrial Intelligence \\
OSI      & Open Systems Interconnection \\
OVRO     & Owens Valley Radio Observatory \\
PAHs     & Polycyclic Aromatic Hydrocarbons \\
PanoSETI & Panoramic SETI \\
pc       & parsec\\
PFAS     & Per- and Polyfluoroalkyl Substances \\
PFCs     & Perfluorocarbons \\
PM       & Phase Modulation \\
POSS     & Palomar Observatory Sky Survey \\
PSETI    & Penn State Extraterrestrial Intelligence (Center) \\
PSK      & Phase Shift Keying \\
PTZ      & Pan-Tilt-Zoom \\
QAM      & Quadrature Amplitude Modulation \\
RFI      & Radio Frequency Interference \\
SAM      & Spin Angular Momentum \\
SEC      & Strong Energy Condition \\
SETA     & Search for Extraterrestrial Artifacts \\
SETI     & Search for Extraterrestrial Intelligence \\
SGL      & Solar Gravitational Lens \\
SKA      & Square Kilometre Array \\
SMBH     & Supermassive Black Hole \\
STACEE   & Solar Tower Atmospheric Cherenkov Effect Experiment \\
TESS     & Transiting Exoplanet Survey Satellite \\
TOI      & TESS Object of Interest \\
UAP      & Unidentified Anomalous (formerly Aerial) Phenomena \\
UFO      & Unidentified Flying Object \\
UHECRs   & Ultra-High Energy Cosmic Rays \\
USGS     & United States Geological Survey \\
UV       & Ultraviolet \\
VERITAS  & Very Energetic Radiation Imaging Telescope Array System \\
VLTI     & Very Large Telescope Interferometer \\
WEC      & Weak Energy Condition \\
WISE     & Wide-field Infrared Survey Explorer \\
XFEL     & X-ray Free Electron Laser \\

\end{longtable}

\subsection{Scientific Prefixes}
\begin{table}[h]
    \centering
    \begin{tabular}{rcl}
        Prefix & Symbol & Numerical Value\\
        \hline\hline\\
        Vocto & v & $10^{-33}$\\
        Quecto & q & $10^{-30}$\\
        Ronto & r & $10^{-27}$\\
        Yocto & y & $10^{-24}$\\
        Zepto & z & $10^{-21}$\\
        Atto & a & $10^{-18}$\\
        Femto & f & $10^{-15}$\\
        Pico & p & $10^{-12}$ \\
        Nano & n & $10^{-9}$\\
        Micro & $\mu$& $10^{-6}$\\
        Milli & m & $10^{-3}$\\
        Centi & c & $10^{-2}$\\
        Deci & d & $10^{-1}$\\
        Deca & da & $10^{1}$\\
        Hecto & h & $10^{2}$\\
        Kilo & k & $10^{3}$\\
        Mega & M & $10^{6}$\\
        Giga & G & $10^{9}$\\
        Tera & T & $10^{12}$\\
        Peta & P & $10^{15}$\\
        Exa & E & $10^{18}$\\
        Zetta & Z & $10^{21}$\\
        Yotta & Y & $10^{24}$\\
        Ronna & R & $10^{27}$\\
        Quetta & Q & $10^{30}$\\
        Votta & V & $10^{33}$\\
    \end{tabular}
    \caption{Scientific prefixes from very small to very large scales}
\end{table}

\end{document}